\documentclass[useAMS,usenatbib,usegraphicx]{mn2e}
\usepackage{epsf}

 \newcommand{\bm}[1]{\mbox{\boldmath$#1$}}

\newcommand{\simgt}{\lower.5ex\hbox{$\; \buildrel > \over \sim \;$}}
\newcommand{\simlt}{\lower.5ex\hbox{$\; \buildrel < \over \sim \;$}}

\title[Combined strong and weak lensing analysis of 28 clusters]
{Combined strong and weak lensing analysis of 28 clusters from the
Sloan Giant Arcs Survey 
\thanks{Based on data collected at Subaru Telescope, which is operated
  by the National Astronomical Observatory of Japan. Based on
  observations obtained at the Gemini Observatory, which is operated
  by the Association of Universities for Research in Astronomy, Inc.,
  under a cooperative agreement with the NSF on behalf of the Gemini
  partnership: the National Science Foundation (United States), the
  Science and Technology Facilities Council (United Kingdom), the
  National Research Council (Canada), CONICYT (Chile), the Australian
  Research Council (Australia), Minist\'{e}rio da Ci\^{e}ncia e
  Tecnologia (Brazil) and Ministerio de Ciencia, Tecnolog\'{i}a e
  Innovaci\'{o}n Productiva (Argentina).} 
}

\author[M.~Oguri et al.]
{Masamune Oguri,$^{1,2}$\thanks{E-mail: masamune.oguri@ipmu.jp} 
Matthew B. Bayliss,$^{3,4,5,6}$
H{\aa}kon Dahle,$^{7}$
Keren Sharon,$^{6}$\newauthor
Michael D. Gladders,$^{5,6}$
Priyamvada Natarajan,$^{8,9}$
Joseph F. Hennawi,$^{10}$\newauthor
and Benjamin P. Koester$^{5,6}$\\
$^1$Institute for the Physics and Mathematics of the Universe,
University of Tokyo, 5-1-5 Kashiwanoha, Chiba 277-8583,
Japan.\\ 
$^2$Division of Theoretical Astronomy, National Astronomical
Observatory of Japan, 2-21-1 Osawa, Tokyo 181-8588, Japan.\\ 
$^3$Department of Physics, Harvard University, 17 Oxford St., Cambridge,
MA 02138, USA.\\
$^4$Harvard-Smithsonian Center for Astrophysics, 60 Garden St.,
Cambridge, MA 02138, USA.\\
$^5$Department of Astronomy \& Astrophysics, University of Chicago, 5640
South Ellis Avenue, Chicago, IL 60637, USA.\\ 
$^6$Kavli Institute for Cosmological Physics, University of Chicago, 5640
South Ellis Avenue, Chicago, IL 60637, USA.\\ 
$^7$Institute of Theoretical Astrophysics, University of Oslo,
P.O. Box 1029, Blindern, N-0315 Oslo, Norway.\\
$^8$Department of Astronomy, Yale University, P.O. Box 208101, New
Haven, CT 06511, USA.\\
$^9$Department of Physics, Yale University, P.O. Box 208120, New Haven, 
CT 06520, USA.\\
$^{10}$Max-Planck-Institut f\"{o}r Astronomie, K\"{o}nigstuhl 17, D-69117
Heidelberg, Germany.\\
} 

\begin{document}

\date{\today}

\voffset- .5in

\pagerange{\pageref{firstpage}--\pageref{lastpage}} \pubyear{}

\maketitle

\label{firstpage}

\begin{abstract}
We study the mass distribution of a sample of 28 galaxy clusters
using strong and weak lensing observations. The clusters are
selected via their strong lensing properties as part of the Sloan
Giant Arcs Survey (SGAS) from the Sloan Digital Sky Survey (SDSS). 
Mass modelling of the strong lensing information from the giant arcs 
is combined with weak lensing measurements from deep
Subaru/Suprime-cam images to primarily obtain robust constraints on 
the concentration parameter and the shape of the mass distribution. 
We find that the concentration $c_{\rm vir}$ is a steep function of
the mass, $c_{\rm vir}\propto M_{\rm vir}^{-0.59\pm0.12}$,
with the value roughly consistent with the lensing-bias-corrected
theoretical expectation for high mass ($\sim 10^{15}h^{-1}M_\odot$)
clusters. However, the observationally inferred concentration
parameters appear to be much higher at lower masses 
($\sim 10^{14}h^{-1}M_\odot$), possibly a consequence of the
modification to the inner density profiles provided by baryon
cooling. The steep mass-concentration relation is also supported from
direct stacking analysis of the tangential shear profiles. 
In addition, we explore the two-dimensional shape of the projected
mass distribution by stacking weak lensing shear maps of individual
clusters with prior information on the position angle from strong lens
modelling, and find significant evidence for a large mean ellipticity
with the best-fit value of $\langle e \rangle=0.47\pm0.06$ for the
mass distribution of the stacked sample.  We find that  the luminous
cluster member galaxy distribution traces the overall mass
distribution very well, although the distribution of fainter cluster
galaxies appears to be more extended than the total mass.
\end{abstract}


\section{Introduction}
\label{sec:intro}

Gravitational lensing plays a dominant role in determining the mass
distribution of distant galaxies and clusters of galaxies, because it
allows direct measurements of the distribution of dark matter which
accounts for $\sim 90\%$ of the total matter content of the Universe.
Precise measurements of dark matter distributions are important not
only for understanding the formation of galaxies in the context of the
hierarchical structure formation scenario, but also for testing the
properties of the putative dark matter particle, in particular its
cold and collisionless nature. Observations of clusters are suitable
and very apt for the latter purpose, as the long cooling timescale for
hot gas in clusters indicates that the cluster gravitational potential
is mainly determined by the dynamics of the dominant dark matter
component which are well predicted by $N$-body simulations. However,
it is also expected that the baryonic component in clusters, which
dominates the mass in the innermost regions, should play an
important role at some point. 

There are two regimes of gravitational lensing that can be employed to 
measure and map the mass distribution in clusters. One is strong
lensing, i.e., the regime that produces drastic lensing events with
highly elongated arcs or multiple images of background objects
\citep[e.g.,][for reviews]{kochanek06,kneib11}. The other lensing
regime is weak lensing wherein statistical measurements of small
distortions in the shapes of background galaxies are produced by a
massive, foreground cluster \citep[e.g.,][for a review]{bartelmann01}. 
Since strong and weak lensing probe mass distributions at different
radii, the combination of these two is powerful and essential for the
full understanding and detailed mapping of the gravitational potential
of clusters. 

Lensing studies of clusters have indeed confirmed several important
predictions of the standard $\Lambda$-dominated Cold Dark Matter
($\Lambda$CDM) model. For instance, from numerical simulations, the 
radial run of the density of the mass distribution is predicted to
have a universal form, that of the Navarro-Frenk-White profile
\citep[e.g.,][]{navarro96,navarro97}. Detailed lensing measurements of
massive clusters have confirmed this prediction by and large, although
there does appear to be some cluster-to-cluster variations 
\citep[e.g.,][]{okabe10a,umetsu11b}. Another important prediction is
that massive clusters are on average highly non-spherical with a
typical major to minor axis ratio of $\sim$2:1 \citep{jing02}. Weak
lensing measurements of dark matter distributions in a sample of
massive clusters have directly confirmed this prediction as well
\citep{oguri10a}. Furthermore, lensing analysis of several merging 
clusters of galaxies strongly supports the collisionless nature of
dark matter \citep{clowe06,jee07,bradac08,merten11}, although a
possible case against the collisionless nature has also been
discovered \citep{mahdavi07}. 

However, measurements of the concentration parameter, which appears in
the Navarro-Frenk-White profile, from gravitational lensing
observations for clusters have been controversial. The concentration
parameter is defined as the ratio of the virial radius to the scale
radius, and is a dimensionless parameter that quantifies the degree of
the mass accumulation in the innermost regions. The standard
$\Lambda$CDM model makes a prediction about this parameter too,
namely that more massive haloes or haloes at higher redshifts have
smaller values of the concentration parameter on average
\citep[e.g.,][]{bullock01}. However,
measurements of the dark matter distribution in the massive lensing
cluster A1689 for instance, using both strong and weak lensing, have
indicated that the mass distribution is surprisingly highly
concentrated with an estimated concentration parameter of 
$c_{\rm vir}\sim 12$, which is significantly higher than the expected
standard $\Lambda$CDM prediction of $c_{\rm vir}\sim 4$
\citep{broadhurst05,umetsu08}.   

The interpretation of this result of lensing clusters being
over-concentrated requires careful consideration due to the role
played by projection and selection effects. This is because lensing
observables can only measure the projected mass distribution, and hence
the recovery of the three-dimensional mass distribution requires
additional assumptions about the elongation along the line-of-sight
direction. For instance, because of the large triaxiality of dark
matter haloes, the mass and concentration parameter inferred from
gravitational lensing depend strongly on the orientation with respect
to the line-of-sight direction.  This naturally implies  that both the
mass and concentration are significantly overestimated when observed
along the major axis \citep{clowe04,oguri05,gavazzi05,corless07}. The
large orientation dependence in turn suggests that strong lensing
selected clusters represent a highly biased population with their
major axes preferentially aligned with the line-of-sight direction
\citep{hennawi07,oguri09a,meneghetti10}. Thus, the dark matter
distribution of A1689 alone does not pose a severe challenge to the
$\Lambda$CDM model
\citep{oguri05,oguri09a,corless09,sereno10,coe10,morandi11a,sereno11,morandi11b}, 
although such high values of the concentration parameter appear to be
common in the combined  strong and weak lensing analysis of massive
clusters \citep{comerford07,broadhurst08b,oguri09b,umetsu11b,zitrin11b}.  
High concentrations can be tested with the distribution of the
Einstein radii, but the results have still not converged 
\citep{broadhurst08a,oguri09a,richard10,gralla11,zitrin11a,zitrin11c}.
Also puzzling is the fact that such high concentrations have not been
claimed in weak lensing analysis \citep{okabe10a} or X-ray analysis
\citep{buote07,ettori10} for samples of massive clusters. 

While the combined strong and weak lensing analysis allows accurate
and robust measurements of the concentration parameter, the current
main limitation is the small number of clusters available for such a 
detailed combined  analysis. The Sloan Giant Arcs Survey
\citep[SGAS;][]{hennawi08,bayliss11a}, which is a survey of strongly
lensed giant arcs from the Sloan Digital Sky Survey
\citep[SDSS;][]{york00}, has already discovered more than 30 bright
giant arcs and therefore offers an ideal technique to expand the
sample of clusters appropriate for detailed lensing analysis.

In this paper, we present a systematic study of strong and weak
lensing analysis for a sample of 28 clusters from the SGAS, based on
our extensive follow-up imaging observations with Subaru/Suprime-cam
\citep{miyazaki02}. We study the radial dark matter distributions of
the clusters in detail to measure the concentration parameters for
these clusters. For this purpose we conduct a stacked lensing analysis
as well as individual modelling of clusters. With the stacked lensing
technique we also study the two-dimensional mass distribution, and
constrain the ellipticity of the projected mass distribution. 

The structure of this paper is as follows. We describe our cluster
sample and follow-up imaging observations with Subaru/Suprime-cam in
Section~\ref{sec:sample}, and present the strong and weak lensing
analysis in Section~\ref{sec:lens}. The results are combined to
discuss the mass-concentration relation in Section~\ref{sec:swlens}. 
We also conduct a stacked lensing analysis, which is detailed in
Section~\ref{sec:stack}. We then study the cluster member galaxy
distribution in Section~\ref{sec:member}, and summarize the main
conclusions in Section~\ref{sec:conclusion}. In 
Appendix~\ref{sec:conc} we conduct semi-analytic calculations to
evaluate the effect of the lensing bias on various observables. 
Appendix~\ref{sec:app} summarizes the strong and weak lensing analysis
results of individual clusters. Throughout the paper we
assume the standard $\Lambda$-dominated flat cosmological model with
the matter density $\Omega_M=1-\Omega_\Lambda=0.275$, the
dimensionless Hubble constant $h=0.702$, the baryon density 
$\Omega_b h^2=0.02255$, the spectral index $n_s=0.968$, and a
normalization for the matter power spectrum $\sigma_8=0.816$
\citep{komatsu11} where needed.  

\section{Cluster sample and follow-up observations}
\label{sec:sample}

\subsection{Cluster sample}

We draw our sample of clusters for detailed lensing analysis from
SGAS, which involved an initial extensive visual search of giant arcs in
red-sequence selected clusters in the SDSS. The sample is constructed
utilizing two selection methods. One is the SDSS ``Visual'' survey which
selects giant arc candidates from the visual inspection of the SDSS
imaging data (M. D. Gladders et al., in preparation). The other
survey is the SDSS ``Blind'' survey which searches for giant arcs from
$g$-band follow-up imaging of the most massive $\sim 200$ clusters
selected from the SDSS imaging data \citep{hennawi08}. Some of the
clusters have also been reported in \citet{wen11}. We are
conducting massive spectroscopic follow-up observations of these new
giant arcs with the Gemini Multi-Object Spectrograph 
\citep[GMOS;][]{hook04} to measure redshifts for the newly discovered
arcs as well as redshifts for the lensing clusters. See
\citet{bayliss11b} for details of our spectroscopic follow-up and
successful redshift measurements for more than 20 giant arcs.
We note that this paper includes a few new GMOS spectroscopy results
obtained after the publication of \citet{bayliss11b}.
We also include two cluster-scale quasar lenses discovered from 
the SDSS Quasar Lens Search
\citep[SQLS;][]{oguri06,oguri08a,inada08,inada10}, as they satisfy
similar selection criteria to the SGAS lens sample, and have also
been observed at the Subaru telescope.

\begin{table*}
 \caption{Cluster sample and summary of Subaru/Suprime-cam imaging
   observations. The redshifts of the clusters are taken from
   \citet{bayliss11b}. 
\label{tab:sample}}   
 \begin{tabular}{@{}cccccccccc}
  \hline
   Name
   & R.A.
   & Decl.
   & $z$
   & Exp ($g$)
   & Seeing ($g$)
   & Exp ($r$)
   & Seeing ($r$)
   & Exp ($i$)
   & Seeing ($i$) \\
   & (J2000)
   & (J2000)
   &
   & (sec) 
   & (arcsec) 
   & (sec) 
   & (arcsec) 
   & (sec) 
   & (arcsec) \\
 \hline
SDSSJ0851+3331 & 08 51 38.86 & +33 31 06.1 & 0.370   & 1200 & 0.79  & 1800 & 1.05  & 1680 & 0.71  \\
SDSSJ0915+3826 & 09 15 39.00 & +38 26 58.5 & 0.397   & 1200 & 0.89  & 2100 & 0.87  & 1680 & 0.67  \\
SDSSJ0957+0509 & 09 57 39.19 & +05 09 31.9 & 0.448   & 1200 & 1.01  & 2100 & 0.61  & 1680 & 0.81  \\
SDSSJ1004+4112 & 10 04 34.18 & +41 12 43.5 & 0.68    & 810  & 0.71  & 1210 & 0.67  & 1340 & 0.61  \\
SDSSJ1029+2623 & 10 29 12.48 & +26 23 32.0 & 0.584   & 1200 & 0.79  & 2700 & 0.65  & 1920 & 1.05  \\
SDSSJ1038+4849 & 10 38 42.90 & +48 49 18.7 & 0.430   & 900  & 0.75  & 2100 & 0.79  & 1680 & 0.83  \\
SDSSJ1050+0017 & 10 50 39.90 & +00 17 07.1 & 0.60$^a$& 1200 & 0.59  & 2100 & 0.67  & 1680 & 0.65  \\
RCS2J1055+5547 & 10 55 04.59 & +55 48 23.3 & 0.466   & 1200 & 0.95  & 2100 & 0.89  & 1440 & 0.83  \\
SDSSJ1110+6459 & 11 10 17.70 & +64 59 47.8 & 0.659$^b$& 1200 & 1.23  & 2100 & 0.89  & 1680 & 1.05  \\
SDSSJ1115+5319 & 11 15 14.85 & +53 19 54.6 & 0.466   & 1200 & 1.39  & 1500 & 1.13  & 1200 & 1.13  \\
SDSSJ1138+2754 & 11 38 08.95 & +27 54 30.7 & 0.451   & 900  & 0.81  & 2100 & 0.85  & 1440 & 0.85  \\
SDSSJ1152+3313 & 11 52 00.15 & +33 13 42.1 & 0.362   & 1200 & 1.17  & 1800 & 0.55  & 1680 & 1.21  \\
SDSSJ1152+0930 & 11 52 47.38 & +09 30 14.7 & 0.517   & 1200 & 1.13  & 2100 & 1.01  & 1200 & 0.93  \\
SDSSJ1209+2640 & 12 09 23.68 & +26 40 46.7 & 0.561   & 1200 & 1.25  & 2100 & 0.79  & 960  & 1.13  \\
SDSSJ1226+2149$^c$ & 12 26 51.11 & +21 49 52.3 & 0.435   & 1200 & 1.05  & $R_{\rm c}$=2170 & 0.81  & 1680 & 0.91  \\
A1703          & 13 15 05.24 & +51 49 02.6 & 0.277   & 1200 & 0.97  & 2100 & 0.91  & 1200 & 0.87  \\
SDSSJ1315+5439 & 13 15 09.30 & +54 37 51.8 & 0.588$^d$   & 1200 & 0.69  & 1500 & 0.91  & 1680 & 0.71  \\
GHO132029+3155 & 13 22 48.77 & +31 39 17.8 & 0.308   & 1200 & 0.77  & 2100 & 0.81  & 1680 & 0.87  \\
SDSSJ1329+2243 & 13 29 34.49 & +22 43 16.2 & 0.443$^d$   & 1200 & 0.69  & 2100 & 0.85  & 1680 & 0.69  \\
SDSSJ1343+4155 & 13 43 32.85 & +41 55 03.4 & 0.418   & 1200 & 0.83  & 1500 & 0.83  & 1440 & 1.33  \\
SDSSJ1420+3955 & 14 20 40.33 & +39 55 09.8 & 0.607   & 1200 & 1.29  & 1800 & 0.79  & 1440 & 0.73  \\
SDSSJ1446+3032 & 14 46 34.02 & +30 32 58.2 & 0.464   & 1200 & 0.85  & 2100 & 0.83  & 1200 & 0.93  \\
SDSSJ1456+5702 & 14 56 00.78 & +57 02 20.3 & 0.484   & 1200 & 0.71  & 2100 & 0.81  & 1440 & 1.11  \\
SDSSJ1531+3414 & 15 31 10.60 & +34 14 25.0 & 0.335   & 1200 & 0.91  & 1500 & 0.99  & 1200 & 1.01  \\
SDSSJ1621+0607 & 16 21 32.36 & +06 07 19.0 & 0.342   & 1500 & 0.77  & 2100 & 0.85  & 1440 & 1.31  \\
SDSSJ1632+3500 & 16 32 10.26 & +35 00 29.7 & 0.49$^a$& 900  & 0.85  & 2100 & 0.77  & 1440 & 0.77  \\
SDSSJ2111$-$0114&21 11 19.34 &$-$01 14 23.5& 0.638   & 1440 & 0.83  & 2400 & 0.61  & 1680 & 0.53  \\
 \hline
 \end{tabular}
\flushleft{$^a$ Photometric redshifts estimated from the SDSS data, as
  spectroscopic cluster redshifts are not available for these clusters.}
\flushleft{$^b$ Based on the spectroscopy of the brightest cluster
  galaxy at Apache Point Observatory 3.5-meter telescope.} 
\flushleft{$^c$ We use deep $R_{\rm c}$-band images
  retrieved from SMOKA instead of obtaining $r$-band follow-up
  images. This field includes two separate cluster cores, both of
  which act as strong lenses.}
\flushleft{$^d$ Based on the new spectroscopic observation with
  Gemini/GMOS conducted after the publication of \citet{bayliss11b}.}
\end{table*}

\subsection{Observations with Subaru/Suprime-cam}

We observed the SGAS giant arc clusters with Suprime-cam
\citep{miyazaki02} at the Subaru 8.2-meter telescope, primarily for 
the wide-field weak lensing analysis, between 2007 June and 2011
April. Combined with a few images retrieved from the archive system
named SMOKA \citep{baba02}, our sample comprises 28 clusters with
Subaru/Suprime-cam multicolour follow-up imaging.
The Suprime-cam has a large field-of-view of $\sim 34'\times27'$ 
with a pixel scale of $0\farcs202$, and is therefore ideal for weak
lensing studies of massive clusters at intermediate redshifts whose
typical virial radius is $\la 10'$. Our strategy is to obtain deep
images in $g$-, $r$-, and $i$-bands, with the longest exposure in
$r$-band. The deep $r$-band imaging is performed to conduct weak
lensing analysis using $r$-band images, but the additional colour
information from $g$- and $i$-bands are crucial for reliable selection
of background galaxies for the weak lensing analysis (more details
presented in Section~\ref{sec:weaklens}) as well as secure
identifications of multiply imaged arcs in the cluster cores.  We note
that the weak lensing analysis results of the first 4 SGAS clusters
have been published in \citet{oguri09b}. 

The data are reduced using SDFRED and SDFRED2 \citep{yagi02,ouchi04}. 
The reduction procedure includes bias subtraction, flat-fielding, 
distortion correction, sky subtraction, and co-adding to generate the
final mosaic images. When co-adding, we remove some of the frames which
have significantly worse seeing sizes than other frames. We derive the
magnitude zero-points by comparing objects in the reduced images with
the photometric catalogue of the SDSS data. The astrometric calibration
is performed with SCAMP \citep{bertin06}, again using objects in the
SDSS data as astrometric reference sources. The photometric galaxy
catalogue of each cluster is constructed using SExtractor
\citep{bertin96}, with the Galactic extinction correction
\citep{schlegel98}. Our 28 cluster sample and basic parameters of
follow-up images are summarized in Table~\ref{tab:sample}.   

\section{Strong and weak lensing analysis}
\label{sec:lens}

\subsection{Strong lensing analysis}

We mostly follow \citet{oguri09b} for the strong lens modelling
methodology. Basically we assume the elliptical extension of the NFW
profile \citep{navarro96,navarro97} to model the mass distribution of
a dark halo, and add contributions from member galaxies assuming the
pseudo-Jaffe model.The velocity dispersion $\sigma$ and cutoff scale
$r_{\rm cut}$ of the pseudo-Jaffe model are assumed to scale with the
luminosity as $L\propto\sigma^{1/4}$ and $r_{\rm cut}\propto L^{1/2}$. 
The normalization and cut-off radius may be fixed to a typical value
if there are not enough observational constraints. We may
also fix the centre of the main halo to the position of the brightest
cluster galaxy, depending on the number of available constraints and 
the configuration of multiple images.

We identify multiple images based on spectroscopic follow-up results
of \citet{bayliss11b} as well as colours of lensed arcs measured from
the Subaru/Suprime-cam images. While redshifts of arcs are available
for many of our sample clusters, for clusters without any arc
spectroscopy information we assume that the redshift of the main arc
is $z_{\rm arc}=2$ with the conservative 1$\sigma$ error of
$\sigma(z_{\rm arc})=1$, because our extensive spectroscopic
observations \citep{bayliss11a} as well as photometric analysis
\citep{bayliss11c} have convincingly shown that most of the arcs in
our sample fall in this redshift range. The positional uncertainties of
all the arcs are assumed to be $1''$ in the image plane, which is much
larger than measurement uncertainties but is typical of the uncertainties
associated with parametric strong lens modelling of clusters.

We perform strong lens modelling using the software {\it glafic}
\citep{oguri10b}, using the $\chi^2$ minimization in the source plane
\citep[see][]{oguri09b,oguri10b}. From the derived best-fit mass model
we compute the Einstein radius $\theta_{\rm E}$ of the system by
solving the following equation: 
\begin{equation}
\bar{\kappa}(<\theta_{\rm E})=\frac{1}{\pi \theta_{\rm
    E}^2}\int_{|\bm{\theta}|<\theta_{\rm
    E}}\kappa(\bm{\theta})d\bm{\theta} =1, 
\end{equation}
and use only this Einstein radius determination as the constraint from 
our strong lensing analysis. We use the Einstein radius as it is a
robust quantity that is well constrained from the strong lensing
modelling and is fairly insensitive to our model assumptions
\citep[e.g.,][]{jullo07}. We
estimate the error on the Einstein radius by changing one (usually
mass) of the parameters from the best-fit values, optimizing other
parameters, and monitoring the $\chi^2$ differences. However, to be
conservative in our estimates, we always assign a minimum error of
$10\%$ to the Einstein radius, even if the procedure above returns
smaller errors, given the unaccounted complexity of cluster mass
distributions and projections along line-of-sight
\citep[e.g.,][]{daloisio11}. For clusters without arc spectroscopy,
we conservatively estimate the $1\sigma$ error on the Einstein radius 
by changing the arc redshift  by $\pm1$. The Einstein radii are
computed for both the arc redshift $z_{\rm arc}$ and the 
fixed source redshift of $z_s=2$. We use the former value for the
combined strong and weak lensing analysis, whereas the latter value is
used for the statistical, stacked lensing analysis. 

We note that the definition of the Einstein radius adopted here
differs from that in \citet{oguri09b}. In this new definition the
contribution of stellar masses is explicitly included, while in
\citet{oguri09b} the Einstein radius was computed from the dark halo
component alone. The use of this modified definition is because the
separation of dark and luminous components is not obvious, and has to
rely on several assumptions, unless we have additional data such as the
velocity dispersion measurement of the central galaxy
\citep[e.g.,][]{sand08,newman09}. Thus the effect of baryonic
components has been taken into account in interpreting our results.  

\begin{table*}
 \caption{Summary of strong lensing analysis. We tabulate the Einstein
   radii both for arc redshifts and for $z_s=2$, the total number of
   multiple images used for strong lens modelling ($N_{\rm img}$; the
   number of multiple image sets is shown in parentheses), the
   best-fit ellipticity ($e$) and position angle ($\theta_e$) of the
   main dark halo component, and references for arc redshifts.
\label{tab:slens}}   
 \begin{tabular}{@{}cccccccc}
 \hline
     Name
   & $z_{\rm arc}$$^a$
   & $N_{\rm img}$
   & $\theta_{\rm E}$ ($z_s=z_{\rm arc}$)
   & $\theta_{\rm E}$ ($z_s=2$)
   & $e$
   & $\theta_e$$^b$  
   & Ref.$^c$ \\
   &
   & 
   & (arcsec)
   & (arcsec)
   & 
   & (deg)
   &\\
 \hline
SDSSJ0851+3331 & 1.693 & 4(1) & $21.6^{+ 2.2}_{- 2.2}$ & $23.0^{+ 2.3}_{- 2.3}$ & 0.23 & $39.1$ & 1\\  
SDSSJ0915+3826 & 1.501 & 3(1) & $ 9.8^{+ 1.3}_{- 1.0}$ & $11.4^{+ 1.3}_{- 1.1}$ & 0.28 & $-74.9$& 1,2 \\ 
SDSSJ0957+0509 & 1.820 & 3(1) & $ 5.2^{+ 0.5}_{- 0.5}$ & $ 5.4^{+ 0.5}_{- 0.5}$ & 0.82 & $64.2$ & 1,3\\
SDSSJ1004+4112 & 1.734 & 31(8)& $ 7.3^{+ 0.7}_{- 0.7}$ & $ 8.9^{+ 0.9}_{- 0.9}$ & 0.14 & $-28.0$& 4,5 \\
SDSSJ1029+2623 & 2.197 & 3(1) & $10.7^{+ 4.8}_{- 6.1}$ & $ 9.9^{+ 4.8}_{- 5.9}$ & 0.58 & $-81.5$& 6,7 \\
SDSSJ1038+4849 & 2.198 & 9(3) & $12.6^{+ 1.3}_{- 1.6}$ & $11.2^{+ 1.5}_{- 1.4}$ & 0.15 & $-52.8$& 1,8,9\\
SDSSJ1050+0017 &$2\pm1$& 3(1) & $16.1^{+17.9}_{- 2.5}$ & $16.1^{+17.9}_{- 2.5}$ & 0.28 & $-6.9$ &  \\  
RCS2J1055+5548 & 1.250 & 3(1) & $10.0^{+ 1.0}_{- 1.0}$ & $12.7^{+ 1.3}_{- 1.3}$ & 0.82 & $3.2$  & 1\\  
SDSSJ1110+6459 &$2\pm1$& 3(1) & $ 8.4^{+12.7}_{- 1.5}$ & $ 8.4^{+12.7}_{- 1.5}$ & 0.60 & $76.9$ & \\   
SDSSJ1115+5319 &$2\pm1$& 5(1) & $21.9^{+18.2}_{- 3.4}$ & $21.9^{+18.2}_{- 3.4}$ & 0.63 & $-47.4$&  \\  
SDSSJ1138+2754 & 1.334 & 5(2) & $ 9.8^{+ 1.0}_{- 1.5}$ & $12.9^{+ 1.3}_{- 3.6}$ & 0.50 & $-18.7$& 1 \\ 
SDSSJ1152+3313 & 2.491 & 7(2) & $ 8.7^{+ 0.9}_{- 0.9}$ & $ 8.2^{+ 0.8}_{- 0.8}$ & 0.34 & $64.5$ & 1\\  
SDSSJ1152+0930 &$2\pm1$& 3(1) & $ 4.5^{+ 8.2}_{- 0.7}$ & $ 4.5^{+ 8.2}_{- 0.7}$ & 0.80 & $-38.5$& \\  
SDSSJ1209+2640 & 1.021 & 6(2) & $ 8.8^{+ 0.9}_{- 0.9}$ & $21.3^{+ 2.1}_{- 2.1}$ & 0.14 & $66.1$ & 1,10\\
SDSSJ1226+2149 & 1.605 & 3(1) & $14.0^{+ 3.4}_{- 2.6}$ & $15.7^{+ 3.8}_{- 2.8}$ & 0.47 & $-57.5$& 1 \\ 
SDSSJ1226+2152 & 2.923 & 3(1) & $10.0^{+ 2.8}_{- 6.4}$ & $ 8.6^{+ 2.4}_{- 5.5}$ & 0.18 & $12.3$ & 1,11\\
         A1703 & 2.627 & 21(6)& $27.4^{+ 2.7}_{- 2.7}$ & $25.3^{+ 2.5}_{- 2.5}$ & 0.33 & $-26.2$& 1,12,13\\ 
SDSSJ1315+5439 &$2\pm1$& 3(1) & $16.9^{+20.6}_{- 2.9}$ & $16.9^{+20.6}_{- 2.9}$ & 0.18 & $-36.8$&  \\  
GHO132029+3155 &$2\pm1$& 4(1) & $21.5^{+ 7.0}_{- 2.2}$ & $21.5^{+ 7.0}_{- 2.2}$ & 0.10 & $76.5$ & \\   
SDSSJ1329+2243 & 2.040 & 3(1) & $10.9^{+ 1.6}_{- 1.6}$ & $10.8^{+ 1.6}_{- 1.6}$ & 0.22 & $-4.9$ & 14\\   
SDSSJ1343+4155 & 2.091 & 3(1) & $ 5.4^{+ 2.5}_{- 1.6}$ & $ 5.3^{+ 2.5}_{- 1.6}$ & 0.73 & $64.9$ & 1,2,15\\
SDSSJ1420+3955 & 2.161 & 6(2) & $ 9.9^{+ 2.8}_{- 1.0}$ & $ 9.4^{+ 2.7}_{- 0.9}$ & 0.77 & $73.6$ & 1\\  
SDSSJ1446+3032 &$2\pm1$& 4(1) & $16.8^{+13.1}_{- 2.3}$ & $16.8^{+13.1}_{- 2.3}$ & 0.22 & $-41.6$&  \\  
SDSSJ1456+5702 & 0.833 & 6(2) & $13.2^{+ 1.3}_{- 1.3}$ & $30.1^{+ 9.2}_{- 5.0}$ & 0.21 & $-13.6$& 1\\  
SDSSJ1531+3414 & 1.096 & 6(2) & $11.7^{+ 1.2}_{- 1.2}$ & $16.6^{+ 1.7}_{- 1.7}$ & 0.09 & $-37.5$& 1 \\ 
SDSSJ1621+0607 & 4.131 & 3(1) & $12.5^{+ 1.5}_{- 2.5}$ & $10.3^{+ 1.3}_{- 2.2}$ & 0.33 & $-40.0$& 1 \\ 
SDSSJ1632+3500 &$2\pm1$& 4(1) & $14.3^{+12.6}_{- 2.1}$ & $14.3^{+12.6}_{- 2.1}$ & 0.30 & $-61.9$&  \\  
SDSSJ2111$-$0114 & 2.858 & 9(3) & $17.7^{+10.6}_{- 5.8}$ & $14.5^{+ 9.2}_{- 4.8}$ & 0.24 & $10.0$ & 1 \\ 
 \hline
 \end{tabular}
\flushleft{$^a$ Note that for some clusters there are multiple arcs
  with different source redshifts. In this case we show the redshift
  of the main arc.}
\flushleft{$^b$ The position angle $\theta_e$ is measured East of North.}
\flushleft{$^c$ 
 1 -- \citet{bayliss11b}; 
 2 -- \citet{bayliss10};
 3 -- \citet{kubo10}; 
 4 -- \citet{inada03}; 
 5 -- \citet{sharon05}; 
 6 -- \citet{inada06}; 
 7 -- \citet{oguri08b}; 
 8 -- \citet{belokurov09}; 
 9 -- \citet{kubo09}; 
 10 -- \citet{ofek08};
 11 -- \citet{koester10}; 
 12 -- \citet{limousin08}; 
 13 -- \citet{richard09}; 
 14 -- new redshift not yet published in the literature; 
 15 -- \citet{diehl09}.}
\end{table*}

Table~\ref{tab:slens} lists the Einstein radii derived from strong
lensing modelling. SDSSJ1226+2149 and SDSSJ1226+2152 are two nearby
strong lensing cores in a highly complex massive structure, both of
which are covered by our Subaru/Suprime-cam imaging
\citep{bayliss11b}. We perform strong lensing analysis for both cores.  
We also show the best-fit ellipticity $e$ and the position angle
$\theta_e$ of the best-fit main NFW component, as they are used later
for two-dimensional stacking analysis (Section~\ref{sec:stack2d}).
The best-fit critical curves for individual clusters are presented in
Appendix~\ref{sec:app}. 

Strong lensing analysis for some of these clusters has been published
in the literature
\cite[e.g.,][]{oguri09b,koester10,bayliss10,oguri10b,gralla11}. 
Our estimates of the Einstein radii presented here generally agree with
these previous results. For A1703, our best-fit Einstein radius is in
excellent agreement with that of the independent strong lens
modelling by \citet{richard09}. An exception is SDSSJ1343+4155 whose
Einstein radius in our strong lens modelling result is much smaller
than what presented in \citet{bayliss10} and \citet{gralla11} because
a large offset of the halo centre from the brightest cluster galaxy
has been found in the previous analysis, whereas in the present paper
the centre of the halo is fixed to the location of the brightest
cluster galaxy.  

\subsection{Weak lensing analysis}
\label{sec:weaklens}

For weak lensing measurements, we follow the formalism outlined in
\citet[KSB;][]{kaiser95} using the software package
IMCAT\footnote{http://www.ifa.hawaii.edu/~kaiser/imcat/}.  
We first detect objects in the reduced images using a hierarchical
peak finding algorithm. For all the clusters in our sample, we use
$r$-band images for weak lensing shear measurements. We then measure
the shapes of objects by iteratively refining the centroid of each
object. Stars for correcting the distortion of the Point Spread
Function (PSF) are selected in a standard way by identifying the
appropriate branch in the magnitude-half light radius $r_h$ plane, along
with the peak significance cut $\nu>15$. Shapes of the stars are
measured as a function of the size of the weight function, $r_g$, in
order to make PSF corrections with matched $r_g$ values.  We divide
each co-added image into $4\times3$ chunks and fit the PSF in each
chunk independently with second order bi-polynomials. The smear
polarizability is corrected by computing a scalar polarizability $P_s$
from the trace of the matrix, and then fitting $P_s$ as a function of
magnitude, $r_g$, and the galaxy ellipticity. 
For our weak lensing analysis, we only use galaxies with $\nu>15$ and
$r_h>\overline{r_h^*}+2\sigma(r_h^*)$, where $\overline{r_h^*}$ and 
$\sigma(r_h^*)$ are median and root-mean-square dispersion of
half-light radii for the stars selected above. Given the general
tendency of the KSB algorithm to underestimate the weak lensing shear
\citep{erben01,heymans06,massey07}, we include a calibration factor of
$1/0.9$, i.e., this factor is multiplied to all the estimated shear
values. For each object we assign the statistical weight $w_g$ defined
by \citep[e.g.,][]{hamana03,miyazaki07,hamana09,okabe10a,okabe10b,umetsu10,umetsu11b}.  
\begin{equation}
w_g=\frac{1}{\sigma_g^2+\alpha^2},
\label{eq:weight}
\end{equation}
with $\alpha=0.4$ and $\sigma_g$ is the variance of the shear computed
from 20 neighbors in the magnitude-$r_g$ plane. When computing shear by
averaging shear measurements of galaxies in a bin, we also apply a
$3\sigma$ clipping which appears to reduce the shear measurement bias. 

To check the accuracy of the weak lensing shear measurement, we
perform a series of image simulations. Specifically, we generate a
galaxy catalogue using the software Stuff \citep{bertin09}. Each galaxy
is described by the sum of bulge and disk components, which we model
with Sersic profiles with the index $n=4$ and $n=1$, respectively.
We also add stars in the catalogue. We convolve the image with a PSF
which we assume follows the Moffat profile $\Sigma(r)\propto
[1+(r/a)^2]^{-\beta}$ with an elliptical extension. Based on the
catalogue, we generate a number of realistic Subaru/Suprime-cam like
images with different seeing sizes ($0\farcs5$--$1\farcs1$) and
$\beta$ ($3<\beta<12$) using the software {\it glafic}
\citep{oguri10b}. We find that the resulting shear multiplicative
error \citep[the parameter $m$ in][]{heymans06,massey07} depends on
both seeing size and $\beta$ such that $m$ is smaller for larger
seeing sizes or smaller $\beta$, but our algorithm generally yields
$|m|\la 0.05$ for a wide range of PSF parameters examined here. 

\subsection{Galaxy selection in colour-colour space}

A careful selection of background galaxies is essential for cluster
weak lensing studies, because contamination by cluster member galaxies
is known to dilute the detected weak lensing signal significantly,
particularly near the cluster centres
\citep[e.g.,][]{broadhurst05,medezinski07}. Our $gri$-band imaging is
very powerful for reliable background galaxy selection, because we can
select galaxies efficiently in colour-colour space \citep{medezinski10}. 

\begin{figure}
\begin{center}
 \includegraphics[width=0.85\hsize]{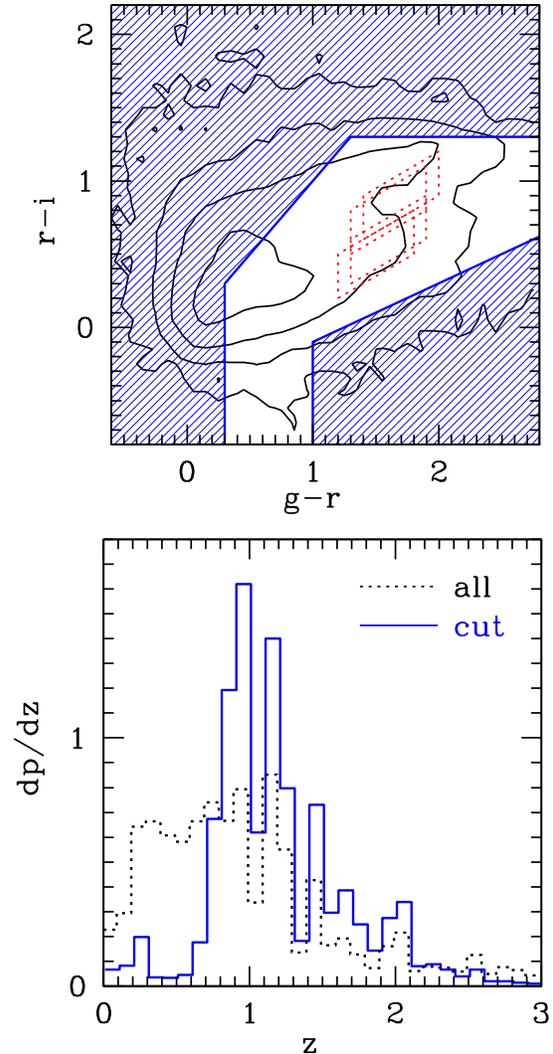}
\end{center}
\caption{{\it Upper:} The colour cut selecting background galaxies for
  weak lensing analysis ({\it shading}). The colour cut for red member
  galaxy selection is indicated by dotted lines. The four regions
  correspond to colour cuts for different cluster redshifts. Solid
  contours indicate galaxy number density in the COSMOS catalogue
  ($i<25$). {\it lower:} Photometric redshift distributions of
  galaxies in the COSMOS catalogue, before ({\it dotted}) and after
  ({\it solid}) the colour cut.  
\label{fig:cosmos}}
\end{figure}

We determine the colour cut that is appropriate for our cluster sample
based on the COSMOS photometric galaxy catalogue \citep{ilbert09}.
Thanks to the wide wavelength coverage from the ultraviolet to 
mid-infrared, the photometric redshifts are very accurate down to
$i\sim 25$. By inspecting the photometric redshift distributions in
each point of the $g-r$ versus $r-i$ colour space, we determine the
colour cut for our analysis as 
\begin{equation}
g-r > 1\;\;\mbox{\&\&}\;\; r-i< 0.4(g-r)-0.5,
\end{equation}
\begin{equation}
g-r<0.3,
\end{equation}
\begin{equation}
r-i>1.3,
\end{equation}
\begin{equation}
r-i > g - r, 
\end{equation}
Figure~\ref{fig:cosmos} shows the cut and resulting COSMOS photometric
redshift distribution. It is seen that our cut efficiently selects
galaxies at $z\ga 0.7$, the redshifts higher than any cluster
redshifts in our sample. In our weak lensing analysis, we also limit
the range of $r$-band magnitude to $21<r<r_{\rm lim}$, where the
limiting magnitude $r_{\rm lim}$ is determined from the galaxy number
counts of each cluster field image (see below).

In addition to background galaxies, we identify cluster (red) member
galaxies by the following criteria: 
\begin{equation}
a_1 -0.3 < g-r< a_1 + 0.3,
\label{eq:gricut_red1}
\end{equation}
\begin{equation}
0.5(g-r)+a_2-0.15<r-i<0.5(g-r)+a_2+0.15,
\label{eq:gricut_red2}
\end{equation}
with ($a_1$, $a_2$) are ($1.5$, $-0.25$) for $z<0.35$, ($1.6$,
$-0.22$) for $0.35<z<0.45$,  ($1.6$, $0$) for  $0.45<z<0.55$, 
and ($1.7$, $0.05$) for $0.55<z$. These cuts are also indicated by the 
upper panel of Figure~\ref{fig:cosmos}.

\begin{figure}
\begin{center}
 \includegraphics[width=0.85\hsize]{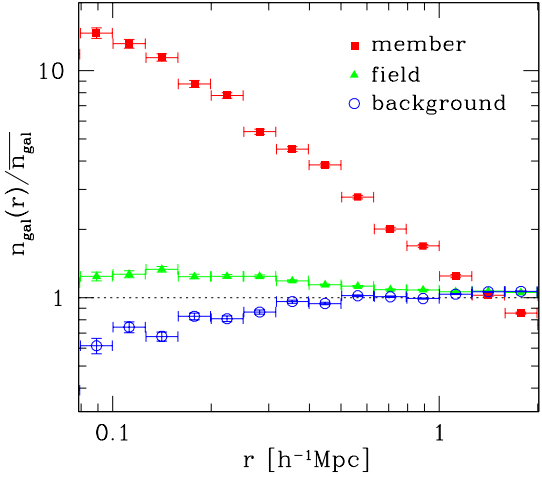}
\end{center}
\caption{The radial number density distributions of galaxies, which are
  obtained by averaging distributions over all 28 clusters. We show
  distributions for background galaxies ({\it open circles}), cluster
  member galaxies ({\it filled squares}), and ``field'' galaxies ({\it
   filled triangles}) that are simply defined as being neither member
  nor background galaxies. Note that the distributions are normalized
  by the overall number densities.
\label{fig:galdist}}
\end{figure}

Figure~\ref{fig:galdist} shows the average radial number density 
distributions of galaxies, background, member, and field (i.e.,
galaxies which are neither background nor member) galaxies in our 28
cluster sample. As expected, the member galaxy density is steeply
rising near the cluster. Field galaxies do not show strong dependence
on the distance from the centre, as that should be dominated by local
foreground galaxies. A slight increase at small radius suggests that
some blue cluster member galaxies are included in the field galaxy
sample. On the other hand, background galaxy density decreases near
the centre, which is in fact expected because of the lensing
magnification and dilution effect of clusters 
\citep[e.g.,][]{broadhurst05,umetsu11a}. These distributions basically
support the successful colour cut for selecting background galaxies. 

\begin{figure}
\begin{center}
 \includegraphics[width=0.85\hsize]{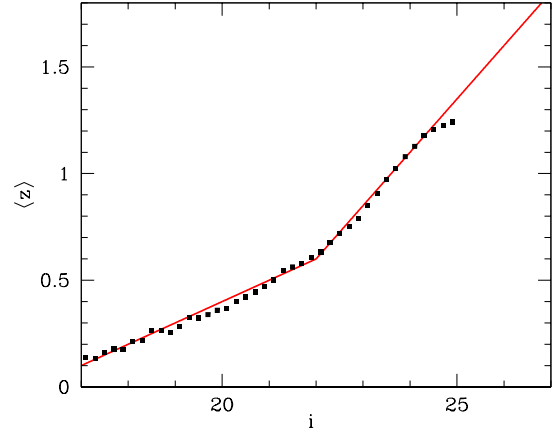}
\end{center}
\caption{The mean galaxy redshift as a function of $i$-band magnitude,
  which is derived from the COSMOS photometric redshift catalogue
  \citep{ilbert09}. The solid line plots our fit given by
  equation~(\ref{eq:zmeanfit}). 
\label{fig:zmean}}
\end{figure}

In order to extract physical quantities for clusters from weak lensing
signals, we need an estimate of the lensing depth of our source galaxy
sample. Again, the COSMOS photometric redshift catalogue is useful for
this purpose, but one problem is that many of our source galaxies are
fainter than the magnitude limit of the COSMOS photometric redshift
catalogue. To overcome this problem, we adopt the following procedure to
estimate the lensing depth. First, from the COSMOS photometric
redshift catalogue \citep{ilbert09} we derive the response function for
our colour cut as $\phi_{\rm cut}(z)=n_{\rm cut}(z)/n_{\rm tot}(z)$,
where $n_{\rm tot}(z)$ and $n_{\rm cut}(z)$ are photometric redshift
distributions of galaxies before and after the colour cut,
respectively. Our weak lensing analysis involves the statistical
weight $w_g$ (equation~\ref{eq:weight}), and hence we compute the mean
weight as a function of $i$-band magnitude as well, 
$\overline{w_g}(i)$, from the background galaxy catalogue. 
Next we derive the redshift distribution of galaxies as a function of
the $i$-band magnitude as follows. First, we adopt the functional form
proposed by \citet{schrabback10}:
\begin{equation}
p(z|i)\propto \left(\frac{z}{z_0}\right)^\alpha
\left(\exp\left[-\left(\frac{z}{z_0}\right)^\beta\right]
+cu^d\exp\left[-\left(\frac{z}{z_0}\right)^\gamma\right]\right),
\label{eq:zgaldist}
\end{equation}
with $u={\rm max}(0,i-23)$ and ($\alpha$, $\beta$, $c$, $d$,
$\gamma$)$=$(0.678, 5.606, 0.581, 1.851, 1.464). We then recompute the
mean galaxy redshift as a function of $i$-band magnitude for the range
$17<i<25$ using the COSMOS photometric redshift catalogue.  The result
shown in Figure~\ref{fig:zmean} suggests that it is well approximated by
\begin{eqnarray}
\langle z \rangle = \left\{
      \begin{array}{ll}
        0.1(i-22) +0.6 &
        \mbox{($17<i<22$)}, \\ 
        0.25(i-22)+0.6 &
        \mbox{($22<i$)} , 
      \end{array}
   \right. 
\label{eq:zmeanfit}
\end{eqnarray}
We extrapolate this linear relation to a fainter magnitude of $i>25$,
because \citet{schrabback10} has explicitly shown that such
extrapolation can explain the redshift distribution of galaxies reasonably
well down to $i\sim 27$. Using this relation, we derive the relation
between $z_0$ in equation~(\ref{eq:zgaldist}) and $i$-band magnitude,
which is approximated by 
\begin{eqnarray}
z_0 = \left\{
      \begin{array}{ll}
        0.16(i-22)+0.97 & \mbox{($17<i<22$)}, \\ 
        0.4(i-22)+0.97 & \mbox{($22<i<23.1$)}, \\
        1.41 & \mbox{($23.1<i<24.3$)}, \\
        0.2(i-24.3)+1.41 & \mbox{($24.3<i$)}.
      \end{array}
   \right. 
\end{eqnarray}
Using this redshift distribution, we derive the mean depth as
\begin{equation}
\left\langle\frac{D_{ls}}{D_{os}}\right\rangle=\left(\int di
\,\overline{D_i}
\frac{dN}{di}\overline{w_g}(i)\right)\left(\int di 
\,\frac{dN}{di}\overline{w_g}(i)\right)^{-1},
\end{equation}
\begin{equation}
\overline{D_i}=\left(\int dz
\frac{D_{ls}}{D_{os}}p(z|i)\phi_{\rm cut}(z)\right)\left(\int dz\,
p(z|i)\phi_{\rm cut}(z)\right)^{-1},
\end{equation}
where $dN/di$ are the $i$-band number counts of background galaxies
used for weak lensing analysis. We define the effective source
redshift $z_{s,{\rm eff}}$ such that it reproduces the mean depth:
\begin{equation}
\frac{D_{ls}}{D_{os}}(z_{s,{\rm eff}})
=\left\langle\frac{D_{ls}}{D_{os}}\right\rangle.
\end{equation}
Throughout this paper, we assume that all the galaxies are
located at $z_{s,{\rm eff}}$ for our weak lensing analysis.

The depth estimate above can be affected by the lensing magnification,
because the magnification enhances the effective lensing depth. However,
the magnification factor tends to decrease rapidly beyond the Einstein
radius such that a typical magnification factor of the innermost
radial bin for our weak lensing analysis is $\sim 20\%$ or so. This
corresponds to the enhancement of the shear amplitude of $\sim 5\%$,
which is not significant compared with other uncertainties. Moreover
the effect is much smaller at radii where weak lensing signals mostly
come from. Thus we conclude that the magnification effect is
insignificant for our analysis. Photometric errors that become larger
for fainter galaxies can smear the distribution in the color-color
space, and hence can affect the lensing depth estimate, but
photometric errors are included in the COSMOS photometric redshift
catalog as well. Photometric errors for the faintest galaxies in our
weak lensing analysis are similar to those for the faintest galaxies
in the COSMOS catalog. Another possible systematic effect is the size
cut, which may systematically eliminate galaxies at high redshifts
and hence bias the lensing depth estimate. While we expect the effect
to be small because of the conservative magnitude limit we adopt and
the narrow redshift distribution after the color cut, we leave the
detailed exploration of this effect for future work. 

\begin{table*}
 \caption{Summary of the weak lensing analysis. We show the median of 
   stellar ellipticities before ($e^*_{\rm raw}$) and after ($e^*_{\rm
     cor}$) the PSF correction (numbers in parentheses are standard
   deviations), the number of stars used for the PSF correction 
   ($N^*$), the number density of source background galaxies ($n_g$),  
   the $r$-band magnitude limit of source background galaxies ($r_{\rm
     lim}$), the mean lensing depth ($\langle D_{ls}/D_{os}\rangle$)
   and the effective source redshift ($z_{s,{\rm eff}}$), the range of
   the radii for tangential shear fitting ($\theta$ range), and the
   total signal-to-noise ratio of the tangential shear profile ($S/N$).
   \label{tab:wlens}}   
 \begin{tabular}{@{}cccccccccc}
 \hline
   Name
   & $e^*_{\rm raw}$
   & $e^*_{\rm cor}$
   & $N^*$
   & $n_g$
   & $r_{\rm lim}$
   & $\langle D_{ls}/D_{os}\rangle$  
   & $z_{s,{\rm eff}}$
   & $\theta$ range
   & $S/N$ \\
   & ($10^{-2}$)
   & ($10^{-2}$) 
   & 
   & (arcmin$^{-2}$)
   & (mag)
   & 
   & 
   & (arcmin)
   & \\
 \hline
SDSSJ0851+3331 & 1.51(0.90) & 0.01(0.47) &  834 & 10.1 & 25.8 & 0.595 & 1.098 &  0.63--15.85 &  6.93 \\ 
SDSSJ0915+3826 & 1.40(0.83) & 0.02(0.54) &  694 & 11.7 & 25.6 & 0.571 & 1.109 &  0.40--6.31  &  4.59 \\ 
SDSSJ0957+0509 & 1.28(1.74) & 0.01(0.93) &  965 & 11.6 & 26.0 & 0.536 & 1.155 &  0.40--6.31  &  3.68 \\ 
SDSSJ1004+4112 & 0.26(2.39) & 0.02(1.15) &  579 &  8.3 & 25.4 & 0.318 & 1.113 &  0.40--6.31  &  2.41 \\ 
SDSSJ1029+2623 & 1.27(1.75) & 0.01(0.64) &  654 & 17.5 & 26.2 & 0.427 & 1.189 &  0.40--15.85 &  7.30 \\ 
SDSSJ1038+4849 & 1.58(1.22) & 0.03(0.54) &  532 & 12.7 & 26.0 & 0.549 & 1.142 &  0.40--6.31  &  3.61 \\ 
SDSSJ1050+0017 & 2.00(1.59) & 0.02(0.72) &  764 & 16.2 & 26.0 & 0.405 & 1.162 &  0.50--12.59 &  7.52 \\ 
RCS2J1055+5548 & 1.78(1.33) & 0.01(0.52) &  569 & 13.2 & 25.8 & 0.513 & 1.133 &  0.63--10.00 &  7.45 \\ 
SDSSJ1110+6459 & 3.20(1.50) & 0.03(0.75) &  924 & 11.3 & 25.8 & 0.407 & 1.124 &  0.40--2.51  &  4.10 \\ 
SDSSJ1115+5319 & 1.10(0.90) & 0.01(0.53) &  644 & 10.1 & 25.6 & 0.501 & 1.095 &  0.50--19.95 &  6.72 \\ 
SDSSJ1138+2754 & 1.27(1.20) & 0.01(0.48) &  513 & 12.9 & 25.8 & 0.527 & 1.135 &  0.68--14.69 &  9.32 \\ 
SDSSJ1152+3313 & 0.46(1.90) & 0.08(1.35) &  847 &  9.1 & 25.8 & 0.606 & 1.112 &  0.50--5.01  &  2.05 \\ 
SDSSJ1152+0930 & 2.56(0.89) & 0.02(0.45) &  748 & 11.0 & 25.2 & 0.457 & 1.106 &  0.40--10.00 &  5.55 \\ 
SDSSJ1209+2640 & 1.50(0.95) & 0.03(0.49) &  581 & 13.7 & 25.8 & 0.430 & 1.139 &  0.40--10.00 &  7.39 \\ 
SDSSJ1226+2149 & 1.50(1.44) & 0.03(1.01) &  754 &  9.9 & 25.4 & 0.510 & 1.033 &  0.32--10.00 &  6.73 \\ 
SDSSJ1226+2152 & 1.50(1.44) & 0.03(1.01) &  754 &  9.9 & 25.4 & 0.510 & 1.033 &  0.40--2.51  &  1.31 \\ 
         A1703 & 0.95(1.59) & 0.02(0.62) &  686 & 15.4 & 26.0 & 0.690 & 1.101 &  0.68--14.69 & 12.62 \\ 
SDSSJ1315+5439 & 1.17(1.14) & 0.01(0.61) &  658 & 10.2 & 25.6 & 0.407 & 1.140 &  0.63--10.00 &  5.90 \\ 
GHO132029+3155 & 1.93(1.10) & 0.02(0.54) &  593 & 14.1 & 26.2 & 0.666 & 1.137 &  0.68--14.69 & 10.32 \\ 
SDSSJ1329+2243 & 3.12(1.86) & 0.03(0.60) &  751 & 17.9 & 26.2 & 0.539 & 1.149 &  0.63--10.00 &  8.00 \\ 
SDSSJ1343+4155 & 1.78(0.99) & 0.01(0.59) &  626 & 12.9 & 25.6 & 0.553 & 1.117 &  0.63--10.00 &  5.22 \\ 
SDSSJ1420+3955 & 2.19(1.16) & 0.01(0.65) &  705 & 10.4 & 25.4 & 0.382 & 1.119 &  0.40--10.00 &  7.45 \\ 
SDSSJ1446+3032 & 1.20(1.03) & 0.04(0.58) &  928 & 13.9 & 25.6 & 0.513 & 1.127 &  0.50--12.59 &  8.31 \\ 
SDSSJ1456+5702 & 0.95(1.35) & 0.01(0.54) &  768 & 13.1 & 25.8 & 0.499 & 1.143 &  0.50--19.95 &  7.30 \\ 
SDSSJ1531+3414 & 0.77(1.31) & 0.01(0.57) & 1106 & 11.7 & 25.4 & 0.623 & 1.072 &  0.63--15.85 &  7.08 \\ 
SDSSJ1621+0607 & 0.03(1.13) & 0.00(0.50) & 2726 &  7.1 & 25.2 & 0.616 & 1.071 &  0.40--15.85 &  6.39 \\ 
SDSSJ1632+3500 & 2.40(1.87) & 0.03(0.61) & 1679 & 10.8 & 25.8 & 0.494 & 1.142 &  0.50--12.59 &  4.95 \\ 
SDSSJ2111$-$0114 & 1.76(1.85) & 0.04(0.90) & 2645 & 13.4 & 25.8 & 0.371 & 1.157 &  0.50--12.59 &  4.48 \\ 
 \hline
 \end{tabular}
\end{table*}

Table~\ref{tab:wlens} summarizes our weak lensing analysis. It is seen
that our PSF correction algorithm successfully reduces the stellar
ellipticities to the level that are much smaller than typical lensing
shear amplitudes for our cluster sample ($\gamma \ga 0.01$). The
relatively small number density of $n_g\sim 10$~arcmin$^{-2}$ is due
to the colour cut for selecting background galaxies. From the shear
catalogues we compute tangential shear profiles. The tangential shear
$g_+$ is computed from the reduced shear $\bm{g}=(g_1,\,g_2)$ as
\begin{equation}
g_+=-g_1\cos2\phi-g_2\sin2\phi,
\label{eq:g1}
\end{equation}
where $\phi$ is the polar angle. Throughout the paper, we assume that the
position of the brightest galaxy in the strong lensing region is
coincident with the cluster centre. Although our strong lensing
analysis suggests that the position of the mass peak can differ
slightly from the central galaxy position, we find that any such
offset is much smaller than the typical inner radial boundary of our
weak lensing analysis. Hence the effect of off-centreing on our
results should be negligibly small. The average shear value in each
radial bin is estimated by the weighted mean of the tangential shear
as follows:  
\begin{equation}
\bar{g}_+ = \left(\sum_i w_{g,i}g_{+,i}\right)
\left(\sum_i w_{g,i}\right)^{-1},
\end{equation}
where the summation runs over all galaxies in the bin. Similarly the
statistical error in the tangential shear measurement in each bin is
computed from the weighted average of the variance of the shear,
$\sigma_g^2$ \citep[e.g.,][]{okabe10a}. The total signal-to-noise ratio
of the tangential shear profile is shown in Table~\ref{tab:wlens}
and indicates that weak lensing signals are detected significantly
($S/N\ga 5$) for most of our clusters. The mass maps reconstructed
from the weak lensing analysis are shown in Appendix~\ref{sec:app}. 

\section{Combining strong and weak lensing}
\label{sec:swlens}

\subsection{Methodology}

We combine constraints from the tangential shear profile with
constraints from strong lensing. As in \citet{oguri09b}, we combine
both sets of constraints by summing up $\chi^2$:
\begin{equation}
\chi^2=\chi^2_{\rm SL}+\chi^2_{\rm WL}.
\end{equation}
We include strong lensing constraints from the Einstein radius at the arc
redshift:
\begin{equation}
\chi^2_{\rm SL}=\frac{\left[\bar{\theta}_{\rm E}-\theta_{\rm E}(M_{\rm
    vir},\,c_{\rm vir})\right]^2}{\sigma_{\rm E}^2}, 
\end{equation}
where $\bar{\theta}_{\rm E}$ and $\sigma_{\rm E}$ are the best-fit
Einstein radius at $z_s=z_{\rm arc}$ and its error presented in
Table~\ref{tab:slens}, and $\theta_{\rm E}(M_{\rm vir},\,c_{\rm vir})$
is the predicted Einstein radius assuming the NFW profile.
On the other hand, weak lensing constraints come from binned
tangential shear measurements: 
\begin{equation}
\chi^2_{\rm WL}=\sum_i\frac{\left[\bar{g}_{+,i}-g_+(\theta_i;\,M_{\rm
    vir},\,c_{\rm vir})\right]^2}{\sigma_i^2},
\label{eq:c2wl}
\end{equation}
where $\bar{g}_{+,i}$ and $\sigma_i$ are observed reduced shear and
its error at $i$-th radial bin, and $-g_+(\theta_i;\,M_{\rm
  vir},\,c_{\rm vir})$ is the predicted reduced shear by the NFW
 model. The NFW profile, which we adopt as an analytical model for the
 radial mass distribution, has the three-dimensional density profile
 of   
\begin{equation}
\rho(r)=\frac{\rho_s}{(r/r_s)(1+r/r_s)^2},
\label{eq:nfw}
\end{equation}
where 
\begin{equation}
\rho_s=\frac{\Delta_{\rm vir}(z)\bar{\rho}_m(z)c_{\rm vir}^3}
{3\left[\ln(1+c_{\rm vir})-c_{\rm vir}/(1+c_{\rm vir})\right]},
\end{equation}
where $\Delta(z)$ is the nonlinear over-density predicted by the
spherical collapse model. We parametrize the profile with two
parameters, the virial mass $M_{\rm vir}$ 
\begin{equation}
M_{\rm vir}=\frac{4\pi}{3}r_{\rm vir}^3\Delta_{\rm vir}(z)
\bar{\rho}_m(z),
\end{equation}
and the concentration parameter $c_{\rm vir}$
\begin{equation}
c_{\rm vir}=\frac{r_{\rm vir}}{r_s}.
\end{equation}

\begin{figure}
\begin{center}
 \includegraphics[width=0.85\hsize]{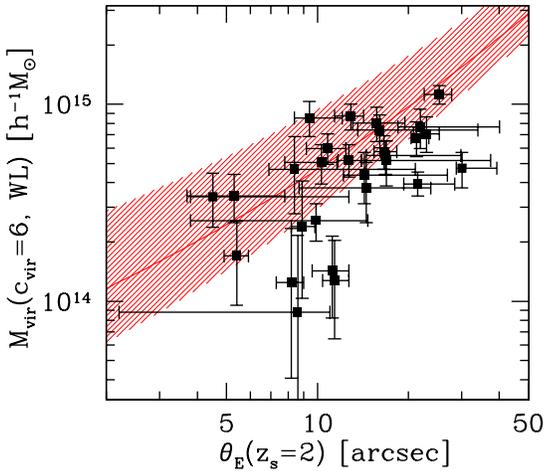}
\end{center}
\caption{Correlation between strong and weak lensing observables. 
The best-fit mass $M_{\rm vir}$ for a fixed concentration of $c_{\rm
vir}=6$ from just weak lensing is compared with the Einstein radius
for the source redshift $z_s=2$ from strong lensing. The red line with
shading shows the expected $M_{\rm vir}$-$\theta_E$ relation assuming
the NFW profile with the concentration of $c_{\rm vir}=6\pm2$. 
\label{fig:mein}}
\end{figure}

\subsection{Correlation between strong and weak lensing observables}

Before deriving best-fit parameters for individual clusters, we first
compare strong and weak lensing observables, namely the Einstein
radius and the virial mass, which should serve as a sanity check for
results of more detailed analysis presented in the following sections. 
For the strong lensing observable, we use the Einstein radius for the
fixed source redshift  
of $z_s=2$ in order to compare results for different clusters. Given
the strong degeneracy between $M_{\rm vir}$ and $c_{\rm vir}$ in weak
lensing analysis, we quantify the strength of the weak lensing signal
by fitting the tangential shear profile for a fixed concentration
parameter of $c_{\rm vir}=6$, a typical value for the concentration 
when the lensing selection effect is taken into account (see
below). Figure~\ref{fig:mein} shows the correlation between the
Einstein radius from strong lensing and the virial mass from weak
lensing. We confirm that these two measurements are indeed correlated
with each other such that the more massive clusters detected via their
weak lensing have larger Einstein radii on average.  

Assuming the NFW profile, the relation between mass and Einstein
radius is determined by the concentration parameter. We also plot the
expected correlation for the concentration assuming $c_{\rm vir}=6$,
as well as the scatter of the relation originating from the shift of 
the concentration by $\pm 2$, for the lens redshift $z=0.46$ which
corresponds to the median redshift of our cluster sample. We find that
the observed correlation roughly follows the expectation for $c_{\rm
  vir}=6$, although there is a tendency for the Einstein radii to be
larger than expected, particularly for low-mass clusters. This
analysis suggests that our sample of clusters are slightly more
concentrated than $c_{\rm vir}\sim 6$, and that clusters with lower
masses on average have higher concentration values. 

\subsection{Fitting results}

Table~\ref{tab:fit} summarizes the results of the combined analysis,
i.e., simultaneous fitting of the observed Einstein radius and
tangential shear profile with the NFW model predictions. The
tangential shear profiles for individual clusters are compared with
best-fit models in Appendix~\ref{sec:app}.  Fitting is
performed in the parameter range $10^{13}h^{-1}M_\odot<M_{\rm
vir}<10^{16}h^{-1}M_\odot$ and $0.01<c_{\rm vir}<39.81$. For four
clusters studied previously in \citet{oguri09b}, our new results are
fully consistent with the old result. In addition, the best-fit mass
and concentration parameter for A1703 are in good agreement with the
result of an independent strong and weak lensing analysis by
\citet{zitrin10}. 

\begin{table*}
 \caption{Constraints on the mass $M_{\rm vir}$ and concentration
   parameter $c_{\rm vir}$ from weak lensing and combined strong and 
   weak lensing analysis. Errors indicate $1\sigma$ errors on each 
   parameter after marginalizing over the other parameter. 
   \label{tab:fit}}   
 \begin{tabular}{@{}ccccccccc}
 \hline
   & \multicolumn{4}{c}{Weak lensing}
   & & \multicolumn{3}{c}{Strong and weak lensing}\\ 
   \cline{2-5} 
   \cline{7-9} 
   Name
   & $\chi^2$/dof
   & $M_{\rm vir}$
   & $c_{\rm vir}$
   & $M_{\rm vir}(c_{\rm vir}=6)$
   & 
   & $\chi^2$/dof
   & $M_{\rm vir}$
   & $c_{\rm vir}$ \\
   &
   & ($10^{14}h^{-1}M_\odot$)
   & 
   & ($10^{14}h^{-1}M_\odot$)
   & 
   &
   & ($10^{14}h^{-1}M_\odot$)
   & \\
 \hline
SDSSJ0851+3331 &  4.2/7 & $ 7.33^{+ 2.44}_{- 1.96}$ & $ 5.62^{+ 3.39}_{- 2.03}$ & $ 7.08^{+ 1.53}_{- 1.39}$ & &  5.8/8 & $ 6.24^{+ 1.80}_{- 1.61}$ & $ 9.44^{+ 3.15}_{- 1.85}$ \\
SDSSJ0915+3826 &  7.0/6 & $ 0.91^{+ 0.30}_{- 0.28}$ & $39.81^{+ 0.00}_{-17.16}$ & $ 1.27^{+ 0.77}_{- 0.63}$ & &  8.0/7 & $ 0.80^{+ 0.50}_{- 0.27}$ & $26.92^{+12.90}_{-10.88}$ \\
SDSSJ0957+0509 &  1.2/6 & $ 0.97^{+ 0.60}_{- 0.31}$ & $39.81^{+ 0.00}_{-27.65}$ & $ 1.70^{+ 0.87}_{- 0.74}$ & &  2.9/7 & $ 1.29^{+ 0.85}_{- 0.61}$ & $ 9.02^{+ 4.47}_{- 2.18}$ \\
SDSSJ1004+4112 &  0.8/6 & $ 2.82^{+ 4.34}_{- 1.92}$ & $ 4.42^{+30.26}_{- 3.74}$ & $ 2.40^{+ 1.67}_{- 1.36}$ & &  1.0/7 & $ 2.21^{+ 2.41}_{- 1.43}$ & $ 8.32^{+11.87}_{- 3.13}$ \\
SDSSJ1029+2623 &  9.7/8 & $ 2.00^{+ 0.73}_{- 0.60}$ & $11.48^{+14.52}_{- 5.02}$ & $ 2.57^{+ 0.56}_{- 0.55}$ & &  9.7/9 & $ 2.02^{+ 0.67}_{- 0.57}$ & $11.09^{+ 9.56}_{- 4.17}$ \\
SDSSJ1038+4849 &  1.2/6 & $ 0.86^{+ 0.71}_{- 0.39}$ & $20.89^{+18.92}_{-13.56}$ & $ 1.43^{+ 0.71}_{- 0.61}$ & &  1.4/7 & $ 0.74^{+ 0.52}_{- 0.12}$ & $39.81^{+ 0.00}_{-21.61}$ \\
SDSSJ1050+0017 &  3.2/7 & $ 6.84^{+ 1.97}_{- 1.71}$ & $ 7.24^{+ 5.34}_{- 2.67}$ & $ 7.41^{+ 1.40}_{- 1.39}$ & &  3.2/8 & $ 6.84^{+ 1.97}_{- 1.65}$ & $ 7.16^{+ 4.86}_{- 2.09}$ \\
RCS2J1055+5548 &  2.1/6 & $ 5.13^{+ 1.71}_{- 1.33}$ & $ 6.17^{+ 4.07}_{- 2.23}$ & $ 5.19^{+ 1.05}_{- 0.92}$ & &  2.2/7 & $ 4.79^{+ 1.31}_{- 1.07}$ & $ 7.41^{+ 1.40}_{- 1.10}$ \\
SDSSJ1110+6459 &  3.2/4 & $ 2.07^{+ 2.15}_{- 0.67}$ & $35.89^{+ 3.92}_{-27.48}$ & $ 4.68^{+ 2.16}_{- 1.89}$ & &  3.7/5 & $ 2.26^{+ 2.41}_{- 0.96}$ & $22.39^{+17.42}_{-15.70}$ \\
SDSSJ1115+5319 &  8.3/8 & $11.61^{+ 3.52}_{- 2.90}$ & $ 2.66^{+ 1.23}_{- 0.90}$ & $ 7.67^{+ 1.77}_{- 1.72}$ & & 13.8/9 & $10.59^{+ 3.05}_{- 2.74}$ & $ 5.25^{+ 1.51}_{- 0.98}$ \\
SDSSJ1138+2754 &  1.6/8 & $11.22^{+ 2.58}_{- 2.31}$ & $ 3.55^{+ 1.52}_{- 1.09}$ & $ 8.71^{+ 1.29}_{- 1.30}$ & &  2.1/9 & $10.35^{+ 2.09}_{- 1.84}$ & $ 4.47^{+ 0.60}_{- 0.53}$ \\
SDSSJ1152+3313 &  0.2/5 & $ 0.73^{+ 1.33}_{- 0.44}$ & $27.54^{+12.27}_{-24.27}$ & $ 1.24^{+ 1.10}_{- 0.84}$ & &  0.2/6 & $ 0.82^{+ 0.94}_{- 0.48}$ & $17.38^{+22.43}_{- 7.38}$ \\
SDSSJ1152+0930 &  6.9/7 & $ 7.24^{+ 3.59}_{- 2.57}$ & $ 1.66^{+ 1.33}_{- 0.87}$ & $ 3.39^{+ 1.08}_{- 1.02}$ & &  8.9/8 & $ 5.75^{+ 2.56}_{- 1.95}$ & $ 3.55^{+ 0.92}_{- 0.66}$ \\
SDSSJ1209+2640 &  9.5/7 & $ 6.92^{+ 2.52}_{- 2.02}$ & $ 5.75^{+ 3.69}_{- 2.25}$ & $ 6.76^{+ 1.37}_{- 1.33}$ & &  9.9/8 & $ 6.03^{+ 1.83}_{- 1.45}$ & $ 7.85^{+ 1.59}_{- 1.25}$ \\
SDSSJ1226+2149 &  0.5/6 & $ 8.81^{+ 3.63}_{- 2.64}$ & $ 5.25^{+ 2.51}_{- 1.74}$ & $ 8.04^{+ 1.63}_{- 1.58}$ & &  0.5/7 & $ 8.61^{+ 3.28}_{- 2.44}$ & $ 5.56^{+ 1.69}_{- 1.14}$ \\
SDSSJ1226+2152 &  0.1/4 & $ 0.80^{+75.05}_{- 0.70}$ & $ 6.84^{+32.97}_{- 6.83}$ & $ 0.88^{+ 1.28}_{- 0.78}$ & &  0.3/5 & $ 0.39^{+ 1.27}_{- 0.25}$ & $39.81^{+ 0.00}_{-33.13}$ \\
         A1703 &  6.3/8 & $12.88^{+ 2.61}_{- 2.17}$ & $ 4.79^{+ 1.24}_{- 1.03}$ & $11.22^{+ 1.22}_{- 1.22}$ & & 10.9/9 & $10.96^{+ 1.92}_{- 1.63}$ & $ 7.08^{+ 1.14}_{- 0.84}$ \\
SDSSJ1315+5439 &  6.7/6 & $ 4.42^{+ 1.82}_{- 1.46}$ & $ 9.44^{+15.97}_{- 4.60}$ & $ 5.19^{+ 1.34}_{- 1.34}$ & &  6.7/7 & $ 4.37^{+ 1.66}_{- 1.38}$ & $ 9.66^{+14.33}_{- 2.82}$ \\
GHO132029+3155 &  5.3/8 & $ 3.43^{+ 0.65}_{- 0.58}$ & $ 8.81^{+ 3.35}_{- 2.35}$ & $ 3.94^{+ 0.58}_{- 0.51}$ & &  9.7/9 & $ 2.95^{+ 0.52}_{- 0.50}$ & $15.67^{+ 4.52}_{- 2.79}$ \\
SDSSJ1329+2243 &  1.1/6 & $ 4.90^{+ 1.34}_{- 1.14}$ & $ 9.89^{+ 7.29}_{- 3.58}$ & $ 5.96^{+ 1.12}_{- 1.00}$ & &  2.7/7 & $ 5.62^{+ 1.38}_{- 1.21}$ & $ 5.82^{+ 1.18}_{- 0.81}$ \\
SDSSJ1343+4155 &  1.8/6 & $ 3.89^{+ 2.07}_{- 1.46}$ & $ 4.57^{+ 5.66}_{- 2.33}$ & $ 3.43^{+ 0.99}_{- 0.92}$ & &  1.9/7 & $ 3.76^{+ 1.55}_{- 1.25}$ & $ 5.07^{+ 1.69}_{- 1.00}$ \\
SDSSJ1420+3955 &  8.9/7 & $ 6.92^{+ 2.20}_{- 1.79}$ & $ 9.55^{+ 6.30}_{- 3.31}$ & $ 8.51^{+ 1.84}_{- 1.67}$ & & 13.5/8 & $ 7.59^{+ 2.53}_{- 2.03}$ & $ 4.57^{+ 1.32}_{- 0.98}$ \\
SDSSJ1446+3032 & 14.8/7 & $ 4.07^{+ 1.17}_{- 0.98}$ & $12.59^{+ 9.04}_{- 4.37}$ & $ 5.50^{+ 1.04}_{- 1.08}$ & & 14.9/8 & $ 4.12^{+ 1.19}_{- 0.99}$ & $12.02^{+ 8.39}_{- 3.80}$ \\
SDSSJ1456+5702 &  7.6/8 & $ 6.68^{+ 2.03}_{- 1.67}$ & $ 2.92^{+ 1.65}_{- 1.16}$ & $ 4.73^{+ 0.96}_{- 0.97}$ & & 21.4/9 & $ 2.69^{+ 0.86}_{- 0.76}$ & $22.65^{+14.51}_{- 6.24}$ \\
SDSSJ1531+3414 &  0.6/7 & $ 5.75^{+ 1.83}_{- 1.44}$ & $ 5.96^{+ 3.27}_{- 2.07}$ & $ 5.75^{+ 1.16}_{- 1.08}$ & &  1.3/8 & $ 5.13^{+ 1.33}_{- 1.19}$ & $ 8.32^{+ 1.57}_{- 1.16}$ \\
SDSSJ1621+0607 &  2.6/8 & $ 6.68^{+ 2.54}_{- 2.01}$ & $ 3.94^{+ 1.89}_{- 1.39}$ & $ 5.07^{+ 1.17}_{- 1.13}$ & &  3.8/9 & $ 5.89^{+ 2.05}_{- 1.67}$ & $ 5.56^{+ 1.44}_{- 1.04}$ \\
SDSSJ1632+3500 &  2.7/7 & $ 4.22^{+ 1.74}_{- 1.40}$ & $ 6.53^{+ 7.27}_{- 3.06}$ & $ 4.37^{+ 1.32}_{- 1.24}$ & &  2.9/8 & $ 3.98^{+ 1.58}_{- 1.26}$ & $ 8.51^{+ 5.94}_{- 2.05}$ \\
SDSSJ2111$-$0114 &  2.2/7 & $ 6.03^{+ 2.58}_{- 2.14}$ & $ 1.91^{+ 1.68}_{- 1.01}$ & $ 3.76^{+ 1.31}_{- 1.25}$ & &  6.0/8 & $ 5.25^{+ 2.43}_{- 1.94}$ & $ 4.79^{+ 3.16}_{- 1.62}$ \\
 \hline
 \end{tabular}
\end{table*}

With a large sample of clusters with measurements of the concentration
parameter from combined strong and weak lensing, we can study the
mass-concentration relation quite well. A caveat is that our sample
of clusters are selected as those having prominent arcs. It has been
noted that clusters selected by strong lensing (e.g., by giant arcs or
large Einstein radii) represent a strongly biased population such that
the concentration parameter inferred from the projected mass
distribution is on average much larger mostly due to the halo
triaxiality \citep{hennawi07,oguri09a,meneghetti10}. Although in
\citet{oguri09b} we assumed the constant enhancement of the
concentration parameter due to the lensing bias, simple considerations
suggest that the lensing bias of the mass-concentration relation
should depend strongly on the mass. To derive more accurate
theoretical predictions based on the $\Lambda$CDM model, in
Appendix~\ref{sec:conc} we conduct a series of semi-analytic
calculations with ray-tracing of extended sources to estimate the
effect of the lensing bias, based on a triaxial halo model of
\citet{jing02}. For reference, we find that the mean
mass-concentration relation at $z=0.45$ with the lensing bias
predicted by this model is roughly described by  
\begin{equation}
\bar{c}_{\rm vir}(z=0.45)\approx 6.3\left(\frac{M_{\rm vir}}
{5\times 10^{14}h^{-1}M_\odot}\right)^{-0.2},
\end{equation}
which show relatively strong dependence on the halo mass, simply
because of the mass dependence of the lensing bias (see
Appendix~\ref{sec:conc}). The scatter of the relation is estimated to
be $\sigma_{\log c}\simeq 0.12$.

\begin{figure}
\begin{center}
 \includegraphics[width=0.96\hsize]{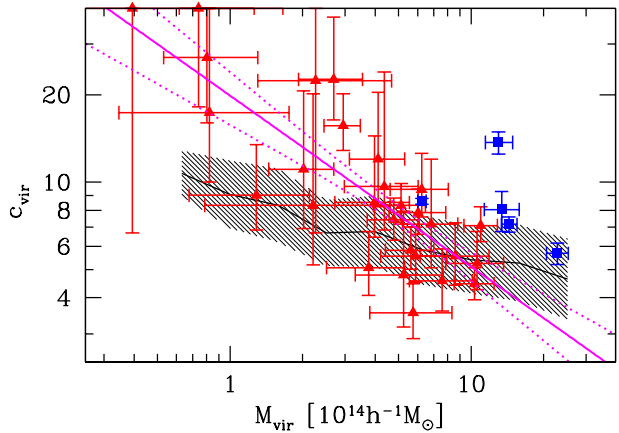}
\end{center}
\caption{The mass-concentration relation obtained from combined strong
  and weak lensing analysis. Filled triangles show our results
  presented in this paper, whereas filled squares show results from
  literature; A1689, A370, CL0024, RXJ1347 \citep{umetsu11b}, and A383
  \citep{zitrin11b}. The black shaded region indicates the predicted
  concentration parameters as a function of the halo mass with the
  lensing bias taken into account (see Appendix~\ref{sec:conc}  for
  details). The solid line is the best-fit mass-concentration relation
  from fitting of our cluster sample (i.e., filled triangles), with the  
  1$\sigma$ range indicated by dotted lines.
\label{fig:clupar}}
\end{figure}

Figure~\ref{fig:clupar} shows the mass-concentration relation obtained
from our lensing analysis for 28 systems. For comparison, we also show
accurate lensing measurements of the concentration parameters for 5
massive clusters from the literature \citep{umetsu11b,zitrin11b}. 
The Figure clearly indicates that measured concentrations are
correlated well with the mass. More massive clusters have on average
smaller concentrations, which is consistent with the theoretical
expectation, although the slope is obviously much steeper. 

To put this on a more quantitative footing, we fit the
mass-concentration parameter with both the normalization and the mass
slope as free parameters. Here we ignore the redshift dependence of
the mass-concentration relation, given the predicted little evolution
of the concentration of massive haloes with redshift
\citep[e.g.,][]{zhao09,prada11}. We use the following estimator for
fitting: 
\begin{equation}
\chi^2=\frac{\left[\log\left(c_{\rm vir,obs}\right)-
\log\left(c_{\rm vir,fit}\right)\right]^2}
{\sigma_{\rm st}^2+\sigma_{\rm in}^2},
\end{equation}
where $\sigma_{\rm st}$ is the $1\sigma$ measurement error on
$\log(c_{\rm vir})$ for individual clusters from the lensing analysis
(see Table~\ref{tab:fit}), and $\sigma_{\rm in}=0.12$ is the intrinsic
dispersion predicted by our calculations. We find that the best-fit
mass-concentration relation from our lensing sample of 28 clusters 
is
\begin{equation}
\bar{c}_{\rm vir}=(7.7\pm0.6)\left(\frac{M_{\rm vir}}
{5\times 10^{14}h^{-1}M_\odot}\right)^{-0.59\pm0.12},
\label{eq:mcfit}
\end{equation}
where we also included $1\sigma$ errors on the normalization and the
slope. 

We detect a strong mass dependence of the concentration parameter with
the slope of $-0.59\pm 0.12$ in our lensing sample, which should be
compared with the predicted slope of $\approx -0.2$ for the strong
lensing selected sample of clusters. We note that steeper
mass-concentration relations than theoretical expectations were also
suggested by previous weak lensing \citep{okabe10a} and X-ray
\citep{schmidt07,buote07,ettori10} analysis \citep[see
also][]{biviano08}.  Our result suggests that the observed
mass-concentration relation is in reasonable agreement with the
simulation results for very massive haloes of 
$M_{\rm vir}\sim 10^{15}h^{-1}M_\odot$. The agreement may be even
better if we adopt recent results of $N$-body simulations by
\citet{prada11}, who argued that previous simulation work
underestimated the mean concentrations at high mass end (see also
Appendix~\ref{sec:conc}).  In contrast, we find that observed
concentrations are much higher than theoretical expectations for less
massive haloes of $M_{\rm vir}\sim 10^{14}h^{-1}M_\odot$, even if we
take account of the mass dependence of the lensing bias. 

A possible concern is the correlation of $M_{\rm vir}$ and $c_{\rm vir}$ 
for fitting of individual clusters which has been ignored in deriving
the mass-concentration relation. We examine the possible effect of the
degeneracy between $M_{\rm vir}$ and $c_{\rm vir}$ by the following
Monte Carlo simulation. For each cluster, we change best-fit values
of $M_{\rm vir}$ and $c_{\rm vir}$ by randomly picking up a point
within the $1\sigma$ confidence region in the $M_{\rm vir}$-$c_{\rm
 vir}$ plane. After choosing new best-fit parameters for all the
clusters, we re-fit the mass-concentration relation to derive best-fit
values of the normalization and slope in equation~(\ref{eq:mcfit}). 
We repeat this simulation for 300 times to check how the correlated
errors between $M_{\rm vir}$ and $c_{\rm vir}$ can affect our
conclusion. We find that the resulting distribution of the best-fit
normalization value is $7.6\pm0.2$, and that of the slope is
$-0.56\pm0.05$, which are small compared with the statistical errors
shown in equation~(\ref{eq:mcfit}), suggesting that the effect of the
correlated error is not very significant. We note that \citet{okabe10a}
also explored potential impacts of the degeneracy between  $M_{\rm
  vir}$ and $c_{\rm vir}$  on their mass-concentration measurement
very carefully, and concluded that the effect is insignificant, which
is consistent with our finding. 

There are a few possible explanations for the excess concentration
for small mass clusters. Perhaps the most significant effect is baryon
cooling. The formation of the central galaxy, and the accompanying
adiabatic contraction of dark matter distribution, enhances the core
density of the cluster and increases the concentration parameter value
for the total mass distribution. This effect is expected to be mass
dependent such that lower mass haloes are affected more pronouncedly,
simply because the fraction of the mass of the central galaxy to the
total mass is larger for smaller halo masses. Indeed, simulations with
radiative cooling and star formation indicate that the concentration
can be significantly enhanced by baryon physics particularly for
low-mass haloes \citep[e.g.,][]{rudd08,mead10}, although the effect
strongly depends on the efficiency of feedback \citep{duffy10,mccarthy10}.
Thus baryon cooling appears to be able to explain the observed strong
mass dependence at least qualitatively. More quantitative estimates of
this effect need to be made using a large sample of simulated clusters
with the baryon physics as well as the proper feedback model included.  

\section{Stacking analysis}
\label{sec:stack}

\subsection{Stacked tangential shear profile}
\label{sec:stackgt}

We can study the average properties of a given sample by stacking lensing
signals. This stacked lensing analysis has been successful for
constraining mean dark matter distributions of cluster samples
\citep[e.g.,][]{mandelbaum06b,johnston07,leauthaud10,okabe10a}. Here we
conduct stacking analysis of the tangential shear profile for our
lensing  sample for studying the mass-concentration relation from another
viewpoint. Note that the off-centreing effect, which has been known to
be one of the most significant systematic errors in stacked lensing
analysis \citep[e.g.,][]{johnston07,mandelbaum08,oguri11a}, should be
negligible for our analysis, because of the detection of weak lensing
signals for individual clusters and the presence of giant arcs which
assure that selected centres (positions of the brightest galaxies in
the strong lensing region) indeed correspond to that of the mass
distribution.  

We perform stacking in the physical length scale. Specifically, we 
compute the differential surface density $\Delta \Sigma_+(r)$ which is
define by
\begin{equation}
\Delta\Sigma_+(r)\equiv\Sigma_{\rm cr}g_+(\theta=r/D_{ol}),
\label{eq:dsig}
\end{equation}
where $\Sigma_{\rm cr}$ is the critical surface mass density for
lensing. We stack $\Delta \Sigma_+(r)$ for different clusters to
obtain the average differential surface density. We do not include
SDSSJ1226+2149 and SDSSJ1226+2152 in our stacking analysis, because
these fields clearly have complicated mass distributions with two
strong lensing cores separated by only $\sim 3'$. Furthermore, we
exclude SDSSJ1110+6459 as well because the two-dimensional weak
lensing map  suggests the presence of a very complicated mass
distribution around the system. We use the remaining 25 clusters for
our stacked lensing analysis. 

\begin{figure}
\begin{center}
 \includegraphics[width=0.96\hsize]{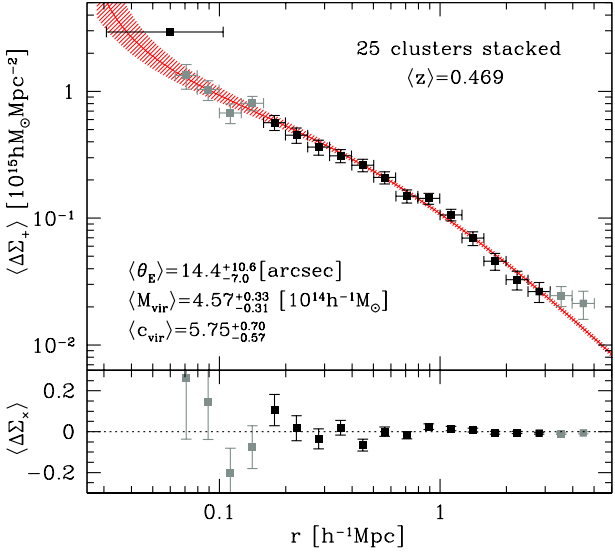}
\end{center}
\caption{The stacked tangential shear profile obtained by combining
 the 25 clusters. The average differential surface density $\langle 
 \Delta\Sigma_+(r)\rangle$ (see equation~\ref{eq:dsig}) is plotted as
 a function of the physical radius $r$. Grey points indicate stacked
 tangential shear measurements from weak lensing that are not used for
 fitting. The upper left point with a horizontal error-bar is the
 constraint from the average Einstein radius. The solid line with
 shading is the best-fit NFW model with $1\sigma$ error range. The
 lower panel plots the stacked profile of the 45$^\circ$ rotated
 component, $\langle\Delta \Sigma_\times(r)\rangle$. 
\label{fig:stack}}
\end{figure}

It should be noted that the reduced shear $g_+$ has a non-linear 
dependence on the mass profile. In fact, the reduced shear is defined
by  $g_+\equiv\gamma_+/(1-\kappa)$, where $\gamma_+$ and $\kappa$ are
tangential shear and convergence. Thus, the quantity defined by
equation~(\ref{eq:dsig}) still depends slightly on the source
redshift via the factor $1/(1-\kappa)$, particularly near the halo
centre. Thus, in comparison with the NFW predictions, we assume the
source redshift of $z_s=1.1$, which is the typical effective source
redshift for our weak lensing analysis (see
Table~\ref{tab:wlens}). Also the non-linear dependence makes it
somewhat difficult to interpret the average profile, and hence our
stacked tangential profile measurement near the centre should be taken
with caution.  

It is known that the matter fluctuations along the line-of-sight
contributes to the total error budget
\citep[e.g.,][]{hoekstra03,hoekstra11,dodelson04,gruen11}. 
While we have ignored this effect for the analysis of individual
clusters presented in Section~\ref{sec:swlens}, here we take into account
the error from the large scale structure in fitting the stacked
tangential shear profile by including the full covariance between
different radial bins \citep[see][for the calculation of the covariance
  matrix]{oguri11a,umetsu11b}. We, however, comment that the error of
the large scale structure is subdominant in our analysis, because of
the relatively small number density of background galaxies after the
colour cut \citep[see also][]{oguri10a}.

In addition to weak lensing, we stack strong lensing constraints
simply by averaging the Einstein radii for the fixed source redshift
$z_s=2$. This constraint is combined with the stacked tangential shear
profile from weak lensing to obtain constraints on the mass and
concentration parameter for the stacked profile. Note that the
Einstein radius is related with the reduced shear as $g_+(\theta_{\rm
  E})=1$. Given the uncertainty
from the non-linearity of the reduced shear and the the possible bias
coming from the uncertainty of the outer mass profile
\citep{oguri11b,becker11}, we restrict tangential shear fitting in the
range $0.158h^{-1}{\rm Mpc}<r<3.16h^{-1}{\rm Mpc}$. However we note
that our results are not largely changed even if we conduct fitting
in the whole radius range. 

\begin{figure}
\begin{center}
 \includegraphics[width=0.96\hsize]{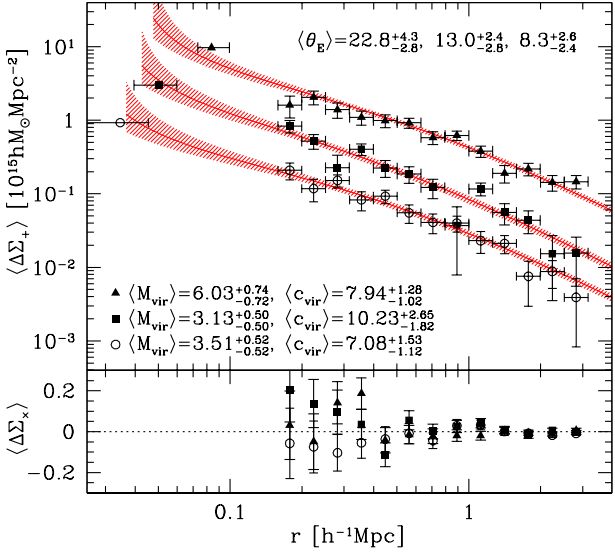}
\end{center}
\caption{Similar to Figure~\ref{fig:stack}, but the stacked lensing
  analysis in three $\theta_{\rm E}$ bins is presented. From top to
  bottom, results for largest to smallest $\theta_{\rm E}$ bins are
  shown. Curves and points for the largest and smallest $\theta_{\rm
    E}$ bins are shifted vertically by $\pm0.5$~dex respectively for
   illustrative purposes. 
\label{fig:stack_ein}}
\end{figure}

\begin{figure}
\begin{center}
 \includegraphics[width=0.96\hsize]{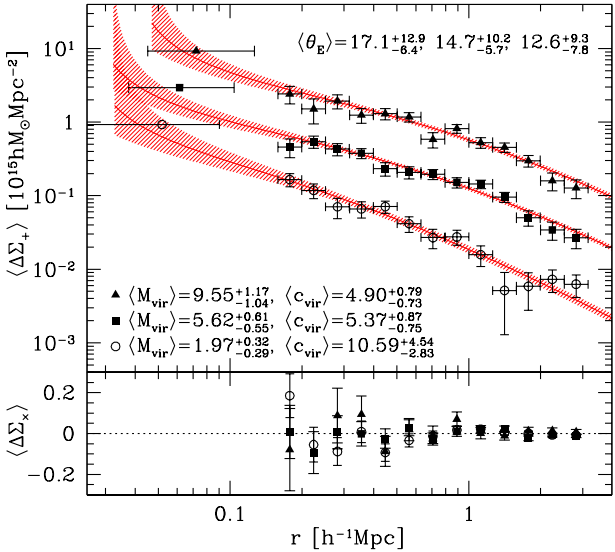}
\end{center}
\caption{Same as Figure~\ref{fig:stack_ein}, but clusters are binned
  in $M_{\rm vir}$. 
\label{fig:stack_mass}}
\end{figure}

\begin{table}
 \caption{Summary of stacked tangential shear analysis
   \label{tab:stack1d}}   
 \begin{tabular}{@{}cccccc}
 \hline
   Sample
   & $N$
   & $\langle z\rangle$
   & $\langle\theta_{\rm E}\rangle$
   & $\langle M_{\rm vir}\rangle$
   & $\langle c_{\rm vir}\rangle$\\
   &
   &
   & (arcsec)
   & ($10^{14}h^{-1}M_\odot$)
   & \\
 \hline
 all & 25 & 0.469 & $14.4^{+10.6}_{-7.0}$ & $4.57^{+0.33}_{-0.31}$ & $5.75^{+0.70}_{-0.57}$\\
 \hline
 $\theta_{\rm E}$-1 & 4 & 0.379 & $22.8^{+4.3}_{-2.8}$ & $6.03^{+0.74}_{-0.72}$ & $7.94^{+1.28}_{-1.02}$\\
 $\theta_{\rm E}$-2 & 5 & 0.416 & $13.0^{+2.4}_{-2.8}$ & $3.13^{+0.50}_{-0.50}$ & $10.23^{+2.65}_{-1.82}$\\
 $\theta_{\rm E}$-3 & 7 & 0.471 & $8.3^{+2.6}_{-2.4}$  & $3.51^{+0.52}_{-0.52}$ & $7.08^{+1.53}_{-1.12}$\\
 \hline
 $M_{\rm vir}$-1 & 5  & 0.480 & $17.1^{+12.9}_{-6.4}$ & $9.55^{+1.17}_{-1.04}$ & $4.90^{+0.79}_{-0.73}$ \\
 $M_{\rm vir}$-2 & 10 & 0.472 & $14.7^{+10.2}_{-5.7}$& $5.62^{+0.61}_{-0.55}$ & $5.37^{+0.87}_{-0.75}$ \\
$M_{\rm vir}$-3  & 10 & 0.460  & $12.6^{+9.3}_{-7.8}$ & $1.97^{+0.32}_{-0.29}$ & $10.59^{+4.54}_{-2.83}$ \\
 \hline
 \end{tabular}
\end{table}

\begin{figure}
\begin{center}
 \includegraphics[width=0.96\hsize]{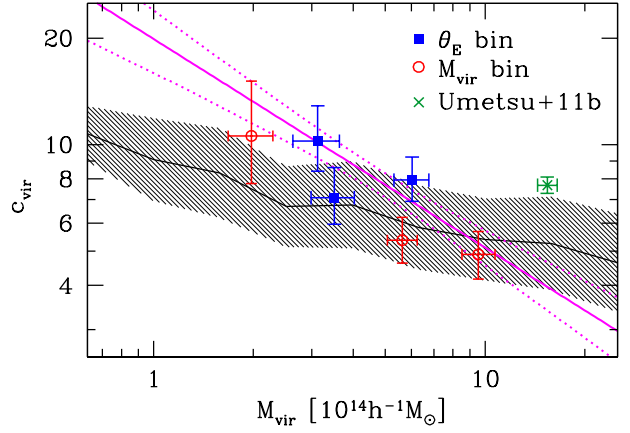}
\end{center}
\caption{The mass-concentration relation obtained from the stacked
 lensing analysis. We show stacking results of 3 $\theta_{\rm E}$ bins
 ({\it filled squares}) and 3 $M_{\rm vir}$ bins ({\it open circles}).
 The mass and concentration measured from stacked strong and weak
 lensing analysis of 4 massive clusters at $z\sim 0.32$
 \citep{umetsu11b} are indicated by a cross. The black shading
 region shows theoretically expected mass-concentration relations with
 the lensing bias (see Appendix~\ref{sec:conc}  for details). The
 solid and dotted lines are bets-fit relation from individual analysis
 shown in Figure~\ref{fig:clupar}. 
 \label{fig:cluspar_stack}} 
\end{figure}

\begin{figure}
\begin{center}
 \includegraphics[width=0.8\hsize]{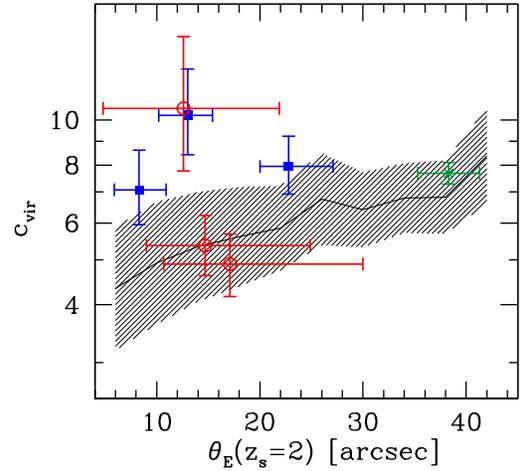}
\end{center}
\caption{Concentration parameters from stacking lensing analysis as a
  function of the Einstein radius for the source redshift
  $z_s=2$. The black shading region indicates the theoretical
  expectation with the selection effect (see Appendix~\ref{sec:conc}
  for details). Symbols are same as Figure~\ref{fig:cluspar_stack}.
 \label{fig:cein}} 
\end{figure}

Figure~\ref{fig:stack} shows the stacking result for all the 25
clusters. The mean cluster redshift for this sample is $\langle
z\rangle=0.469$.  The total signal-to-noise ratio in the whole radius
range of $0.063h^{-1}{\rm Mpc}<r<5.01h^{-1}{\rm Mpc}$ is $S/N=32$.
We find that stacked tangential shear profile from
weak lensing is fitted well by the NFW profile over a wide range in
radius. The average Einstein radius from strong lensing
($\langle\theta_{\rm E}\rangle=14\farcs4^{+10.6}_{-7.0}$) is slightly
larger than the best-fit model predicts ($\theta_{\rm E}=9\farcs1$),
although they are consistent with each other well within $1\sigma$.  
The best-fit mass and concentration are $\langle M_{\rm
  vir}\rangle=4.57^{+0.33}_{-0.31}\times10^{14}h^{-1}M_\odot$ and 
$\langle c_{\rm vir}\rangle=5.75^{+0.70}_{-0.57}$. We note that the 
mean mass measured by the stacking analysis agrees well with the 
mean mass of strong lens selected clusters predicted by ray-tracing
simulations, $\langle M_{\rm vir}\rangle\sim 4.2\times 10^{14}h^{-1}M_\odot$ 
\citep[][see also \citealt{bayliss11b}]{hennawi07}.

The concentration parameter measured in the stacked tangential shear
profile is broadly consistent with the result of individual analysis 
(see Figure~\ref{fig:clupar}), but appears to be slightly smaller 
than the mass-concentration relation constrained from our lensing
sample. Here we estimate the impact of the possible averaging effect
by computing an average shear profile from reduced shear profiles
of the NFW profile with best-fit values of the mass and concentration
from individual cluster analysis (Table~\ref{tab:fit}), and comparing 
it with the reduced shear profile of the NFW profile with the median
values of the mass and concentration. We find that both profiles agree
well near the virial radius, but the averaged profile underestimates
the shear profile toward the centre, with $\sim 10\%$ systematic
difference at around $0.2h^{-1}$Mpc. Both the profiles agree well
again at the innermost radii of $<0.1h^{-1}$Mpc down to the strong
lensing region. The systematic difference translates into the
concentration parameter of $\Delta c_{\rm vir} \sim -1.2$, and hence
it can partly explain the smaller concentration parameter value from
the stacked lensing analysis. 

Another possible reason for the smaller concentration from the stacked
tangential shear is a wide range of
$\theta_{\rm E}$ of our sample, which results in the large error on
the mean Einstein radius and therefore in the much weaker constraints
from strong lensing compared with individual modelling cases. Hence, we
conduct the same stacking analysis by dividing our cluster sample into 3
$\theta_{\rm E}$ bins. In order to assure reasonable constraints from
strong lensing, we remove 9 clusters which have large errors on
$\theta_{\rm E}$ mostly because of the lack of arc redshift
information. To test the mass dependence of the concentration, in
addition to $\theta_{\rm E}$ bins we consider 3 $M_{\rm vir}$ bins
too. We use all the 25 clusters for the mass bin analysis.  

Results of our stacking analysis in different $\theta_{\rm E}$ and
$M_{\rm vir}$ bins are shown in Figures~\ref{fig:stack_ein} and
\ref{fig:stack_mass}, respectively, and are summarized in
Table~\ref{tab:stack1d}. We find that clusters in the largest
$\theta_{\rm E}$  bin are indeed most massive. However, the second and
third $\theta_{\rm E}$ bins have similar mean virial masses, and the
difference of the Einstein radii appear to be derived by the different
concentrations. On the other hand, different mass bins have similar
Einstein radii, but the concentrations are clearly larger for smaller
masses.  

These results can be used to check the mass-concentration relation
inferred from individual analysis of clusters. 
Figure~\ref{fig:cluspar_stack} shows the mass-concentration relation
similar to Figure~\ref{fig:clupar}, but this time the relation
obtained from stacked lensing analysis. We find that the
mass-concentration relation from stacking analysis is in reasonable
agreement with the best-fit relation constrained from individual
analysis of strong and weak lensing (equation~\ref{eq:mcfit}). 
In particular, the strong mass dependence of the concentration is
clearly seen in the stacking analysis as well. The slightly smaller
normalization compared with individual analysis can partly be ascribed
to the averaging effect as described above. Thus the stacking analysis
further confirms the measurement of the mass-concentration relation
from our sample of clusters. 

In Figure~\ref{fig:cein} we study the dependence of concentration
parameters derived from the stacking analysis with the Einstein
radius. In particular we compare it with semi-analytic calculation
conducted in Appendix~\ref{sec:conc} which predicts that the clusters
with larger Einstein radii are more concentrated. While it is hard to
see this trend in our cluster sample, we find that the high
concentration of massive lensing clusters presented by
\citet{umetsu11b} can be explained in this context. Our result
suggests that the average concentration of the \citet{umetsu11b}
cluster sample is in good agreement with the theoretical expectation
given the very large Einstein radii of $\theta_{\rm E}\sim 40''$.

\begin{figure*}
\begin{center}
 \includegraphics[width=0.47\hsize]{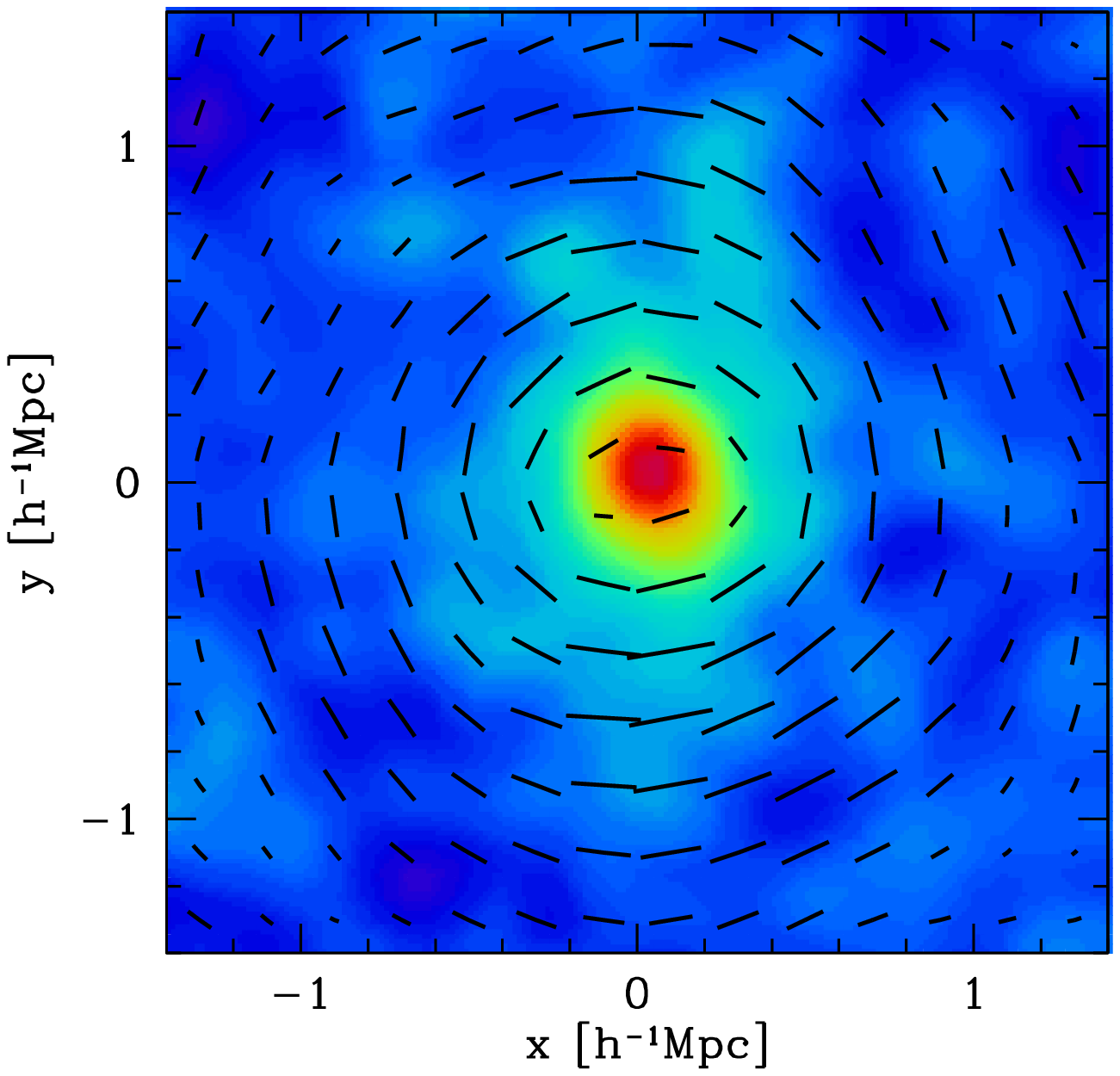}
 \includegraphics[width=0.47\hsize]{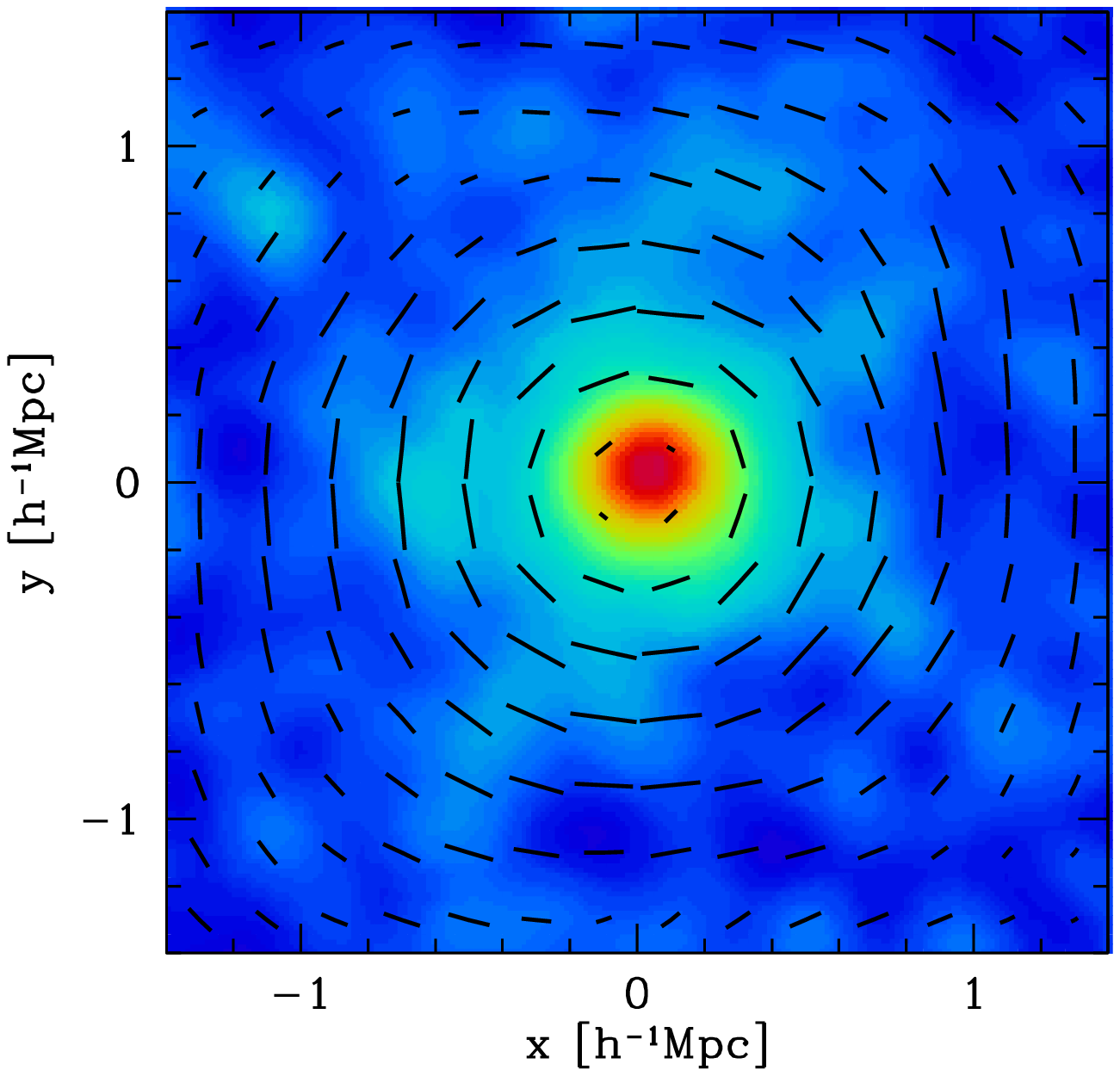}
\end{center}
\caption{The two-dimensional weak lensing shear maps obtained from
  stacking analysis of 25 clusters. The sticks shows observed
  directions and strengths of weak lensing shear distortion. Colour
  contours are the surface density map reconstructed from the shear
  map using the standard inversion technique \citep{kaiser93}. Both
  the shear and density maps are smoothed for illustrative purpose. 
  {\it Left:} The result when the position angle of each cluster is
  aligned to the North-South axis before stacking, by using the
  position angle measured in strong lens modelling. The resulting
  stacked density distribution is clearly elongated along the
  North-South direction. {\it Right:} The result without any alignment
  of the position angle when stacking. The resulting density
  distribution is nearly circular symmetric in this case.
\label{fig:smap}}
\end{figure*}

\subsection{Two-dimensional stacking analysis}
\label{sec:stack2d}

In addition to stacking of the tangential shear profile, we conduct
stacking of two-dimensional (2D) shear maps to study the mean shape of
the projected dark matter distribution in clusters. Such 2D stacking
analysis has been attempted for samples of galaxies
\citep{natarajan00,hoekstra04,mandelbaum06a,parker07} or for clusters
\citep{evans09}. The biggest problem of these 2D stacking analysis has
been that the position angle (orientation) of the projected mass
distribution has to be known for each cluster when stacking. 
In previous work it was assumed that the position angle of the mass
distribution coincides with that of the light distribution, e.g., the
surface brightness distribution of the central or satellite
galaxy distributions, although the assumption has not yet been
fully justified \citep{oguri10a,bett11}. 

Our unique sample of strong and weak lensing clusters provides an
important means of overcoming this difficulty. The idea is that strong
lens modelling can generally constrain the position angle of the dark
halo component quite well, which can be used as a prior information for
the position angle to stack weak lensing signals. This procedure
evades any assumptions on the alignment between mass and light, and
hence should enable much more robust 2D stacking analysis.

As in Section~\ref{sec:stackgt}, we conduct stacking analysis in the
physical length scale. For each cluster, we adopt the position angle
obtained in strong lens mass modelling ($\theta_e$ in
Table~\ref{tab:slens}) to rotate the catalogue of the background galaxies
by $-\theta_e$ such that the the position angle of the dark halo is
aligned with the North-South axis. Specifically, the position of a
galaxy at ($x$, $y$) with respect to the cluster centre is changed to
\begin{eqnarray}
x'&=&x\cos\theta_e+y\sin\theta_e,\\
y'&=&-x\sin\theta_e+y\cos\theta_e,
\end{eqnarray}
and the two shear components ($g_1$, $g_2$) are modified as
\begin{eqnarray}
g_1'&=&g_1\cos2\theta_e+g_2\sin2\theta_e,\\
g_2'&=&-g_1\sin2\theta_e+g_2\cos2\theta_e.
\end{eqnarray}
We stack the rotated shear catalogue in the physical unit, $\Sigma_{\rm
  cr}\bm{g}'(\bm{r}')$, to obtain the average 2D shear map of our
cluster catalogue. The cluster catalogues analyzed in this section is same
as those in Section~\ref{sec:stackgt}, containing 25 clusters in total.

The stacked 2D shear map, as well as the corresponding density map
reconstructed from the shear map, are shown in Figure~\ref{fig:smap}.
As expected, the projected mass distribution from the stacked 2D shear
map is quite elongated along the North-South direction, suggesting the
highly elongated mass distribution of our cluster sample. As a sanity
check, we also compute the 2D shear map without any alignment of the
position angle when stacking. The resulting mass distribution shown in
Figure~\ref{fig:smap} appears to be circular symmetric, which supports
that the highly elongated distribution in our stacked map is not an
artifact.

We constrain the ellipticity of the projected 2D mass distribution 
by  directly fitting the 2D shear map with the elliptical NFW model
prediction. Here we closely follow the procedure detailed in
\citet{oguri10a} for the 2D shear fitting. Briefly, we modify the
convergence $\kappa(r)$ (i.e., the projected surface mass density) of
the spherical NFW profile simply by introducing the ellipticity in the
iso-density contour as $r^2\rightarrow x^2/(1-e)+y^2(1-e)$. With this
procedure our definition of the 
ellipticity is $e=1-b/a$, where $a$ and $b$ are major and minor axis
lengths of the isodensity contour. The corresponding shear pattern is
computed by solving the Poisson equation. We then construct pixelized
distortion field by computing mean shear and errors in each bin, and 
compare it with the elliptical model prediction, adopting the pixel
size of $0.1h^{-1}{\rm Mpc}$. We add the contribution of the large
scale structure to the error covariance matrix \citep[see][]{oguri10a}.
We perform fitting in a $6h^{-1}{\rm Mpc}\times 6h^{-1}{\rm Mpc}$
region, but remove the innermost $4\times4$ pixels considering several
possible systematics that might be affecting signals near the
centre. Unlike \citet{oguri10a}, we fix the mass centre to the assumed
centre (the position of the brightest galaxy in strong lensing
region), because strong lensing available for our cluster sample
allows a reliable identification of the mass centre for each
cluster. Thus we fit the 2D shear map with four parameters 
($M_{\rm vir}$, $c_{\rm vir}$, $e$, $\theta_e$), employing a Markov
Chain Monte Carlo technique. 

\begin{figure}
\begin{center}
 \includegraphics[width=0.9\hsize]{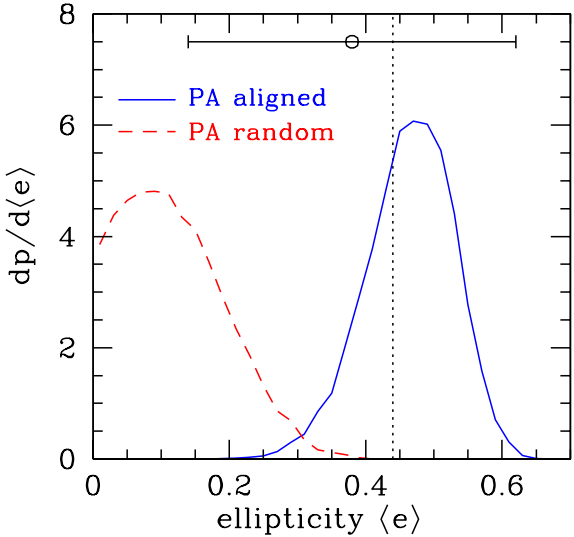}
\end{center}
\caption{The marginalized probability distribution of the mean
  ellipticity $\langle e \rangle$ from stacked weak lensing analysis
  of 25 clusters. The solid line indicates the case when the position
  angles are aligned 
  according to those measured with strong lens modelling, in which the
  mean ellipticity is detected at $5\sigma$ level ($\langle e 
  \rangle=0.47\pm0.06$). The dashed line is the marginalized probability
  distribution for stacking without any alignment of the position
  angles, for which the mass distribution is consistent with the
  circular symmetric distribution ($\langle e \rangle<0.19$). The
  vertical dotted line indicates the theoretical expectation, $\langle e  
  \rangle=0.44$, based on a triaxial halo model of \citet{jing02}. The
  open circle with errorbar shows the average ellipticity and
  $1\sigma$ scatter from strong lens modelling.
\label{fig:ell}}
\end{figure}

In Figure~\ref{fig:ell}, we show the posterior likelihood
distribution of the mean ellipticity $\langle e \rangle$ from the 2D
stacking analysis of all the 25 clusters. When the position angles are
aligned, the resulting density distribution is indeed quite elliptical
with the mean ellipticity of $\langle e \rangle=0.47\pm0.06$.
We find that the elliptical NFW model improve fitting by
$\Delta\chi^2=26.9$ compared with the case $e=0$, which indicates that
the detection of the elliptical mass distribution is significant at the
$5\sigma$ level. The measured mean ellipticity is consistent with the
average ellipticity from strong lens modelling $\langle e 
\rangle=0.38\pm0.24$, although the latter involves large scatter.
The best-fit position angle of
$\theta_e=9.1^{+3.9}_{-4.1}$~deg slightly deviates from the expected
position angle of $\theta_e=0$, but they are consistent with each
other within $2\sigma$ ($\Delta\chi^2<4$). In contrast, if the
position angles are not aligned in stacking shear signals, the
resulting constraint on the mean ellipticity is 
$\langle e \rangle<0.19$, i.e., it is fully consistent with the
circular symmetric mass distribution $e=0$ within $1\sigma$. 

\begin{figure}
  \begin{center}
    \includegraphics[width=0.96\hsize]{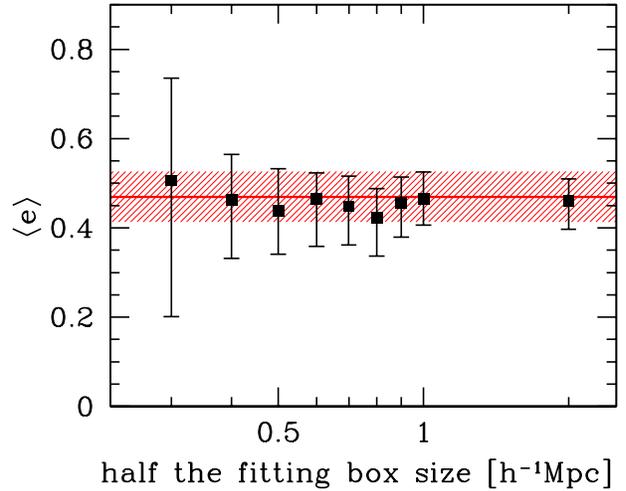}
  \end{center}
  \caption{The mean ellipticities measured in stacked 2D shear map as a
    function of the box size for fitting. Our fiducial result and its
    $1\sigma$ range adopting half the box size of $3h^{-1}{\rm Mpc}$ are
    indicated by the horizontal line with shading. 
    \label{fig:ell_box}}
\end{figure}

We compare this result with the theoretical prediction in the
$\Lambda$CDM model. For this purpose we employ a triaxial model of
\citet{jing02}. Assuming that the halo orientation is random, we
compute the probability distribution of the ellipticity by projecting
the triaxial halo along arbitrary directions \citep{oguri03,oguri04}. 
In this analysis we fix the mass and redshift of the halo to $M_{\rm
  vir}=4.6\times 10^{14}h^{-1}M_\odot$ and $z=0.469$, which are mean
mass and redshift of the 25 clusters. We find that the mean
ellipticity predicted by this model is $\langle e \rangle=0.44$, in
excellent agreement with the measured ellipticity. The analysis
presented in Appendix~\ref{sec:conc} indicates that the effect of the
lensing bias on the mean ellipticity is small, with a possible shift
of the mean ellipticity of $\la 0.05$ at most, and hence it does not
affect our conclusion. Our result is also in good agreement with the
previous lensing measurement of the ellipticity by \citet{oguri10a} in
which 2D shear maps of individual clusters are fitted with the
elliptical NFW profile, rather than examining the stacked shear map.

\begin{figure}
\begin{center}
 \includegraphics[width=0.96\hsize]{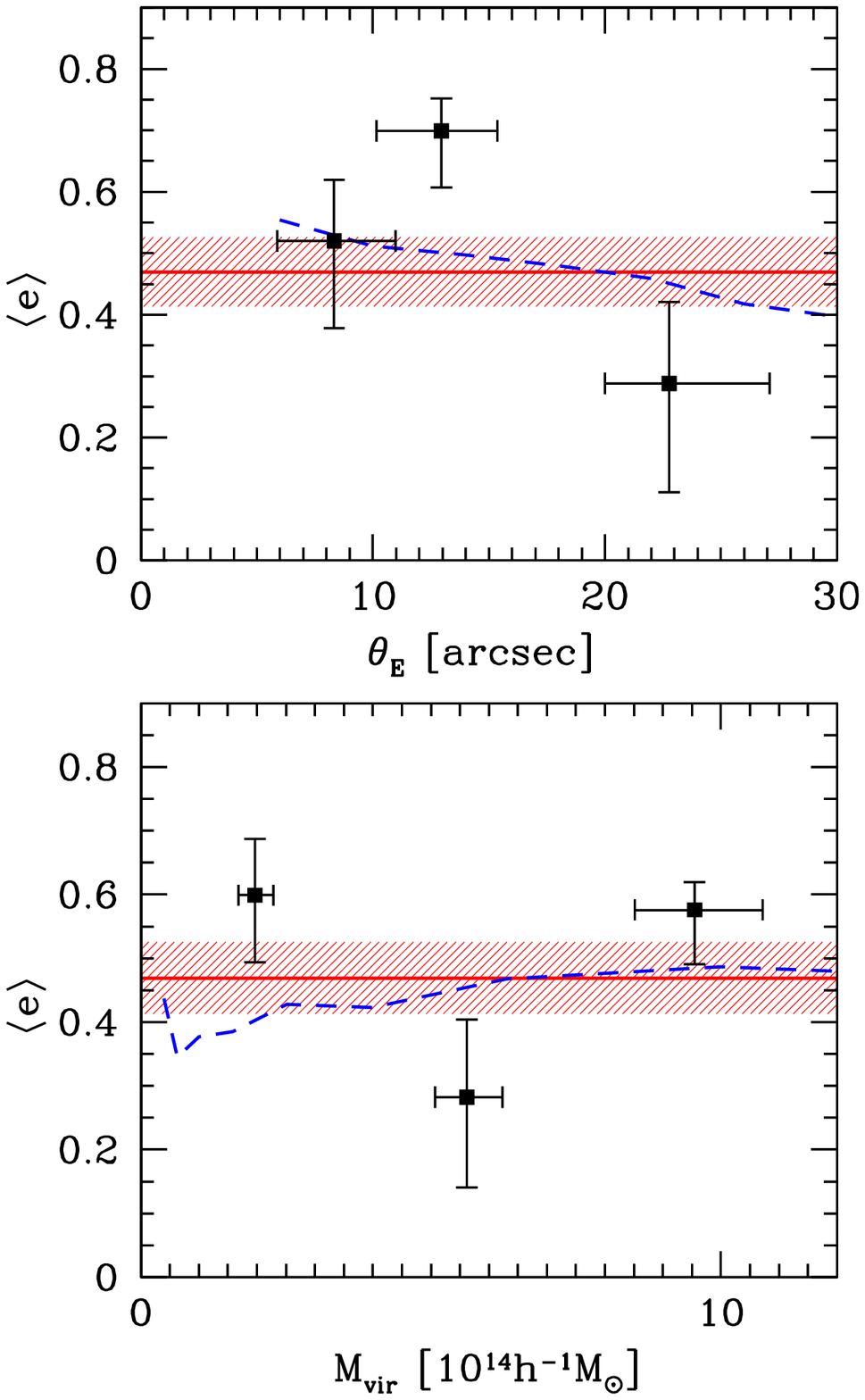}
\end{center}
\caption{Mean ellipticities from stacked shear maps in three different
  $\theta_{\rm E}$ ({\it upper panel}) and $M_{\rm vir}$ ({\it lower
    panel}) bins. The horizontal line with shading indicate best-fit
  and 1$\sigma$ range of the full sample result. Dashed lines show the
  predicted mean ellipticity from semi-analytic calculations with the
  lensing bias (see Appendix~\ref{sec:conc} for details).
  \label{fig:fit2d}} 
\end{figure}

\begin{table}
 \caption{Summary of the two-dimensional stacking analysis
   \label{tab:stack2d}}   
 \begin{tabular}{@{}ccc}
 \hline
   Sample
   & $\langle e\rangle$
   & $\langle \theta_e\rangle$\\
   &
   & (deg)\\
 \hline
 all   & $0.47^{+0.06}_{-0.06}$ & $9.1^{+3.9}_{-4.1}$ \\
 \hline
 $\theta_{\rm E}$-1 & $0.29^{+0.13}_{-0.18}$ & $14.1^{+13.9}_{-18.8}$ \\
 $\theta_{\rm E}$-2 & $0.70^{+0.05}_{-0.09}$ & $13.0^{+4.4}_{-4.3}$ \\
 $\theta_{\rm E}$-3 & $0.52^{+0.10}_{-0.14}$ & $6.7^{+12.2}_{-9.2}$ \\
 \hline
 $M_{\rm vir}$-1 & $0.58^{+0.04}_{-0.09}$ & $5.2^{+4.4}_{-4.5}$ \\
 $M_{\rm vir}$-2 & $0.28^{+0.12}_{-0.14}$ & $9.7^{+11.3}_{-17.6}$ \\
 $M_{\rm vir}$-3  & $0.60^{+0.09}_{-0.11}$ &  $16.7^{+7.0}_{-8.7}$ \\
 \hline
 \end{tabular}
\end{table}

We check the sensitivity of our ellipticity result on the size of the
fitting region, as one possible concern is that infalling matter
associated with the filamentary structure outside clusters might boost
the mean ellipticity. Figure~\ref{fig:ell_box} shows how the
constraint on the mean ellipticity changes by making the size of
the fitting region smaller from our fiducial choice (half the box size
of $3h^{-1}{\rm Mpc}$). The Figure indicates that the detection of the
mean ellipticity of $\sim 0.45$ is robust against the choice of the
fitting size, as the results are consistent down to half the box size
of $\sim 0.3h^{-1}{\rm Mpc}$ where the constraint become significantly
weak. The analysis also implies that the ellipticity of the projected
mass distribution does not change very much with radius. Theoretically,
dark haloes are expected to be more elongated near the centre
\citep{jing02}, although the effect of 
baryon cooling and star formation can make the shape rounder
particularly near the centre \citep{kazantzidis04,lau11}. Our
detection of the elliptical mass distribution down to small radii 
may therefore help constraining the amount of cooling in clusters. 

Finally we check the mean ellipticities for different $\theta_{\rm E}$
and $M_{\rm vir}$ bins. Figure~\ref{fig:fit2d} shows the
results, and Table~\ref{tab:stack2d} gives the summary of constraints
on the mean ellipticities. We find a possible decrease of $\langle e
\rangle$ at the largest Einstein radius bin, although it is not very
significant given the large errorbars. On the other hand, there
appears no simple trend in the mass bin result. 

While results for all subsamples are almost consistent with the full
sample result, the dependence on $\theta_{\rm E}$ might be suggestive
of conflicting selection effects on our sample. There are two main
lensing biases that can affect the results on the mean ellipticity
measurement. One is the orientation bias, i.e., clusters with larger
Einstein radii appear to be rounder than normal clusters because of
the alignment of the major axis with the line-of-sight direction
\citep{oguri09a}. The other selection effect is directly related to
the ellipticity of the projected mass distribution. Because the
ellipticity significantly enhances the giant arc cross section
\citep[e.g.,][]{meneghetti03,meneghetti07}, a sample of clusters with
prominent lensed arcs should have more elliptical projected mass
distribution. These two selection effects apparently conflict with
each other, although a naive expectation is that the orientation bias
dominates for large Einstein radii, whereas the ellipticity bias is
more significant for small Einstein radii. Indeed, our semi-analytic
calculations presented in Appendix~\ref{sec:conc} shows a clear
dependence of the mean ellipticity on the Einstein radius, as plotted
in Figure~\ref{fig:fit2d}. The observed trend appears to be
consistent, at least qualitatively, with the theoretical prediction. 
A large sample is needed to confirm this trend more robustly.

\section{Relation between mass and member galaxy distributions}
\label{sec:member}

Understanding the relation between the mass and cluster member
galaxy distributions is important for the cosmological use of optical
clusters \citep[e.g.,][]{koester07,rozo09,rozo11,rykoff11} as well as
detailed stacking analysis for a large sample of clusters
\citep[e.g.,][]{evans09}.  The spatial distribution of galaxies, in
comparison with that of subhaloes in an $N$-body simulation, is crucial
for exploring the formation history of cluster member galaxies
\citep[e.g.,][]{nagai05}. 
Here we examine spatial distributions of cluster member galaxies
using the stacking technique, and compare them with accurate
measurements of mass distributions from stacked lensing analysis. 

While the colour information from $gri$-band images allows us to select
the galaxy population around a given redshift efficiently, the
selection is not perfect in the sense that there is some
contamination from foreground and background galaxies. Thus, in this
paper we take advantage of the high number density of galaxies after
stacking to subtract the foreground and background galaxies
statistically. Specifically, we assume that the stacked number density
distribution is described by the sum of cluster member galaxies and
background galaxies as:
\begin{equation}
\Sigma_g(\bm{r})=a_g\Sigma_{\rm nfw}(\bm{r})+b_g,
\label{eq:dgal}
\end{equation}
where $a_g$ is the normalization, $\Sigma_{\rm nfw}$ is the surface
density distribution of the NFW profile, and $b_g$ is the number
density of foreground/background galaxies which is assumed to be
constant across the field.

In this paper, we focus on two parameters that describe the number
density distribution of member galaxies. One is the scale radius $r_s$
of the NFW profile, which is used to check whether or not the galaxy
distribution is more extended than the mass distribution. The other
is the ellipticity of the projected number density distribution, as our
stacking analysis enables detailed analysis of 2D distributions for
both mass and member galaxy distributions. We constrain the scale
radius $r_s$ by fitting azimuthally averaged radial profile of the
galaxy number density, whereas the ellipticity is derived from the 2D
fitting of the density distribution. The total number of parameters
for the radial profile fitting is 3 ($r_s$, $a_g$, and $b_g$), and the
that of 2D fitting is 5 ($e$, $\theta_e$, $r_s$, $a_g$, and $b_g$).

We consider two distinct galaxy populations, one is red member
galaxies selected by the colour cut (equations~\ref{eq:gricut_red1} and 
\ref{eq:gricut_red2}), and the other is all galaxies without any colour
cut. For each galaxy population, we consider galaxies that are more
luminous than the luminosity cut, i.e., $L>L_{\rm cut}$. The
luminosity cut is defined in terms of $L_*$, where $L_*$ is the
luminosity corresponding to the $i$-band absolute magnitude
$M_*^i=-21.22+5\log h$ \citep{rykoff11}. For the highest cluster
redshift in our sample, a luminosity $0.1L_*$ corresponds to $i\sim
24.5$, which is well above the magnitude limit of our Subaru imaging.

\begin{figure}
\begin{center}
 \includegraphics[width=0.96\hsize]{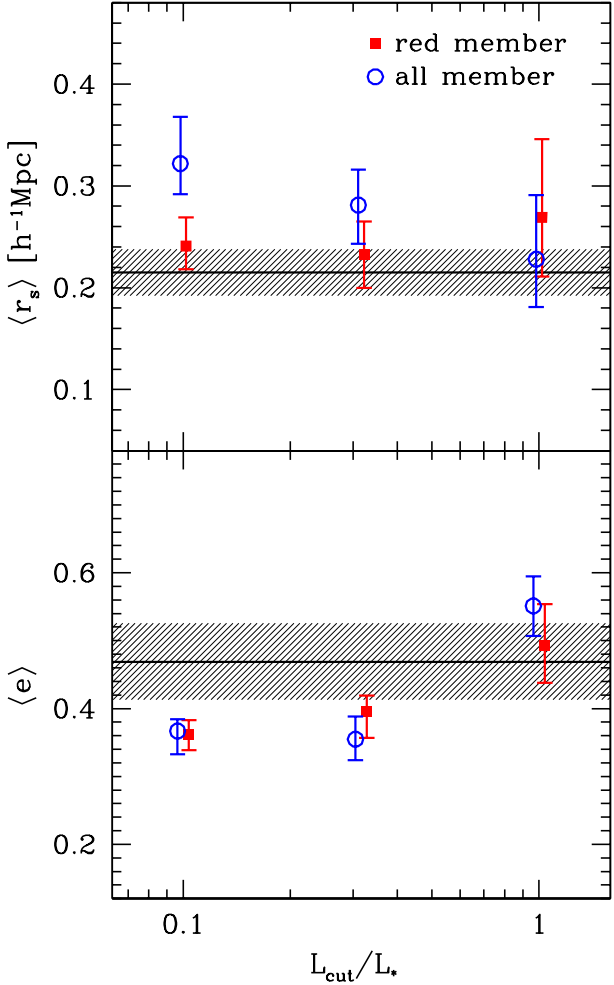}
\end{center}
\caption{The mean scale radius $\langle r_s\rangle$ ({\it upper}) and
  mean ellipticity $\langle e\rangle$ ({\it lower}) of the member
  galaxy distribution from stacking 25 clusters, as a function of the
  luminosity cut $L_{\rm cut}$. Filled squares and open circles show
  the results for red member galaxies and all member galaxies,
  respectively. Horizontal lines with shading are the scale radius and
  ellipticity for the mass distribution, measured by stacked lensing,
  for the same cluster sample. 
  \label{fig:fitmem}} 
\end{figure}

\begin{table}
 \caption{analysis results on the member galaxy distribution
   \label{tab:mem}}   
 \begin{tabular}{@{}cccccc}
 \hline
   $L_{\rm lim}$
   &  $\langle r_s\rangle$ (red)
   &  $\langle e\rangle$ (red)
   &  $\langle r_s\rangle$ (all)
   &  $\langle e\rangle$ (all)\\
   ($L_*$)
   & ($h^{-1}{\rm Mpc}$)
   & 
   & ($h^{-1}{\rm Mpc}$)
   & \\
 \hline
  1 &  $0.27^{+0.08}_{-0.06}$ & $0.49^{+0.06}_{-0.05}$ &  $0.23^{+0.06}_{-0.05}$ & $0.55^{+0.04}_{-0.04}$\\
  $10^{-0.5}$&  $0.23^{+0.03}_{-0.03}$ & $0.40^{+0.02}_{-0.04}$&  $0.28^{+0.04}_{-0.04}$ & $0.36^{+0.03}_{-0.03}$ \\
  $10^{-1}$&  $0.24^{+0.03}_{-0.02}$ & $0.36^{+0.02}_{-0.02}$&  $0.32^{+0.05}_{-0.03}$ & $0.37^{+0.02}_{-0.03}$ \\
 \hline
 \end{tabular}
\end{table}

We show our fitting results in Figure~\ref{fig:fitmem} and
Table~\ref{tab:mem}. We find that the mass and galaxy number
distributions agree reasonably well for the luminous galaxies, $L\ga
L_*$, for both the scale radius and ellipticity. However, for fainter
galaxies of $L\sim 0.1L_*$ the scale radius of the galaxy distribution
tends to be larger (i.e., smaller concentration parameter values), and 
the average ellipticity is smaller. 

The radial distribution has also been studied in previous work.
Measurements of the projected number density distributions suggest
low concentration values of $c\sim 2-5$ 
\citep{lin04,katgert04,hansen05,lin07,biviano09}.
\citet{lin07} found that more luminous galaxies tend to be distributed
with large concentrations, which is consistent with our results
in Figure~\ref{fig:fitmem}. Simulation work by \citet{nagai05} also
predicts that the galaxy number density distribution is slightly more
extended than the mass distribution, which is again consistent with
our result, although no significant dependence of the concentration on
the stellar mass was reported from the simulations. 

There have been measurements of the ellipticity of the projected
galaxy number density distributions
\citep[e.g.,][]{paz06,niederste10}, but its connection with the
underlying mass distribution has not yet fully been explored. 
The similar ellipticity for luminous member galaxies is reasonable,
but the origin of the smaller mean ellipticity for fainter galaxies
is unclear. Essentially, the smaller mean ellipticity for the stacked
cluster sample indicates either the galaxy distribution is on average
rounder or there is a large misalignment (i.e., difference in the
position angles) between the mass and galaxy distribution. In either
case careful simulation work is crucial to check if the observed
distribution is in agreement with current understanding of cluster
galaxy formation. 

\section{Conclusion}
\label{sec:conclusion}

We have performed a combined strong and weak lensing analysis for a
sample of 28 clusters in the redshift range $0.28<z<0.68$. The
cluster sample is based on the SGAS, a large survey of giant arcs
amongst SDSS clusters. In this paper, we have presented extensive
follow-up observations with Subaru/Suprime-cam, which enables reliable
weak lensing measurements out to large radii. Combined with the giant
arcs with significant amount of spectroscopic information, we can
constrain mass distributions of these clusters quite well to measure
the concentration and shape of the mass distribution. 

The mass-concentration relation derived from our lensing analysis has
the slope of $c_{\rm vir}\propto M_{\rm vir}^{-0.59\pm0.12}$, which is
significantly steeper than the slope predicted by the theoretical
expectation of $-0.2$ which includes the effect of the lensing
bias. Concentrations measured by the combined lensing analysis are in
reasonable agreement with the $\Lambda$CDM prediction for massive
clusters ($M_{\rm vir}\sim 10^{15}h^{-1}M_\odot$), if we take proper
account of the lensing bias. The result indicates that the anomalously
high concentration 
\citep{broadhurst05,broadhurst08b,oguri09b,umetsu11b,gralla11}, which
has been claimed from analysis of small number of lensing clusters, is
now much less evident, thanks to the much larger number of clusters we
have  analyzed in the paper. On the other hand, observed concentrations
appear to be significantly higher than theoretical expectations
for lower masses of $M_{\rm vir}\sim 10^{14}h^{-1}M_\odot$, which may 
be explained as arising from the effects of cooling baryons in cluster
centres. 

We have also stacked tangential shear profiles for the sample of
clusters to obtain an accurate mean profile, which is seen to be in good
agreement with the NFW profile. Our stacking analysis at different
Einstein radii and mass bins has confirmed our results on the
mass-concentration relation from the individual analysis of these clusters. 
In addition to the radial profile, we have explored the stacked 2D shear
map to study the shape of the projected mass distribution. By aligning
shear maps of individual clusters with the position angles of dark haloes
from strong lens modelling, we were able to detect the 
elliptical shape of the stacked mass distribution at the $5\sigma$ level.
The mean ellipticity of $\langle e \rangle=0.47\pm0.06$ is in
excellent agreement with the $\Lambda$CDM expectation. The significant
detection of the highly elliptical shape of dark matter haloes with
weak lensing has also been reported by \citet{oguri10a}; this work
confirms the previous finding using a different sample and technique. 
Finally, based on stacking analysis we have compared distributions of
cluster member galaxies with mass distributions, finding good
agreement between them for luminous ($L\ga L_*$) member
galaxies. Distributions of fainter galaxies are found to be more
extended.  

Our work has demonstrated the power of combined strong and weak
lensing, not only for analysis of individual clusters but also for
detailed stacked lensing analysis. The techniques described in the
paper can be applied to the unique sample of lensing clusters obtained
by the Cluster Lensing And Supernova survey with Hubble
\citep[CLASH;][]{postman11}. The detailed comparison of SGAS results
as presented in the paper with upcoming CLASH results, as well as
lensing results from the Local Cluster Substructure Survey
\citep[LoCuSS;][]{okabe10a}, will be very useful, particularly because
of markedly different sample selection between these surveys.

\section*{Acknowledgments}
We thank T. Hamana, E. Komatsu, Y.-T. Lin, S. Miyazaki, D. Nagai, 
N. Okabe, J. Schaye, M. Takada, and K. Umetsu for useful discussions.
We also thank an anonymous referee for useful comments and
suggestions. This work was supported in part by the FIRST program
"Subaru Measurements of Images and Redshifts (SuMIRe)", World Premier
International Research Center Initiative (WPI Initiative), MEXT,
Japan, and Grant-in-Aid for Scientific Research from the JSPS
(23740161). P.~N. acknowledges support from the National Science
Foundation's Theory Program via the grant AST10-44455. 


\appendix

\section{Expected properties of strong lensing selected clusters}
\label{sec:conc}

We predict properties of clusters in our strong-lens selected cluster
sample using a semi-analytic model developed by \citet{oguri09a},
which is based on a triaxial halo model of \citet{jing02}. In brief, 
a catalogue of haloes are generated according to the mass function and
axis ratio distribution derived from $N$-body simulations, and each
halo is projected along random direction to compute its lensing
property \citep{oguri03,oguri04}. The projected convergence profile
is compared with that of a spherical NFW profile to estimate the mass 
$M_{\rm vir,\,2D}$, the concentration $c_{\rm vir,\,2D}$, and the
ellipticity $e$ of the projected mass distribution. Here we fix the
lens (cluster) and source redshift to $z_l=0.45$ and $z=2$,
respectively, which are typical for our cluster lens sample analyzed
in the paper.

We compute a giant arc cross section $\sigma_{\rm arc}$ of each halo 
from ray-tracing of extended sources using {\it glafic}
\citep{oguri10b}. The source is assumed to have a Sersic profile with
$r_e=0\farcs4$ and the ellipticity randomly assigned between 0 and
0.5. For each cluster we compute the arc cross section for the
length-to-width ratio of $l/w>5$ by randomly throwing source galaxies
in the source plane. Output images are convolved with the Gaussian
kernel with the FWHM of $0\farcs8$ in order to take account of the
seeing effect. 

We study the impact of the lensing bias by averaging concentrations and
ellipticities of the halo catalogue with an appropriate weight that
mimics our selection criteria. Obviously our cluster sample is
weighted by the arc cross section $\sigma_{\rm arc}$, which can be a
reasonable choice of the weight. Furthermore, we preferentially
conduct follow-up observations for clusters with larger Einstein radii
(i.e., giant arcs located more distant from the cluster centre),
because they are expected to be more massive. We can model this
selection effect, e.g., by multiplying $\sqrt{\theta_{\rm E}}$ to the
weight.  

\begin{figure}
\begin{center}
\includegraphics[width=0.85\hsize]{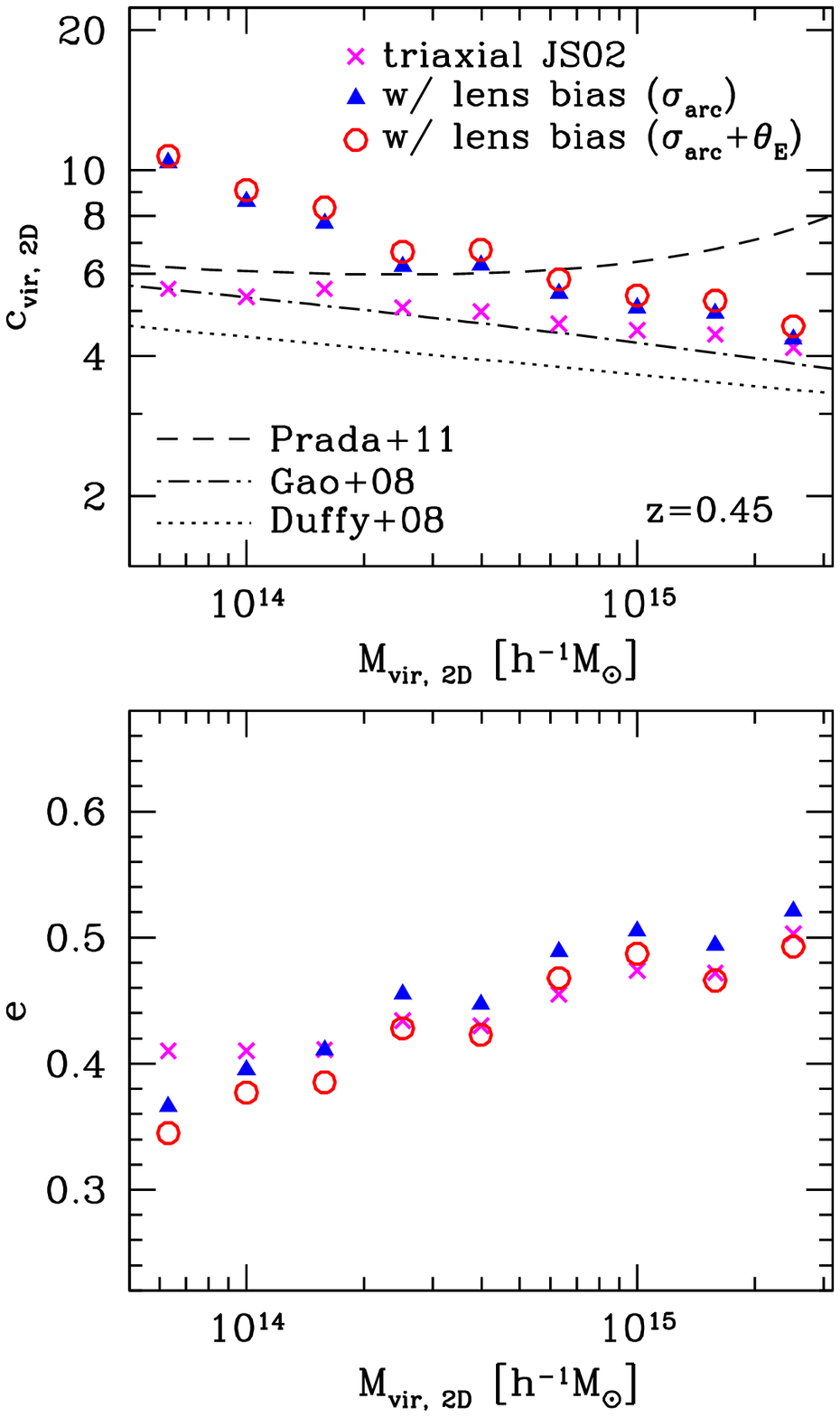}
\end{center}
\caption{{\it Upper:} The mass-concentration relation of the projected
 mass distribution for $z=0.45$, predicted by a triaxial halo model of
 \citet{jing02}. Crosses indicate the average concentration parameter
 from our semi-analytic calculation without any weighting, whereas
 filled triangles and open circles are average concentration
 parameters with the lensing bias, assuming weights from giant arc
 cross sections $\sigma_{\rm arc}$ or arc cross sections plus the
 Einstein radii $\theta_{\rm E}$, respectively. Lines are predicted
 mass-concentration relation from spherical averaging of haloes in
 various $N$-body simulations \citep{gao08,duffy08,prada11}.
 {\it Lower:} Similar to the upper panel, but average ellipticities
 of the project mass distribution as a function of the mass are shown.
 \label{fig:app_conc_comp}} 
\end{figure}

Figure~\ref{fig:app_conc_comp} shows the mass-concentration relation
derived from the projected mass distribution assuming a spherical mass
distribution, which is relevant for the comparison with our lensing
measurement, based on the semi-analytic calculation described above. 
We find that the enhancement of the concentration parameter due to the
lensing bias is a strong function of the mass. For instance, when both
the arc cross section and the Einstein radius are used for computing
the lensing bias, the concentration parameter is enhanced by
$\sim$80\% for $M_{\rm vir}\sim 8\times 10^{13}h^{-1}M_\odot$,
$\sim$30\% for $M_{\rm vir}\sim 4\times 10^{14}h^{-1}M_\odot$, and
$\sim$20\% for $M_{\rm vir}\sim 10^{15}h^{-1}M_\odot$.

We also compare our result based on the triaxial model with other work
studying the mass-concentration relation in $N$-body simulations based
on the spherical NFW profile. We find that our mass-concentration
relation has larger concentration than predicted by the relation of
\citet{duffy08}, which has been adopted for comparisons with lensing
measurements in \citet{oguri09b}.  On the other hand, \citet{prada11}
concluded that concentrations of massive haloes evolve little with the
redshift. As a result, concentrations predicted by the model of
\citet{prada11} are much larger compared with the \citet{duffy08}
relation predicts. We note that such little evolution of massive haloes
have been noted by  \citet{zhao03} and \citet{zhao09}. Our
mass-concentration relation based on the triaxial model resides in
between the \citet{duffy08} and \citet{prada11} relations, and more
resemble the relation presented by \citet{gao08} which was essentially
the modification of the model proposed by \citet{navarro97}. 

\begin{figure}
\begin{center}
\includegraphics[width=0.95\hsize]{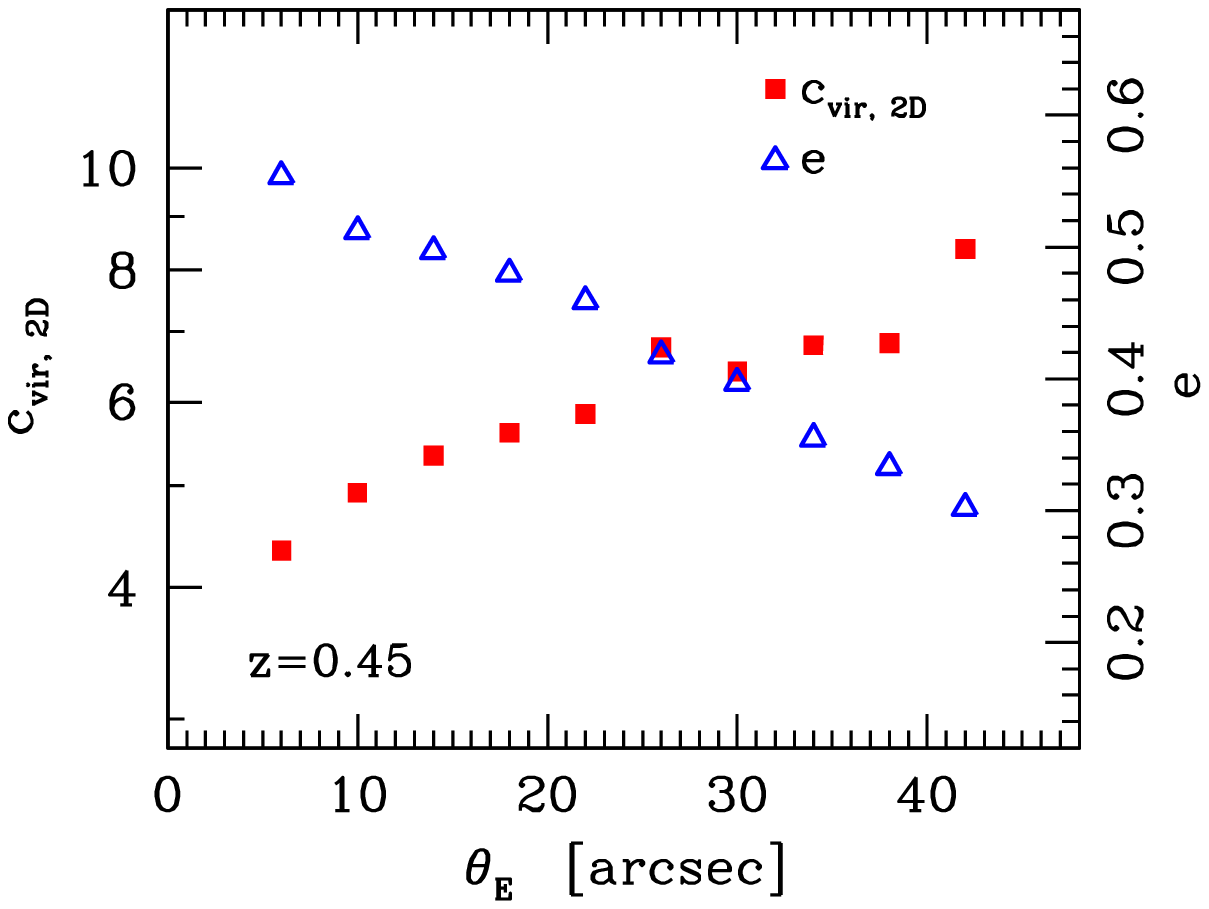}
\end{center}
\caption{The average concentration ({\it filled squares}) and
  ellipticity ({\it open triangles}) of the projected mass
  distributions as a function of the Einstein radius $\theta_{\rm E}$.
 They are derived from a triaxial halo model of \citet{jing02} and
 include the effect of the lensing bias from the arc cross section.
 \label{fig:app_conc_ein}} 
\end{figure}

We also investigate the impact of the lensing bias on the ellipticity
of the projected mass distribution. We find that the lensing bias due
to the arc cross section tends to increase the mean ellipticity, which
is understood by the fact that the ellipticity of the projected mass
distribution significantly enhances the giant arc cross section 
\citep[e.g.,][]{meneghetti03,meneghetti07}. In contrast, the lensing
bias due to the Einstein radius decreases the mean Einstein radius,
because of the alignment of the major axis with the line-of-sight
direction \citep{oguri09a}. Therefore, these two lensing biases
counteract with each other. In either case, however, the effect of the
lensing bias appears to be small, with the change of the mean
ellipticity by $\sim 0.05$ at most. 

\citet{oguri09a} has shown that extreme lensing clusters having
very large Einstein radii are more severely affected by the lensing
bias. Thus we check the dependence of the concentration and projected
ellipticity on the Einstein radius $\theta_{\rm E}$. For the mass
distribution of the cluster, we assume a flat prior of 
$\log M_{\rm vir}$ between $10^{14}h^{-1}M_\odot$ and
$10^{15}h^{-1}M_\odot$, and the number distribution predicted by the
mass function of dark haloes above $10^{15}h^{-1}M_\odot$, which is 
expected to more or less resemble the selection function of our cluster
sample. The result shown in Figure~\ref{fig:app_conc_ein} indicate
that both the concentration and ellipticity depend strongly on the
Einstein radius such that clusters with larger Einstein radii are more
concentrated and spherical. Again, this can be interpreted by the
alignment of the major axis with the line-of-sight direction
\citep{oguri09a}. 

\section{Lensing analysis for individual clusters}
\label{sec:app}

We show for each cluster the critical curve of the best-fit strong
lens modelling plotted on the Subaru/Suprime-cam $gri$-composite
image ($2'\times2'$; squares are positions of multiple images
used for mass modelling), weak lensing mass map plotted on the
Subaru/Suprime-cam $r$-band image (contours are drawn with spacing of
$1\sigma$ noise level, and the cross indicates the position of the
brightest cluster galaxy), and the tangential shear profile as well as 
the best-fit NFW profiles. The shading in the tangential profile plot
indicates the measured Einstein radius and its $1\sigma$ error at the
arc redshift. 

\begin{figure*}
\begin{center}
\includegraphics[width=0.28\hsize]{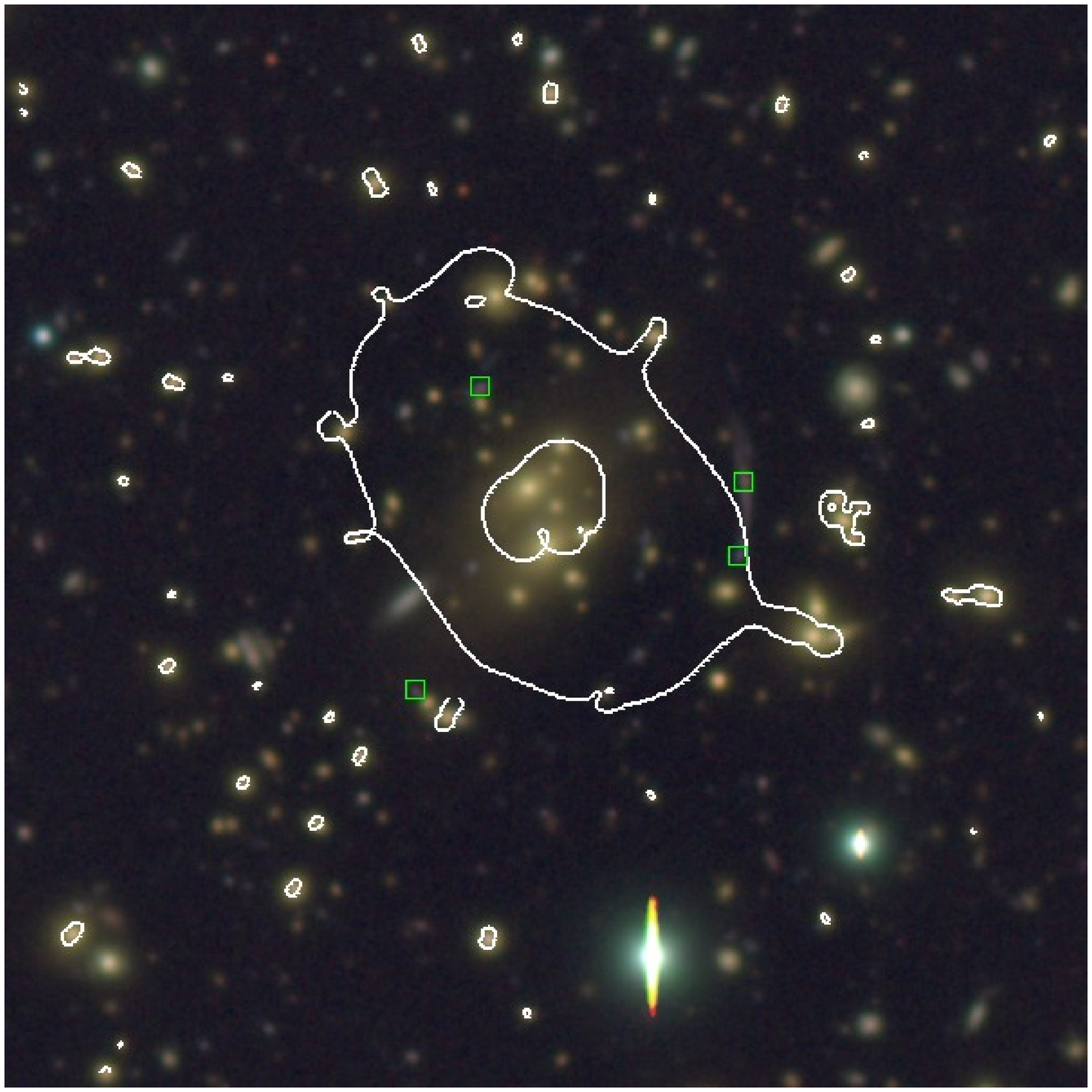}
\includegraphics[width=0.37\hsize]{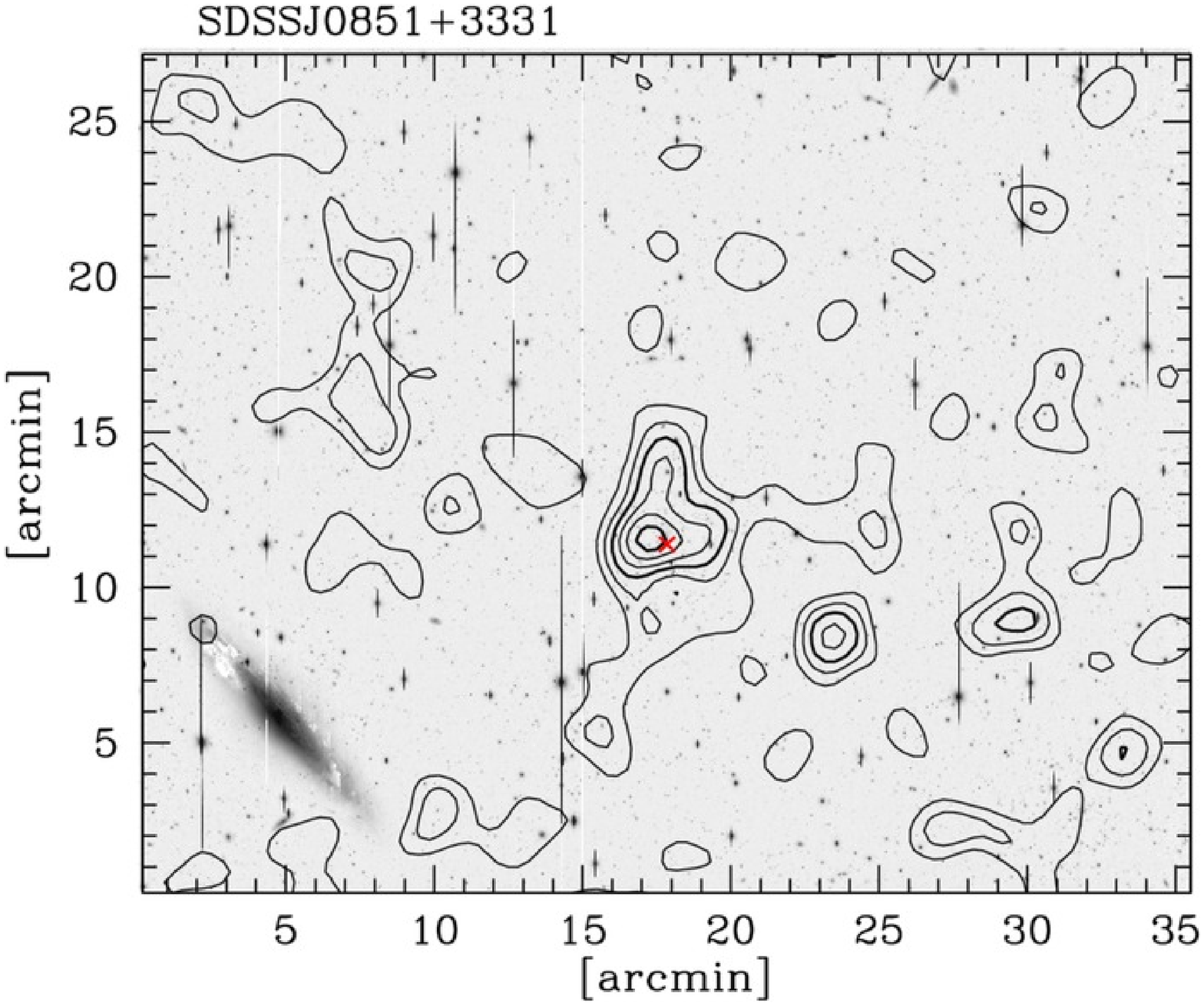}
\includegraphics[width=0.33\hsize]{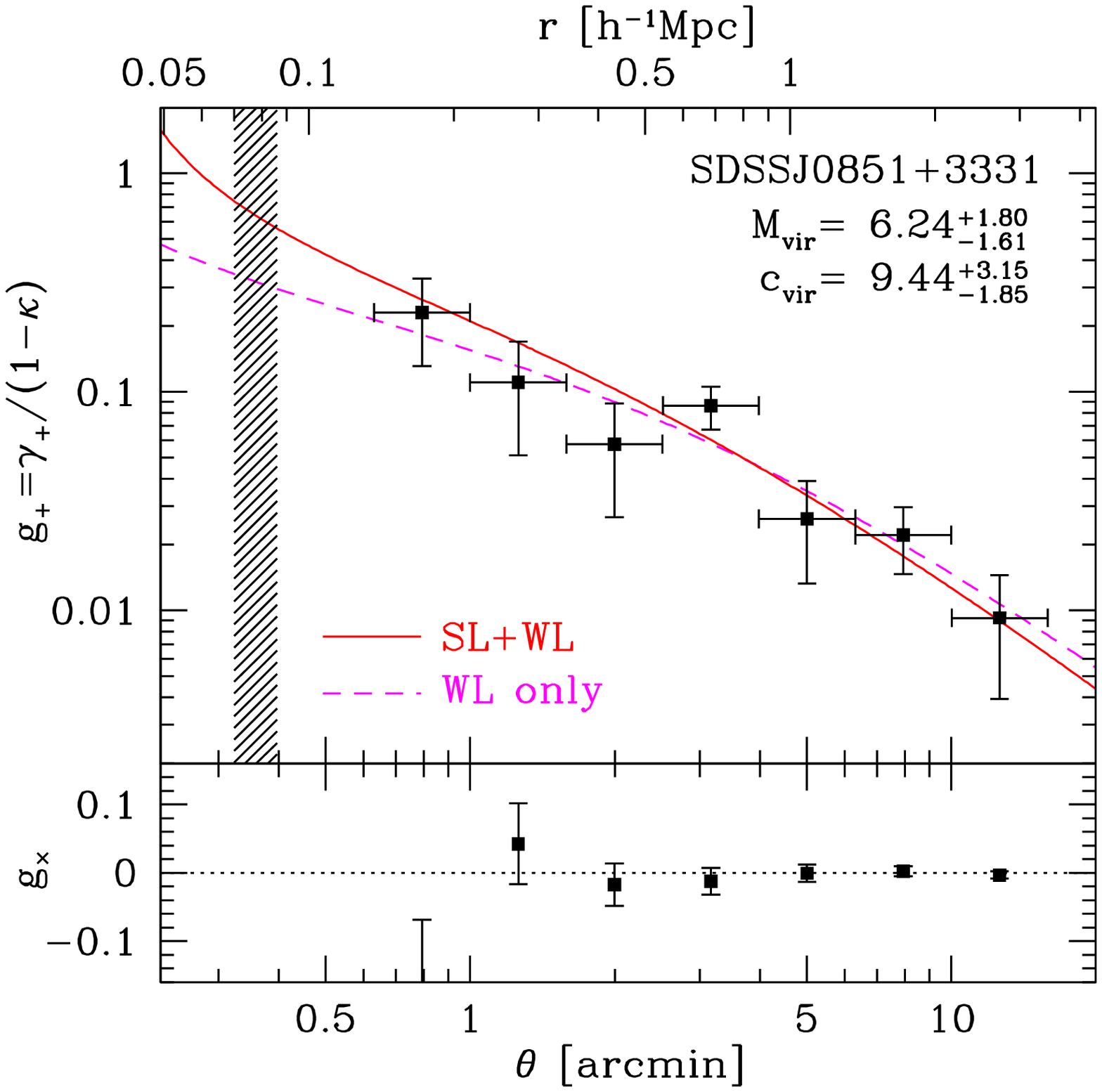}
\includegraphics[width=0.28\hsize]{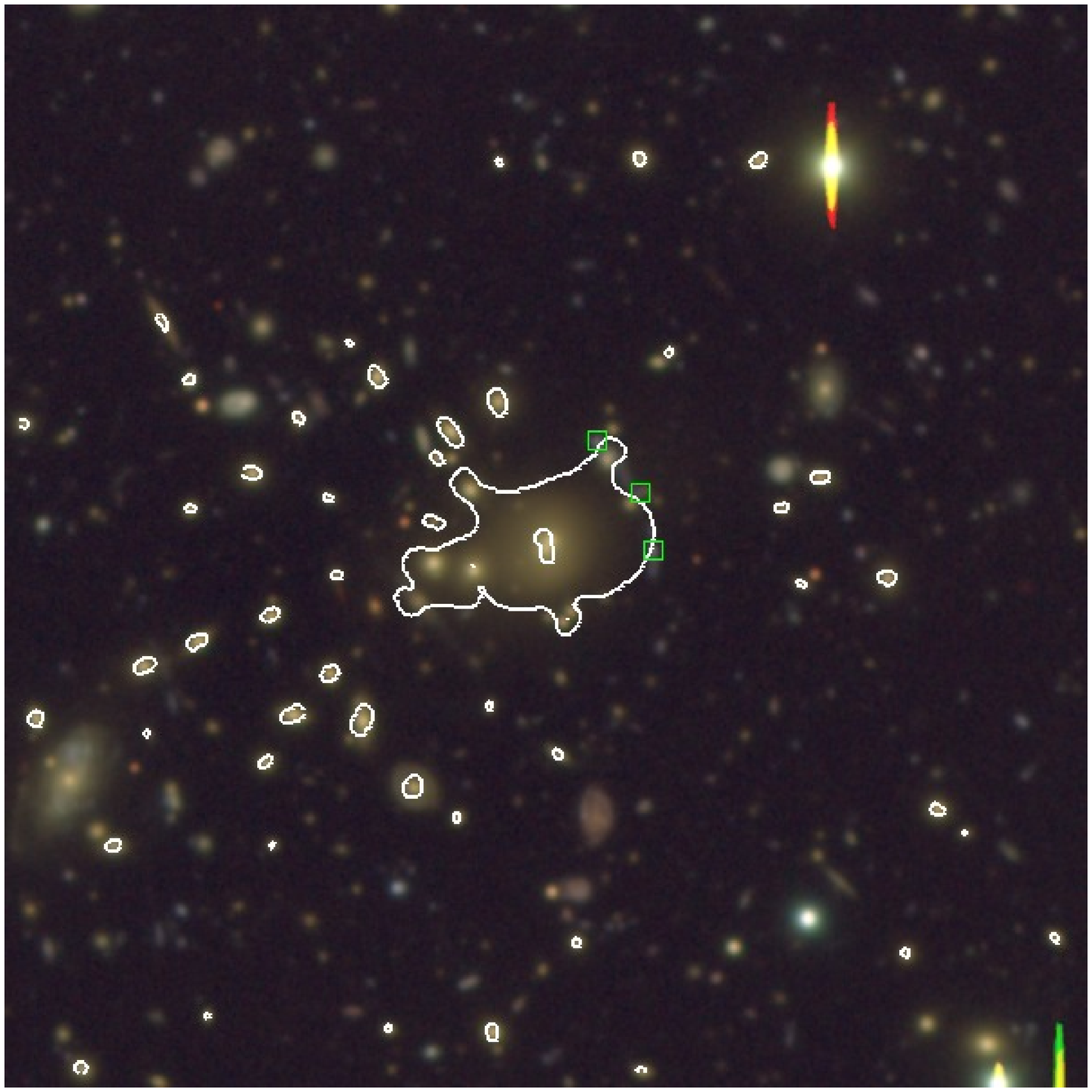}
\includegraphics[width=0.37\hsize]{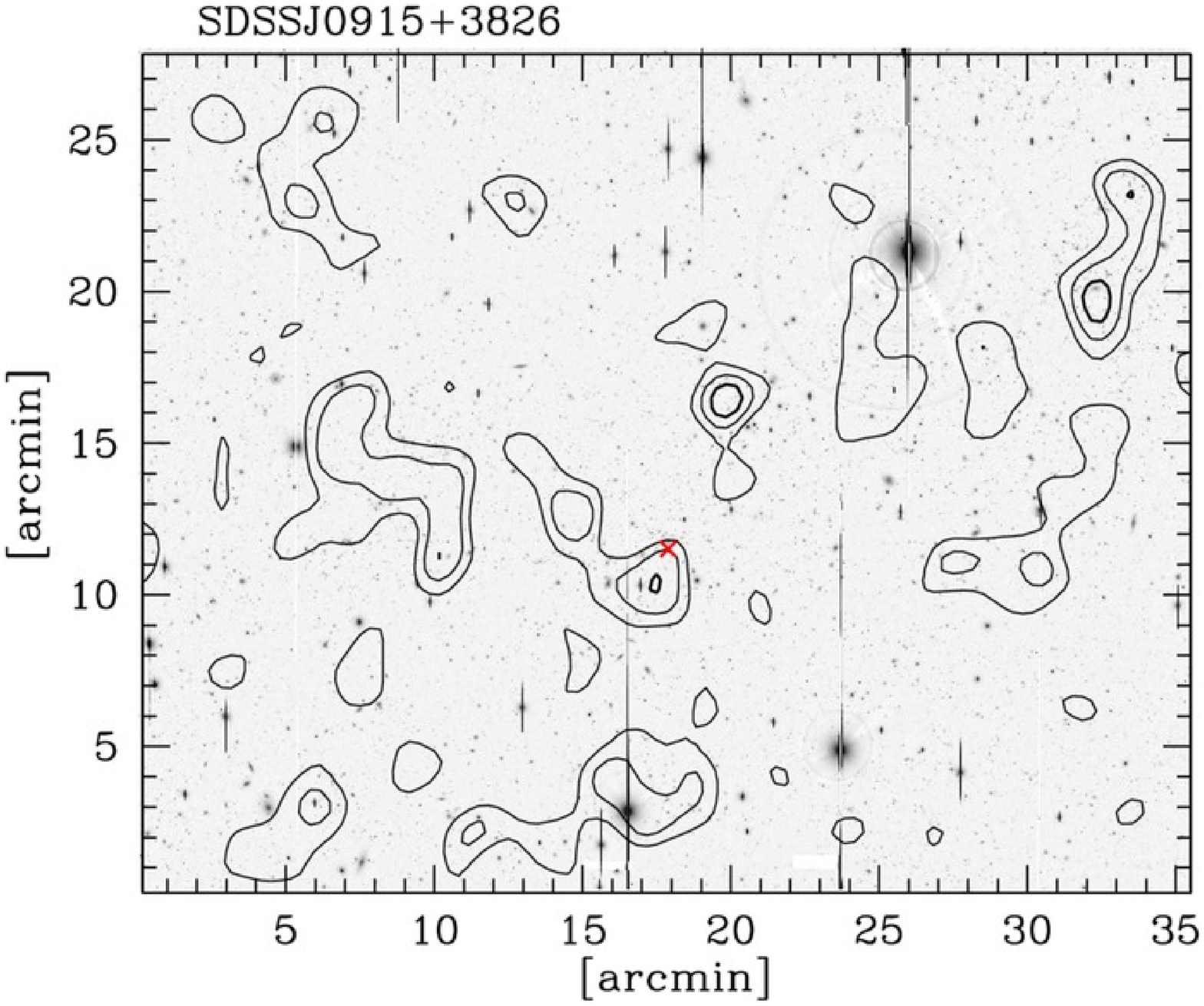}
\includegraphics[width=0.33\hsize]{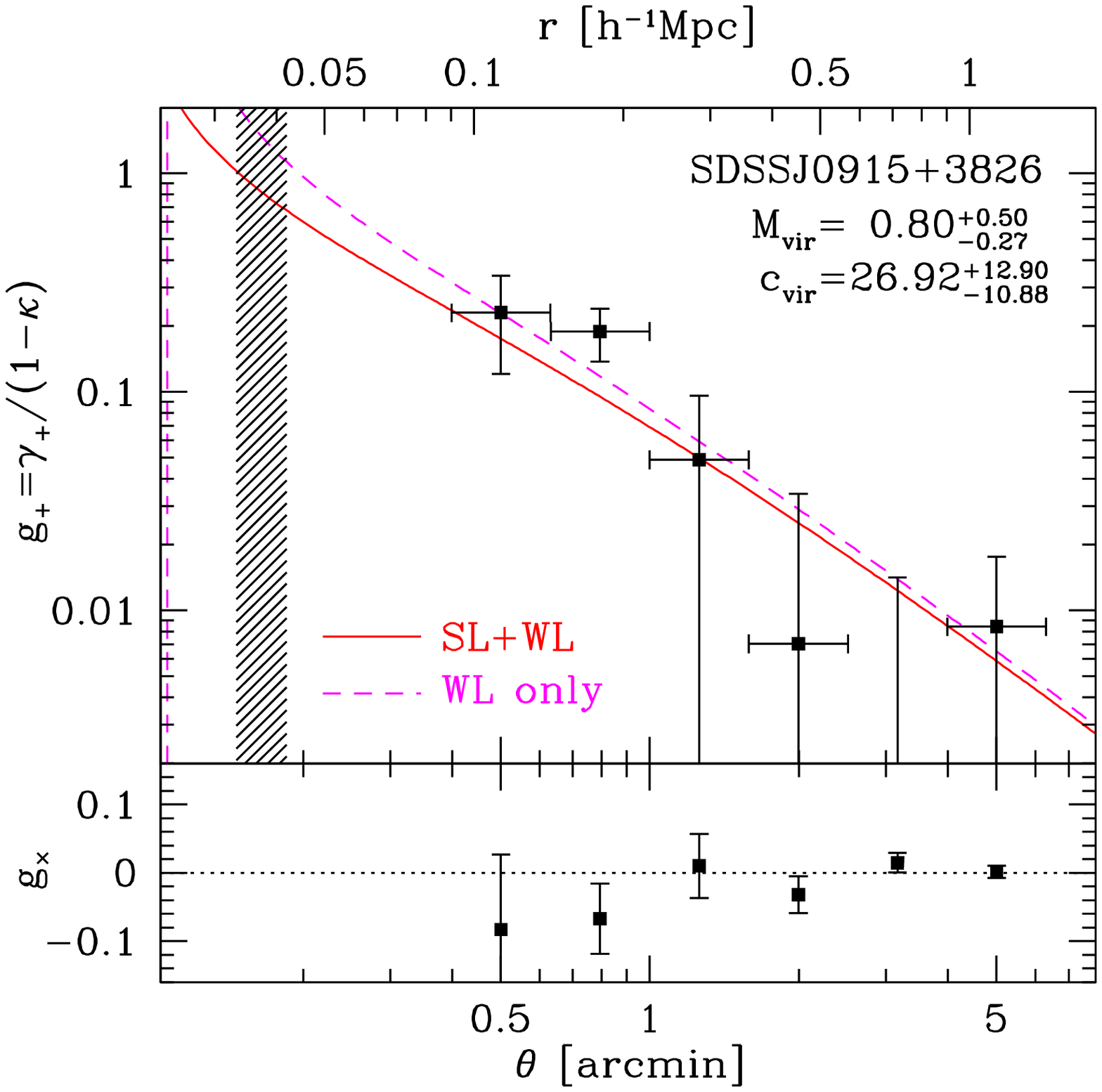}
\includegraphics[width=0.28\hsize]{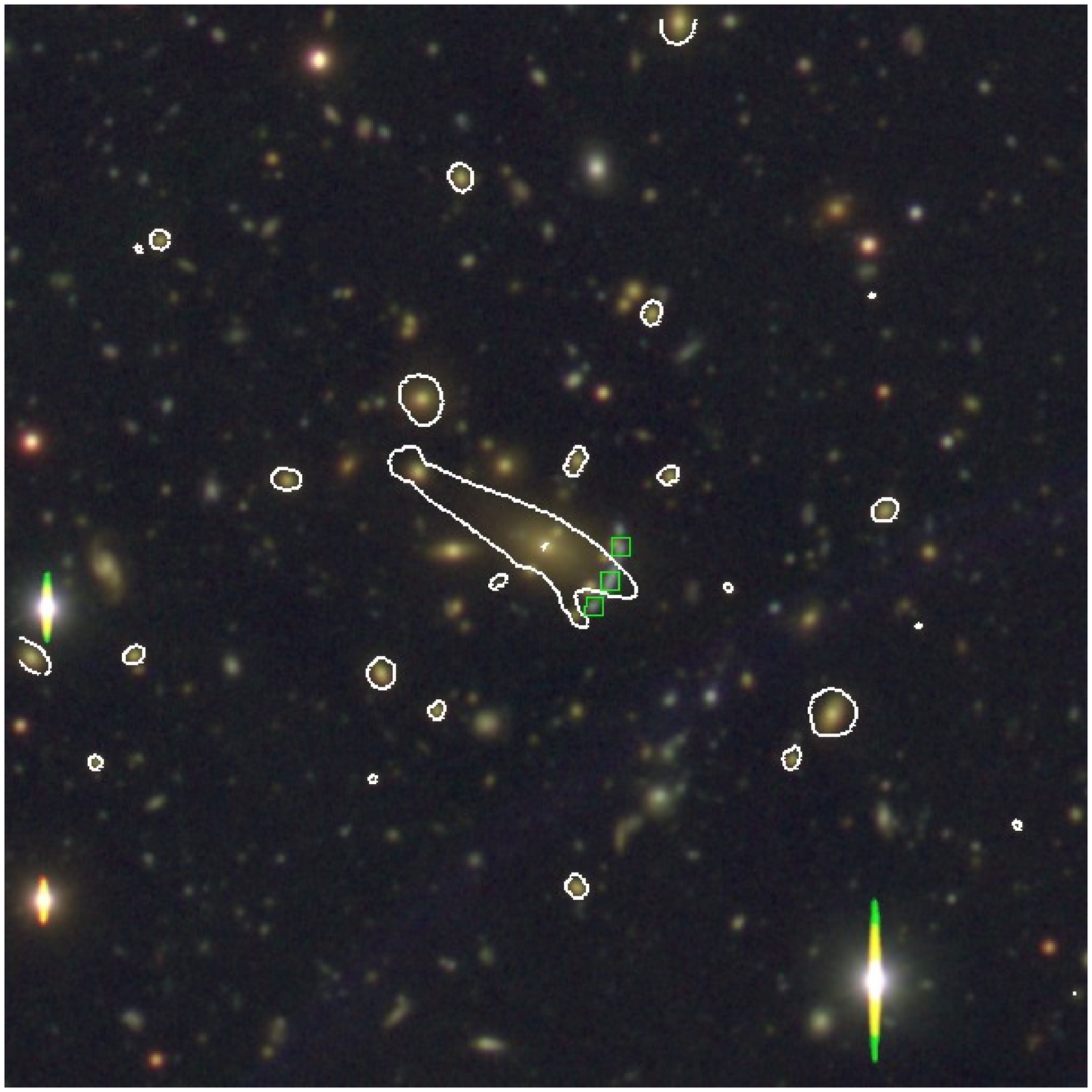}
\includegraphics[width=0.37\hsize]{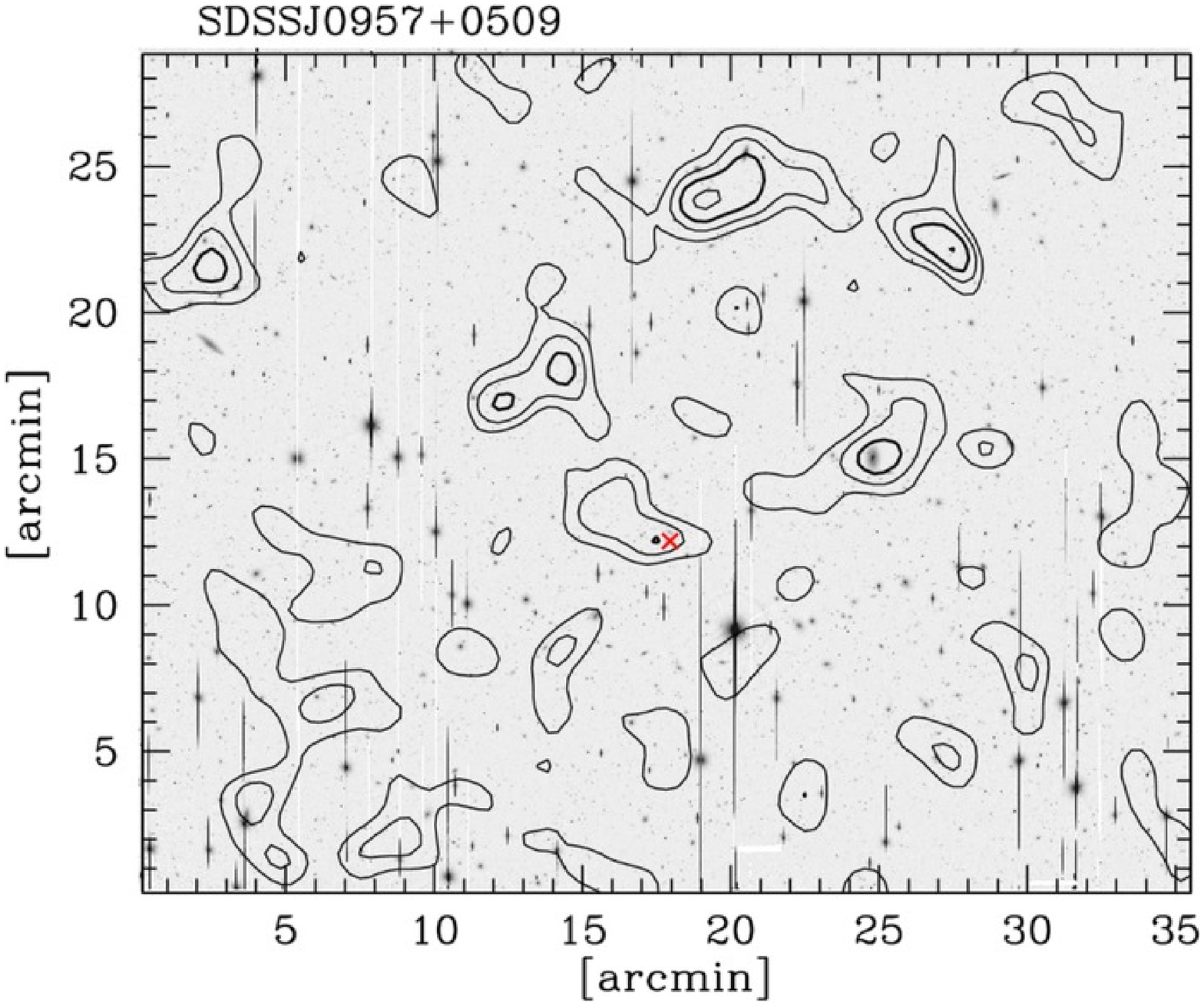}
\includegraphics[width=0.33\hsize]{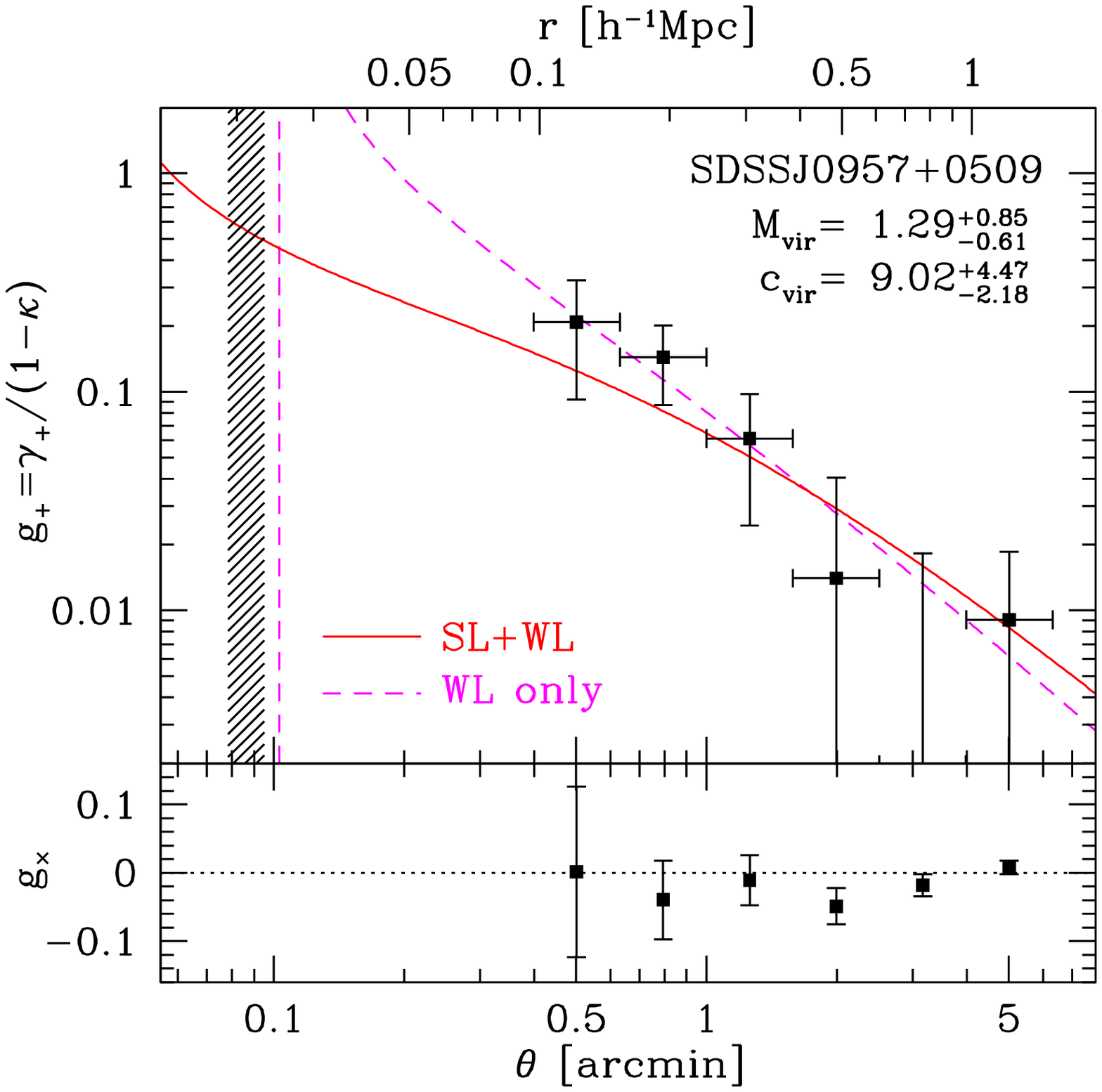}
\includegraphics[width=0.28\hsize]{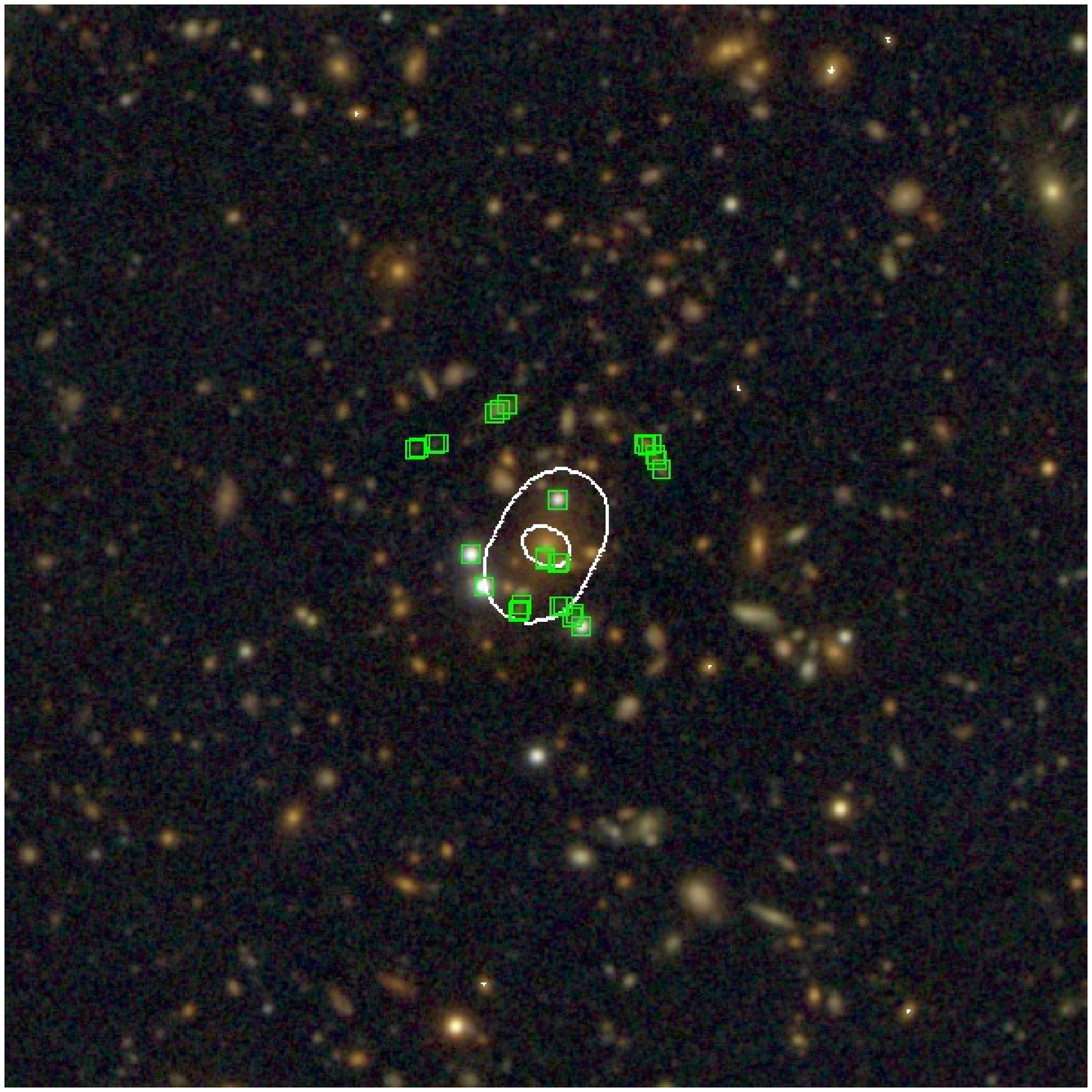}
\includegraphics[width=0.37\hsize]{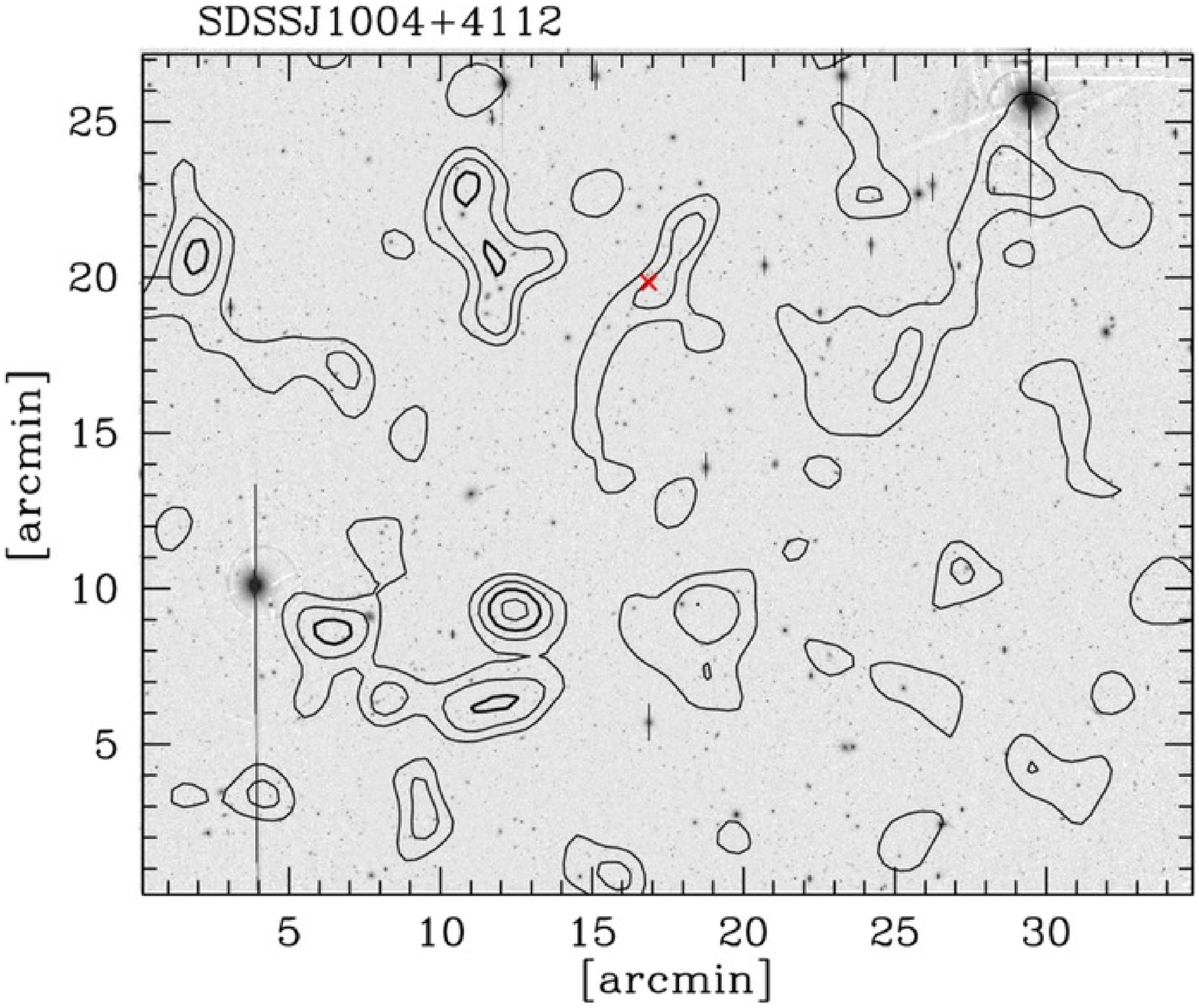}
\includegraphics[width=0.33\hsize]{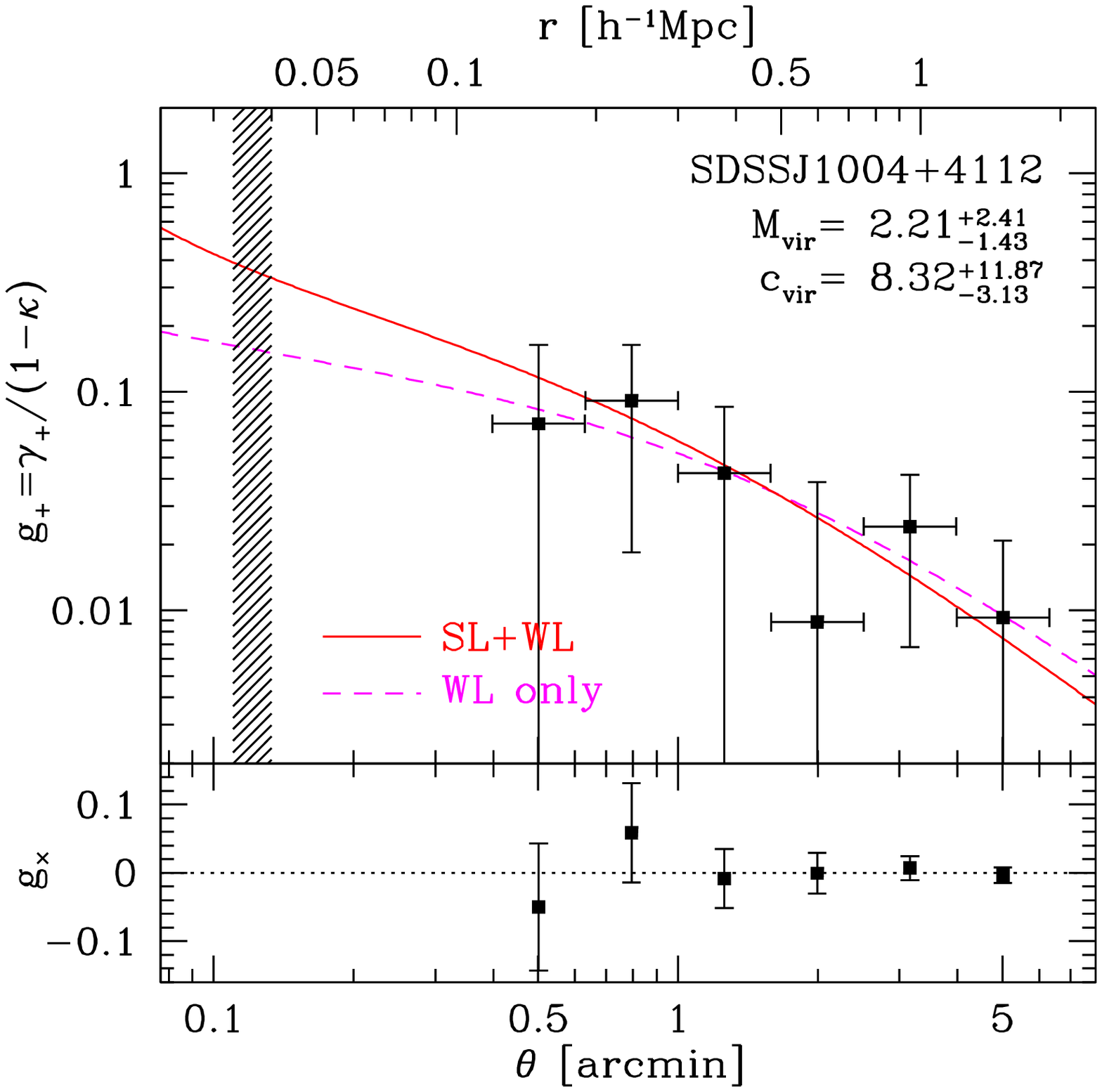}
\end{center}
\caption{SDSSJ0851+3331, SDSSJ0915+3826, SDSSJ0957+0509, SDSSJ1004+4112}
\end{figure*}
\begin{figure*}
\begin{center}
\includegraphics[width=0.28\hsize]{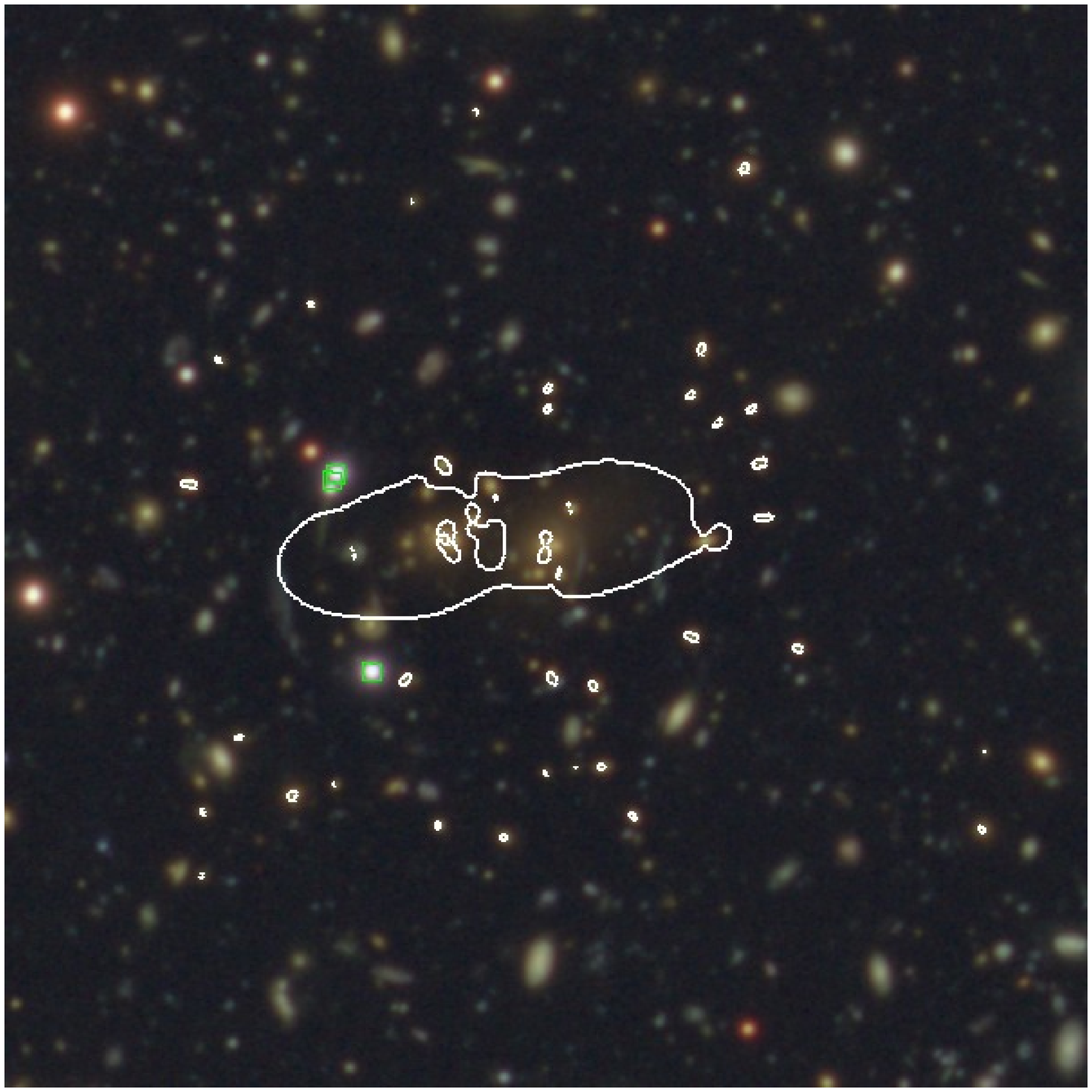}
\includegraphics[width=0.37\hsize]{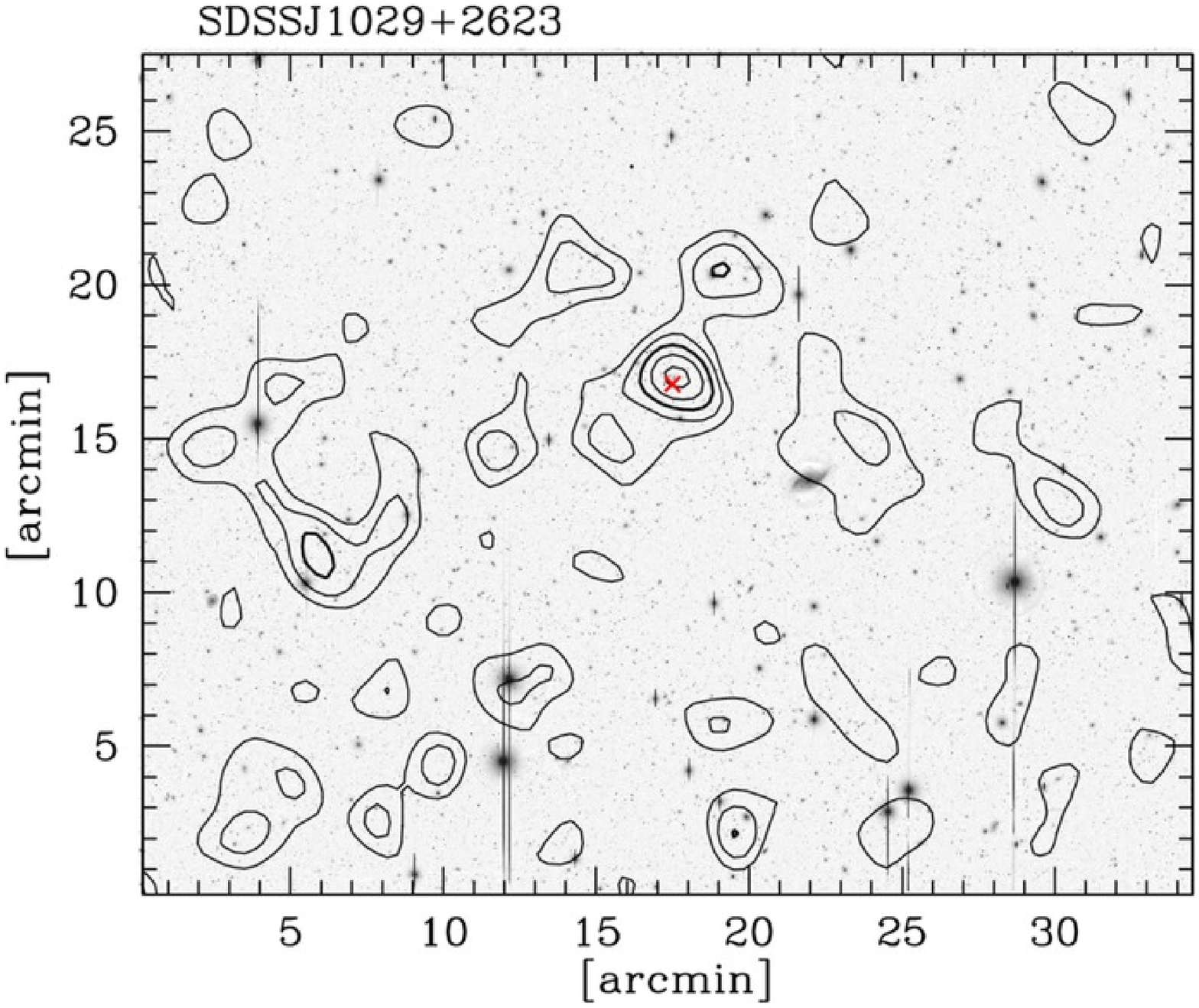}
\includegraphics[width=0.33\hsize]{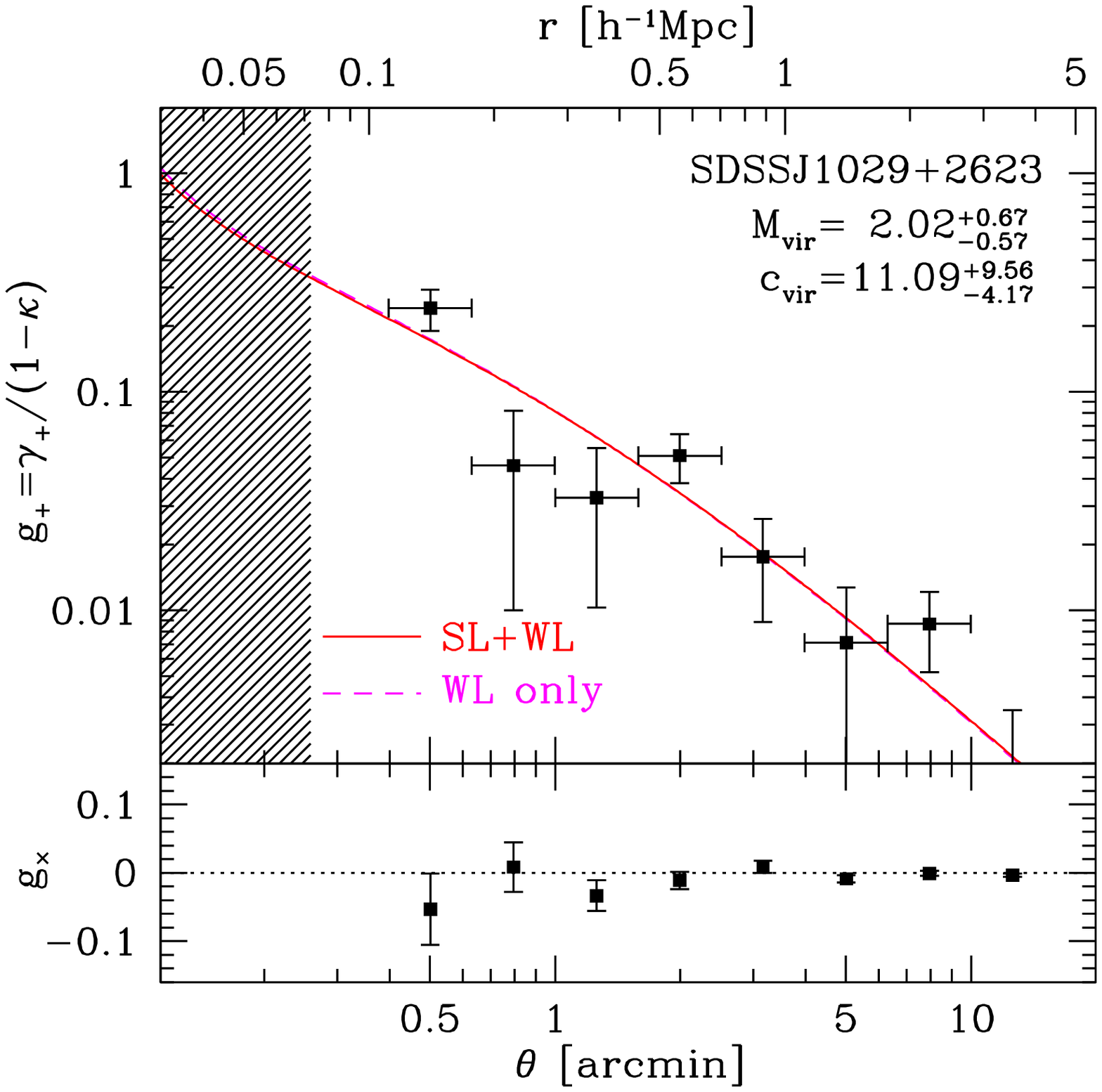}
\includegraphics[width=0.28\hsize]{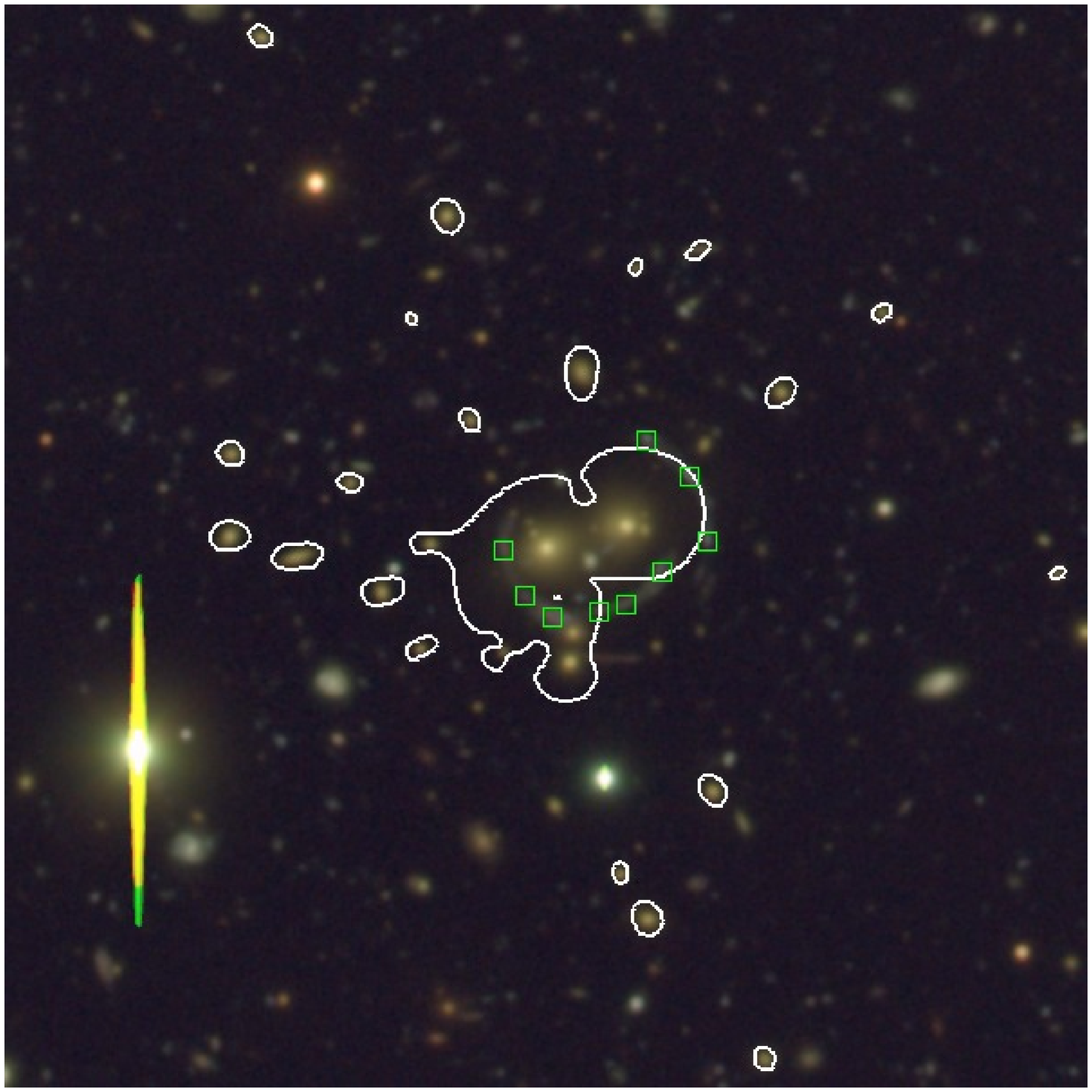}
\includegraphics[width=0.37\hsize]{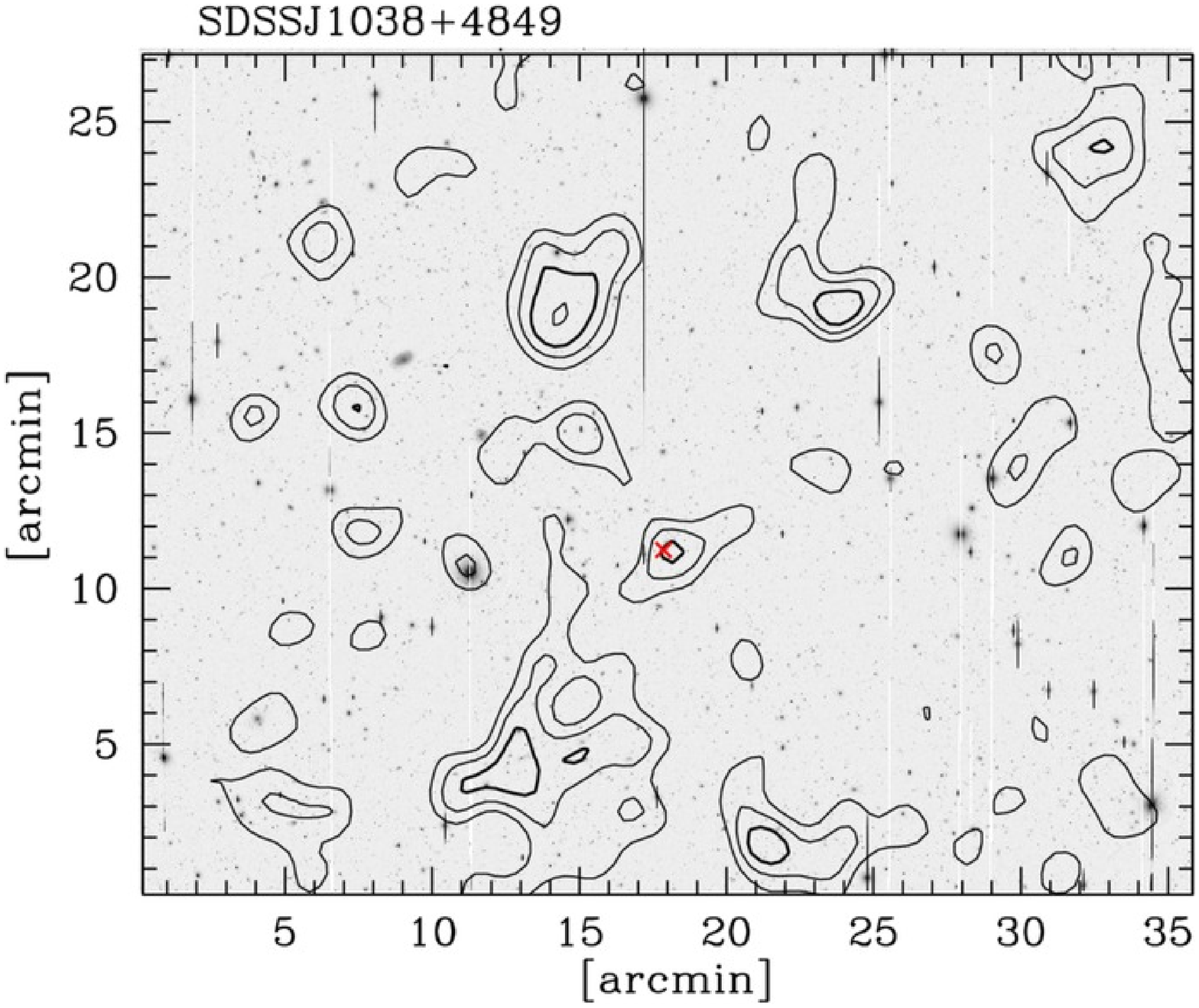}
\includegraphics[width=0.33\hsize]{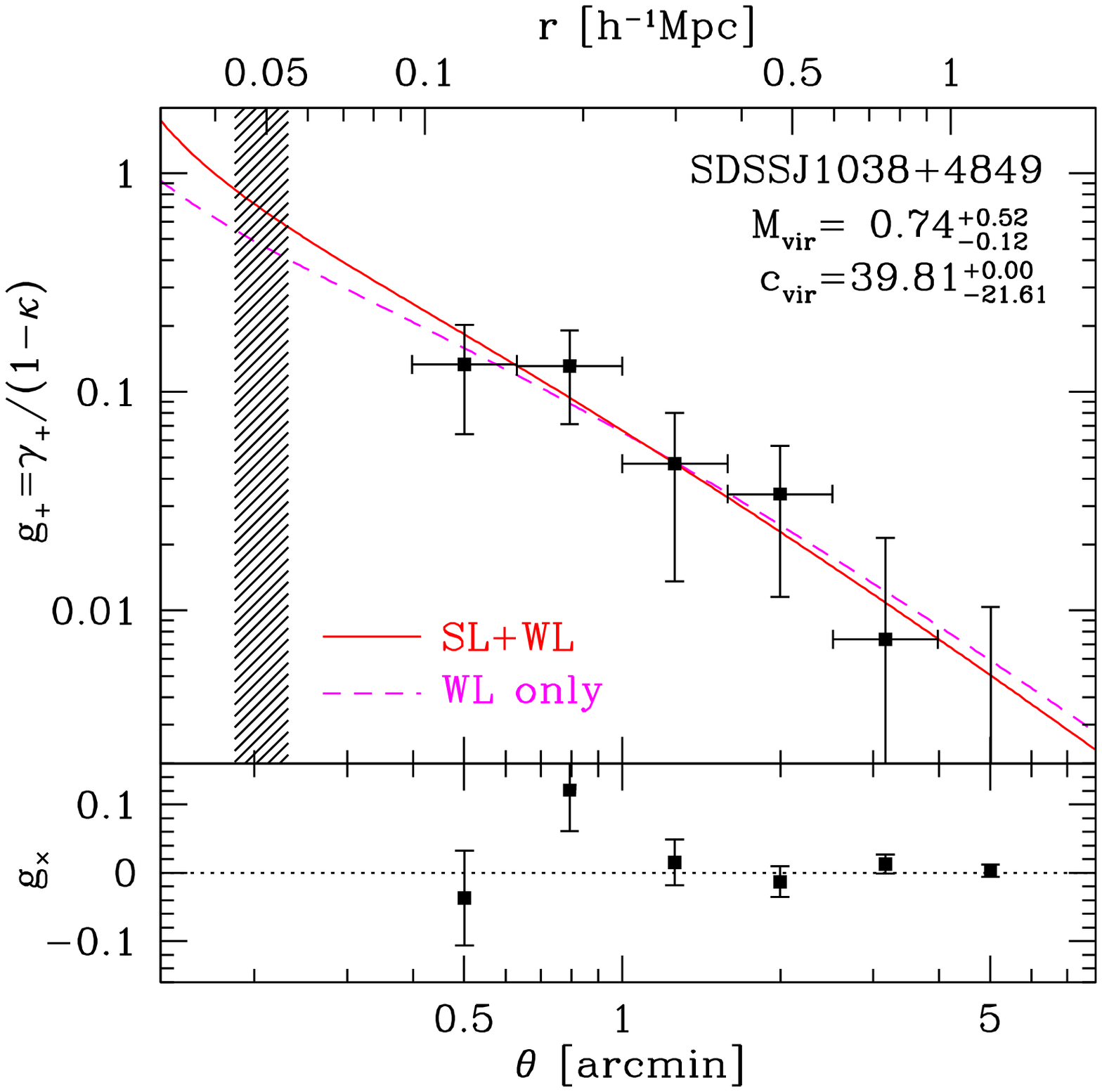}
\includegraphics[width=0.28\hsize]{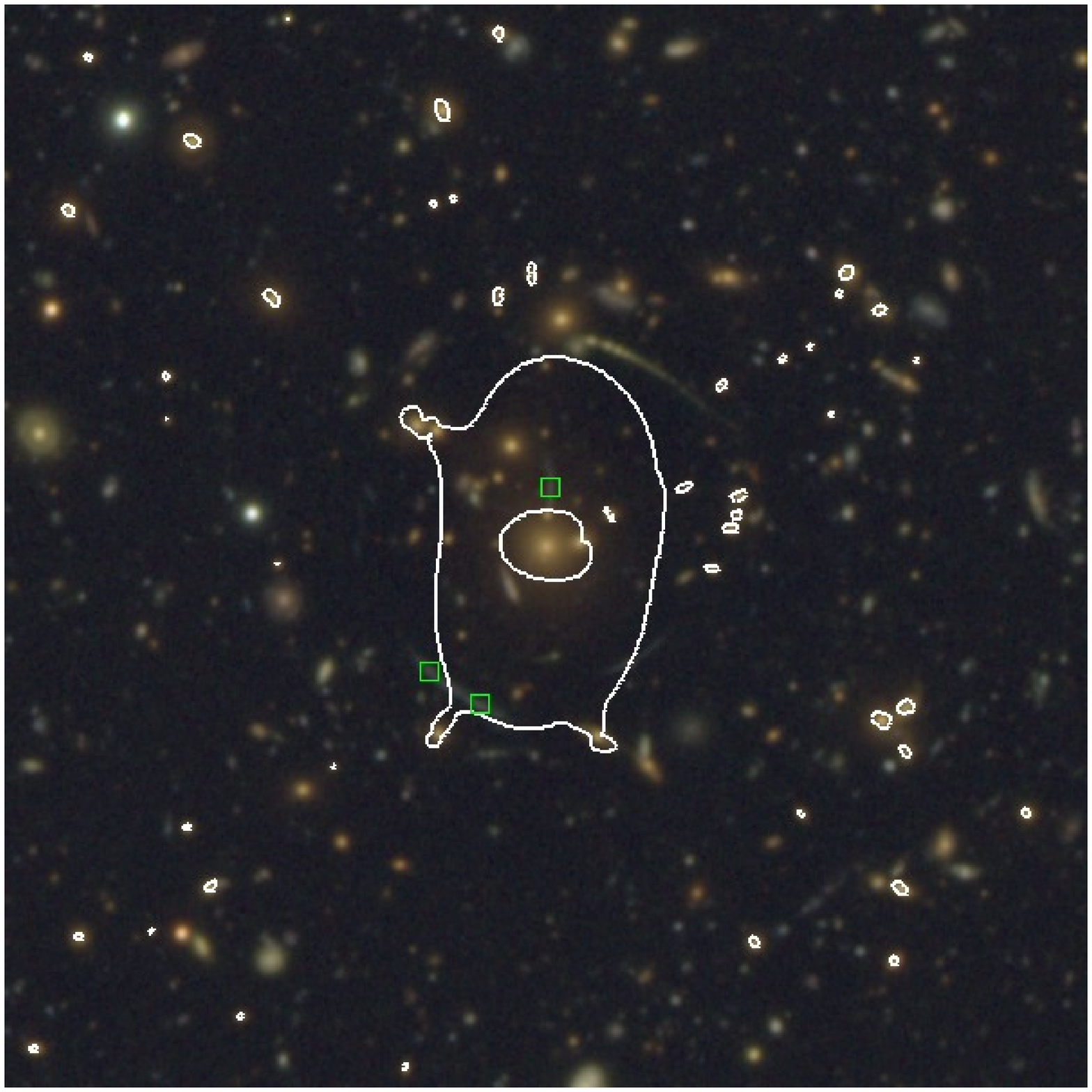}
\includegraphics[width=0.37\hsize]{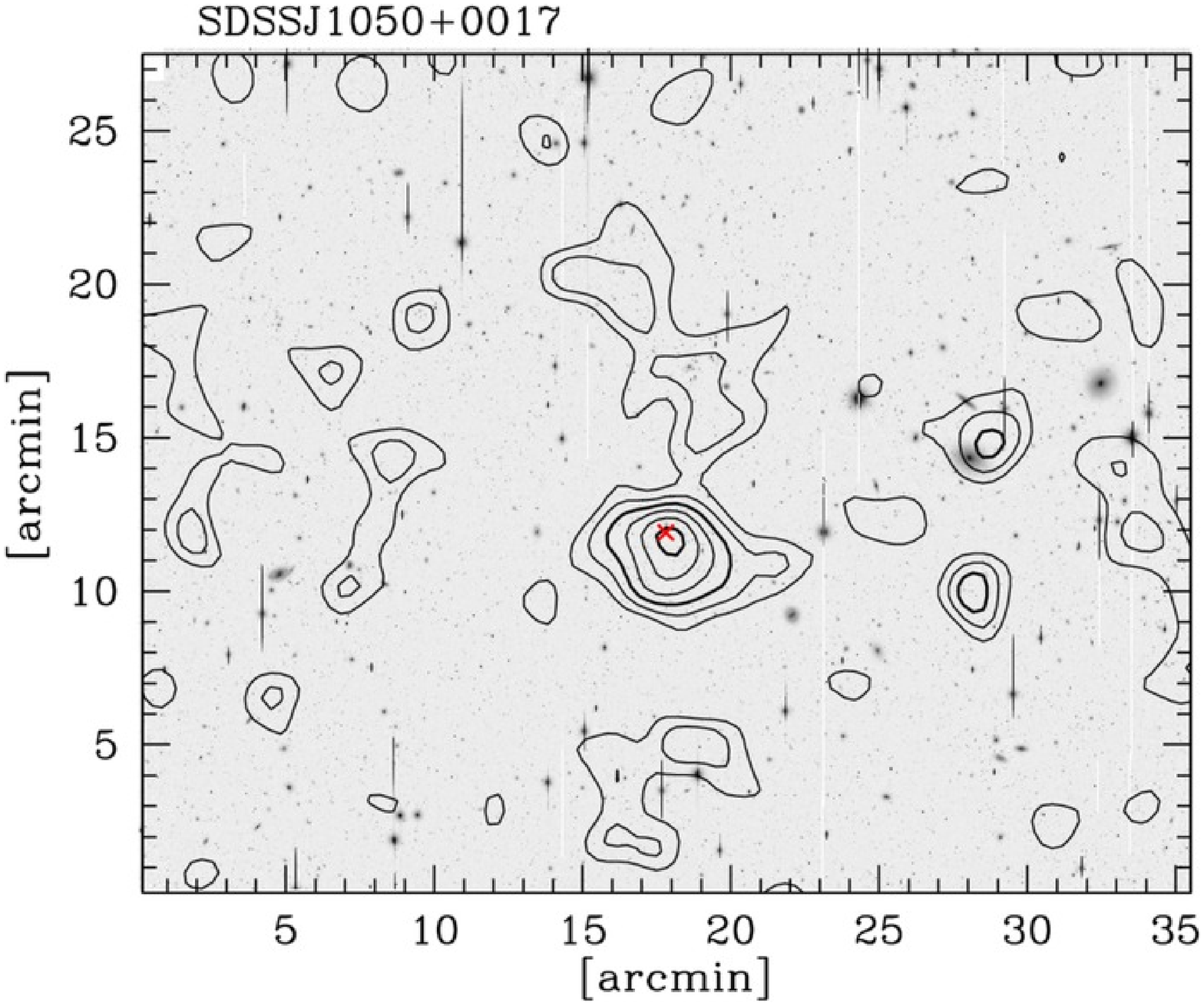}
\includegraphics[width=0.33\hsize]{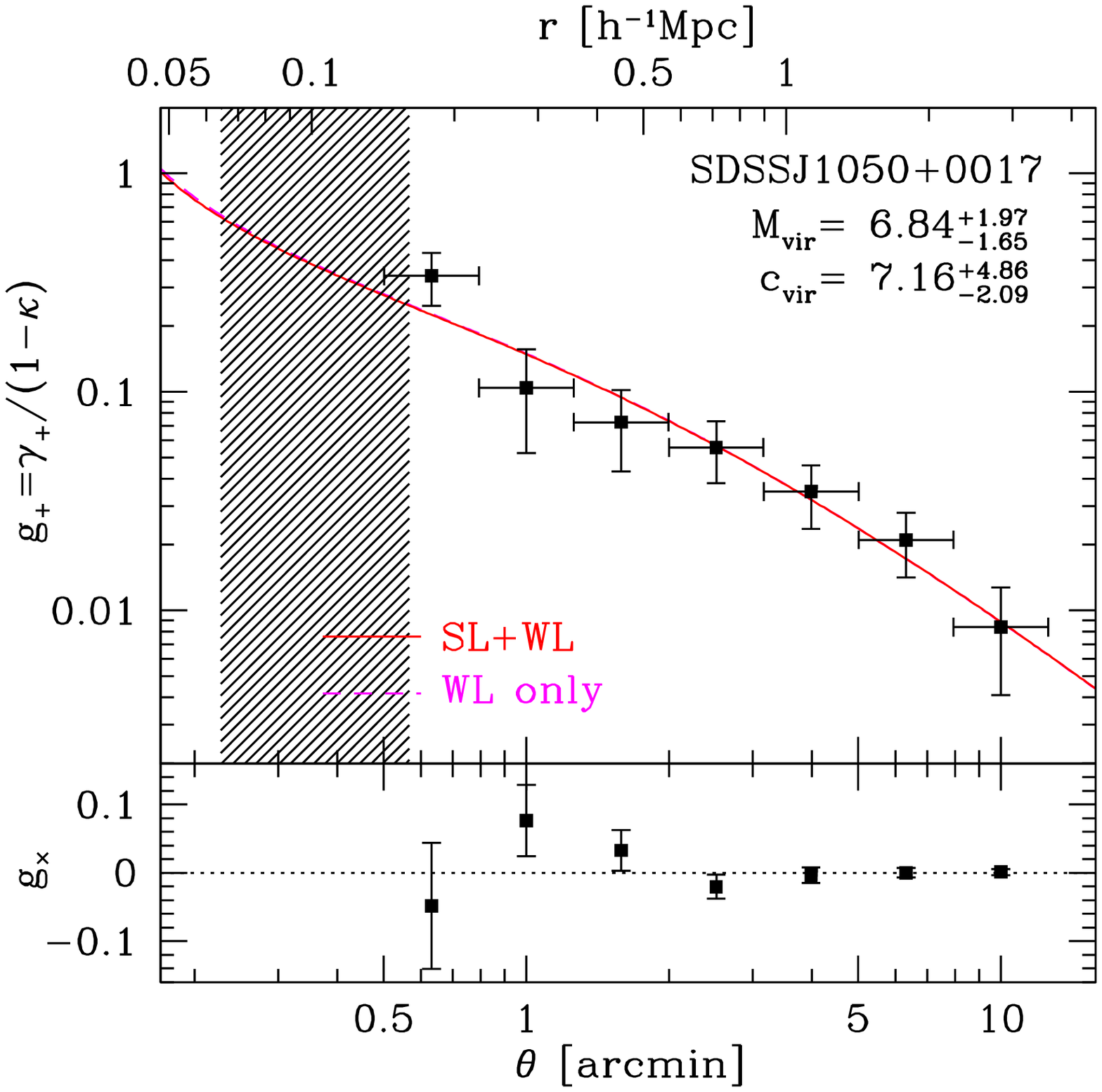}
\includegraphics[width=0.28\hsize]{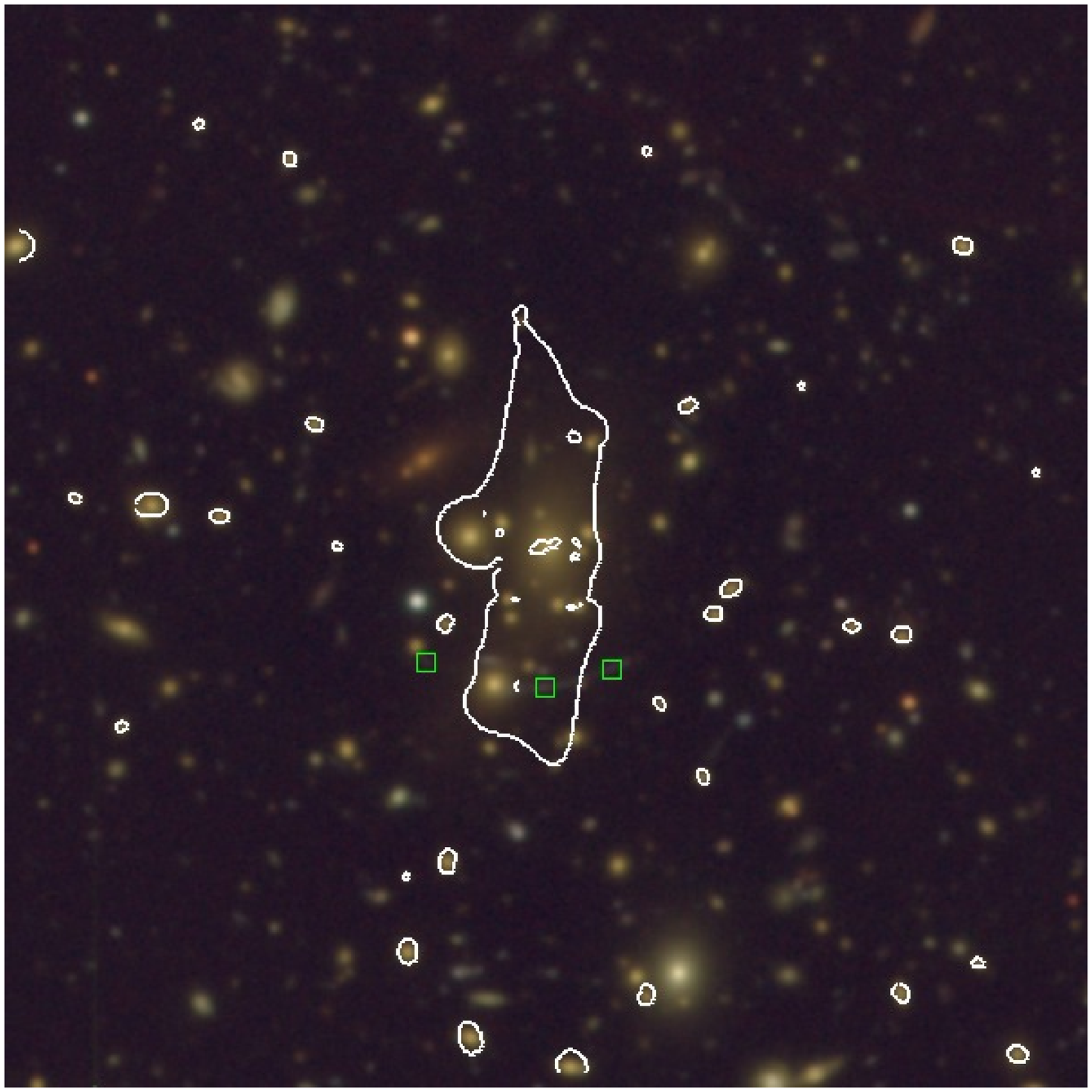}
\includegraphics[width=0.37\hsize]{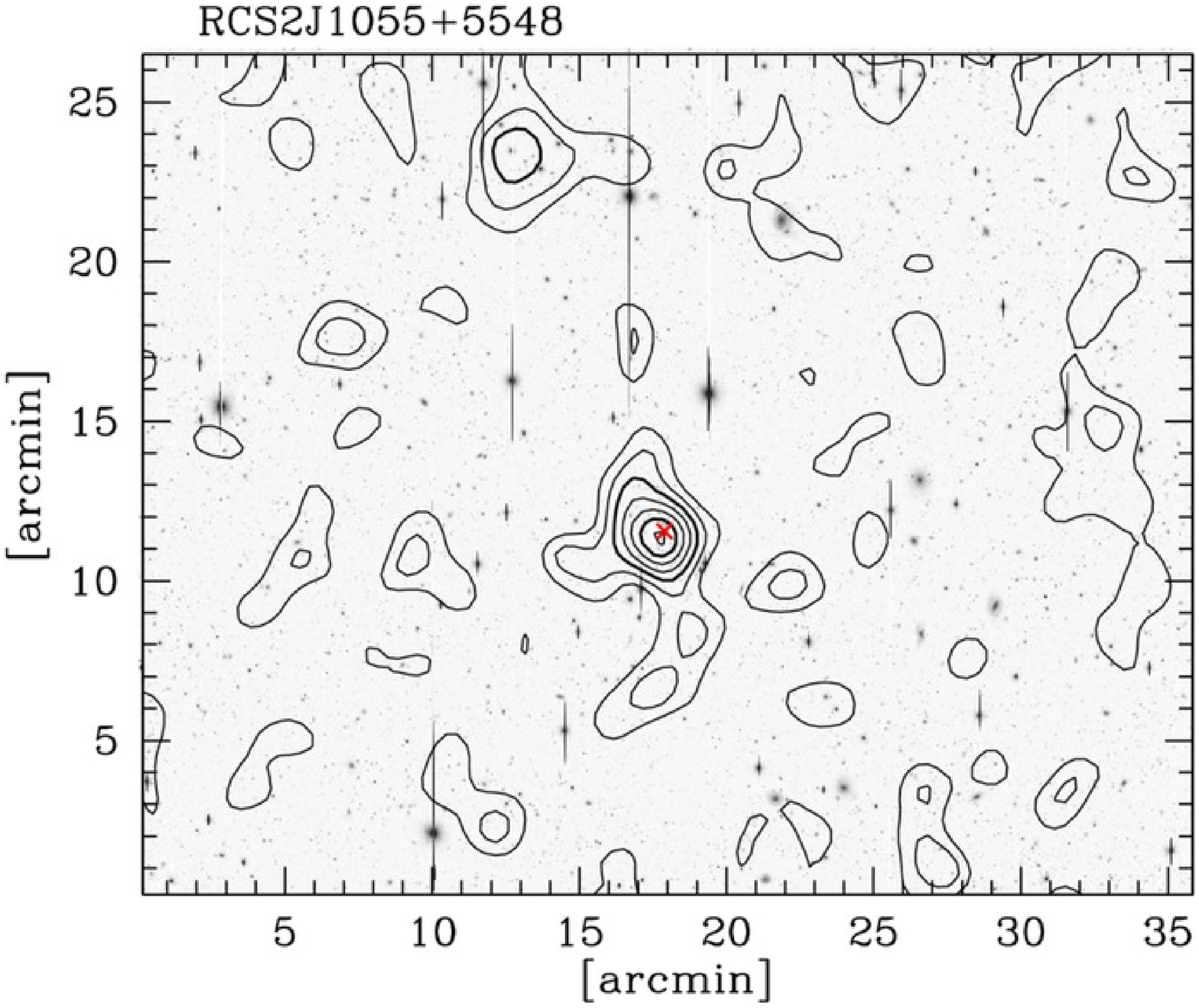}
\includegraphics[width=0.33\hsize]{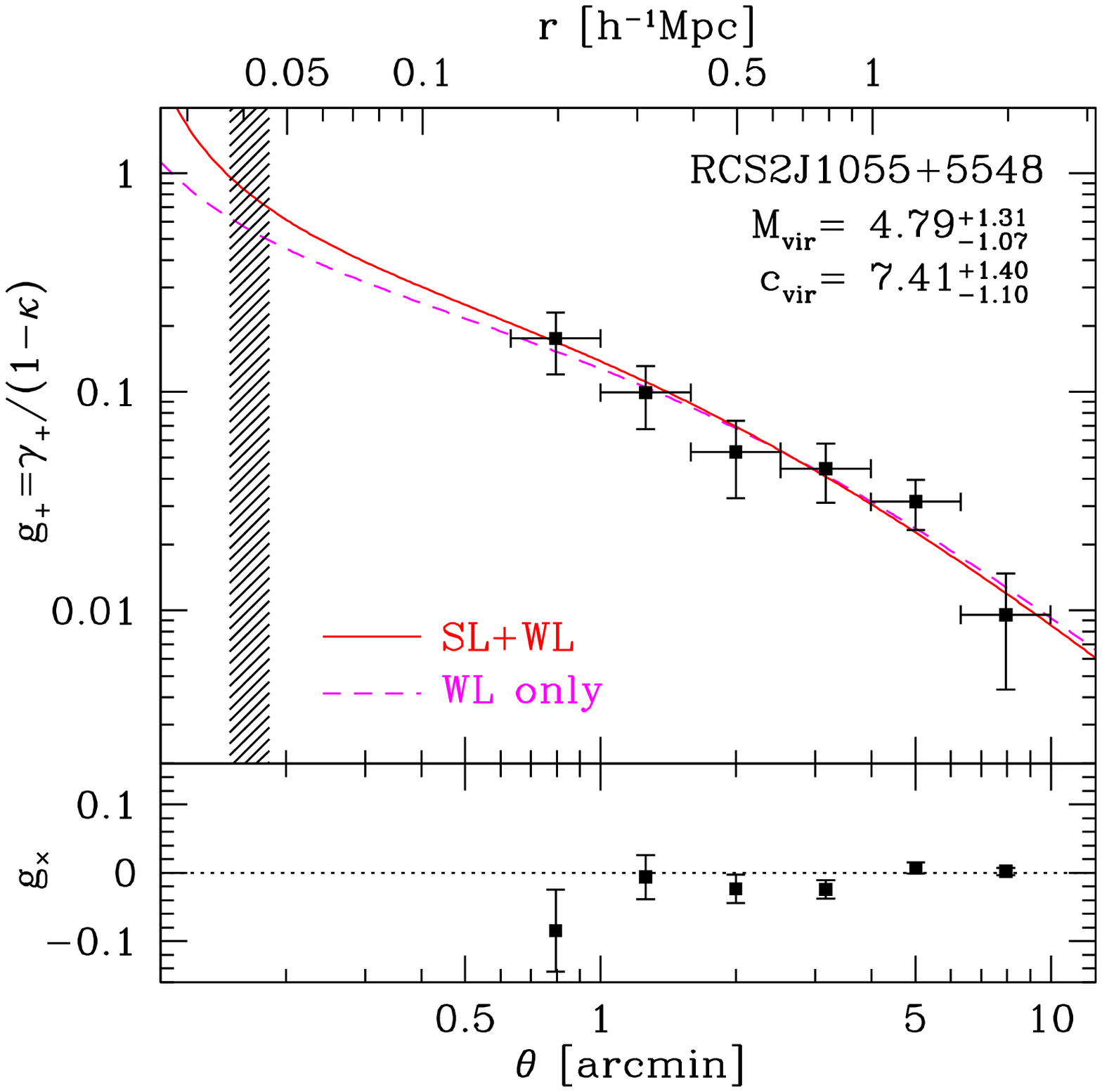}
\end{center}
\caption{SDSSJ1029+2623, SDSSJ1038+4849, SDSSJ1050+0017, RCS2J1055+5548}
\end{figure*}
\begin{figure*}
\begin{center}
\includegraphics[width=0.28\hsize]{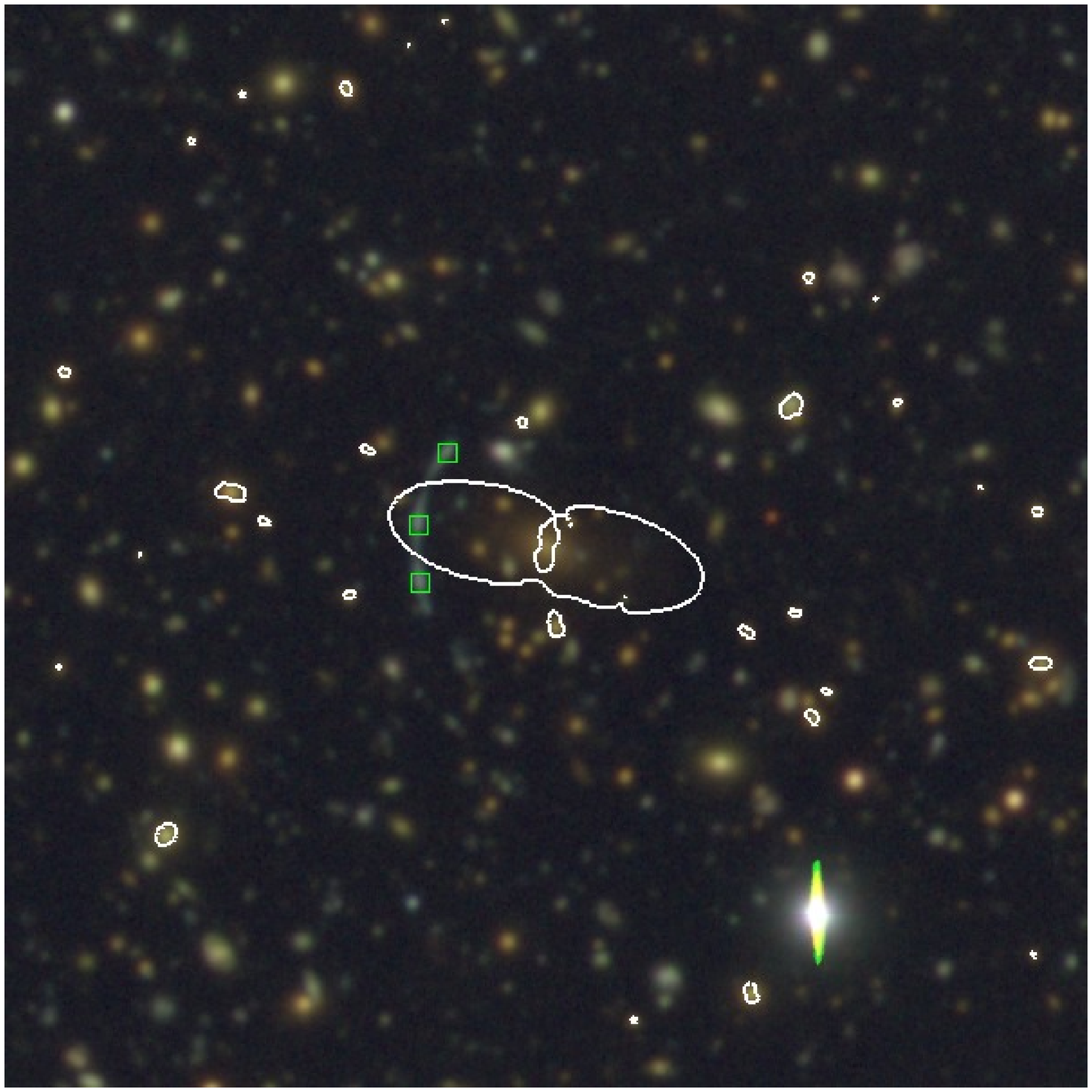}
\includegraphics[width=0.37\hsize]{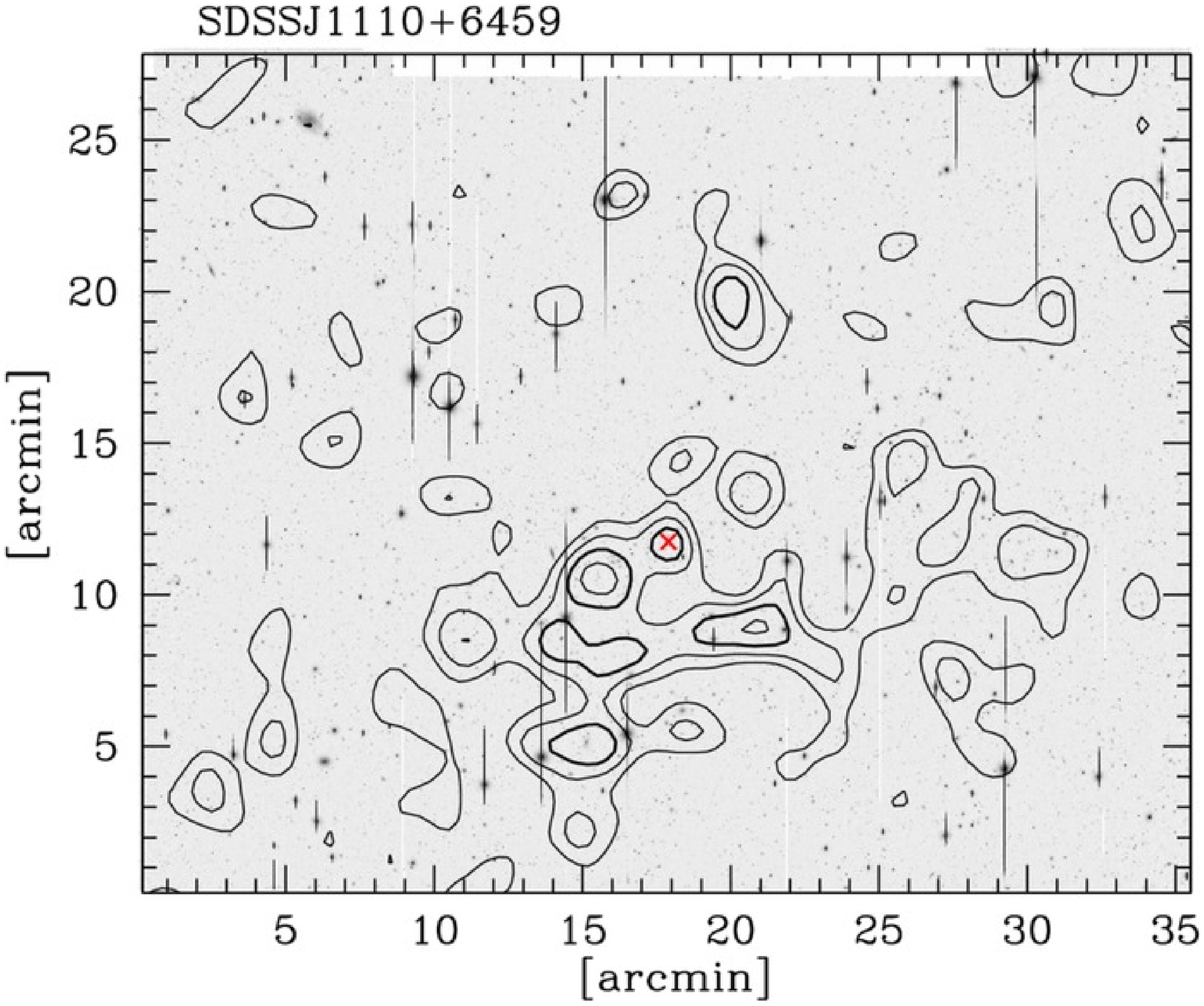}
\includegraphics[width=0.33\hsize]{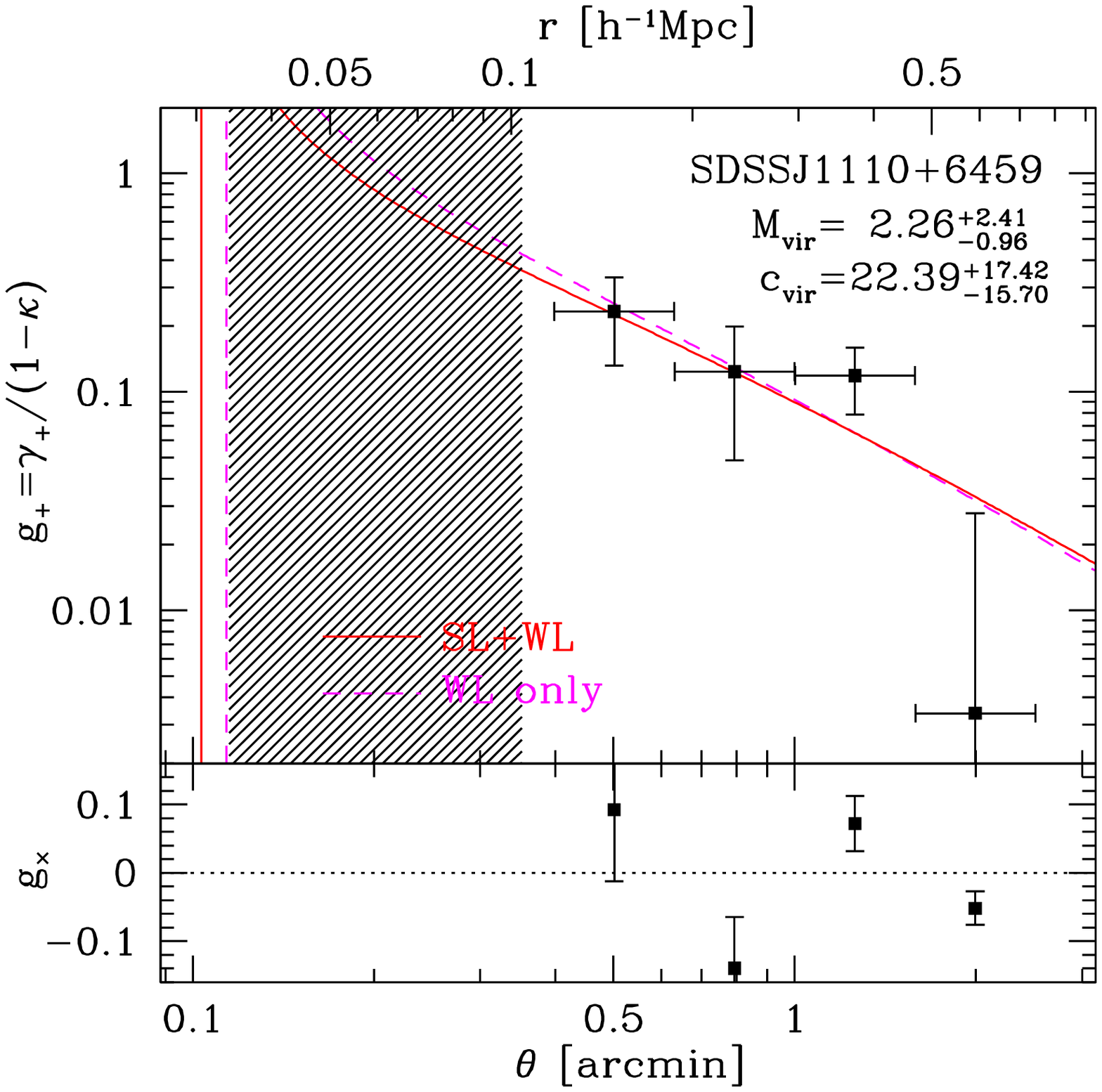}
\includegraphics[width=0.28\hsize]{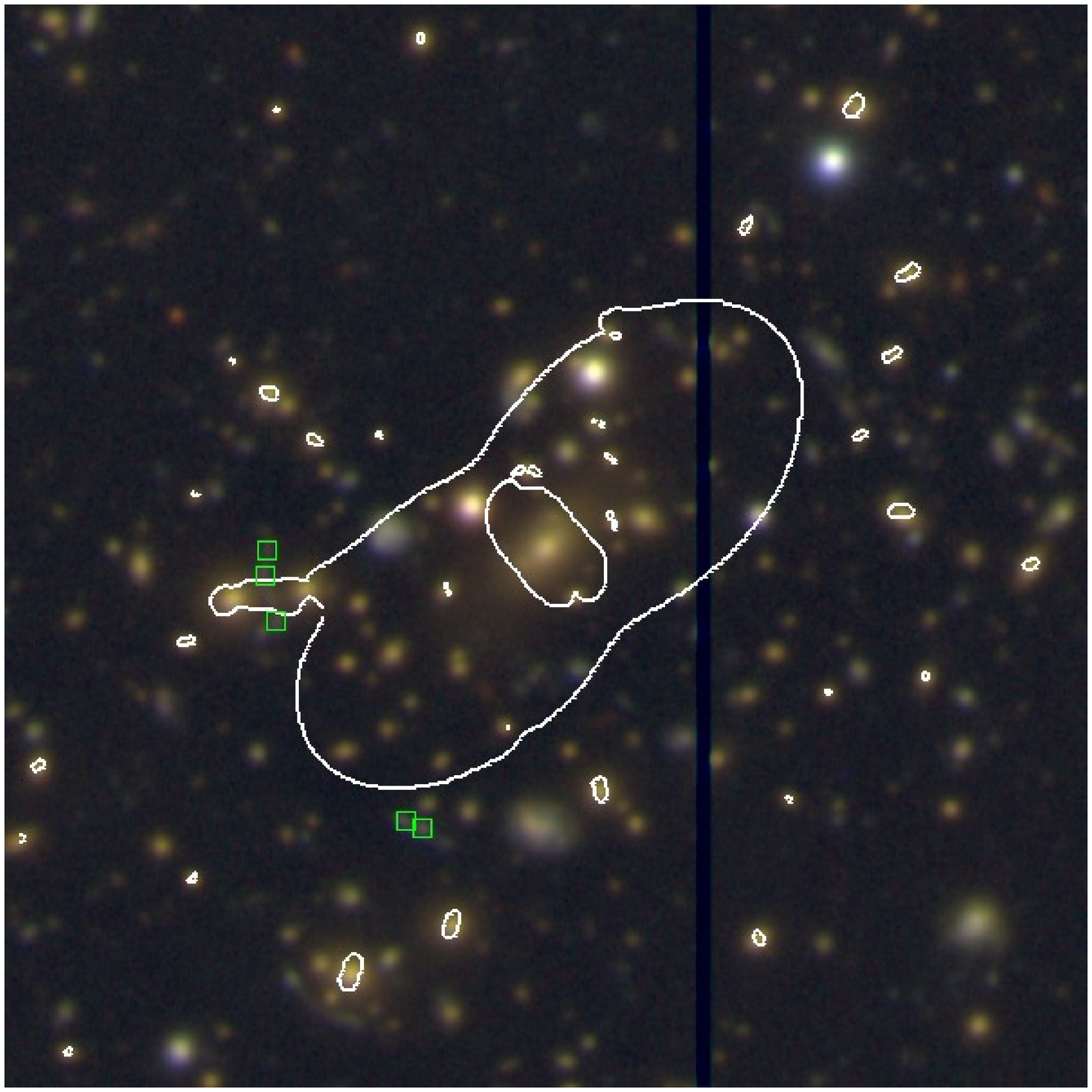}
\includegraphics[width=0.37\hsize]{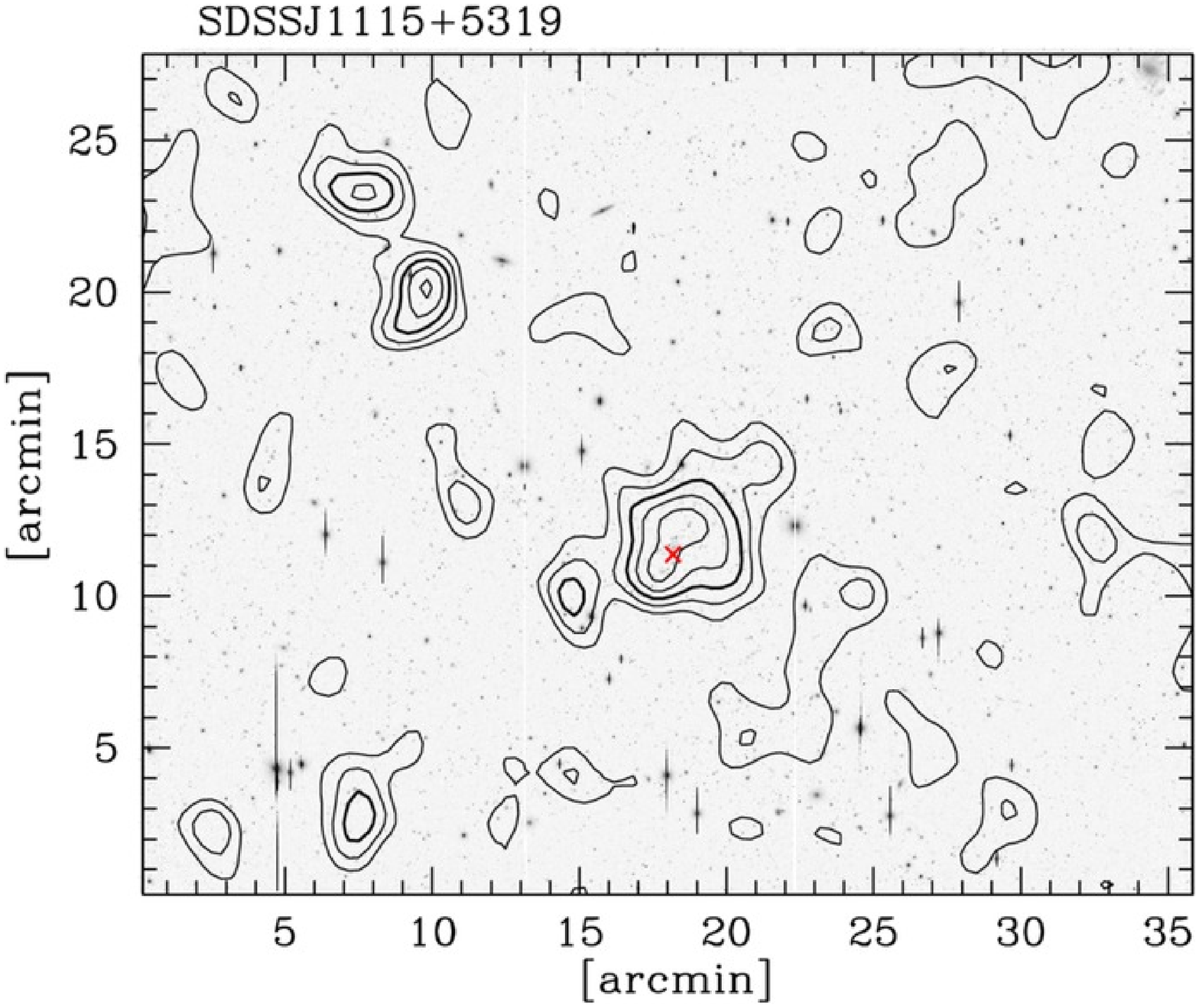}
\includegraphics[width=0.33\hsize]{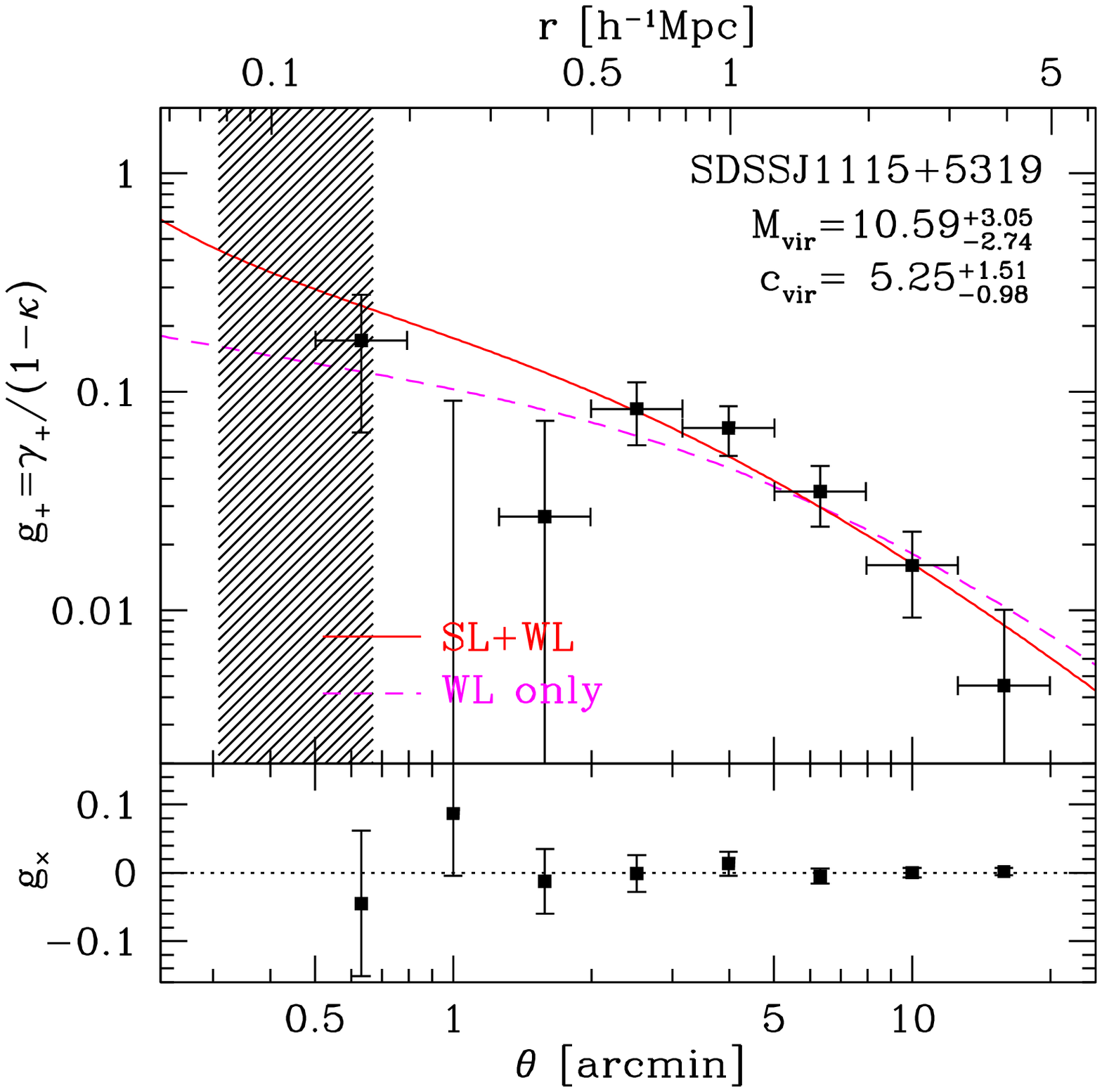}
\includegraphics[width=0.28\hsize]{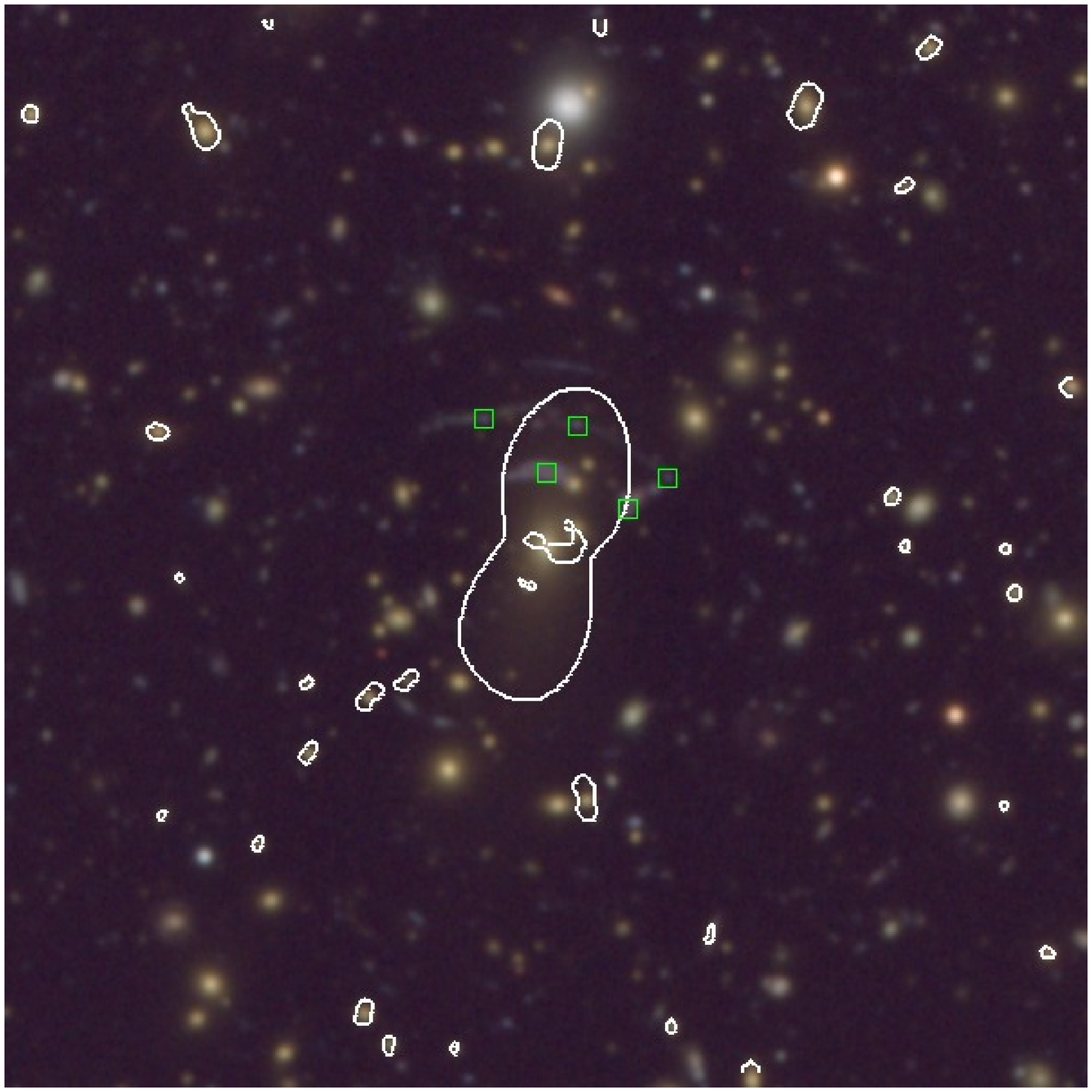}
\includegraphics[width=0.37\hsize]{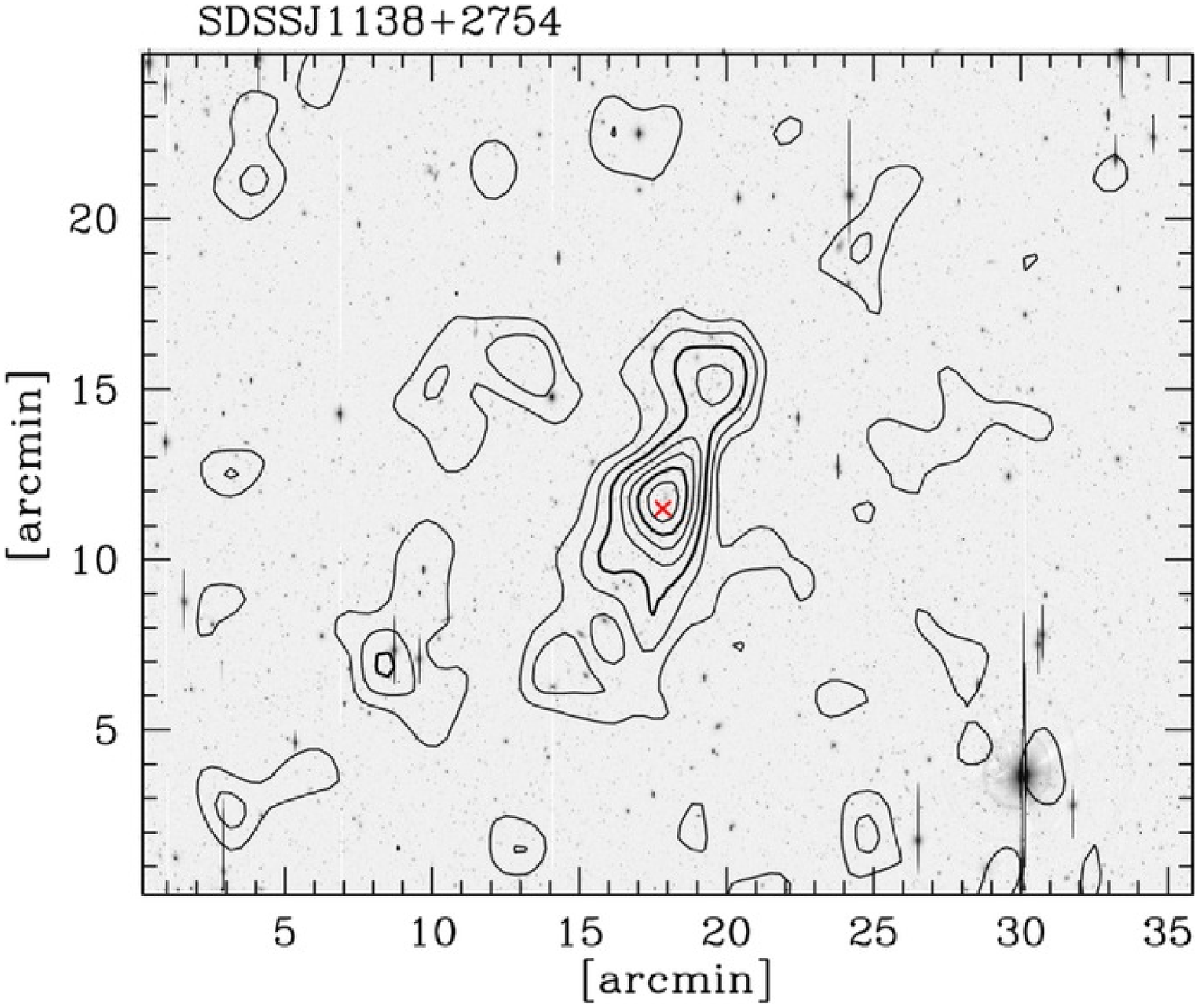}
\includegraphics[width=0.33\hsize]{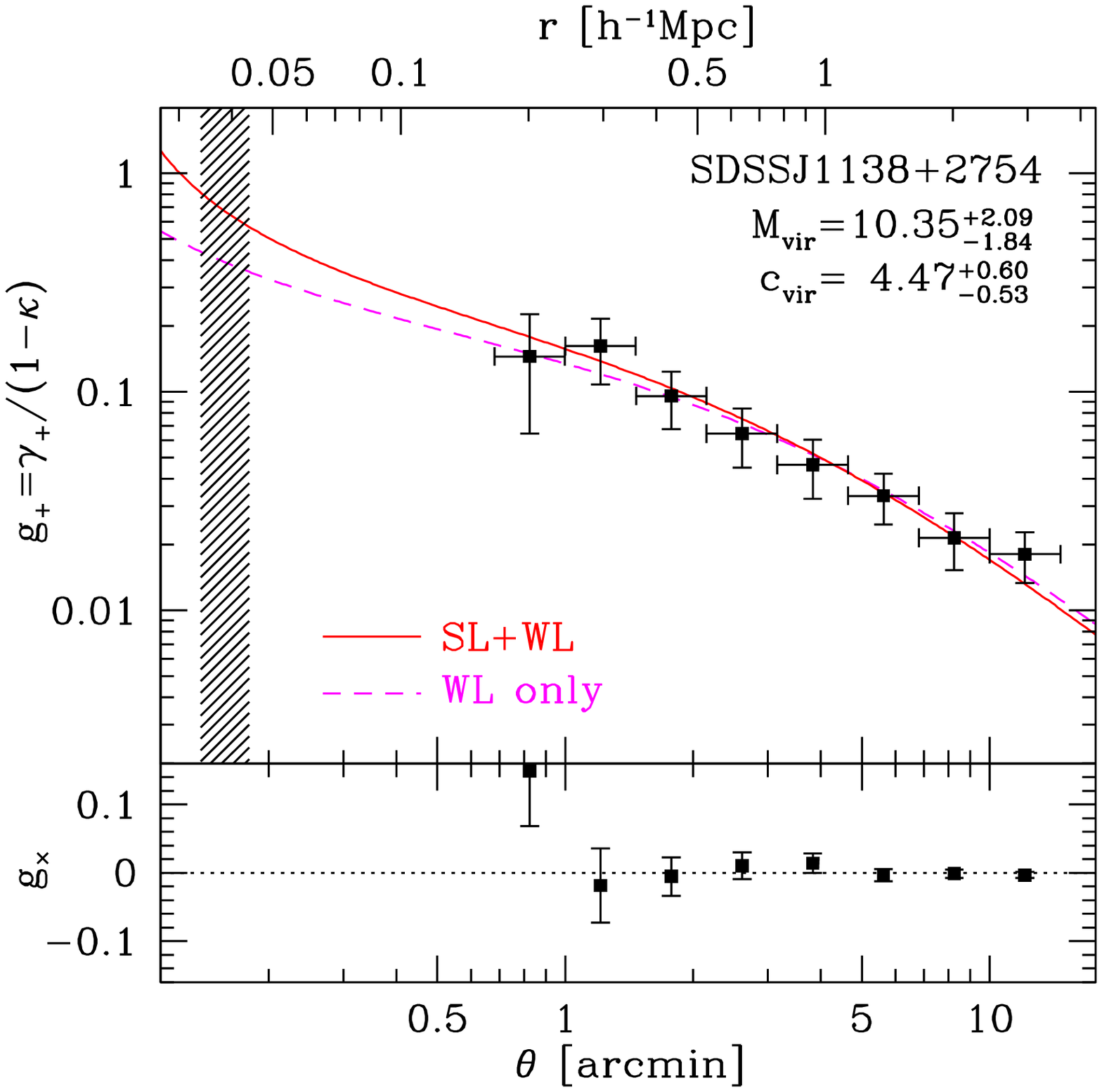}
\includegraphics[width=0.28\hsize]{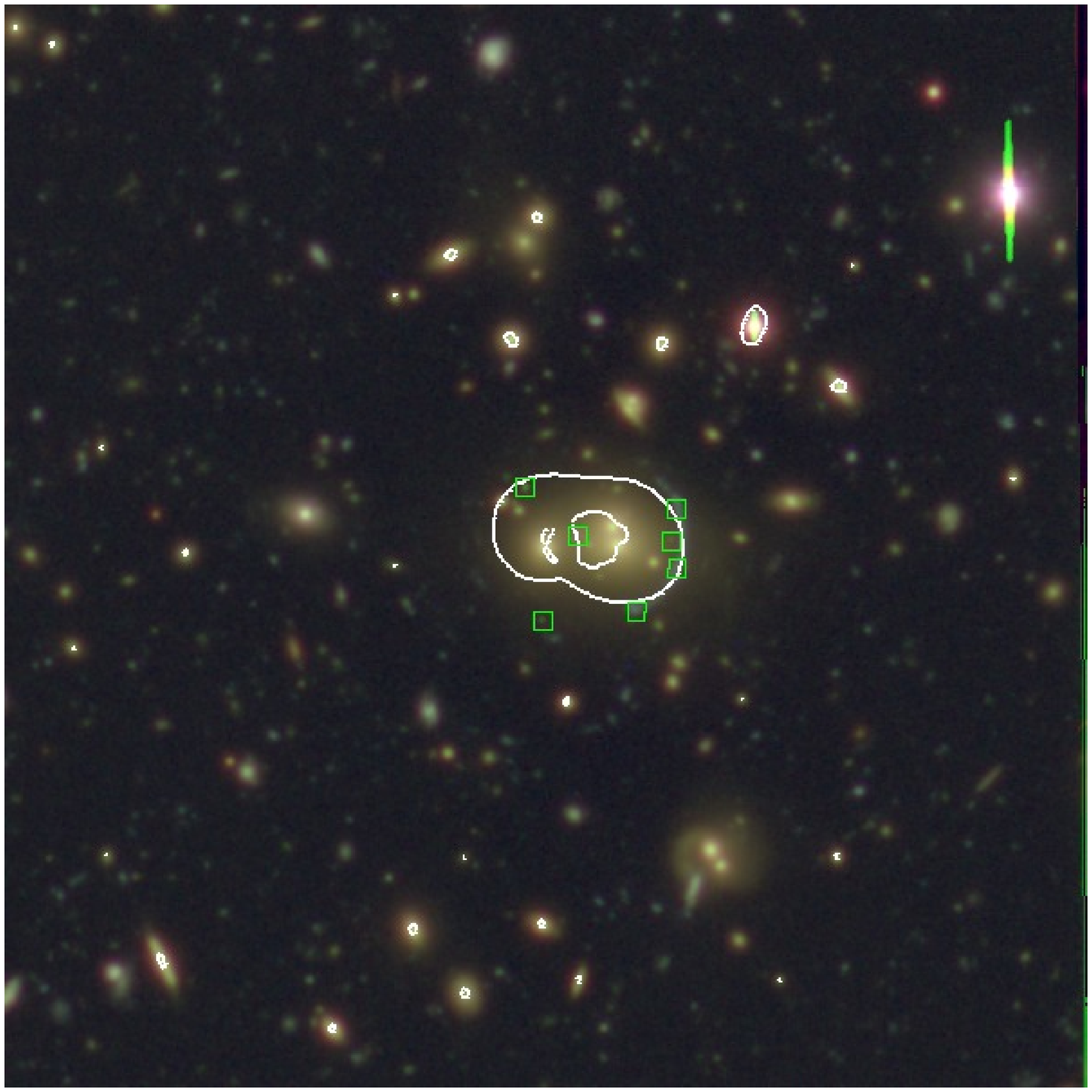}
\includegraphics[width=0.37\hsize]{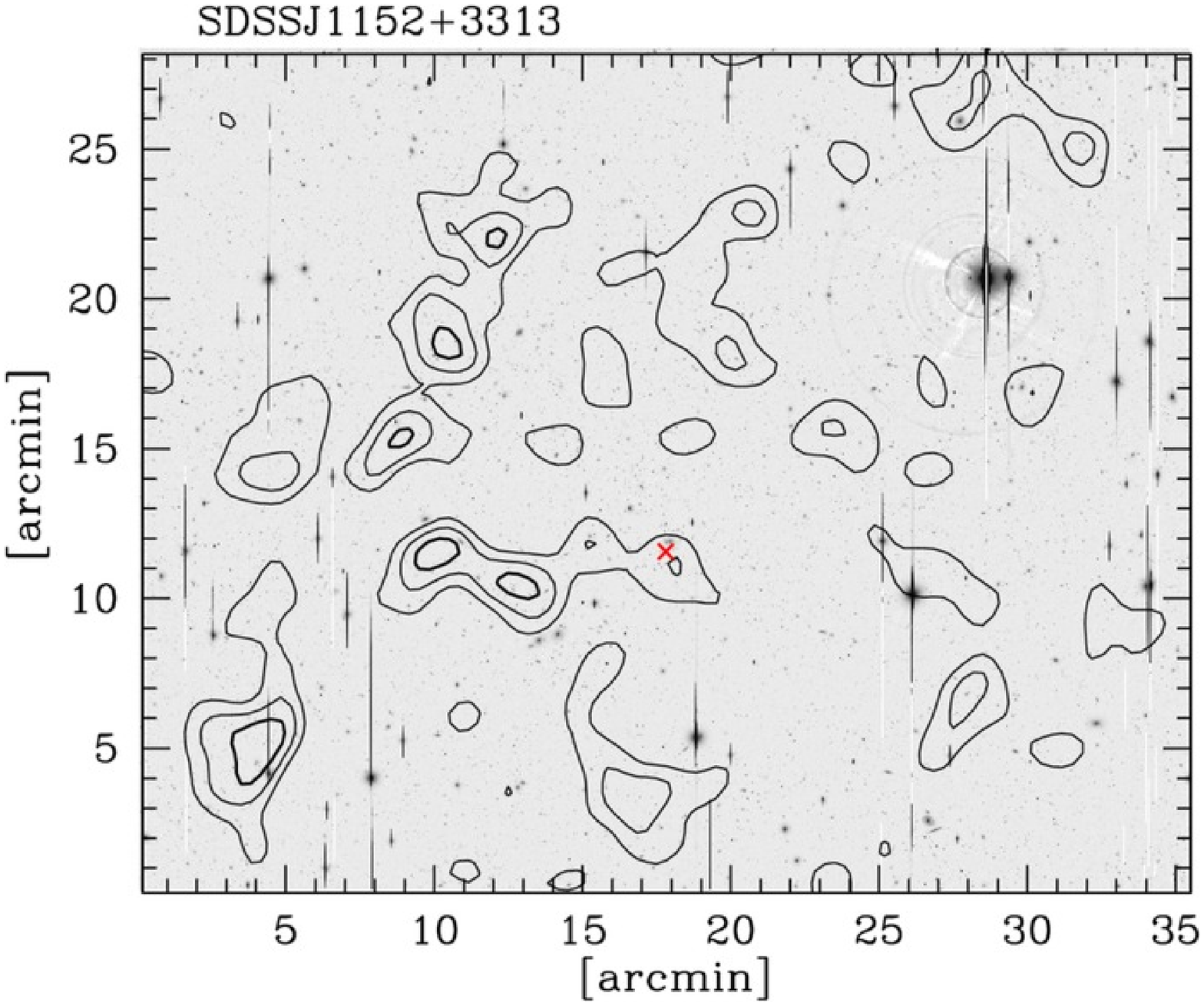}
\includegraphics[width=0.33\hsize]{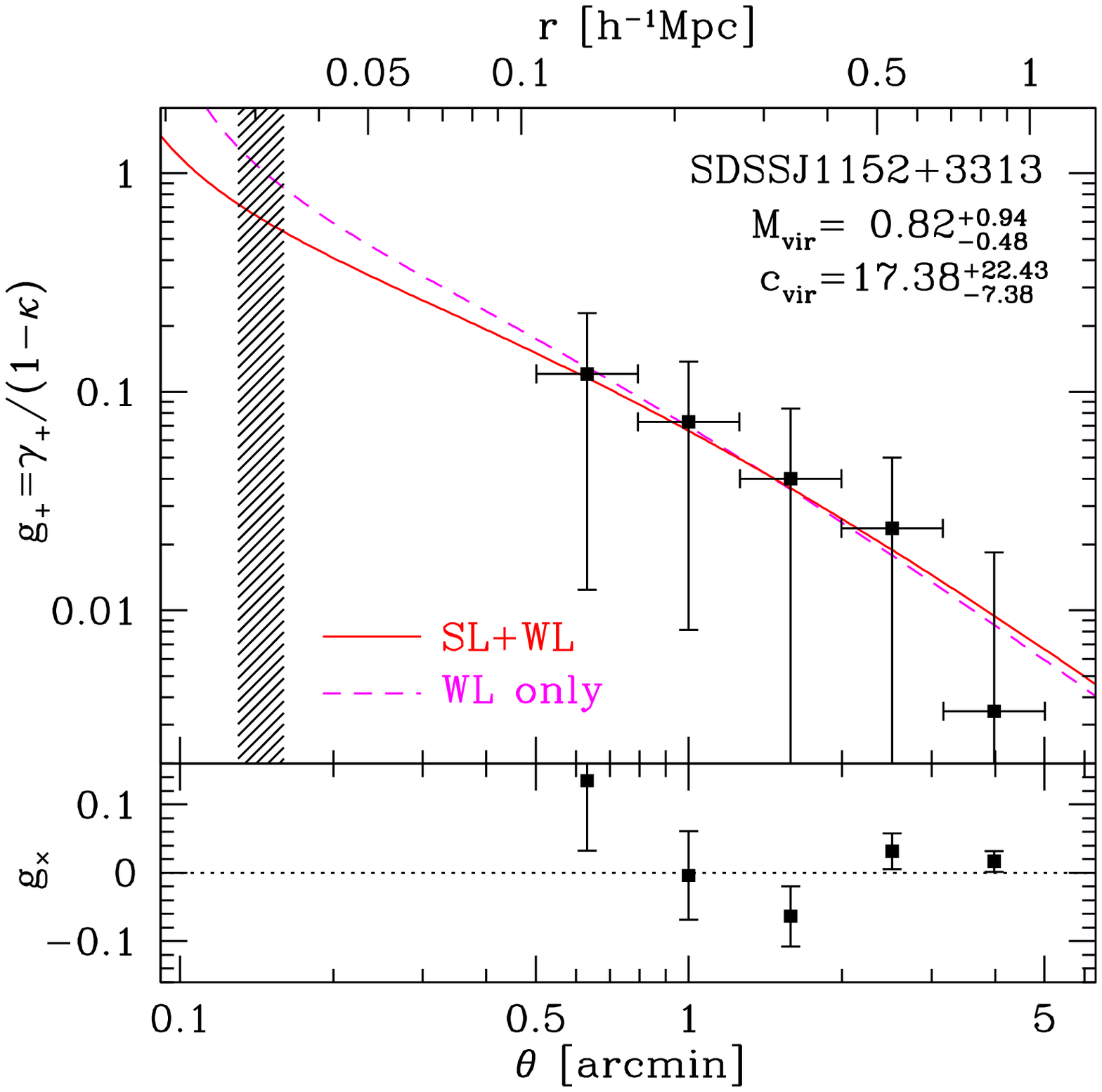}
\end{center}
\caption{SDSSJ1110+6459, SDSSJ1115+5319, SDSSJ1138+2754, SDSSJ1152+3313}
\end{figure*}
\begin{figure*}
\begin{center}
\includegraphics[width=0.28\hsize]{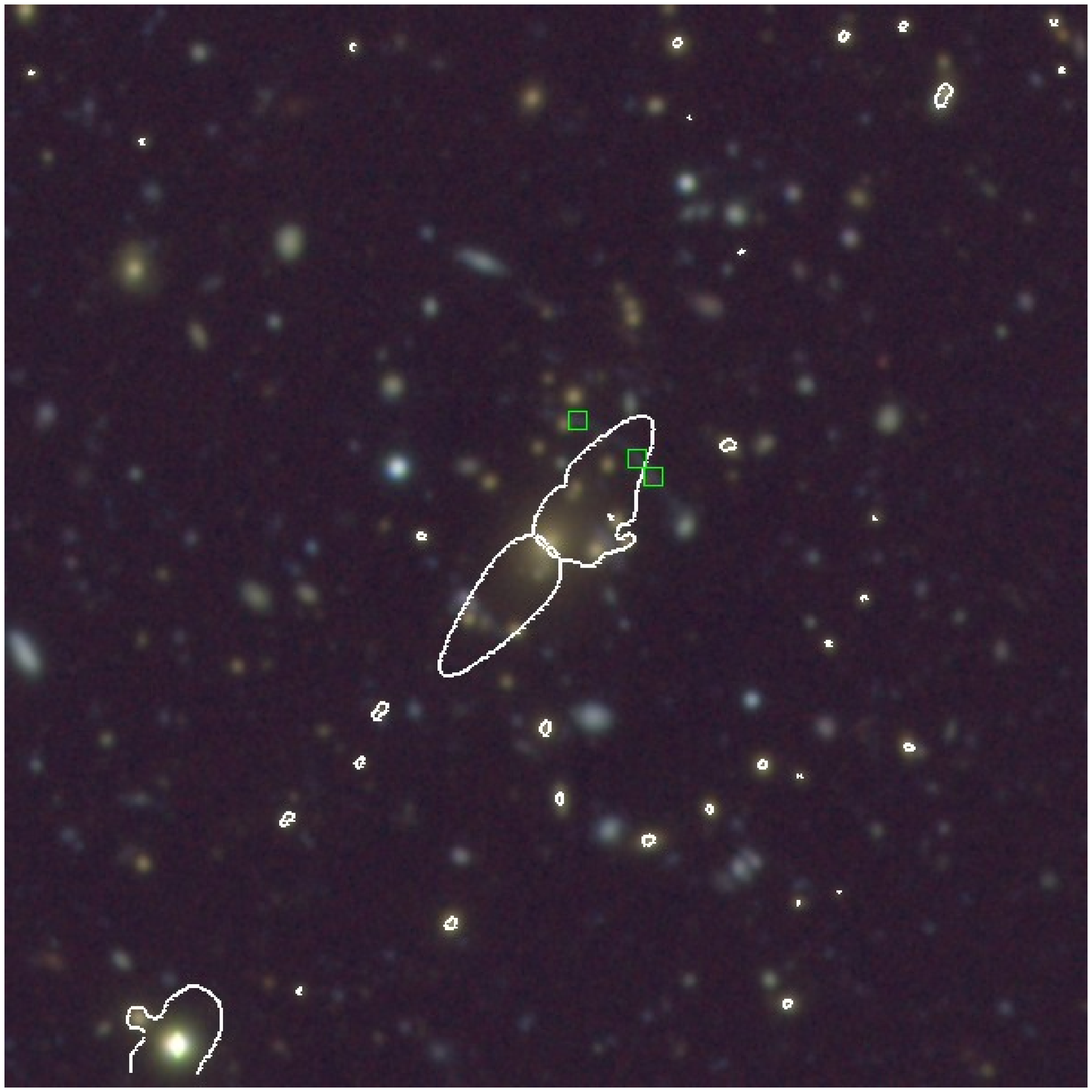}
\includegraphics[width=0.37\hsize]{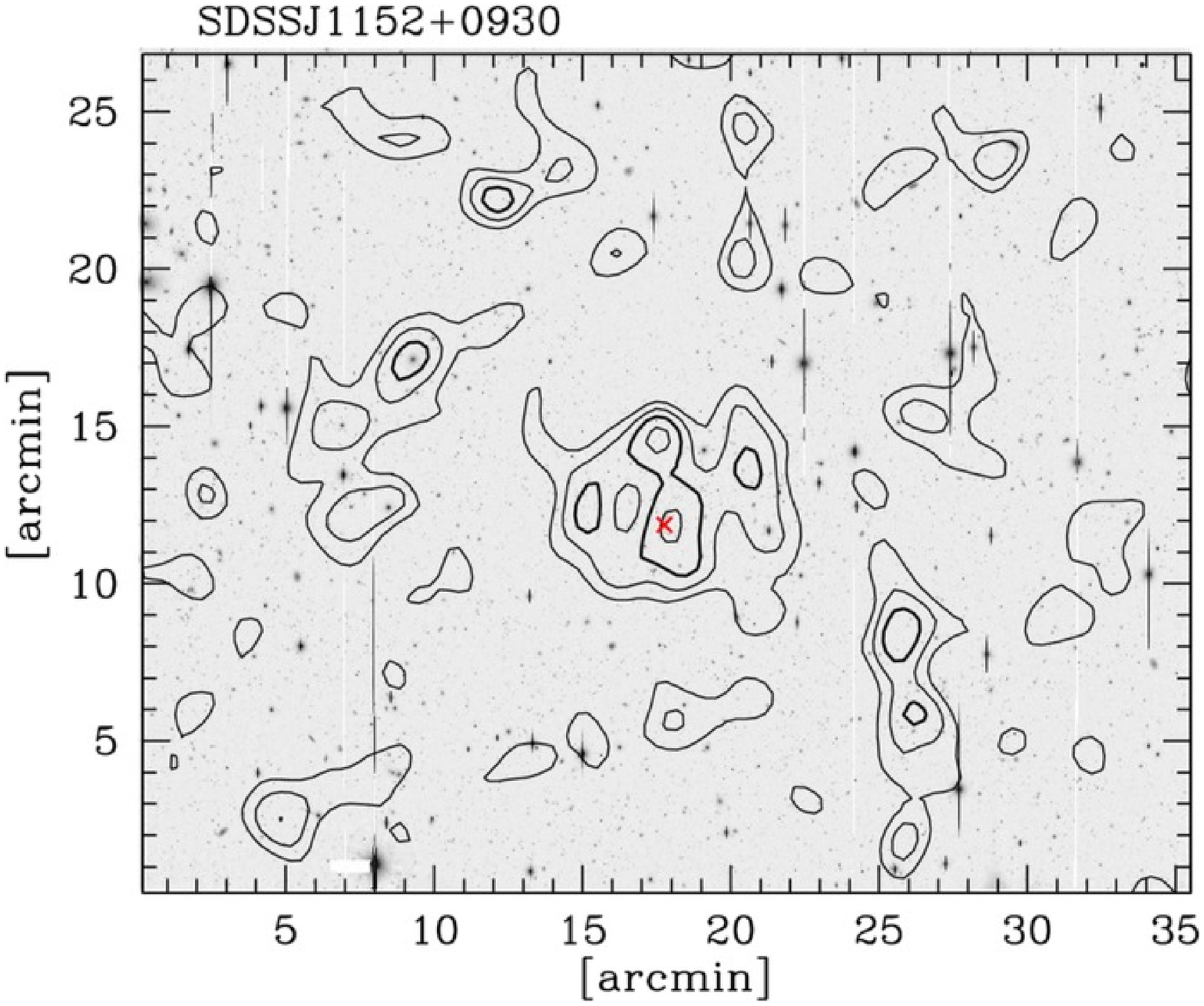}
\includegraphics[width=0.33\hsize]{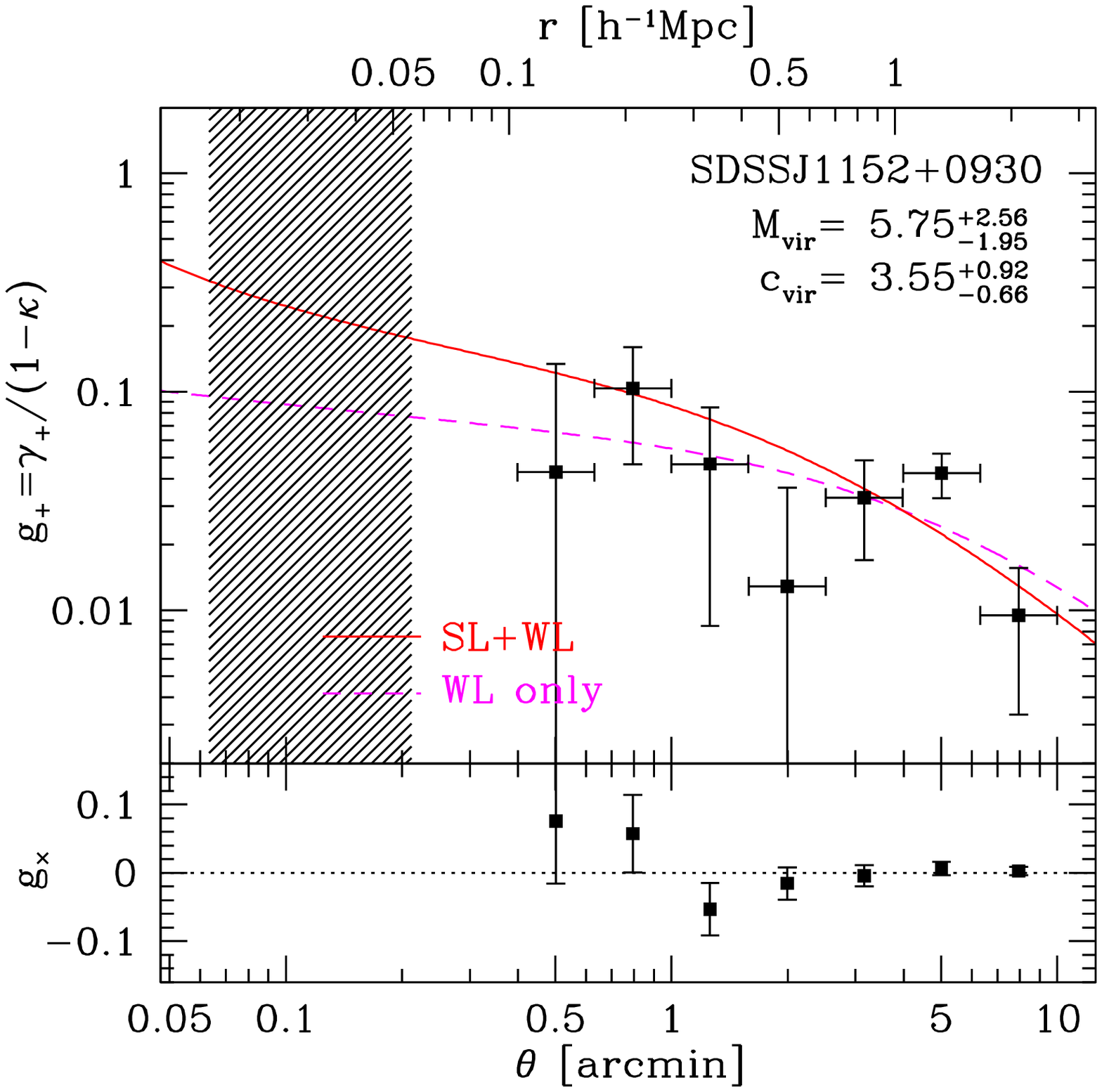}
\includegraphics[width=0.28\hsize]{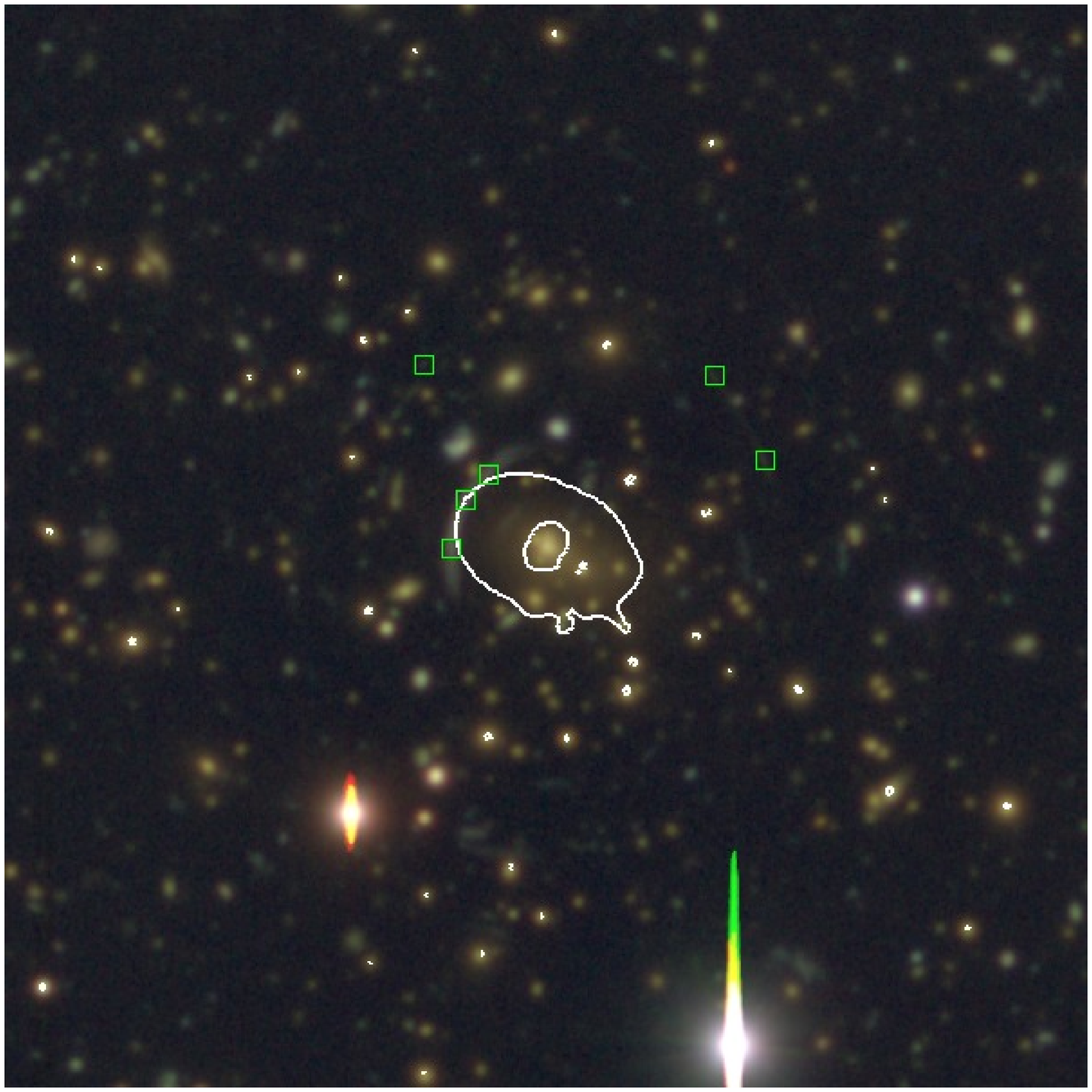}
\includegraphics[width=0.37\hsize]{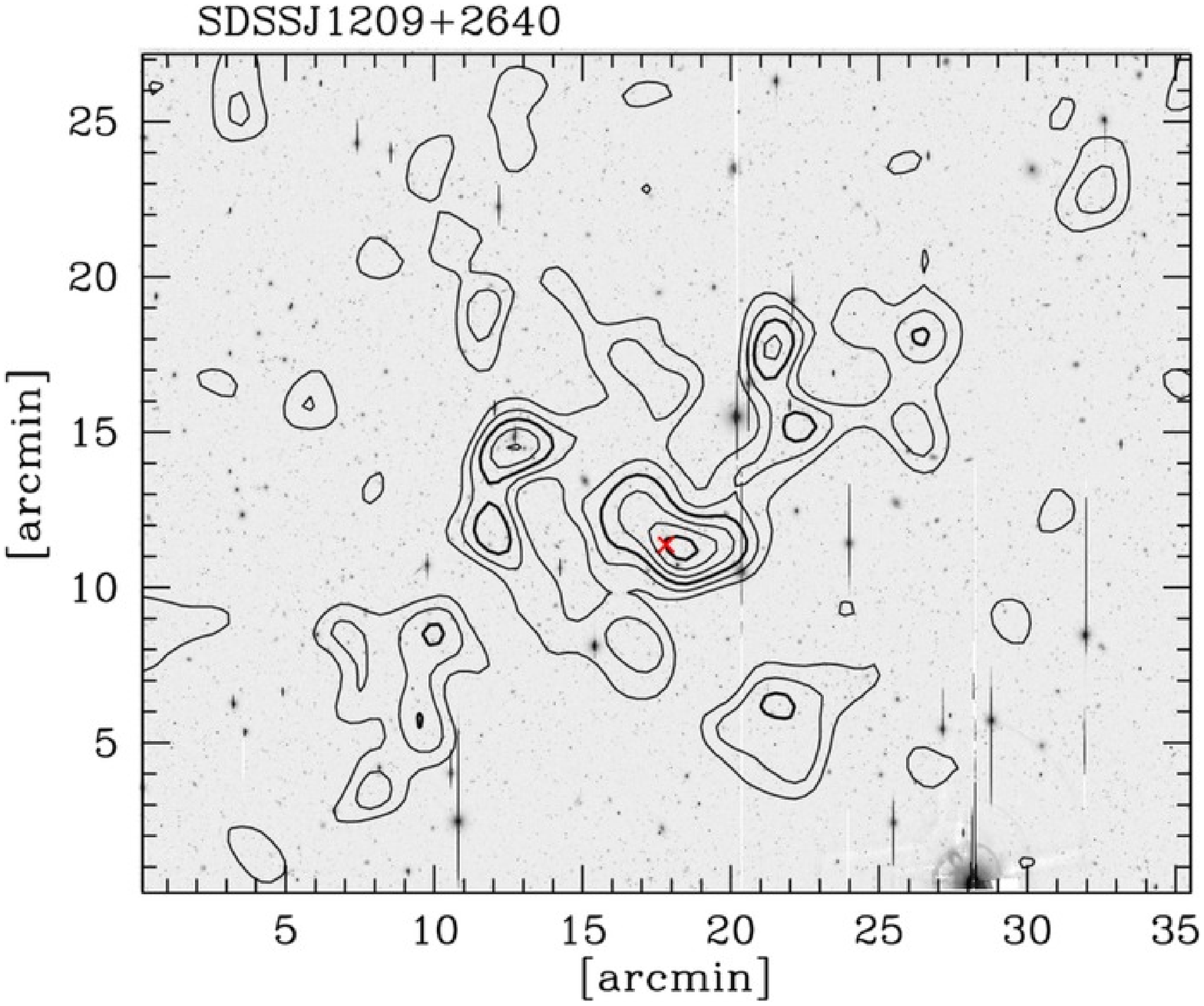}
\includegraphics[width=0.33\hsize]{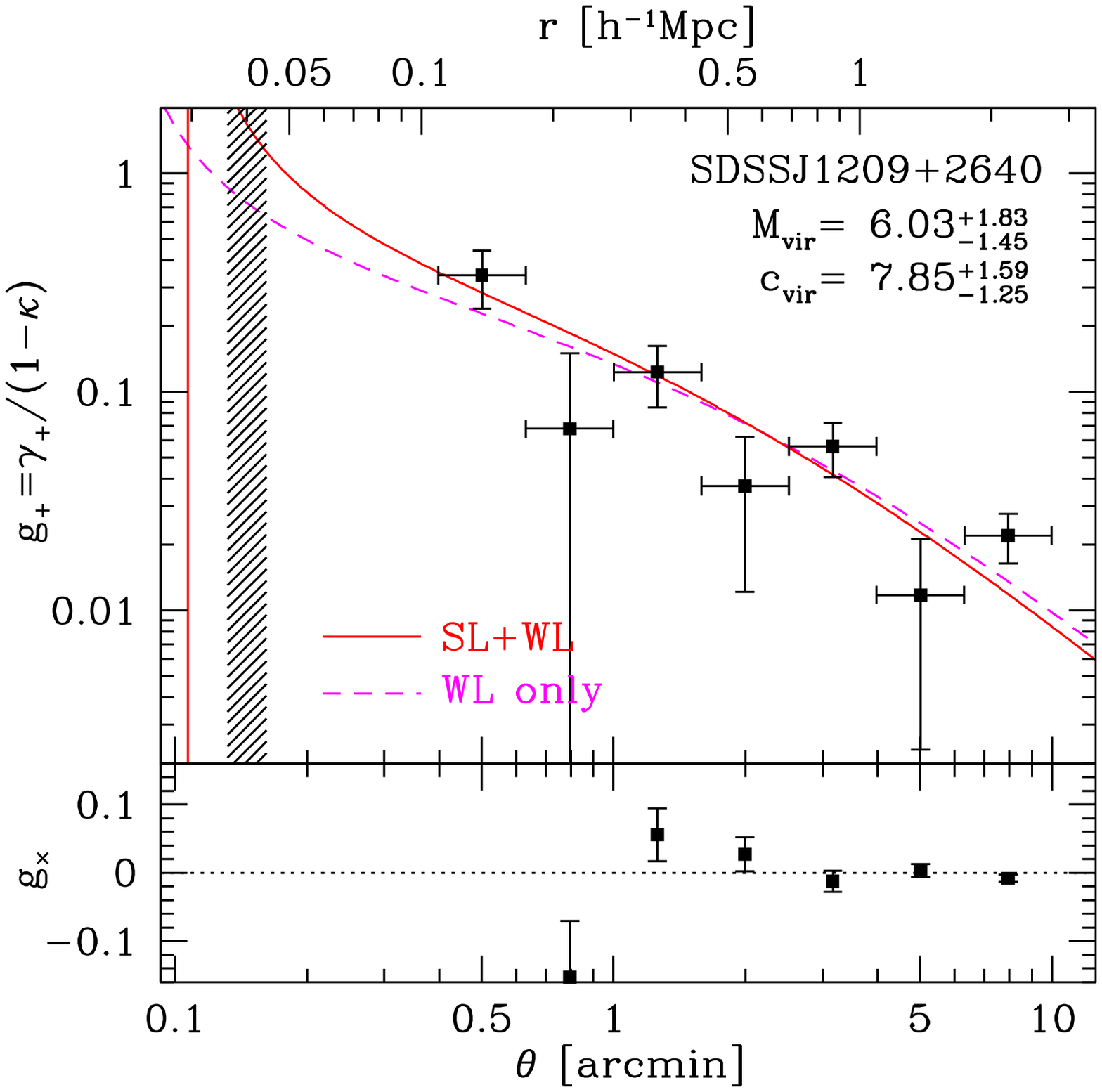}
\includegraphics[width=0.28\hsize]{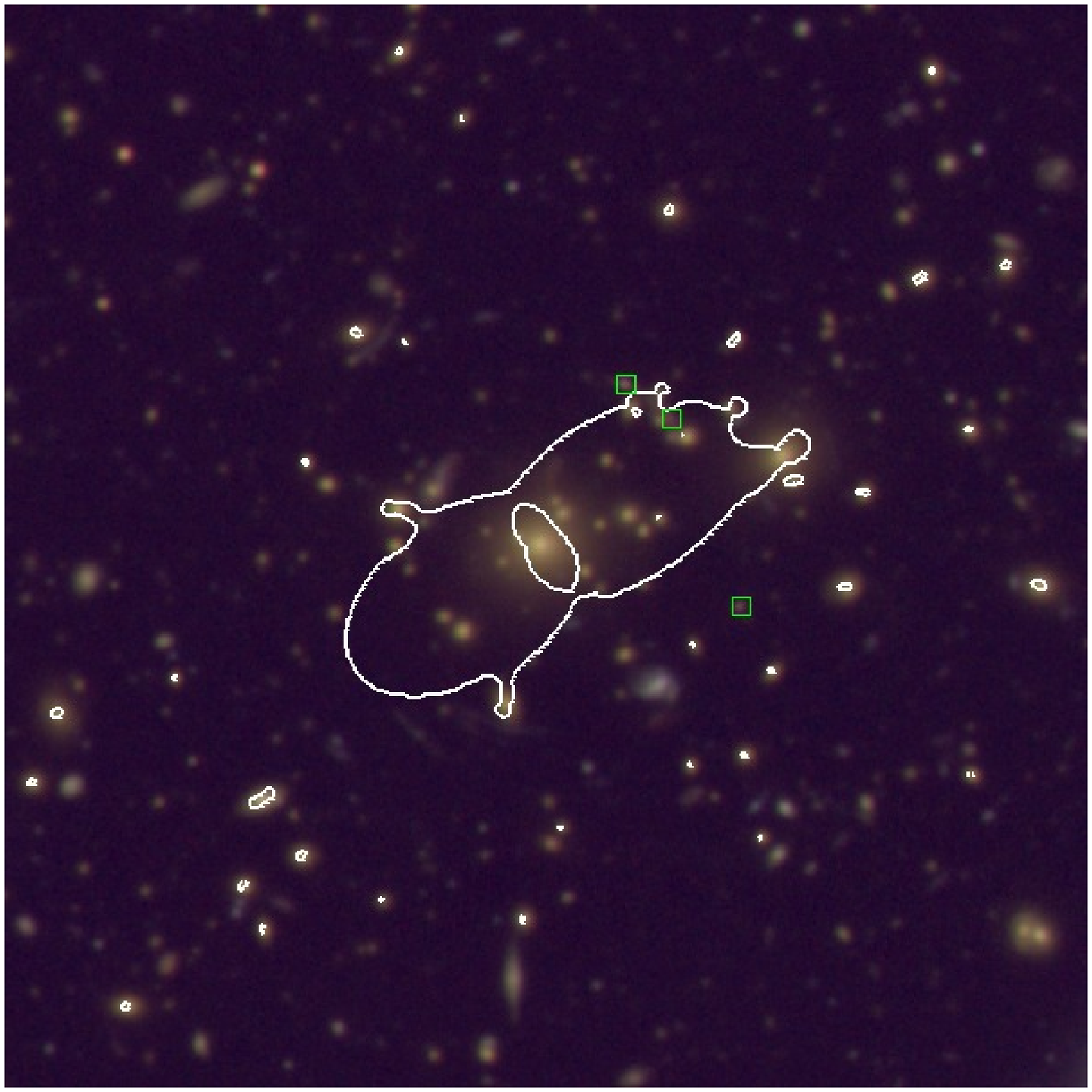}
\includegraphics[width=0.37\hsize]{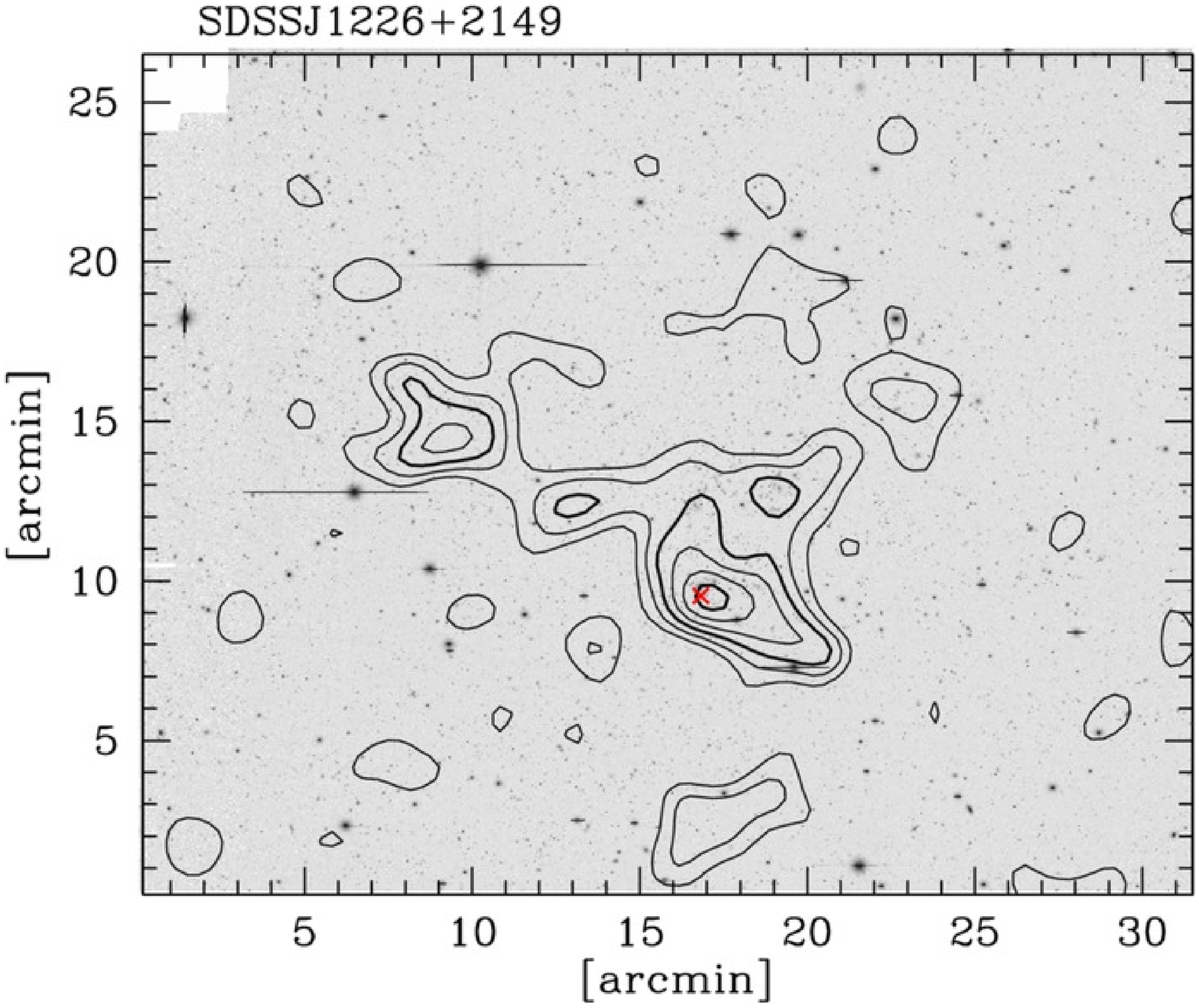}
\includegraphics[width=0.33\hsize]{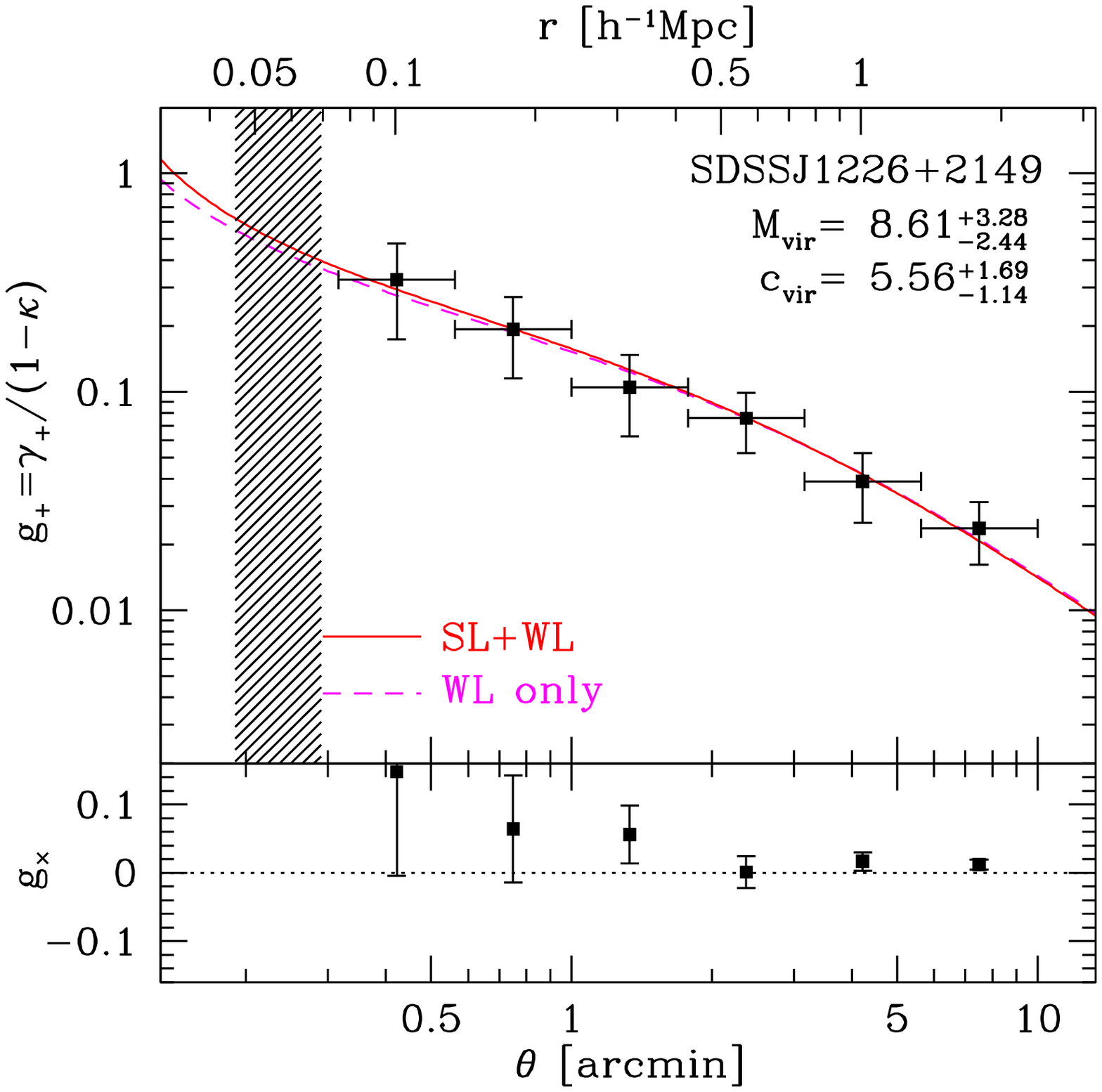}
\includegraphics[width=0.28\hsize]{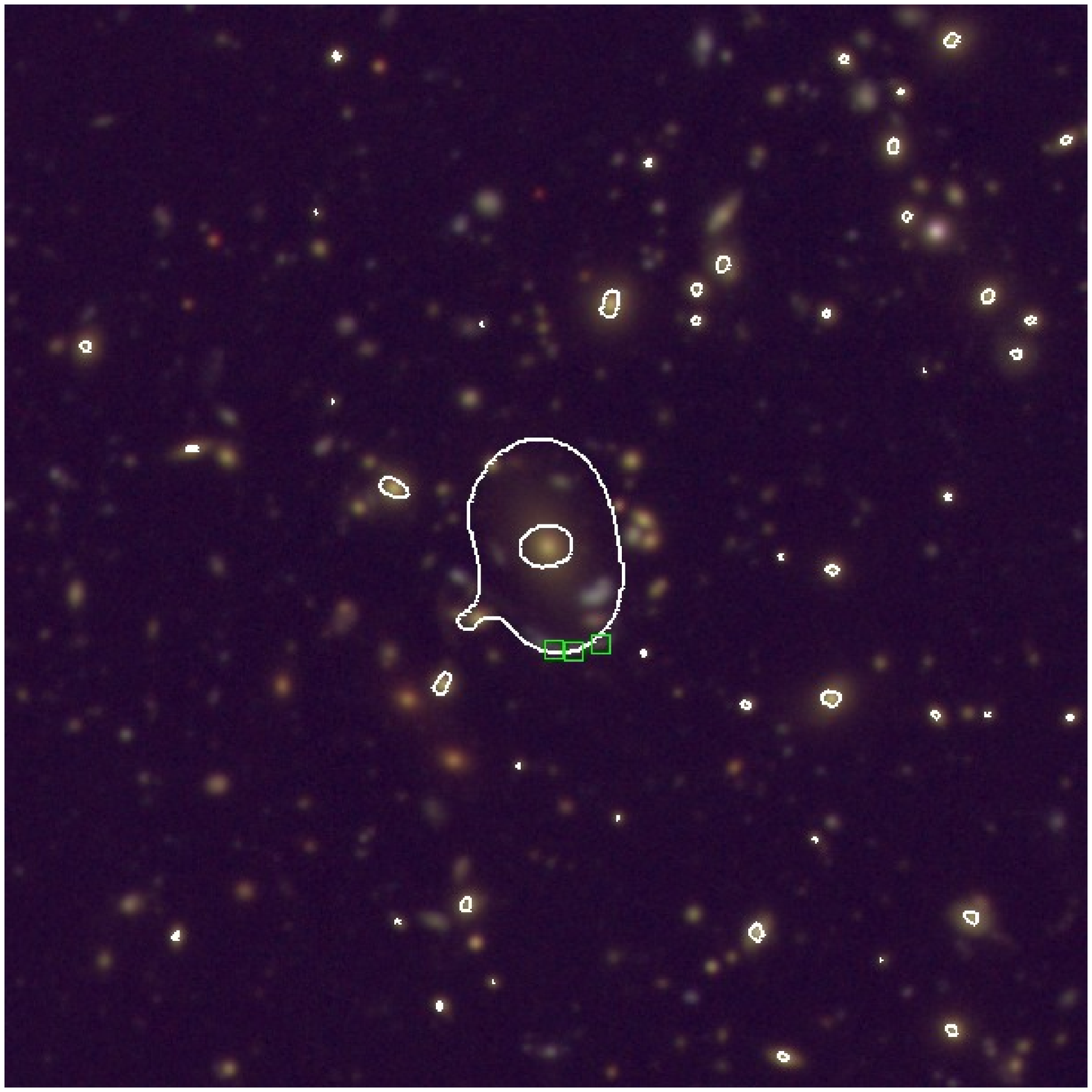}
\includegraphics[width=0.37\hsize]{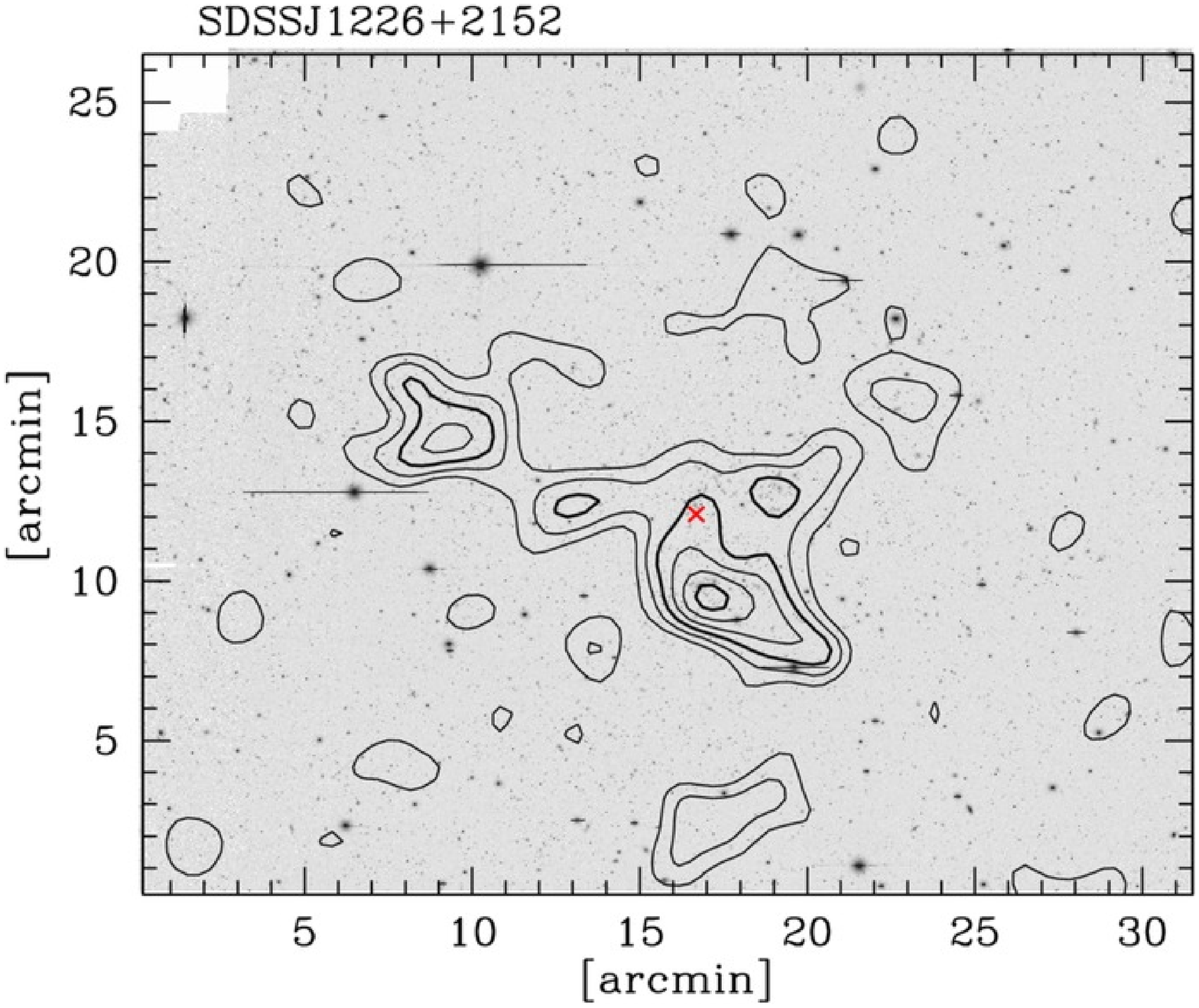}
\includegraphics[width=0.33\hsize]{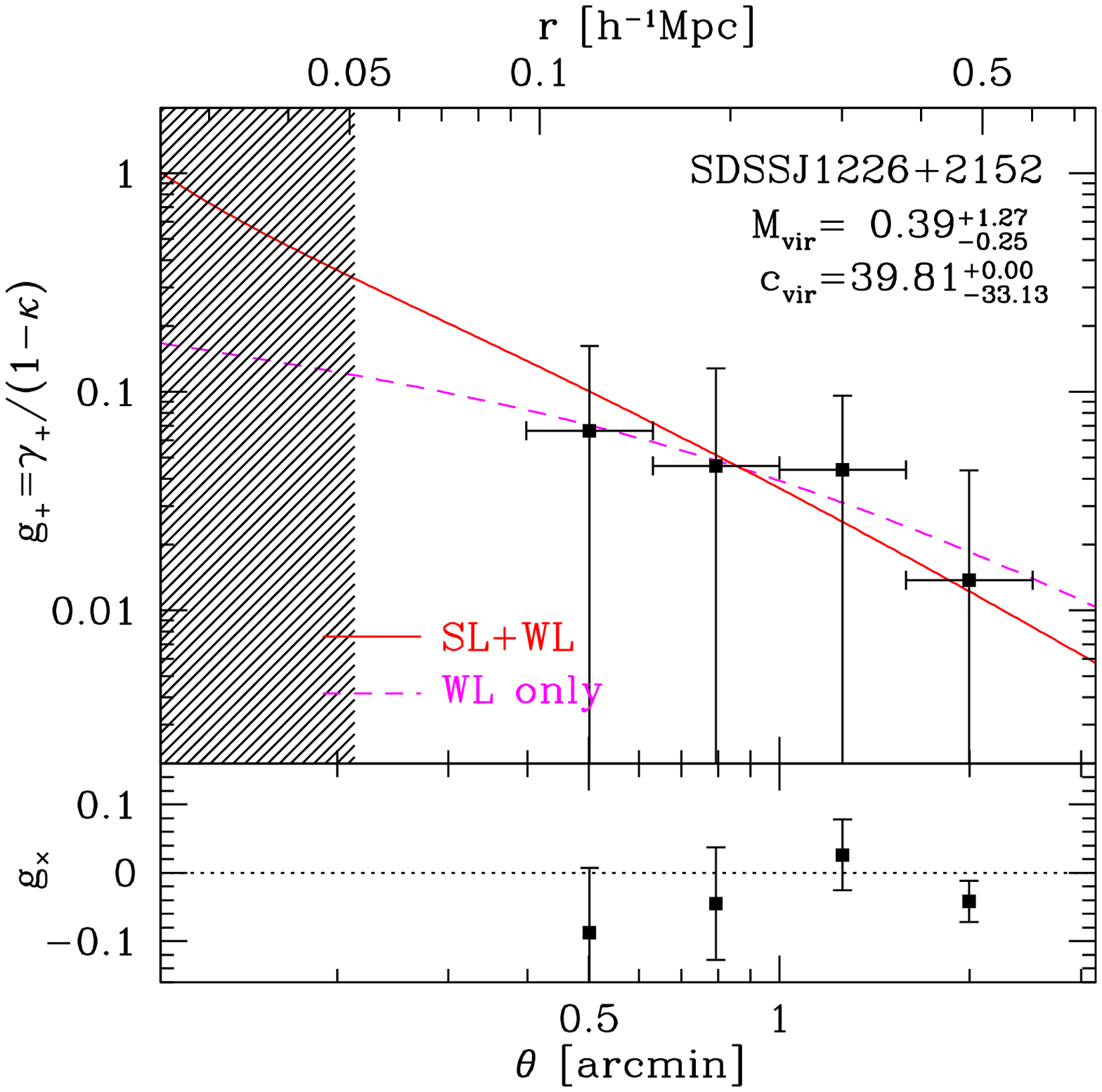}
\end{center}
\caption{SDSSJ1152+0930, SDSSJ1209+2640, SDSSJ1226+2149, SDSSJ1226+2152}
\end{figure*}
\begin{figure*}
\begin{center}
\includegraphics[width=0.28\hsize]{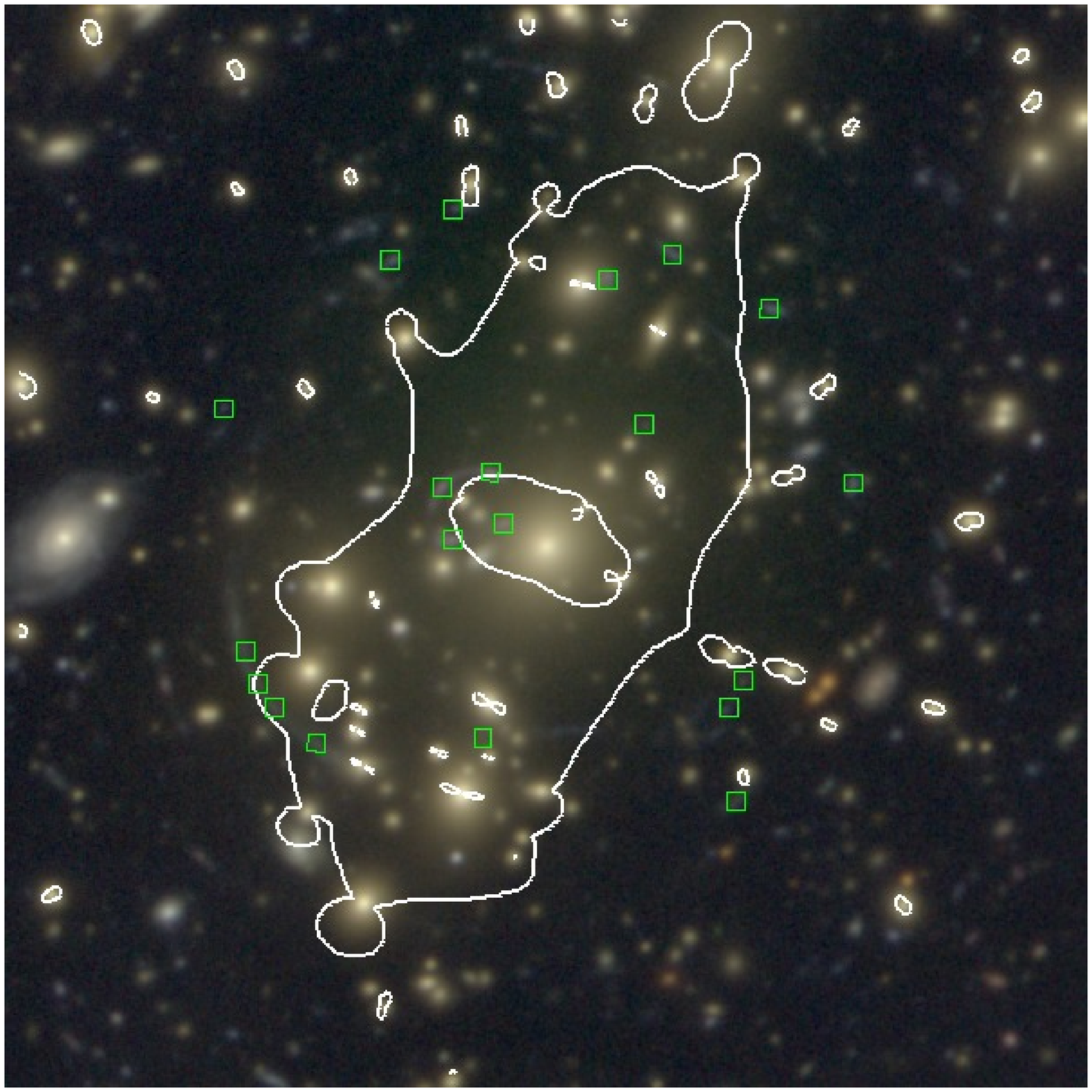}
\includegraphics[width=0.37\hsize]{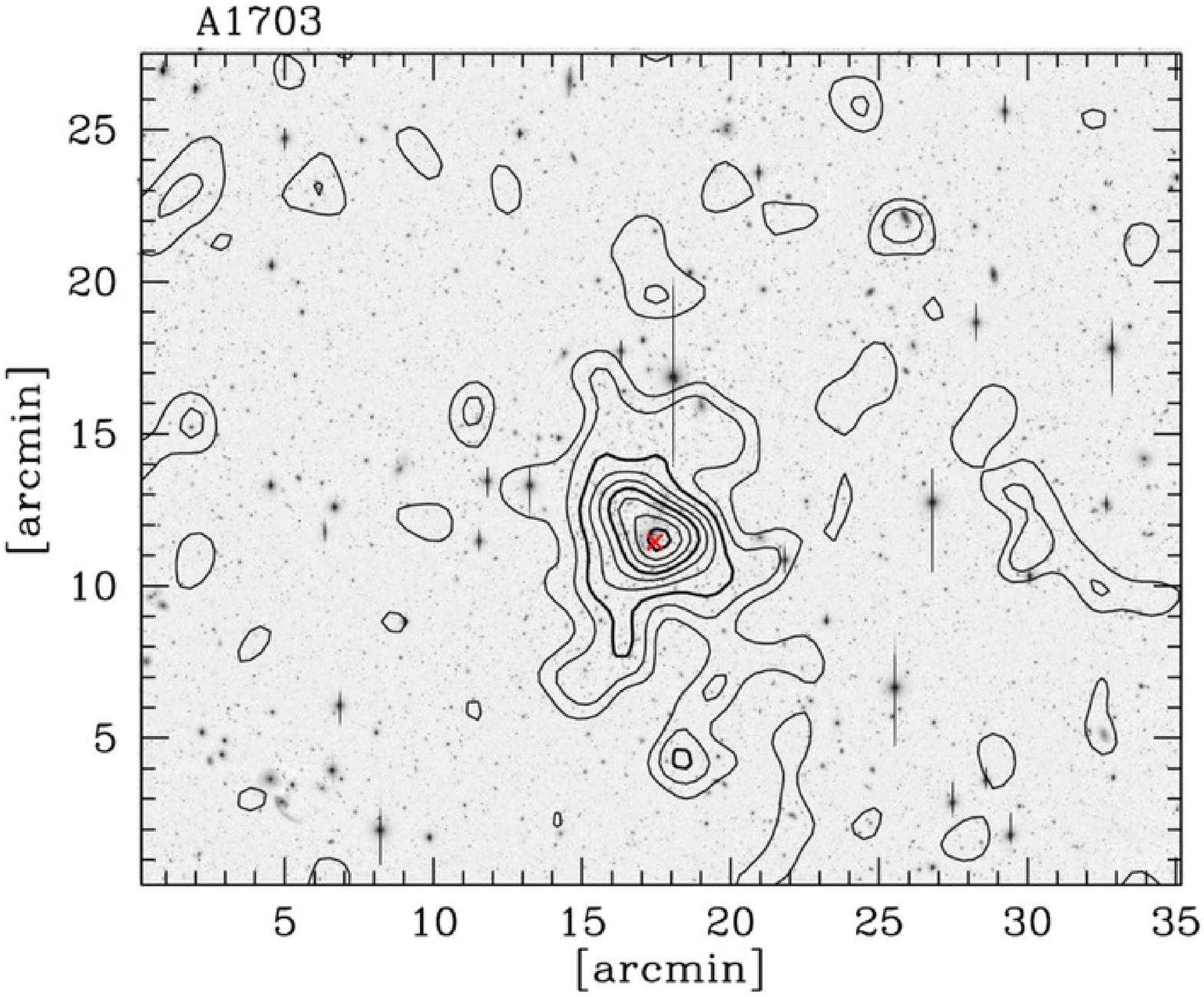}
\includegraphics[width=0.33\hsize]{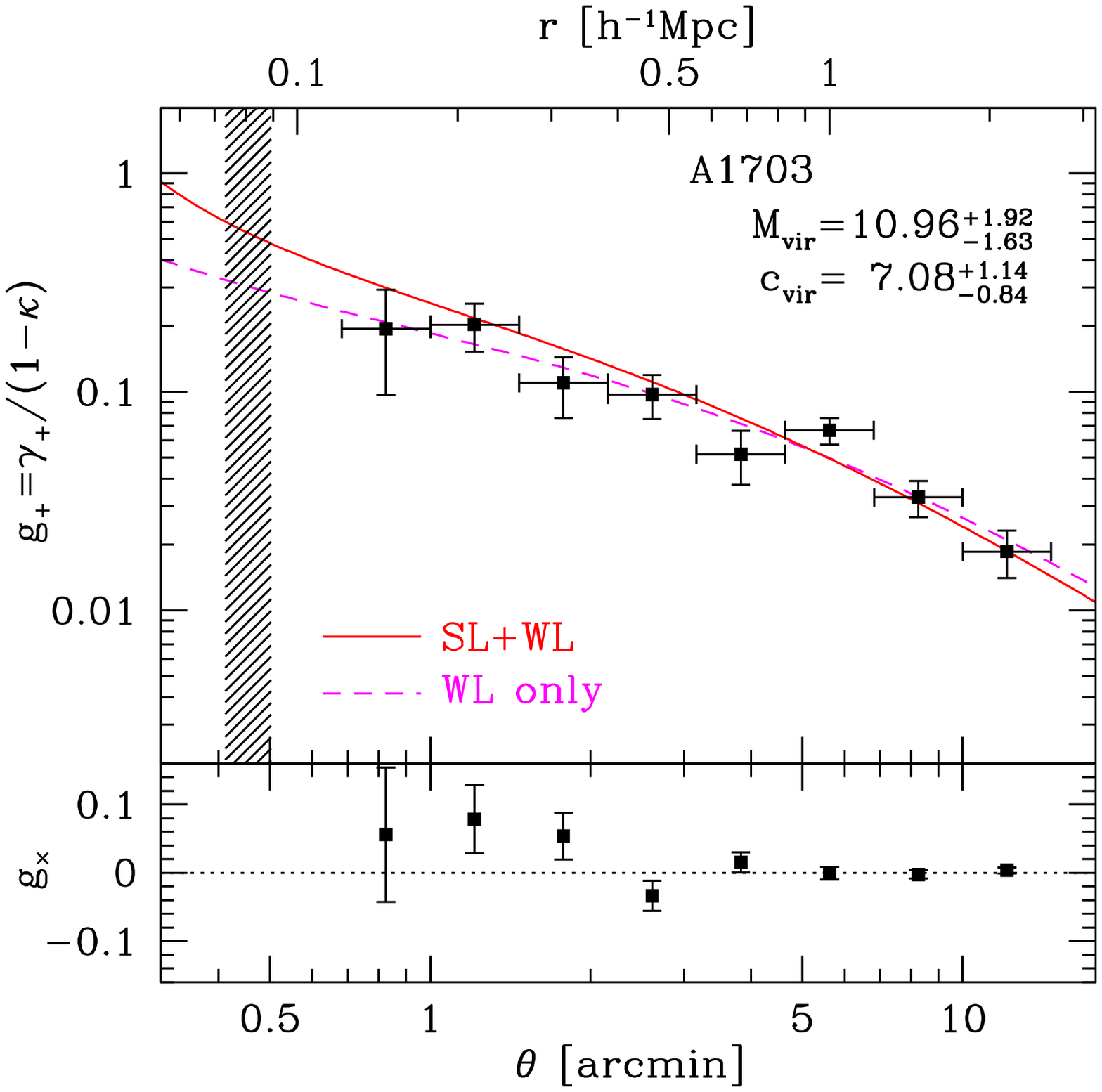}
\includegraphics[width=0.28\hsize]{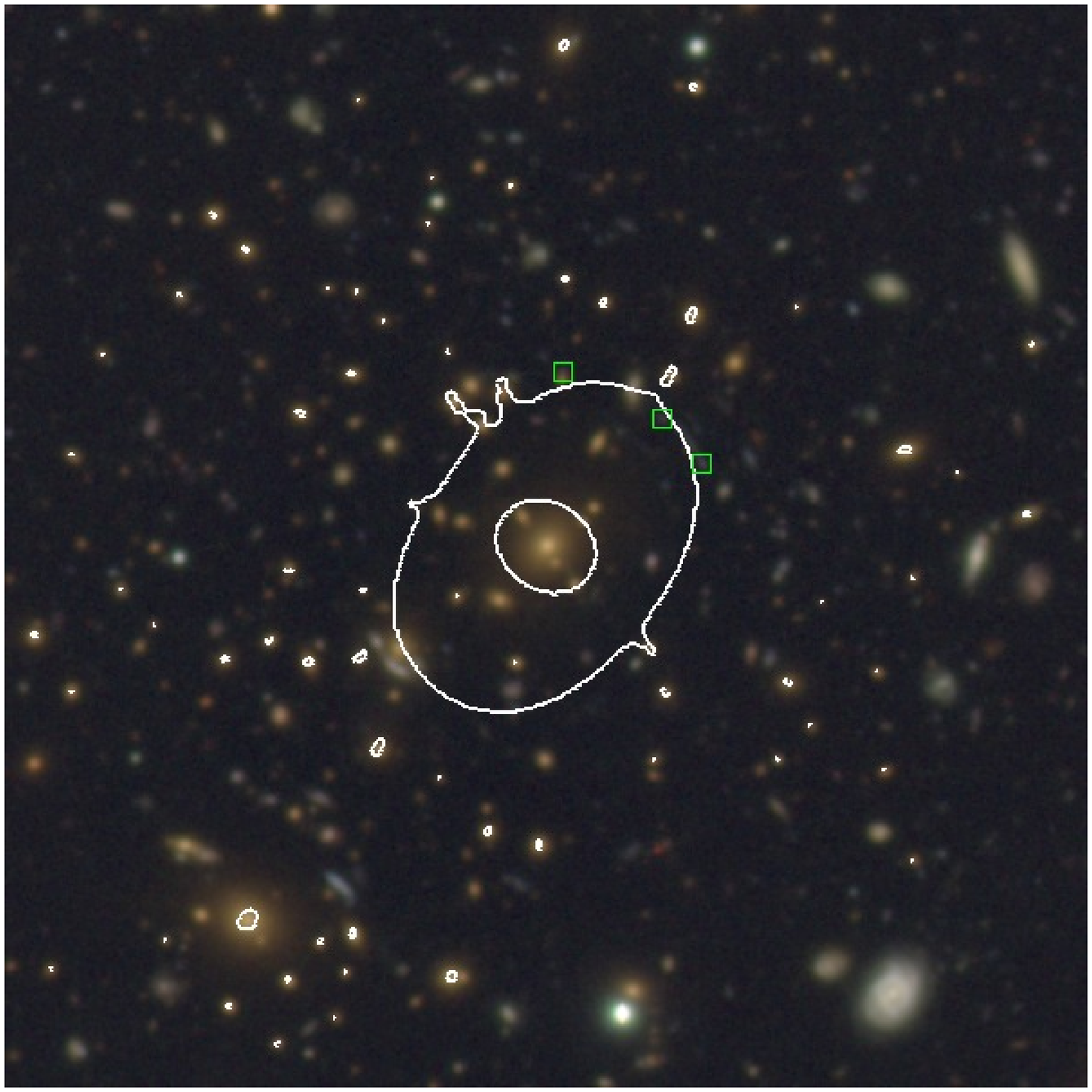}
\includegraphics[width=0.37\hsize]{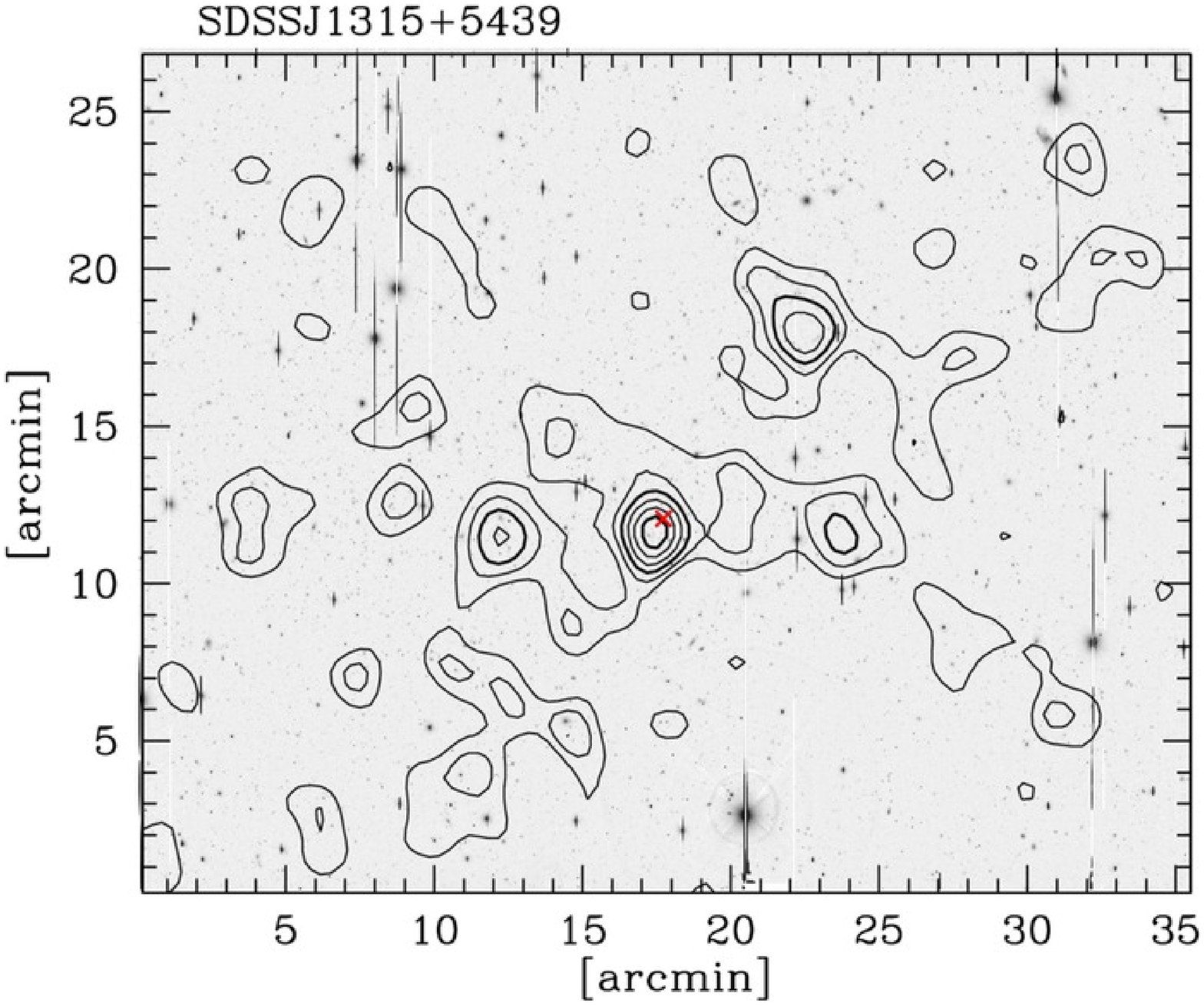}
\includegraphics[width=0.33\hsize]{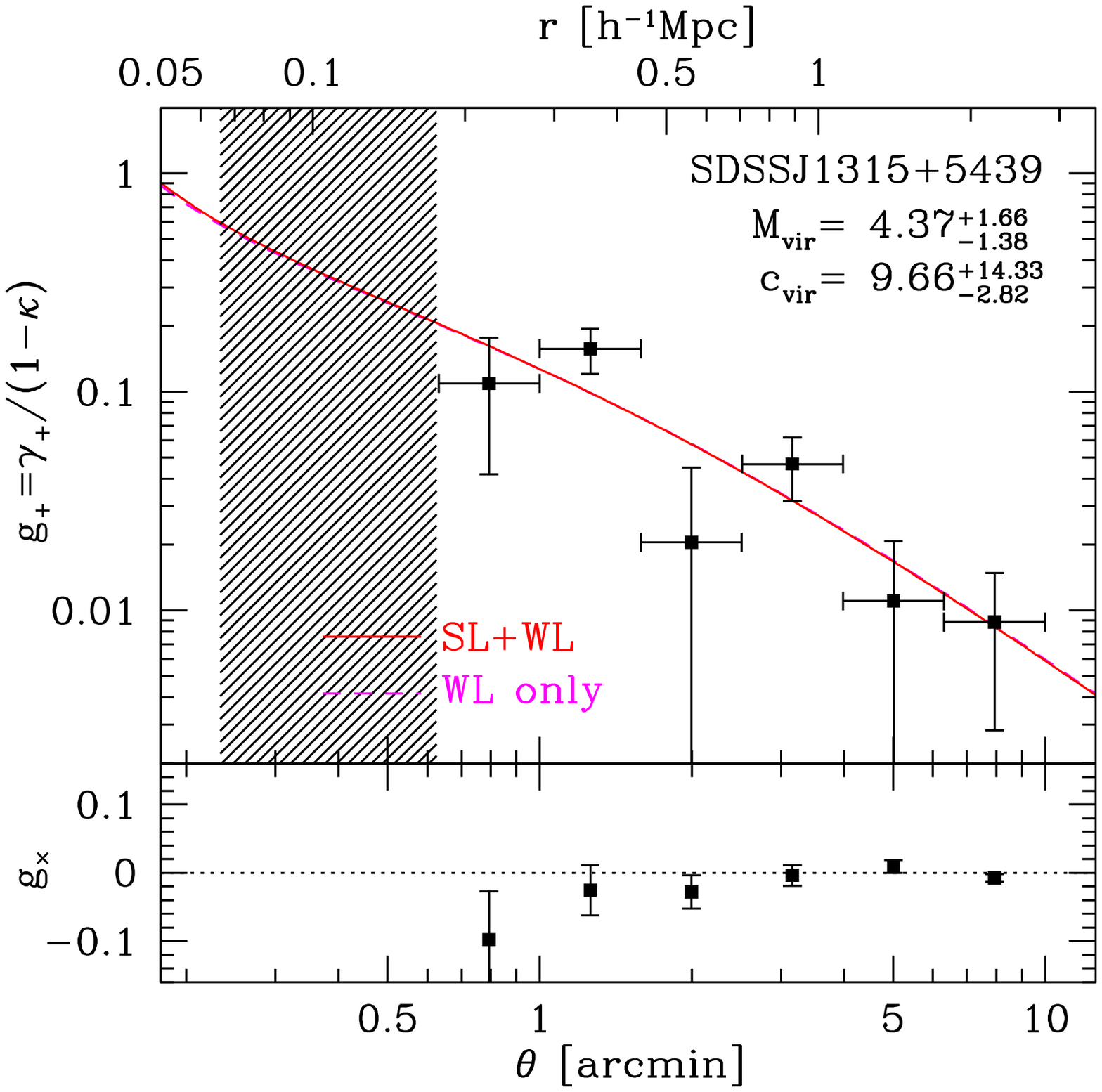}
\includegraphics[width=0.28\hsize]{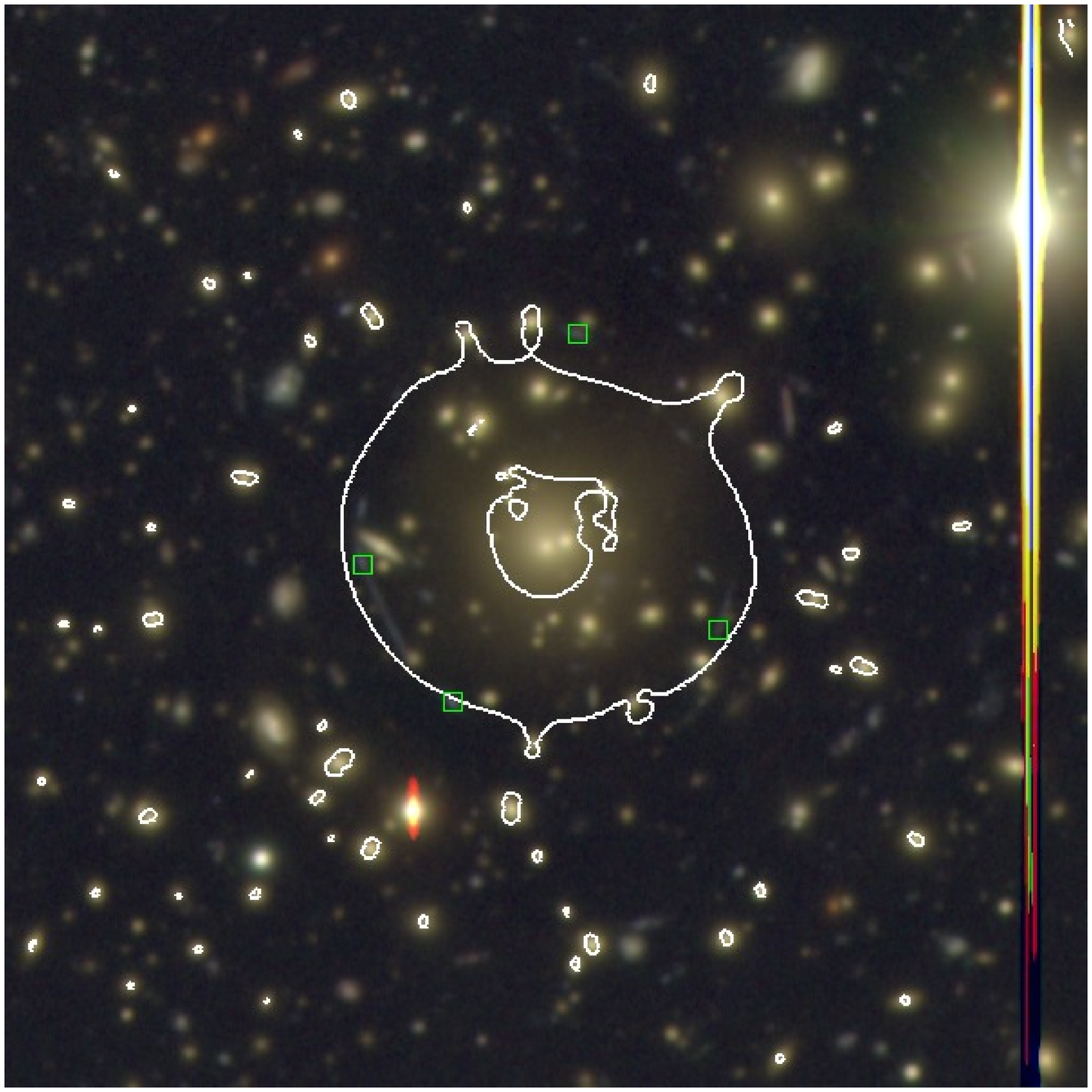}
\includegraphics[width=0.37\hsize]{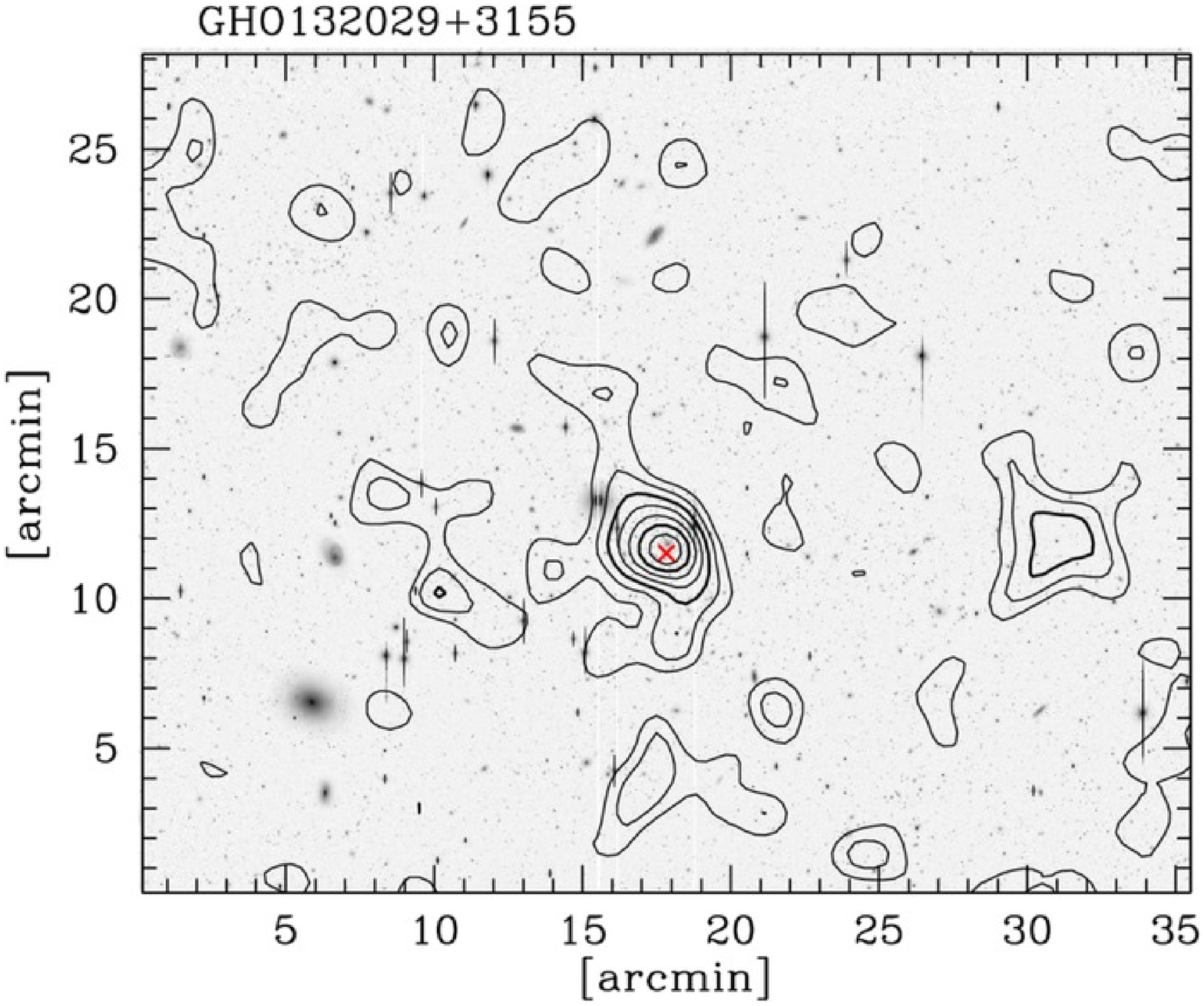}
\includegraphics[width=0.33\hsize]{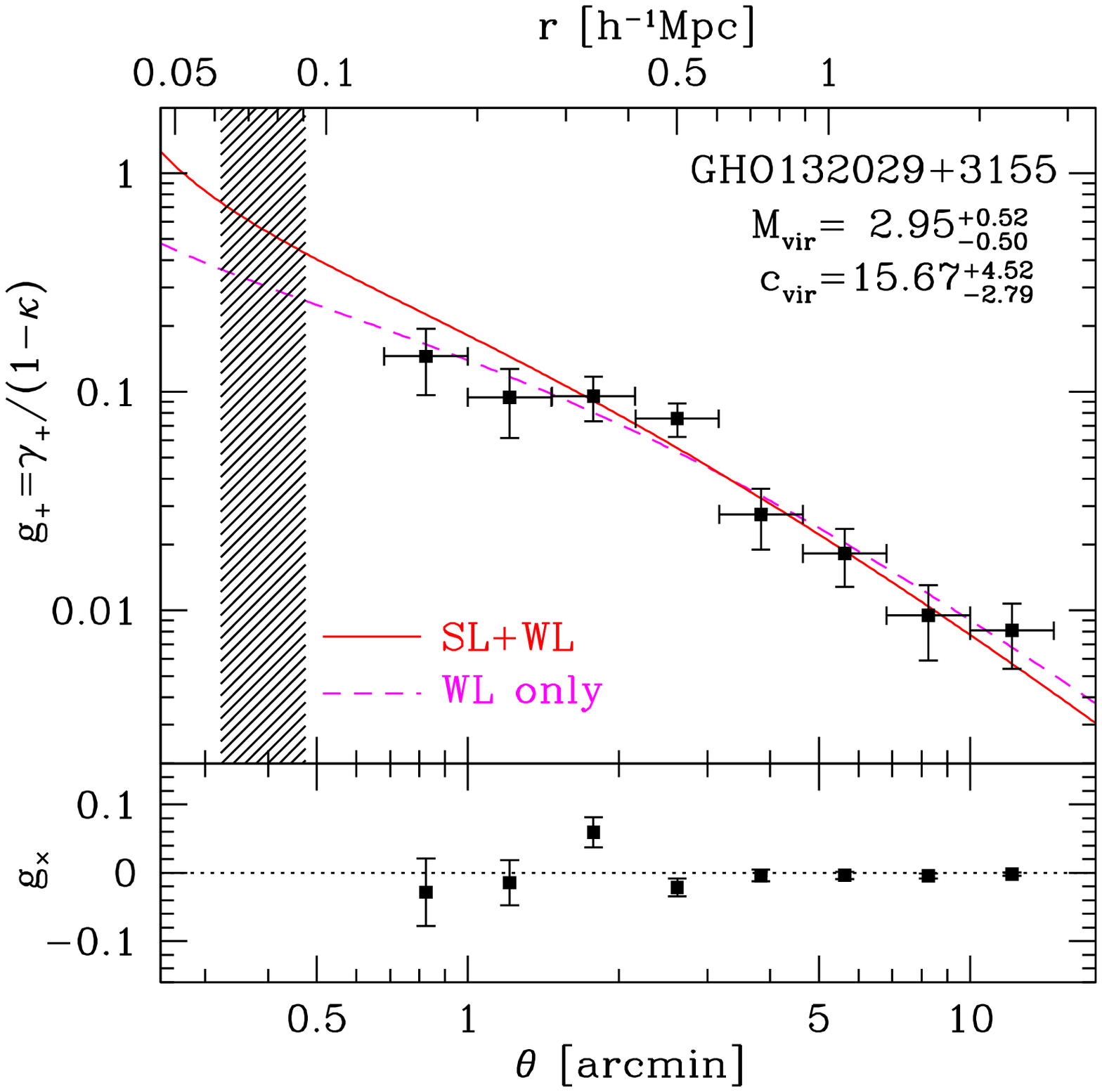}
\includegraphics[width=0.28\hsize]{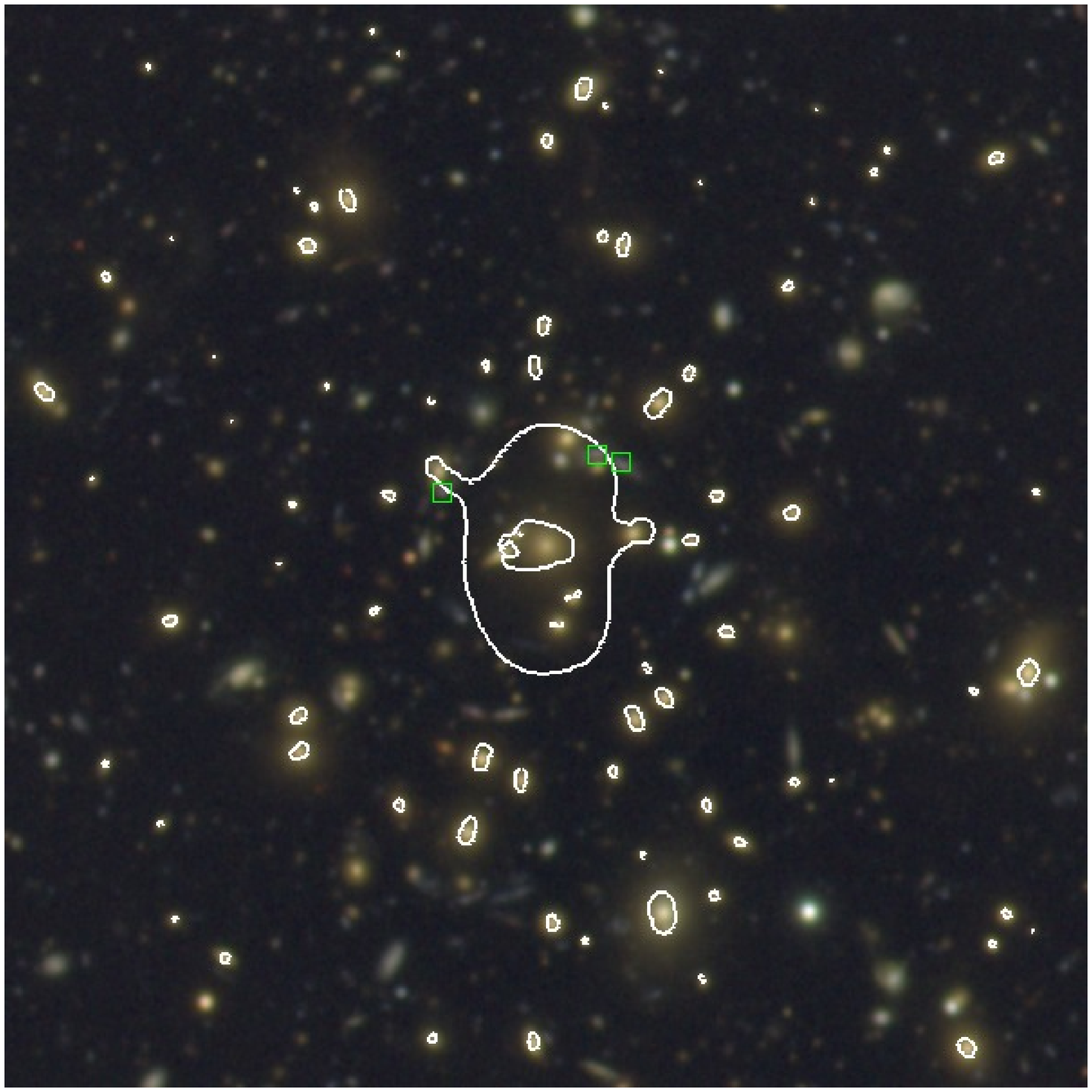}
\includegraphics[width=0.37\hsize]{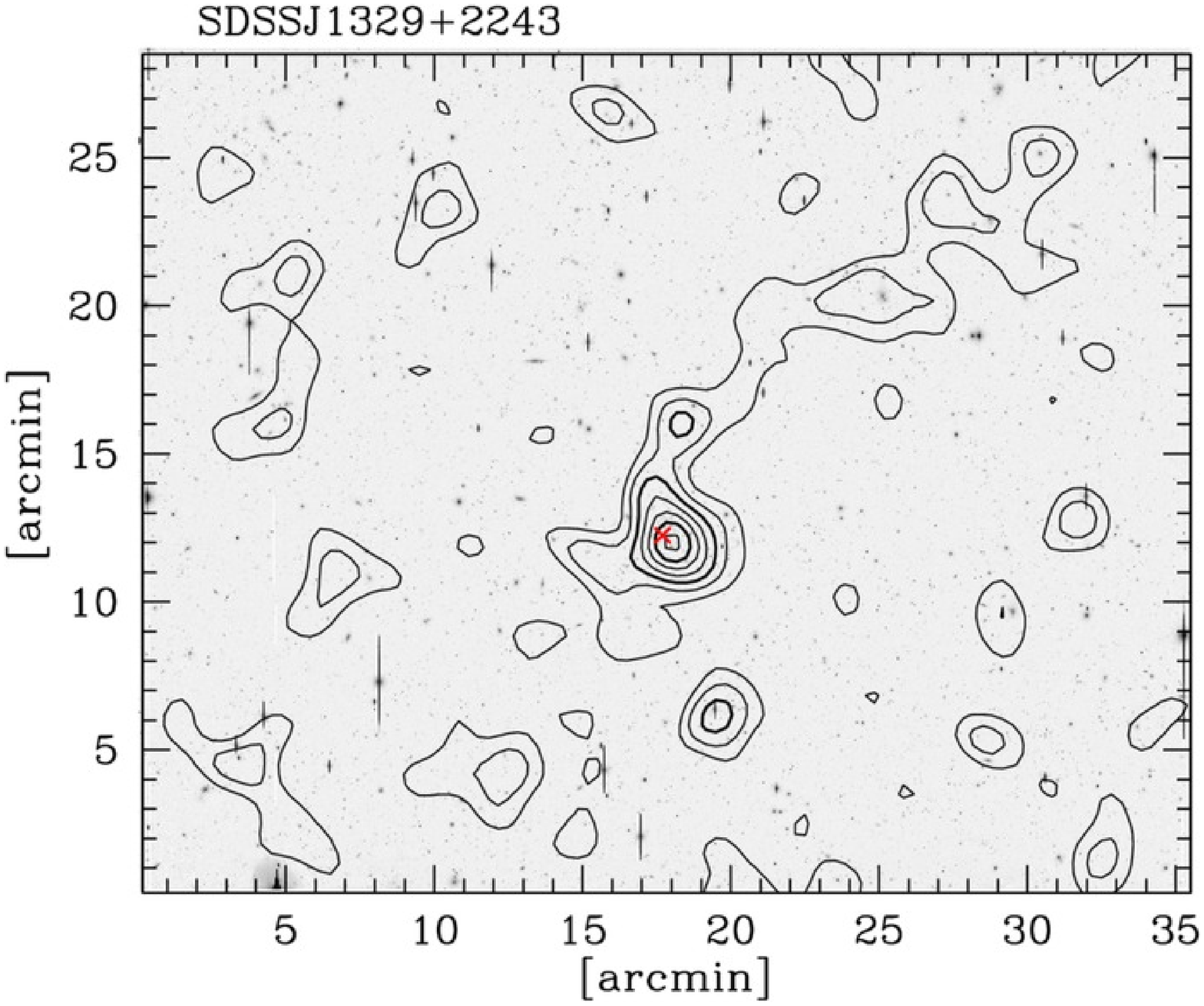}
\includegraphics[width=0.33\hsize]{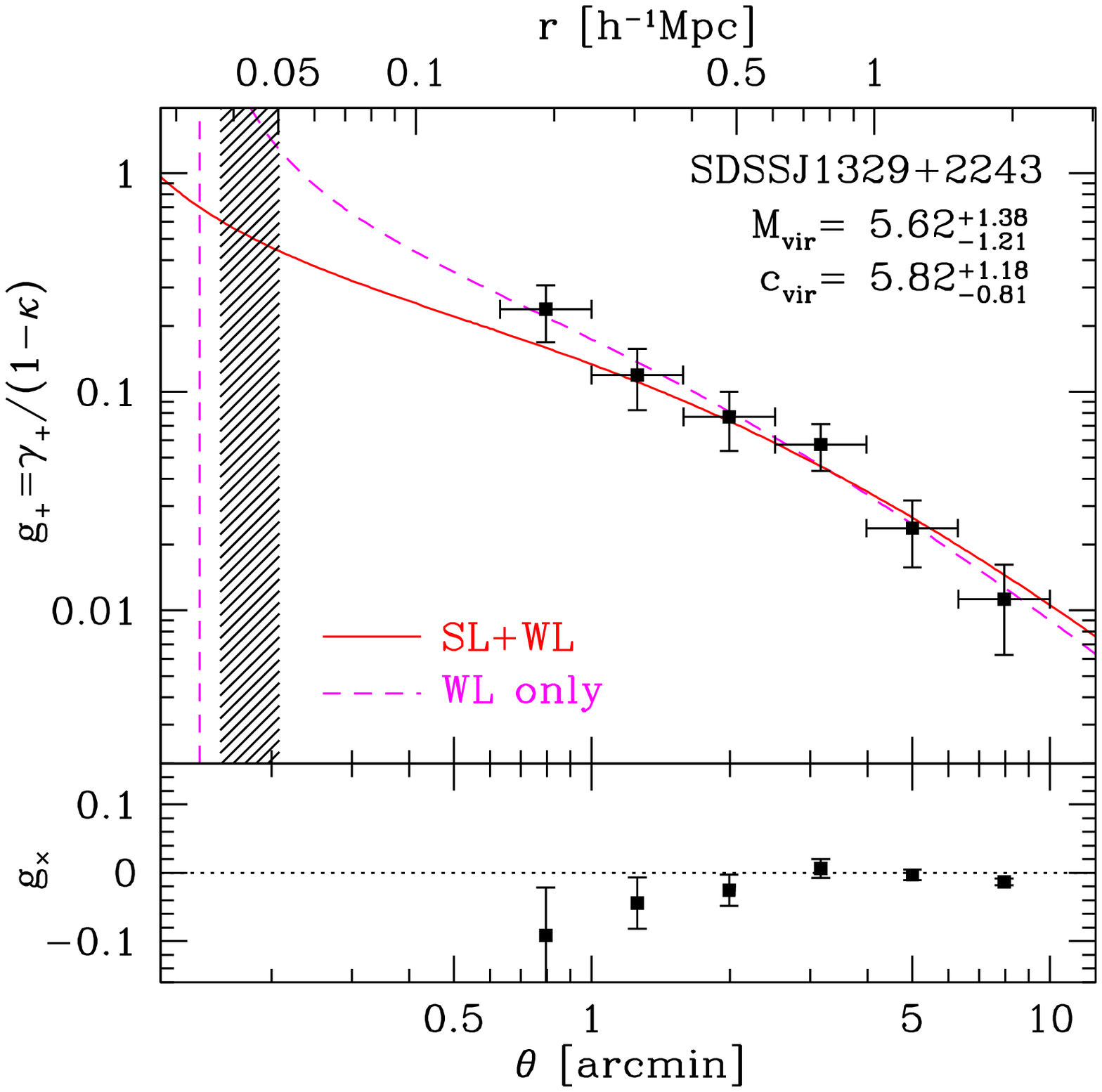}
\end{center}
\caption{A1703, SDSSJ1315+5439, GHO132029+3155, SDSSJ1329+2243}
\end{figure*}
\begin{figure*}
\begin{center}
\includegraphics[width=0.28\hsize]{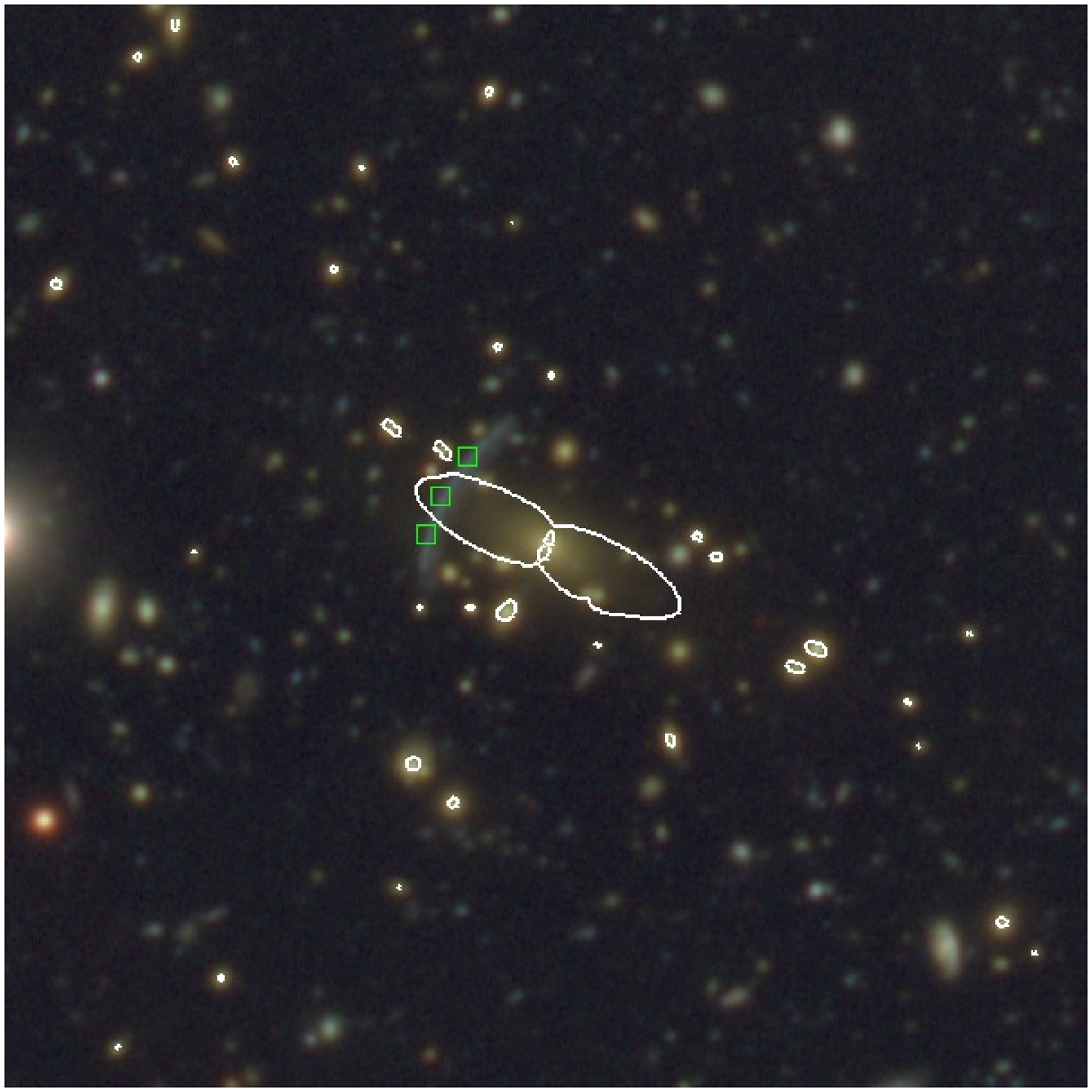}
\includegraphics[width=0.37\hsize]{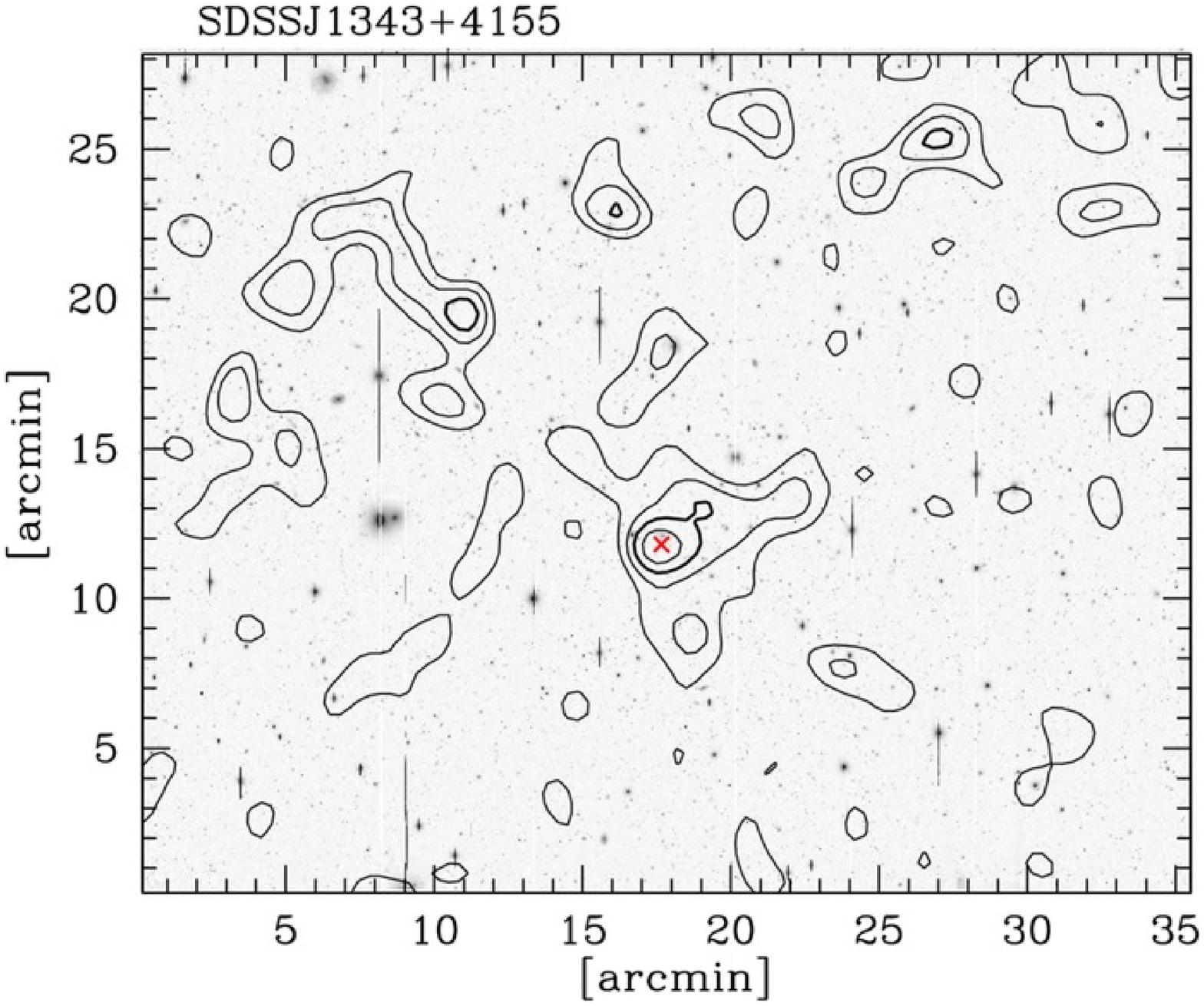}
\includegraphics[width=0.33\hsize]{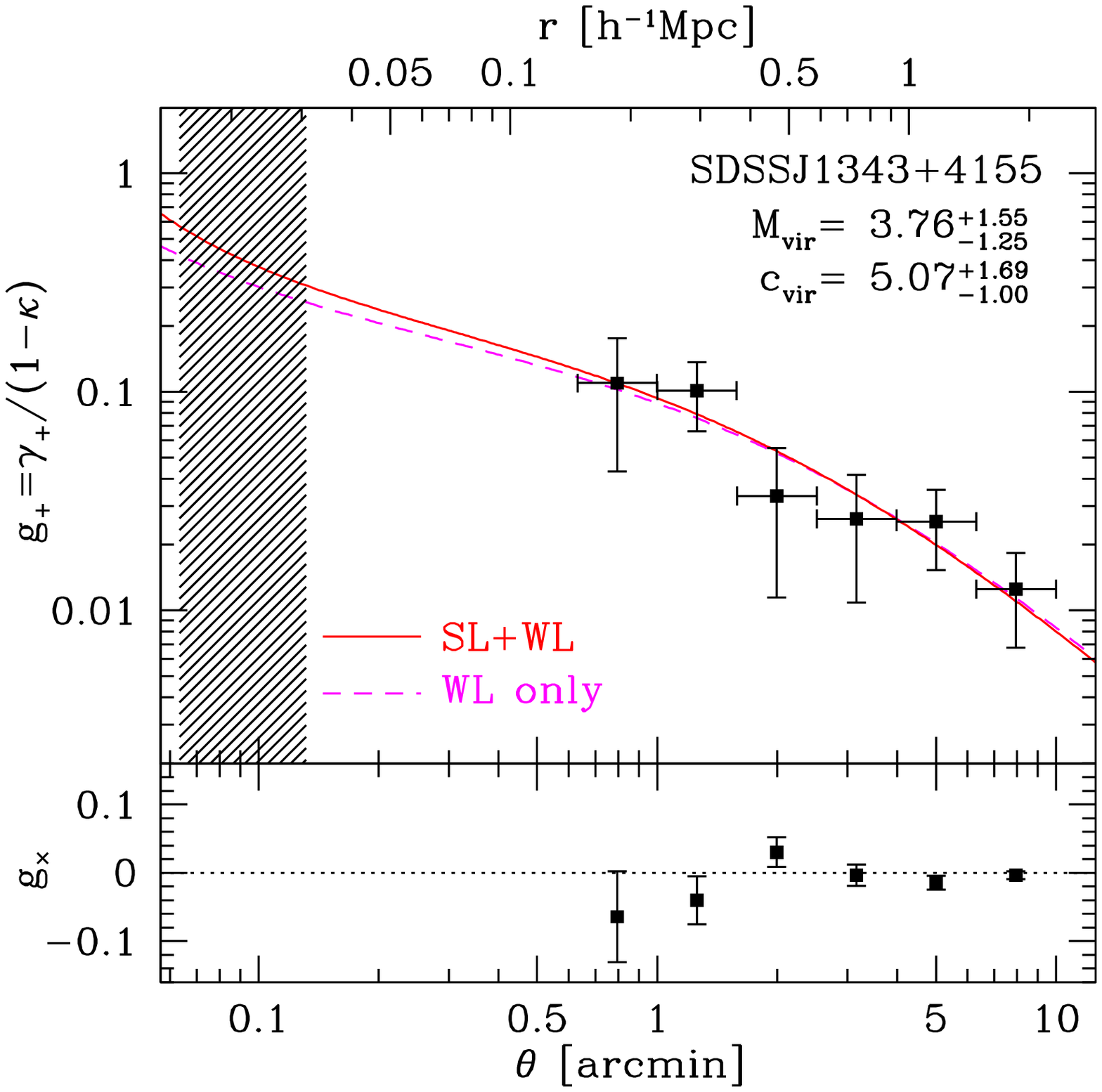}
\includegraphics[width=0.28\hsize]{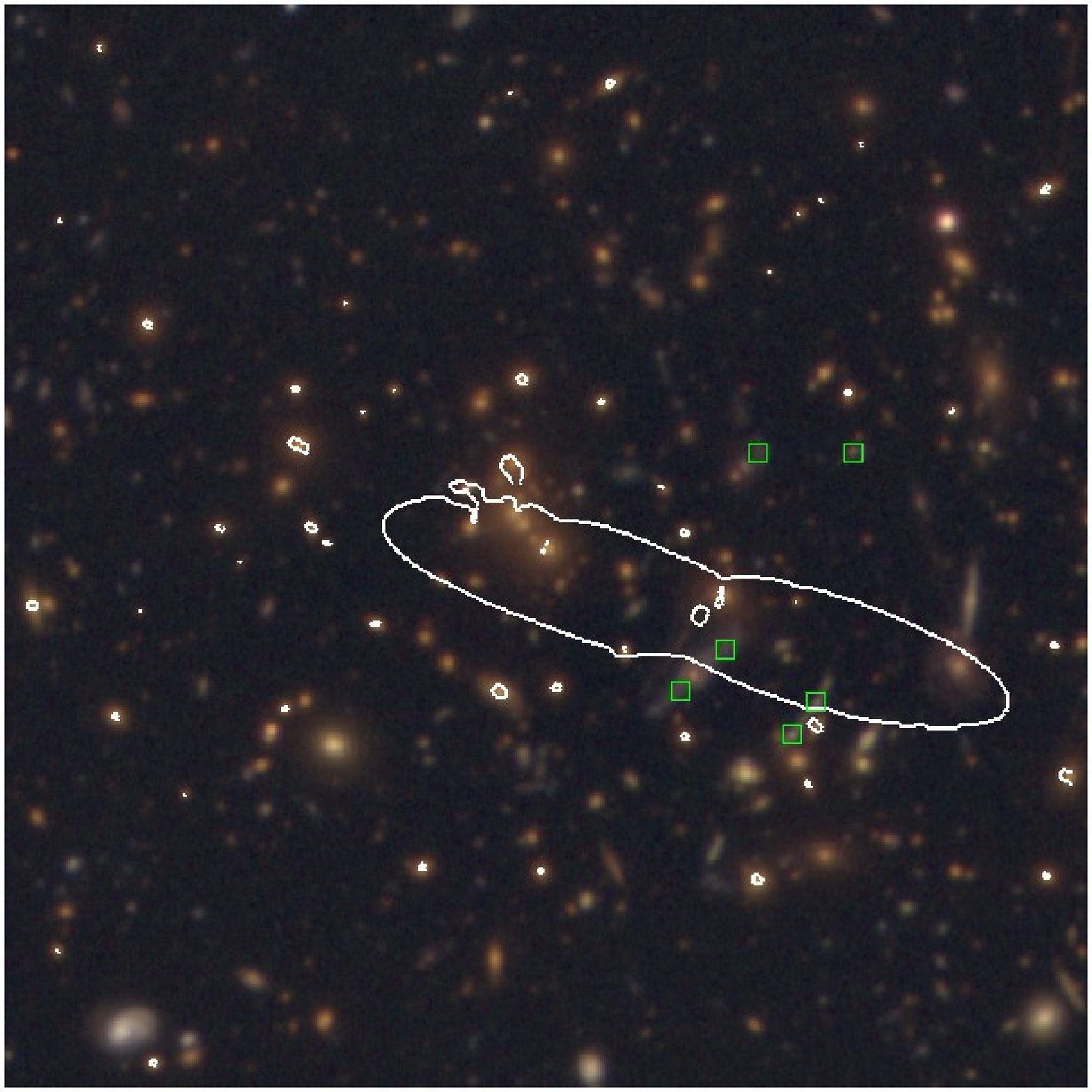}
\includegraphics[width=0.37\hsize]{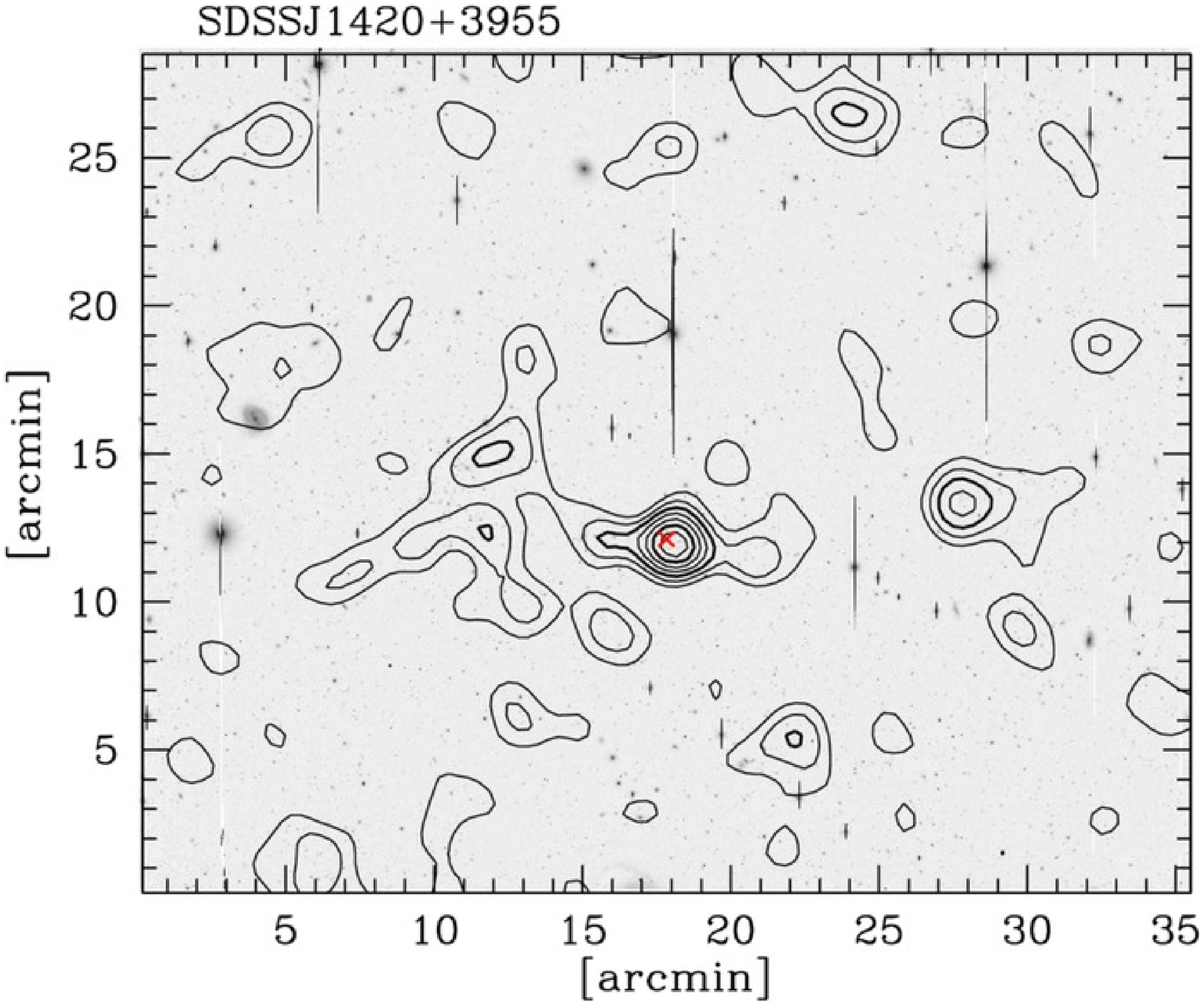}
\includegraphics[width=0.33\hsize]{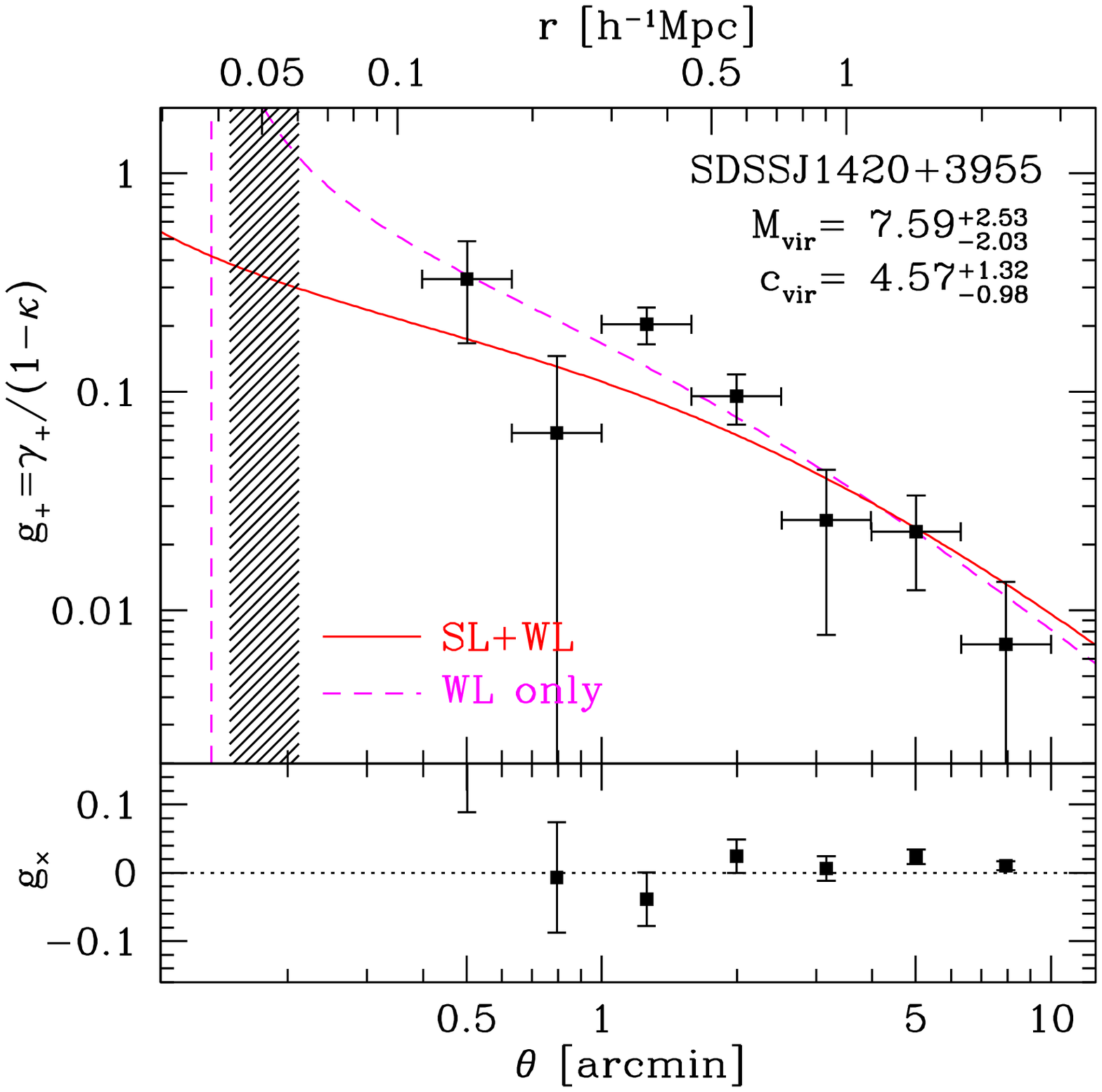}
\includegraphics[width=0.28\hsize]{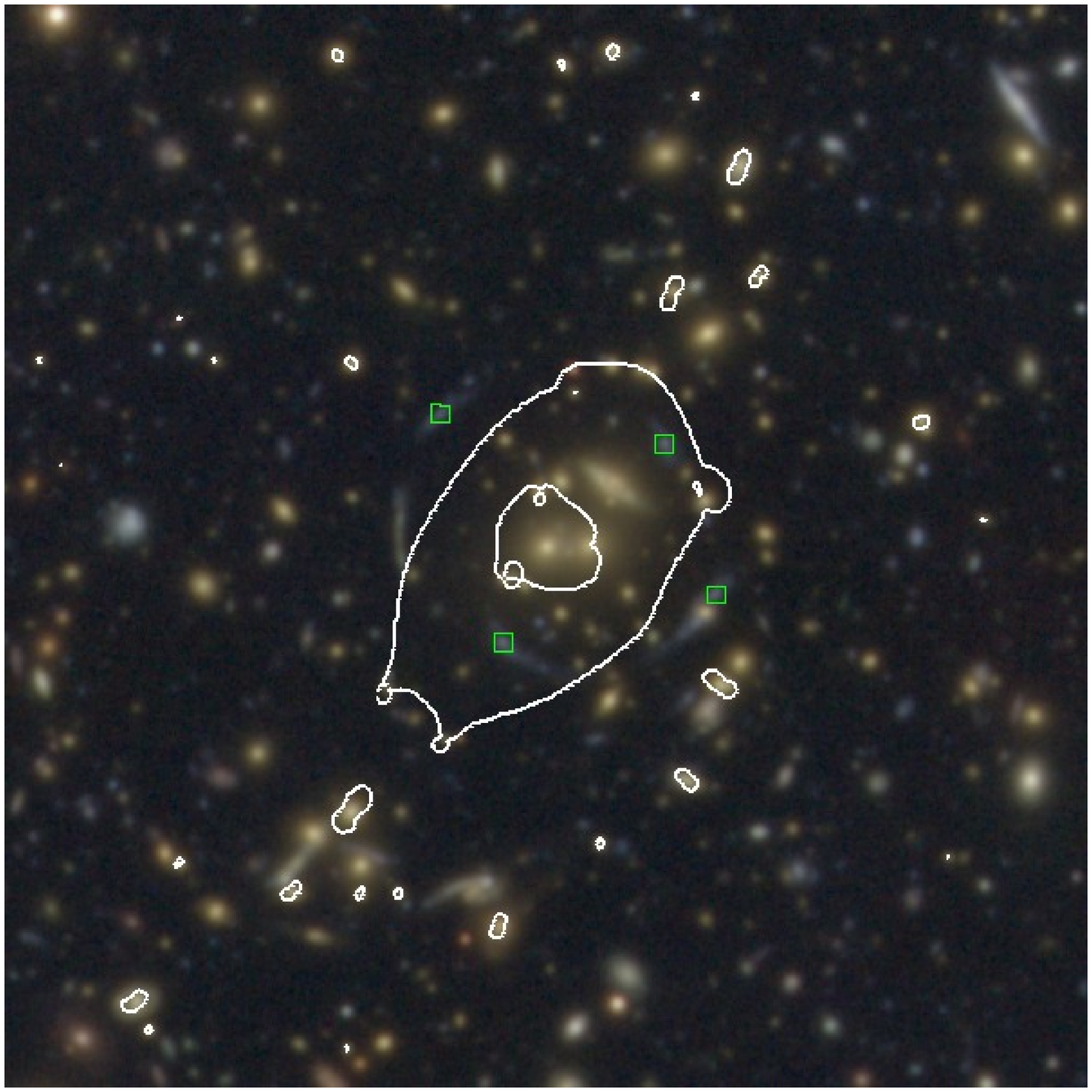}
\includegraphics[width=0.37\hsize]{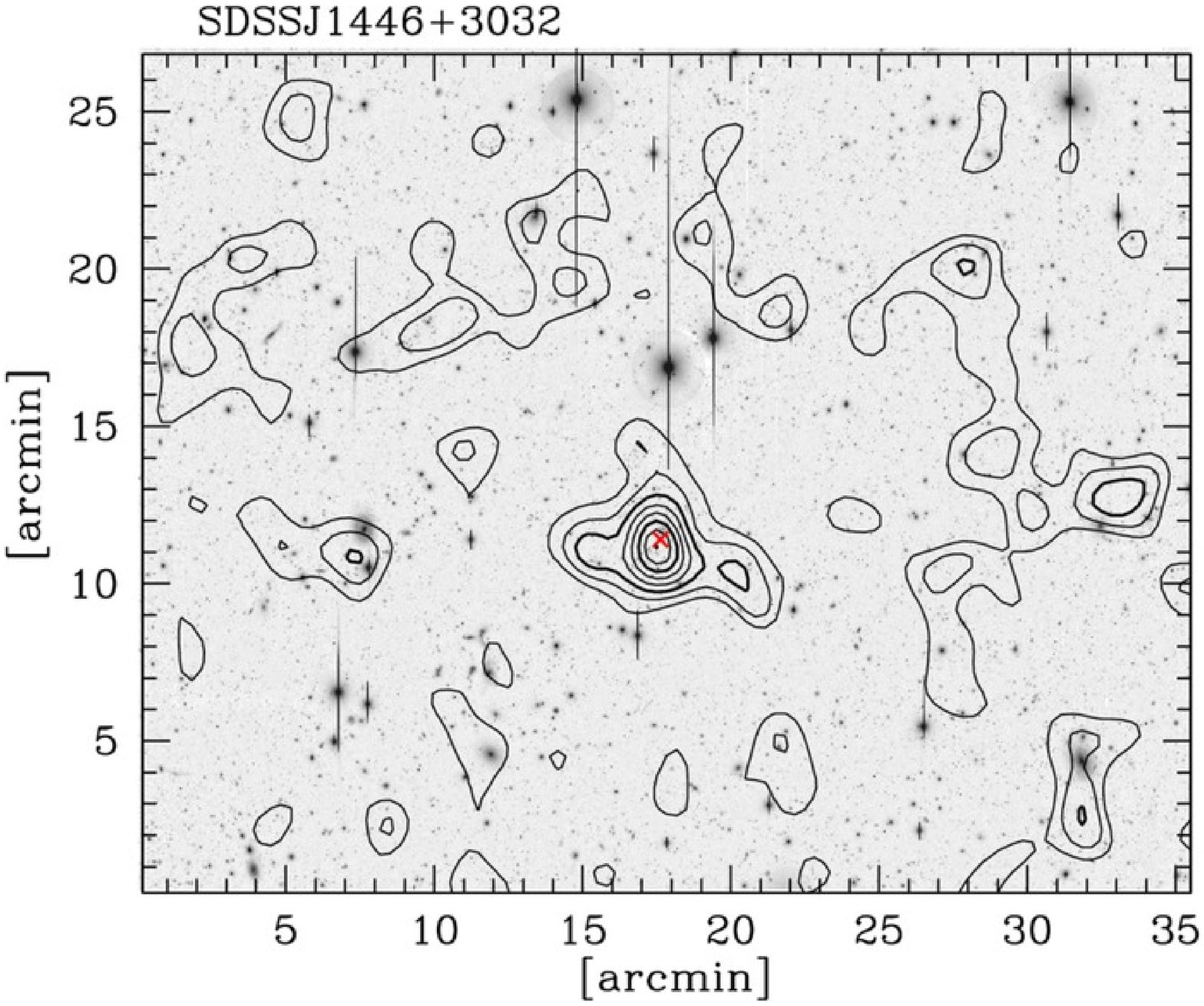}
\includegraphics[width=0.33\hsize]{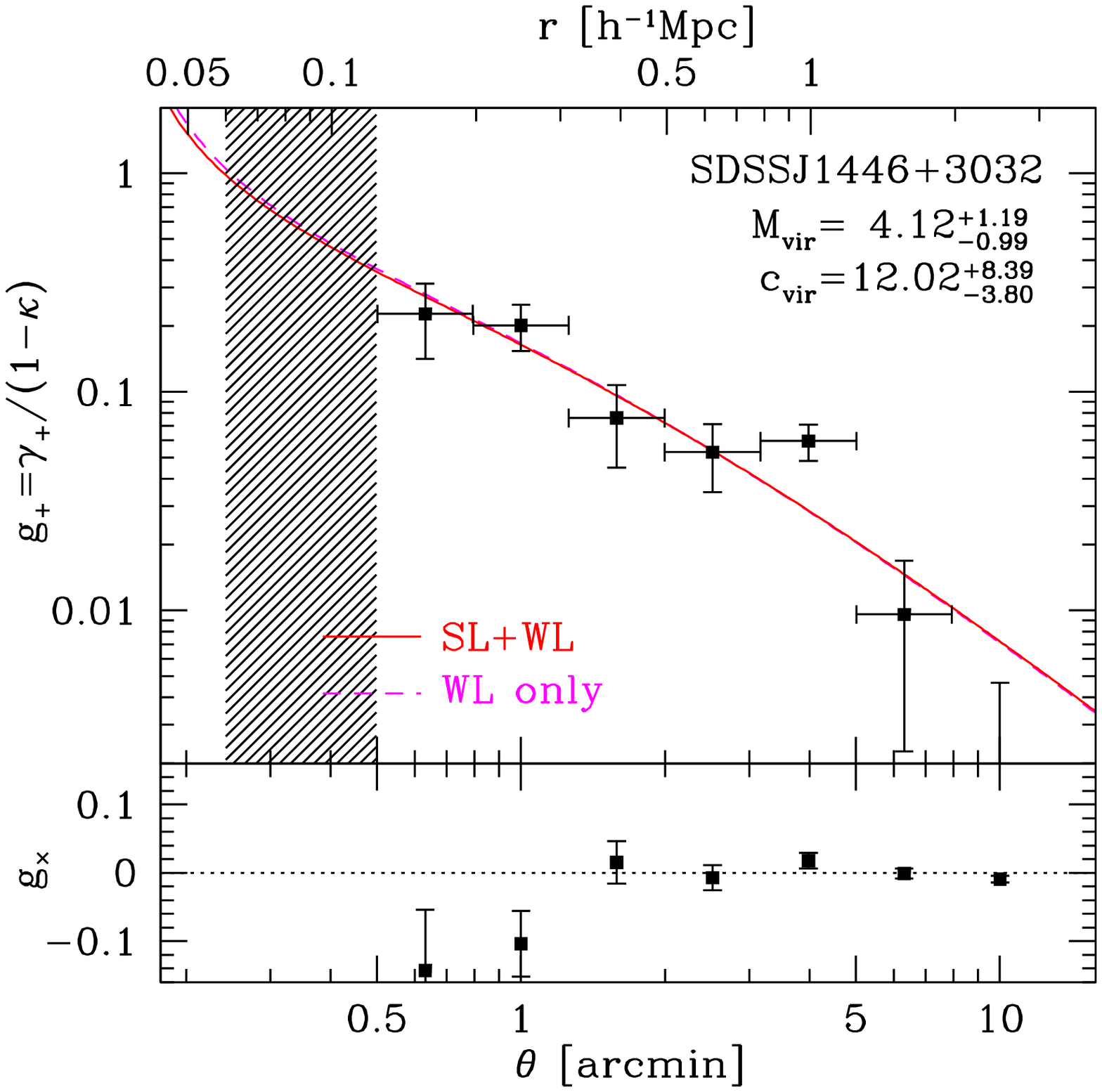}
\includegraphics[width=0.28\hsize]{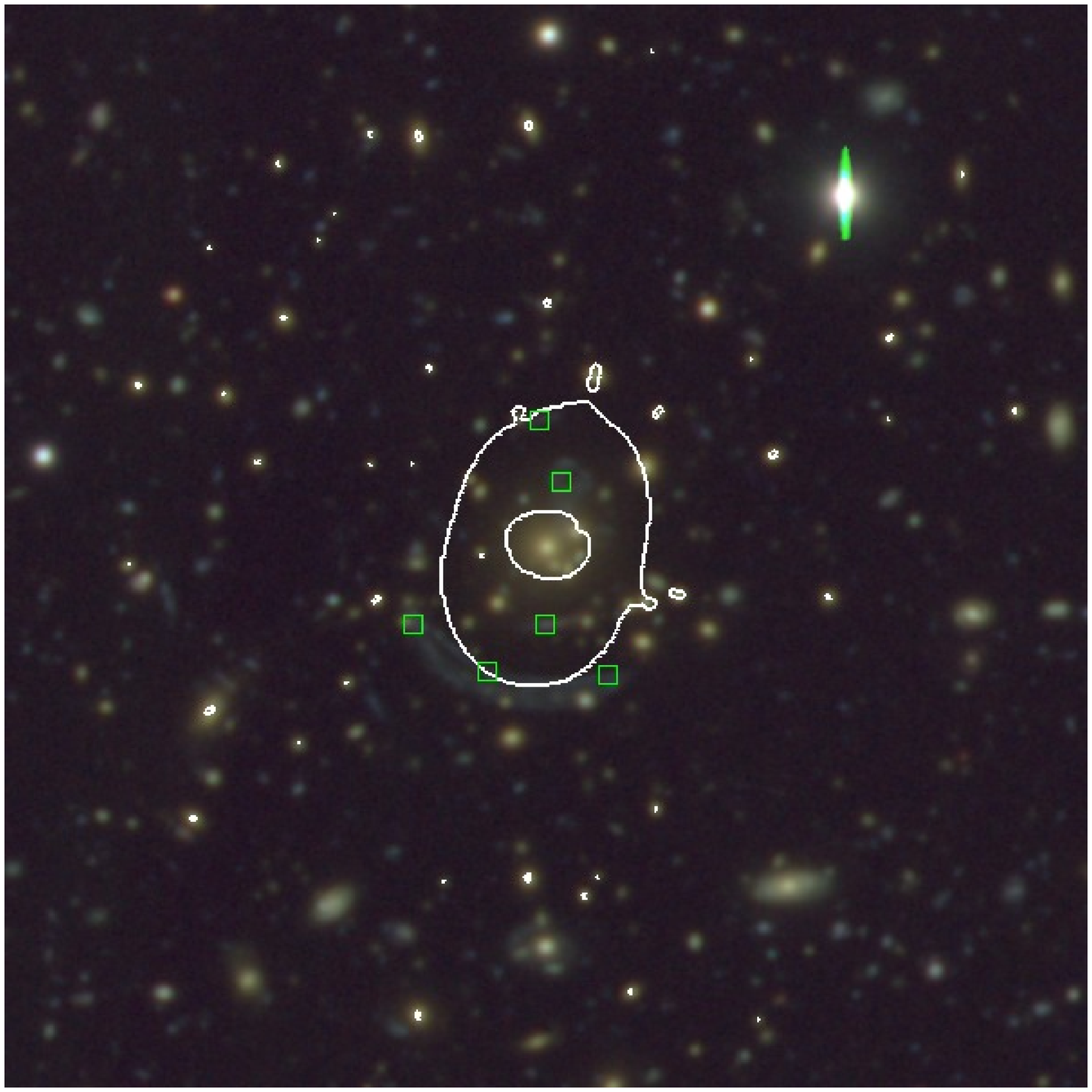}
\includegraphics[width=0.37\hsize]{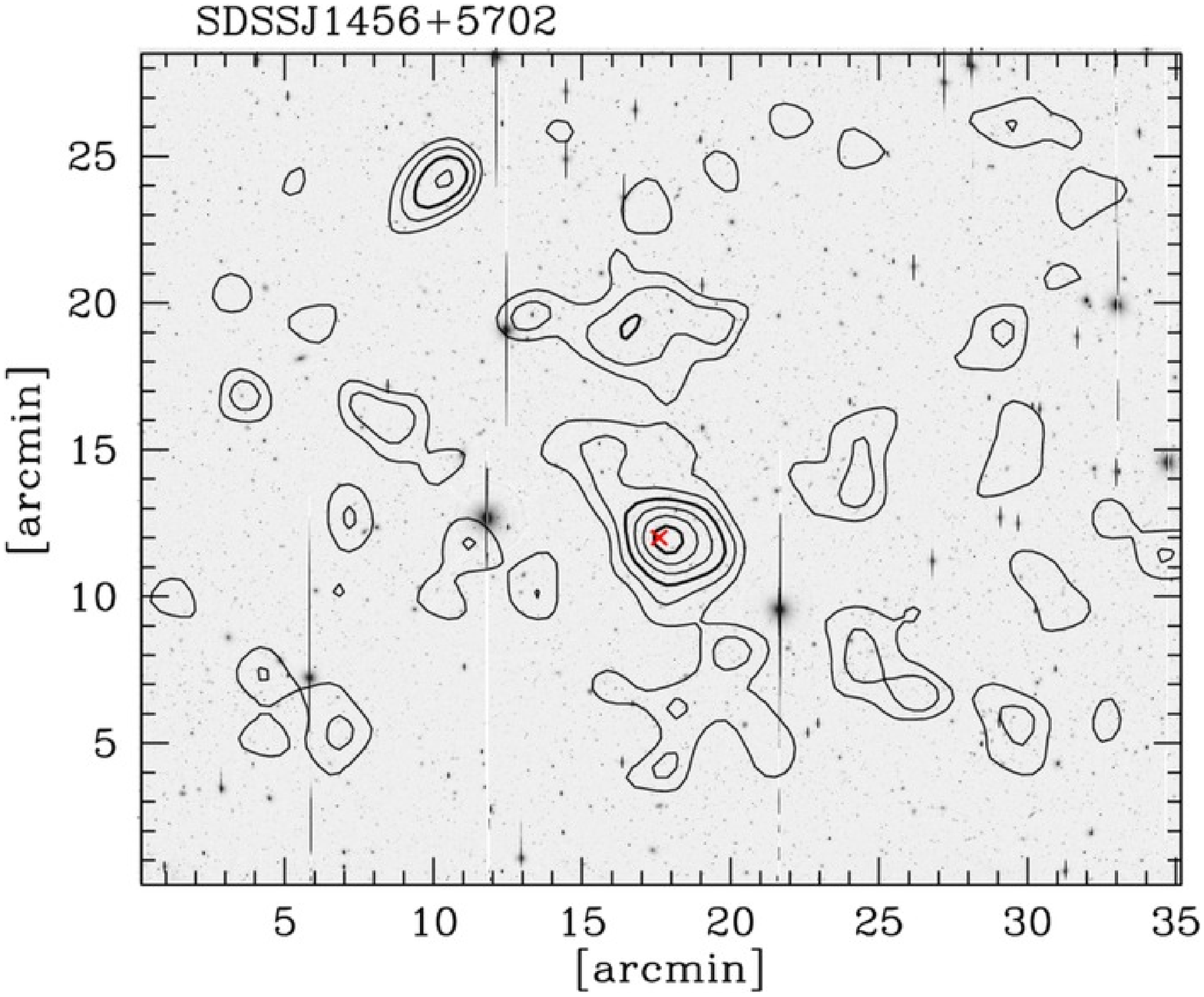}
\includegraphics[width=0.33\hsize]{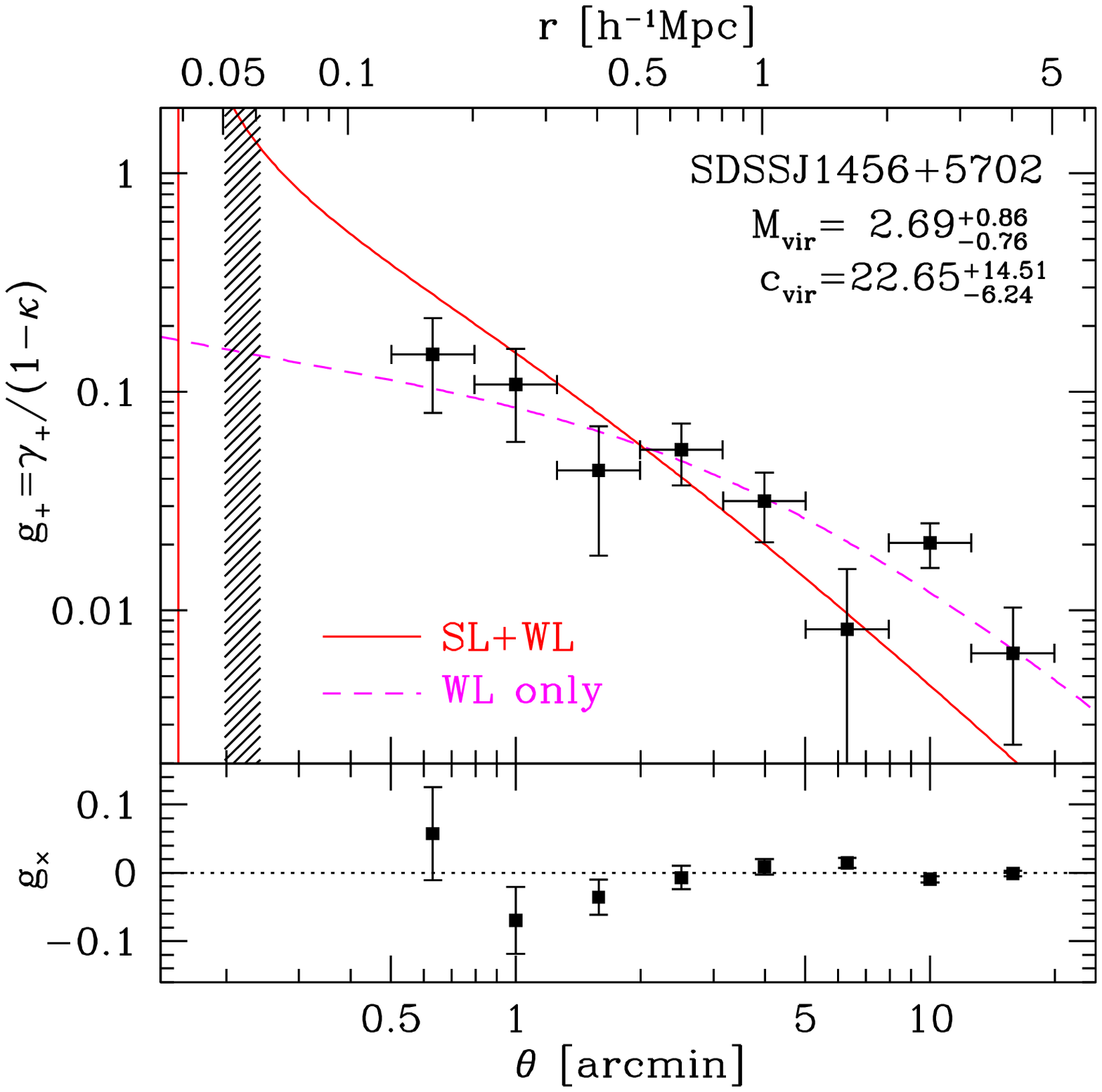}
\end{center}
\caption{SDSSJ1343+4155, SDSSJ1420+3955, SDSSJ1446+3032, SDSSJ1456+5702}
\end{figure*}
\begin{figure*}
\begin{center}
\includegraphics[width=0.28\hsize]{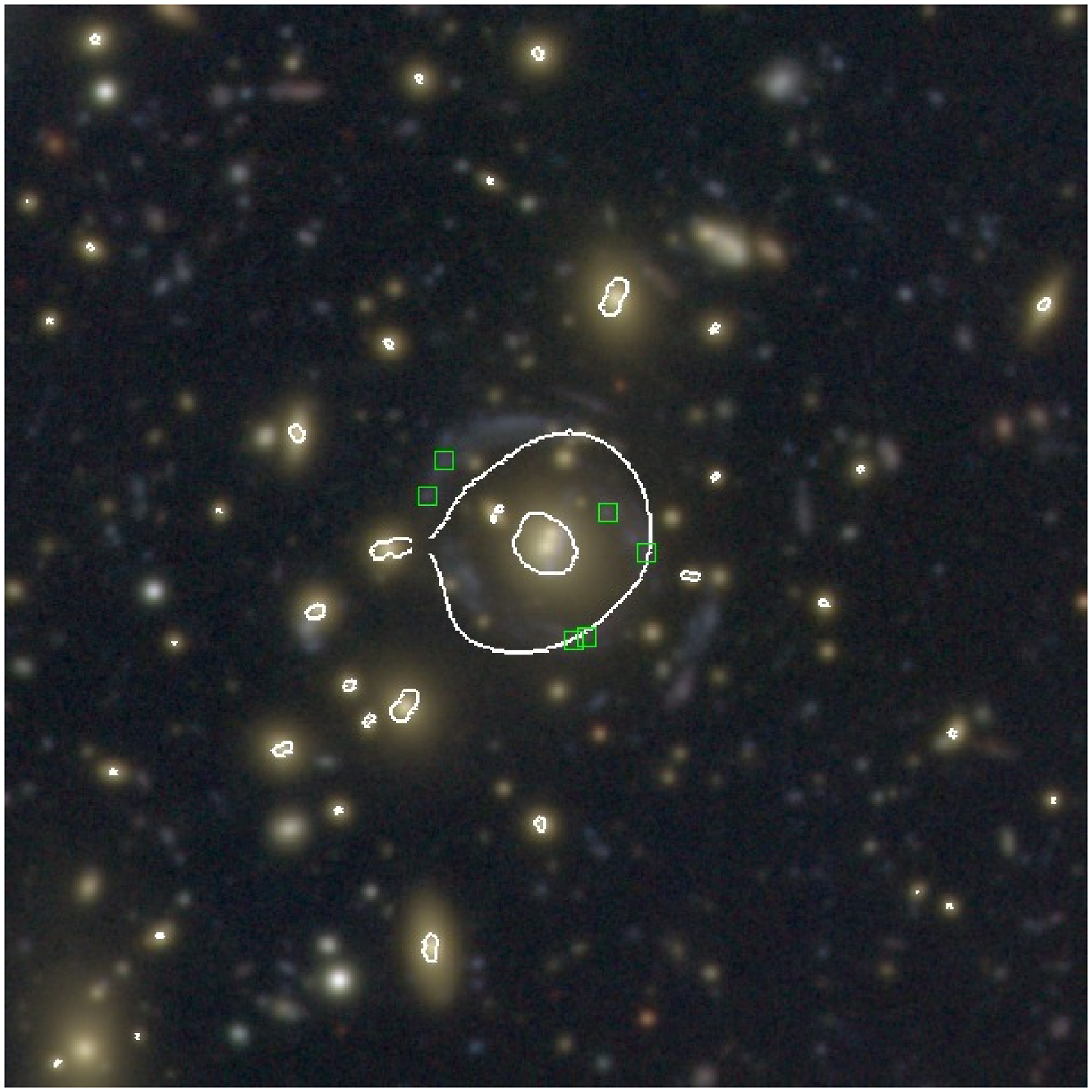}
\includegraphics[width=0.37\hsize]{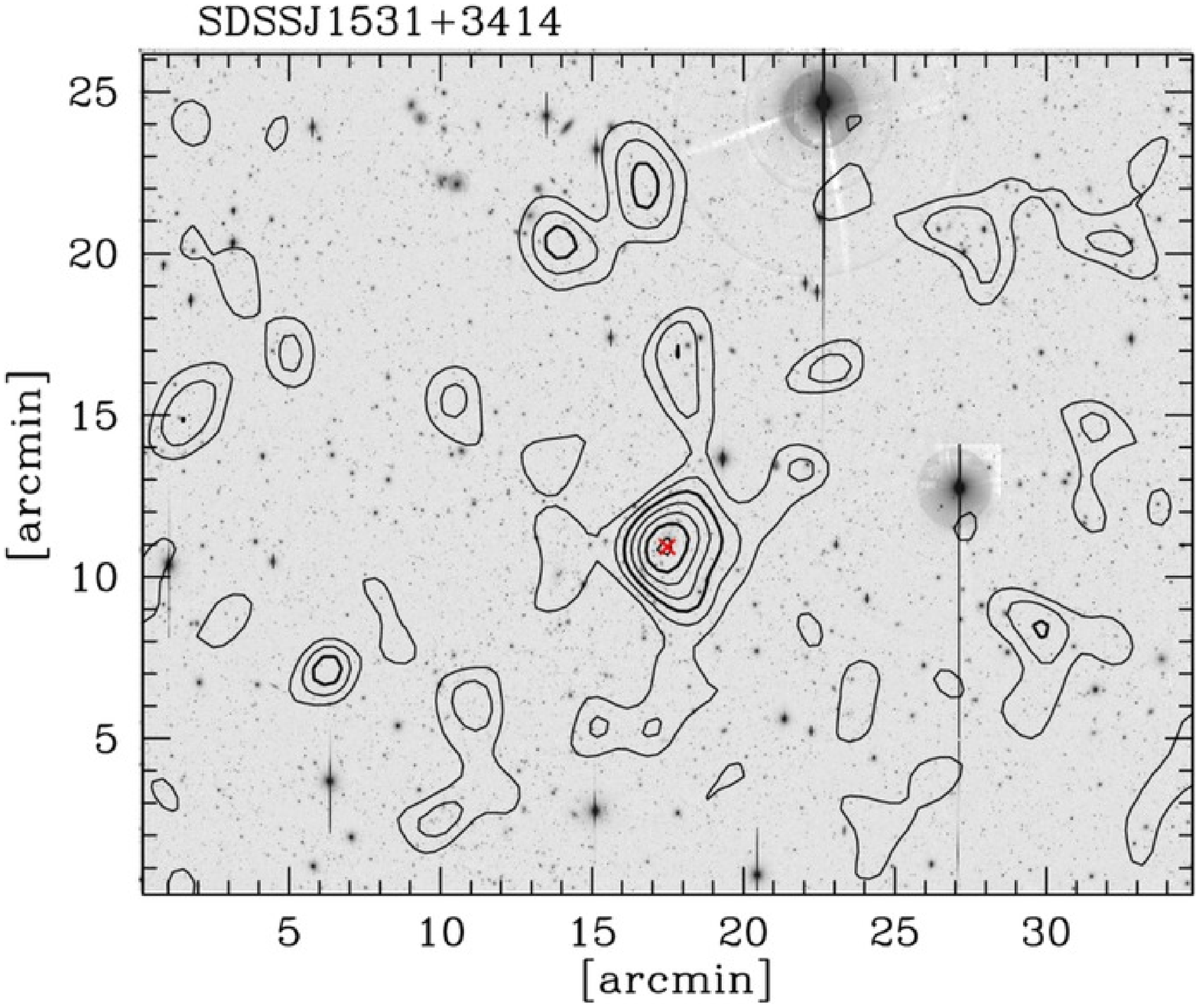}
\includegraphics[width=0.33\hsize]{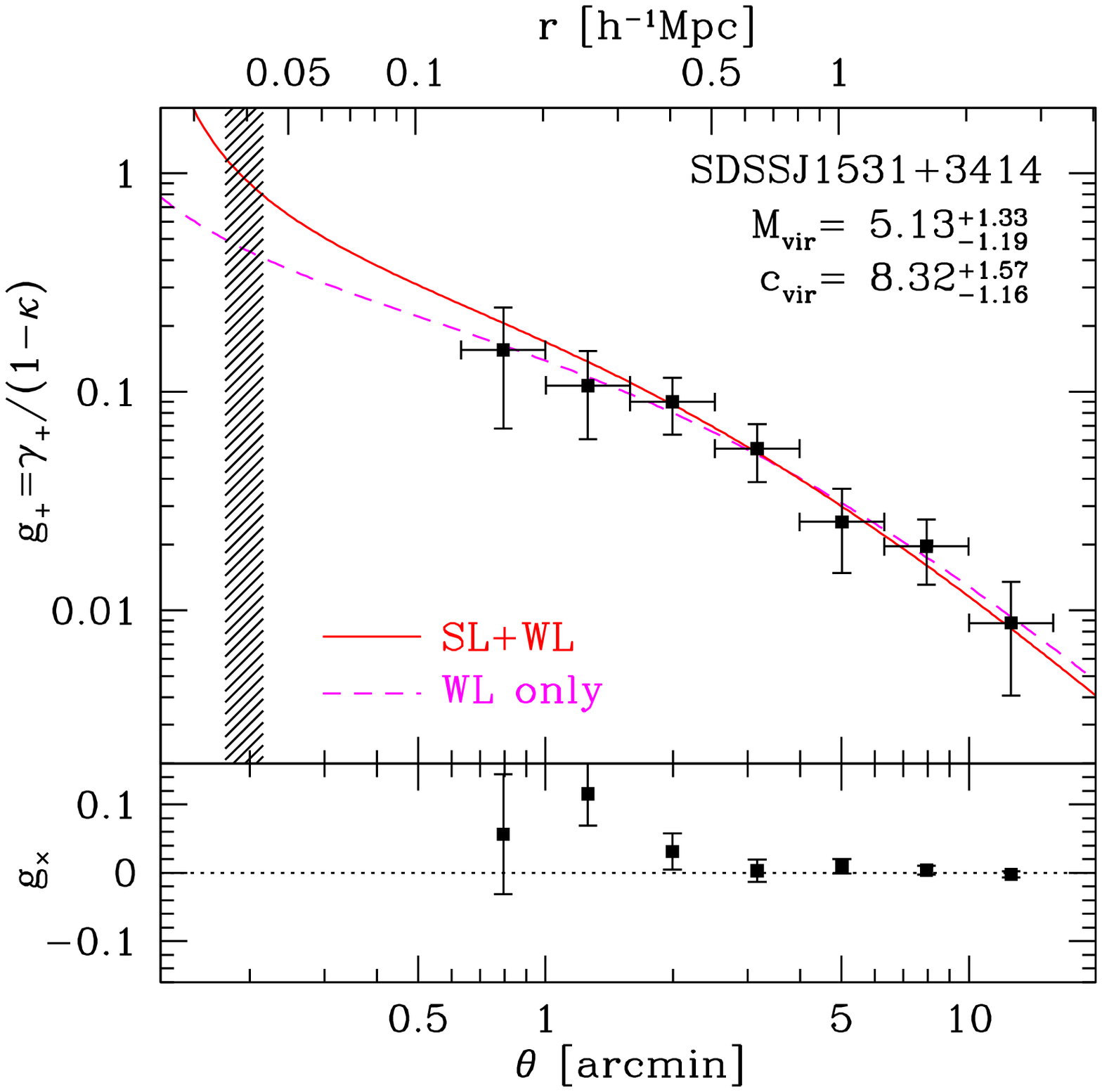}
\includegraphics[width=0.28\hsize]{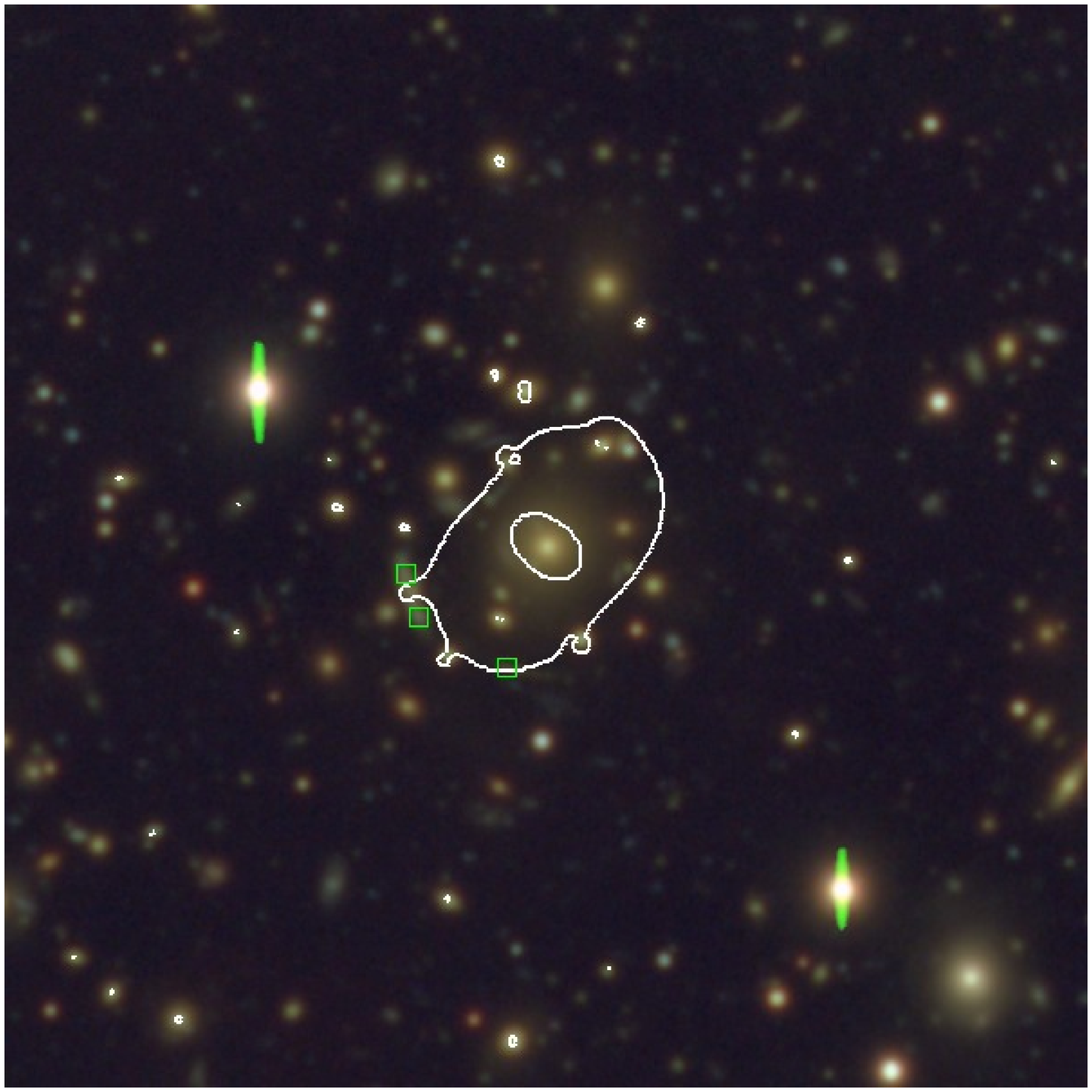}
\includegraphics[width=0.37\hsize]{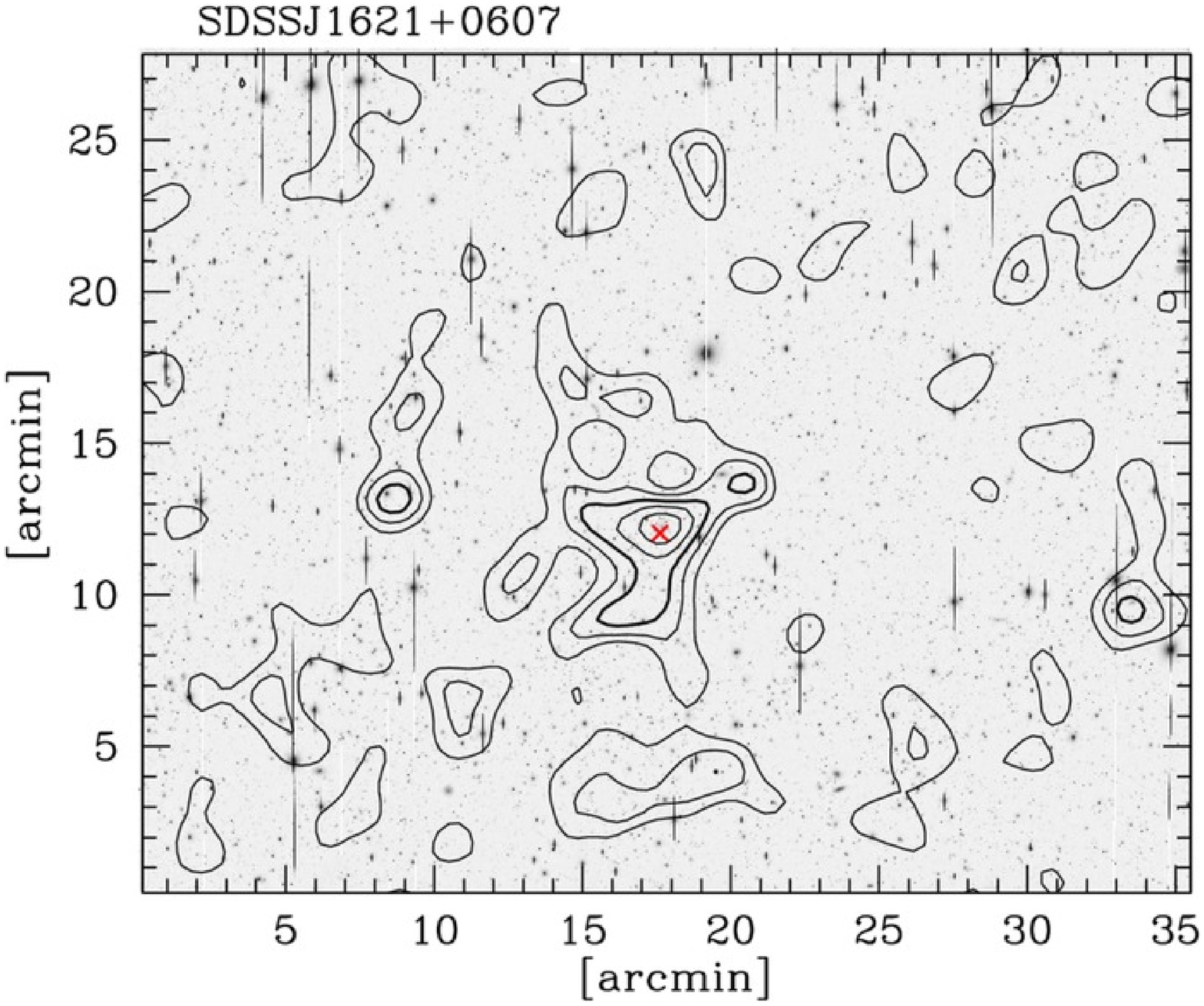}
\includegraphics[width=0.33\hsize]{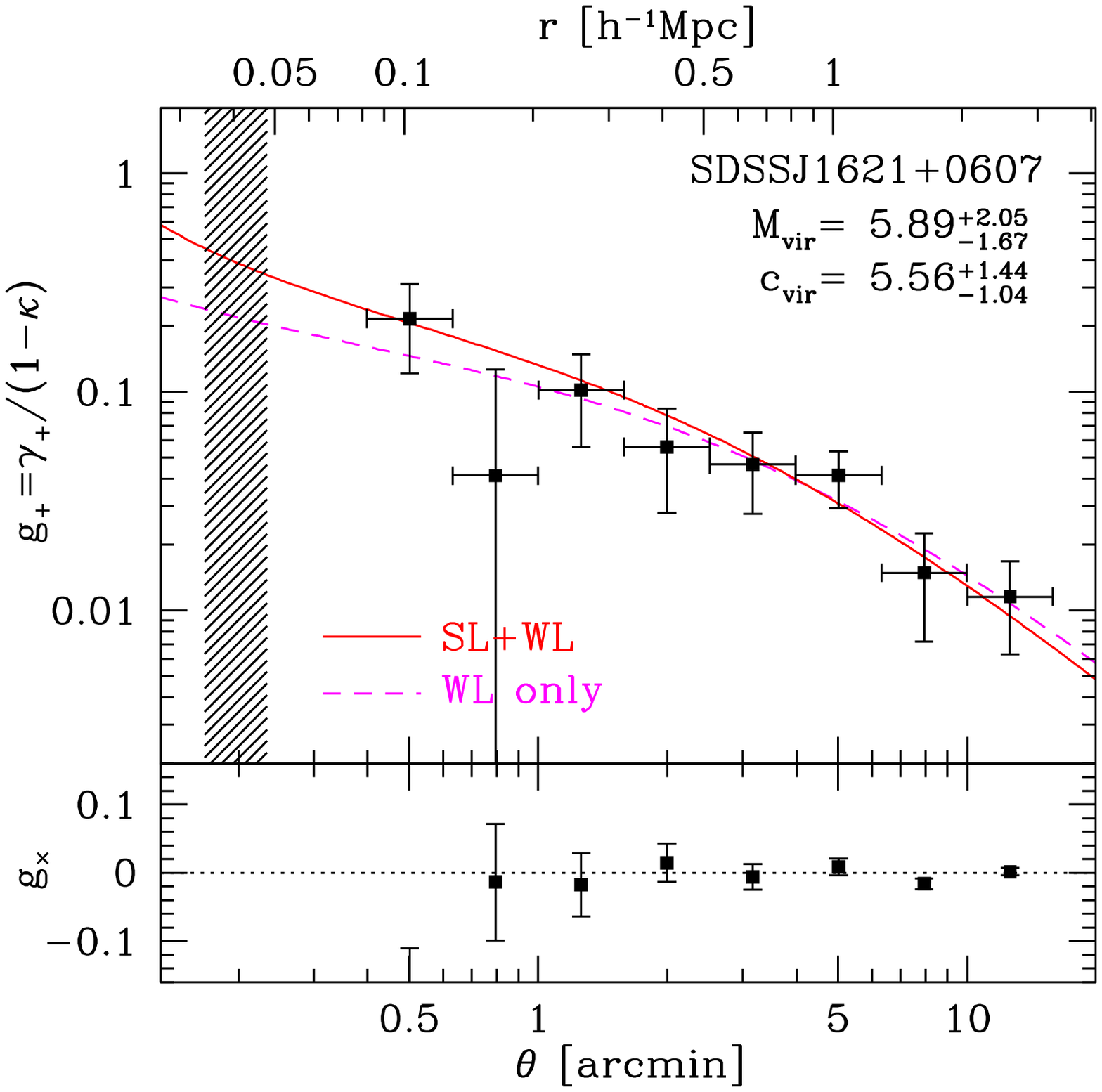}
\includegraphics[width=0.28\hsize]{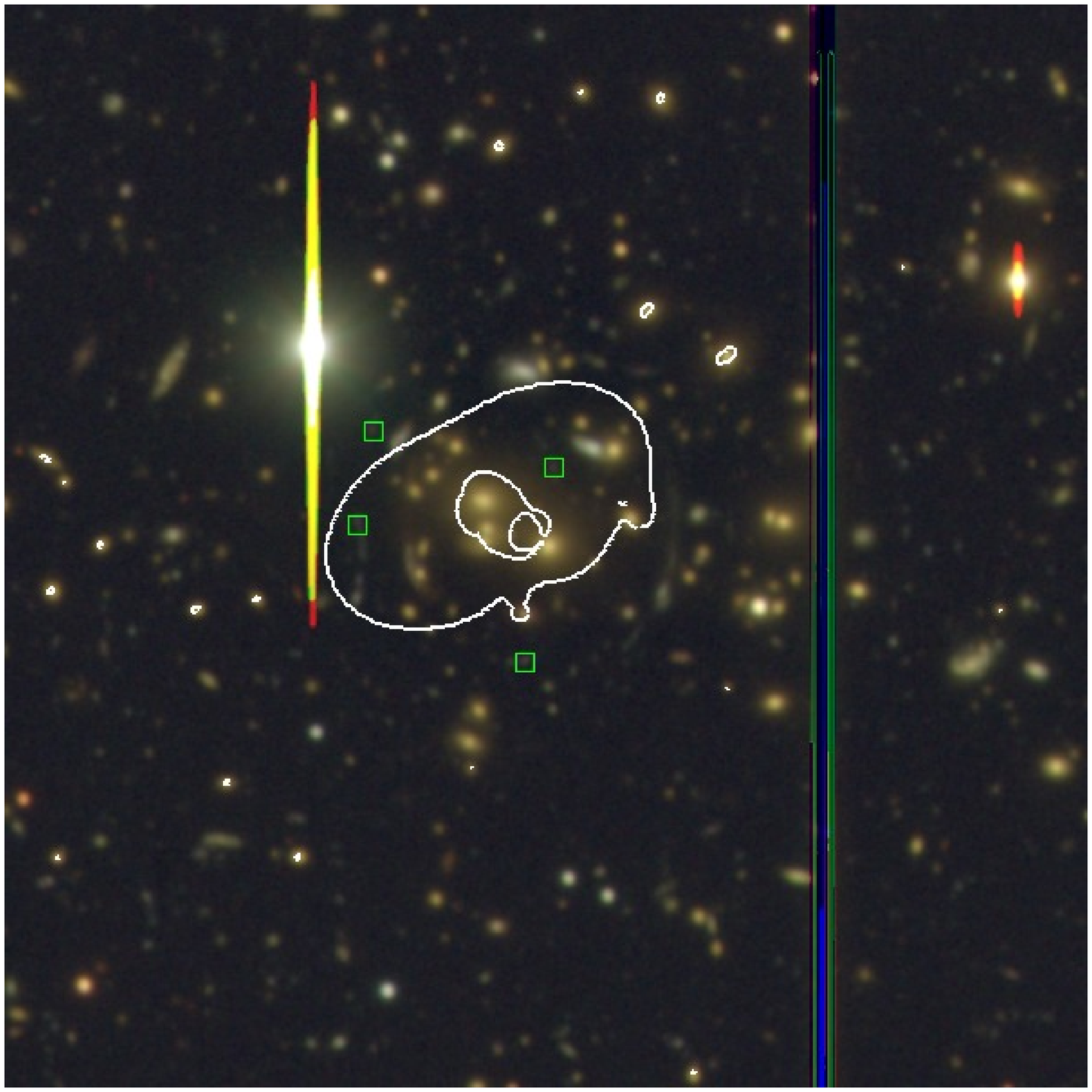}
\includegraphics[width=0.37\hsize]{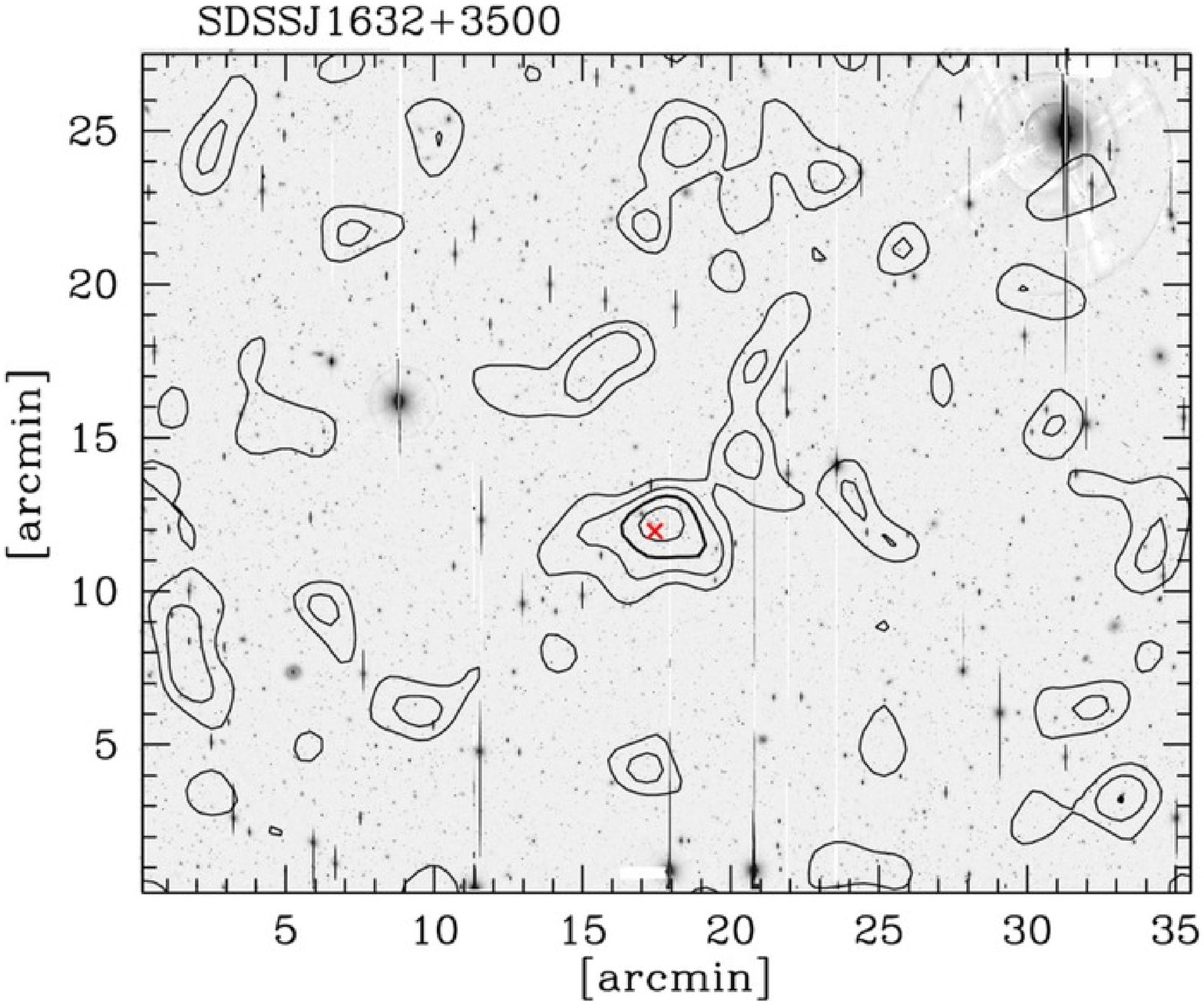}
\includegraphics[width=0.33\hsize]{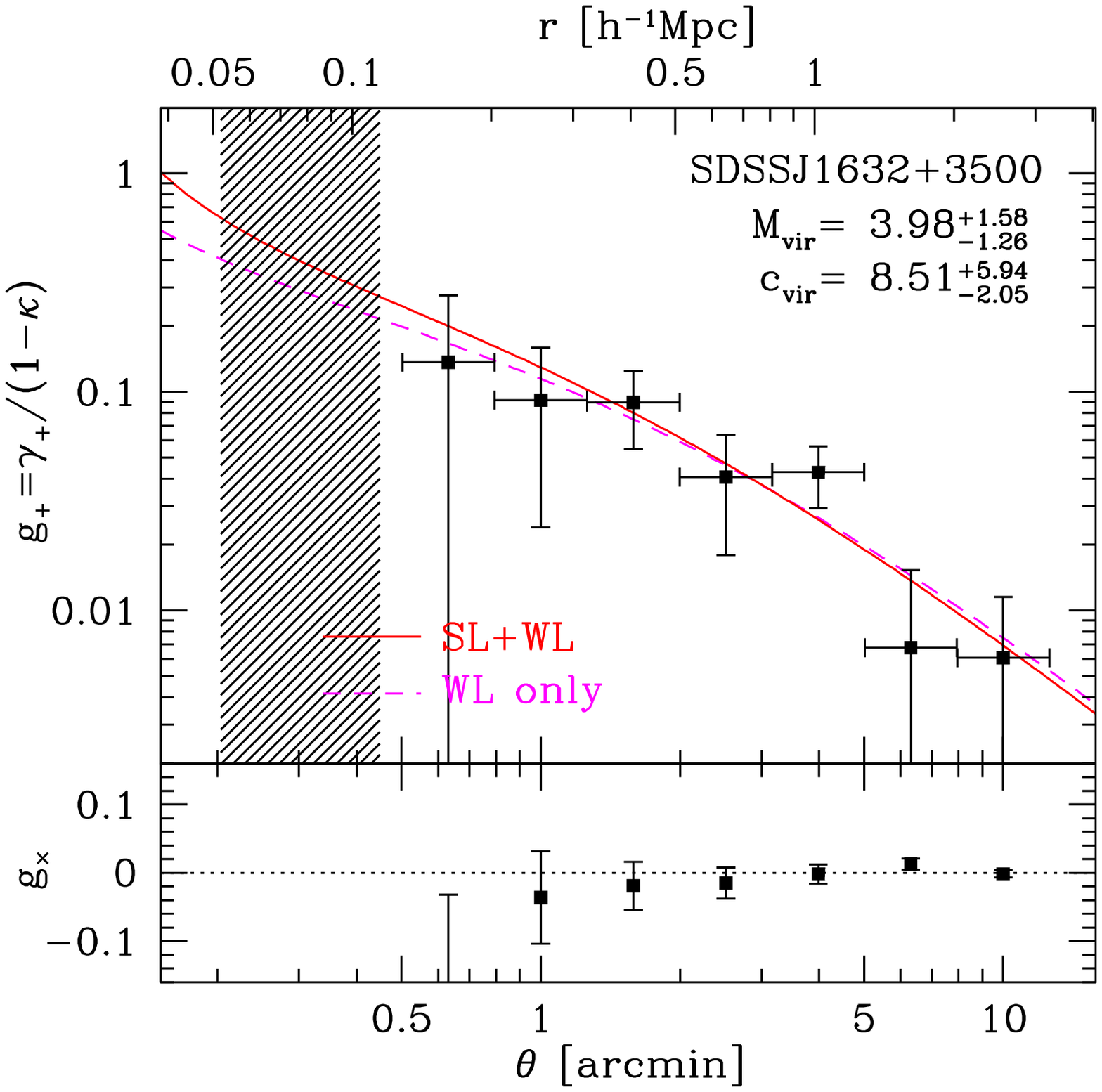}
\includegraphics[width=0.28\hsize]{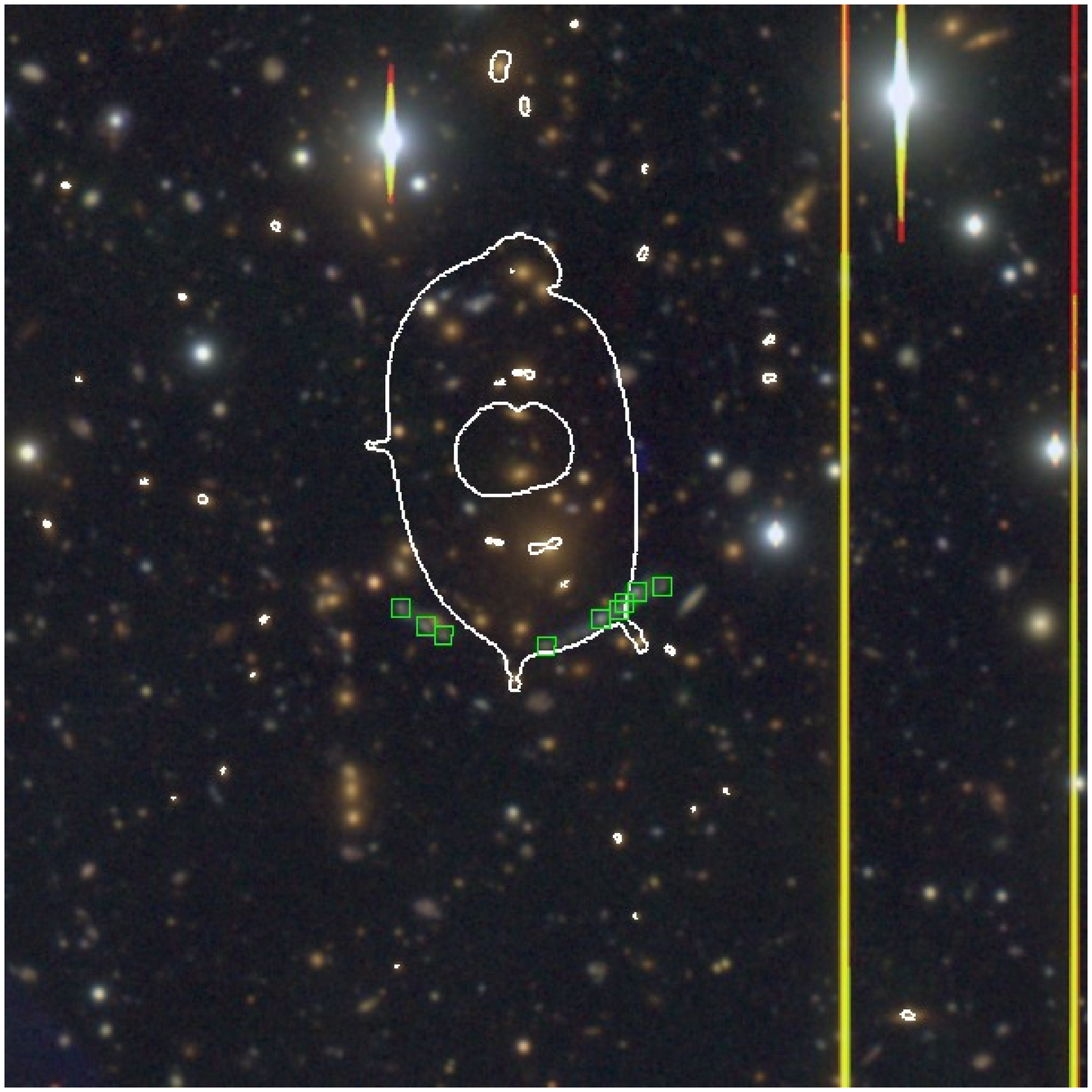}
\includegraphics[width=0.37\hsize]{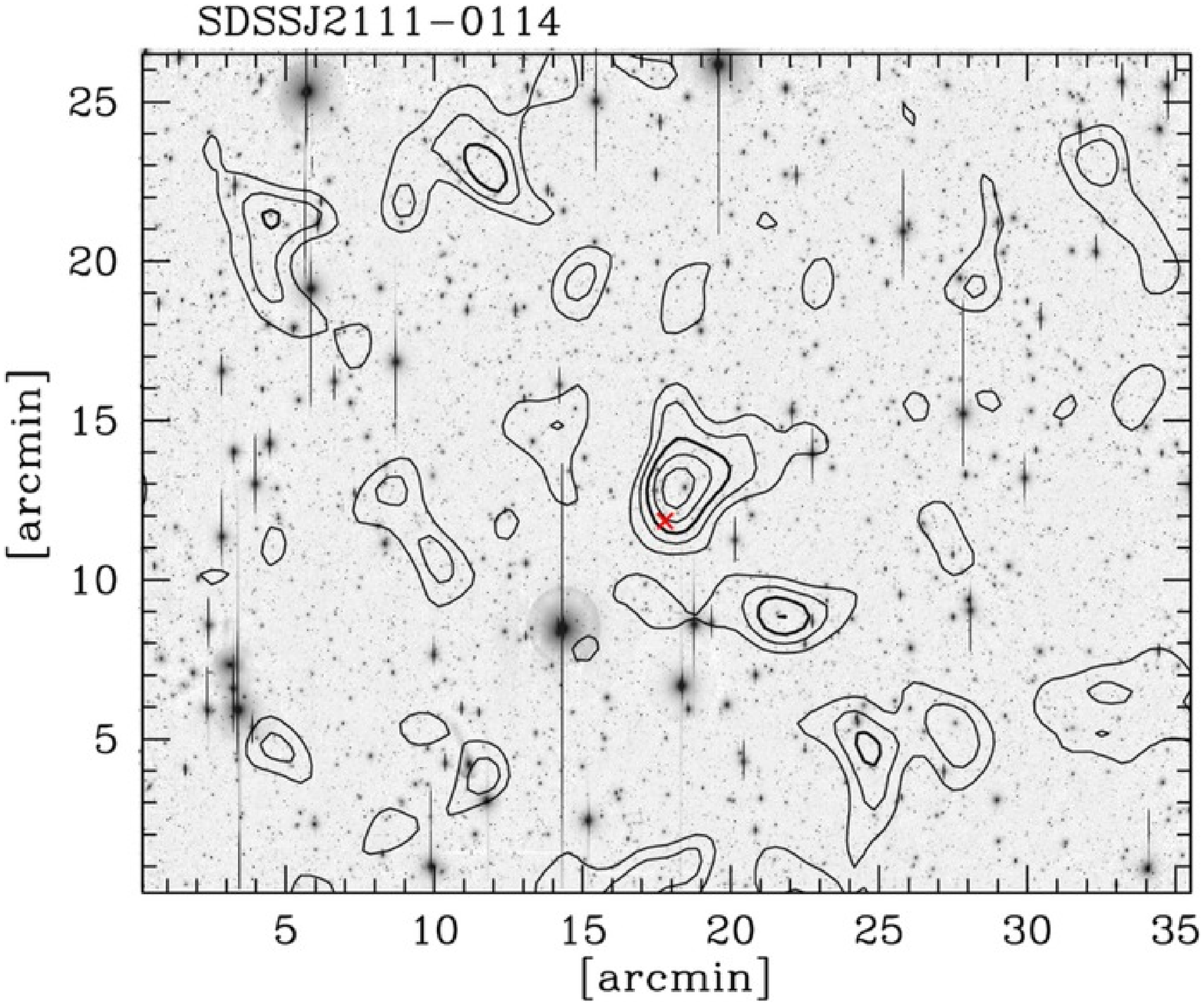}
\includegraphics[width=0.33\hsize]{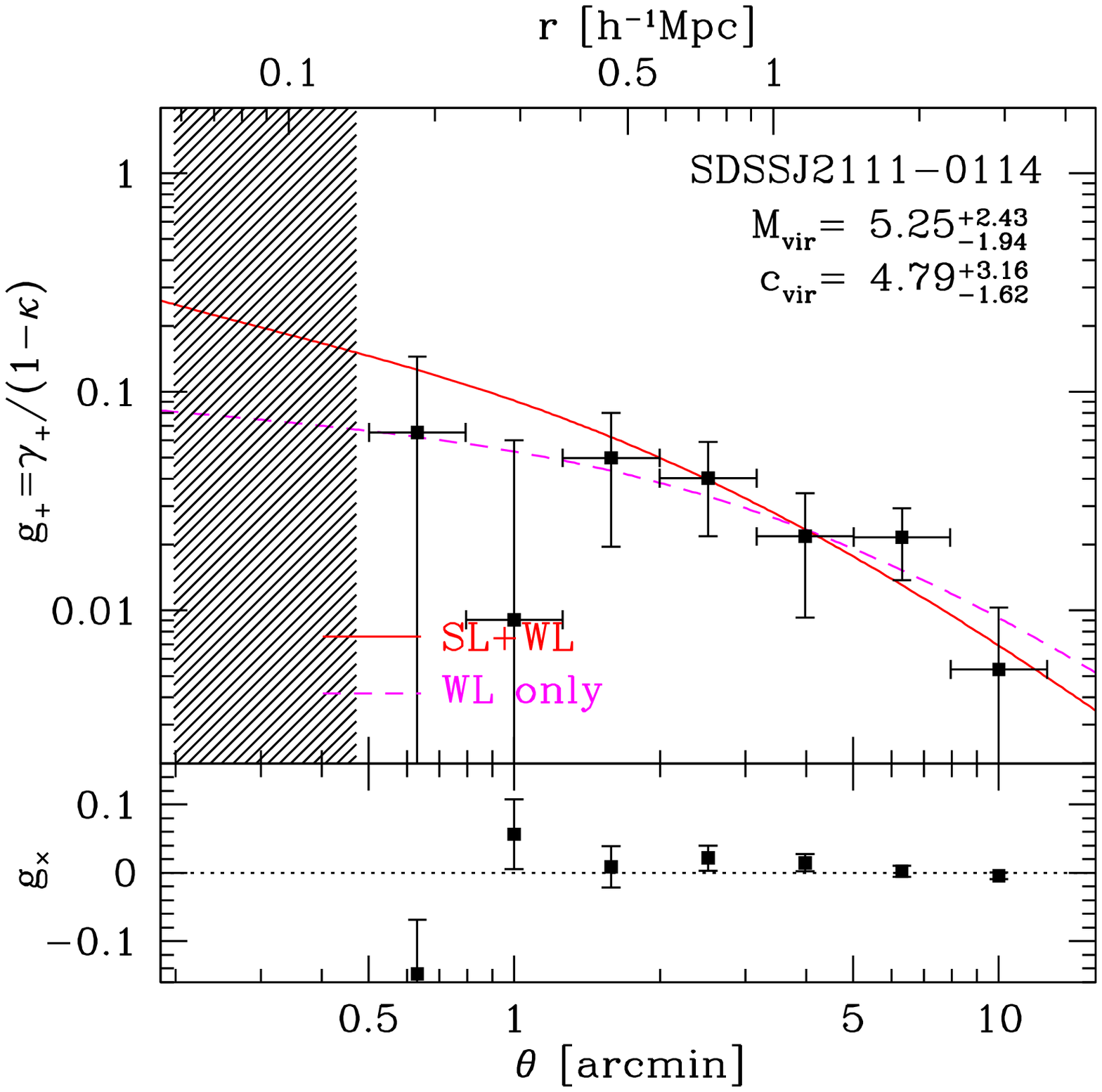}
\end{center}
\caption{SDSSJ1531+3414, SDSSJ1621+0607, SDSSJ1632+3500, SDSSJ2111-0114}
\end{figure*}

\label{lastpage}

\end{document}